\documentclass[12pt,twoside,english,fleqn,twoside,executivepaper]{article}
\usepackage[T1]{fontenc}
\usepackage[latin1]{inputenc}
\usepackage{geometry}
\usepackage{array}
\usepackage{float}
\usepackage{amsmath}
\usepackage{graphicx}
\usepackage{setspace}
\usepackage{amssymb}
\usepackage[authoryear]{natbib}

\makeatletter

\providecommand{\tabularnewline}{\\}

\let\jnl@style=\rmfamily 
\def\ref@jnl#1{{\jnl@style#1}}%
\newcommand\aj{\ref@jnl{AJ\ }}%
\newcommand\araa{\ref@jnl{ARA\&A\ }}%
\newcommand\apj{\ref@jnl{ApJ\ }}%
\newcommand\apjl{\ref@jnl{ApJ\ }}%
\newcommand\apjs{\ref@jnl{ApJS\ }}%
\newcommand\ao{\ref@jnl{Appl.~Opt.\ }}%
\newcommand\apss{\ref@jnl{Ap\&SS\ }}%
\newcommand\aap{\ref@jnl{A\&A\ }}%
\newcommand\aapr{\ref@jnl{A\&A~Rev.\ }}%
\newcommand\aaps{\ref@jnl{A\&AS\ }}%
\newcommand\azh{\ref@jnl{AZh\ }}%
\newcommand\baas{\ref@jnl{BAAS\ }}%
\newcommand\jrasc{\ref@jnl{JRASC\ }}%
\newcommand\memras{\ref@jnl{MmRAS\ }}%
\newcommand\mnras{\ref@jnl{MNRAS\ }}%
\newcommand\pra{\ref@jnl{Phys.~Rev.~A\ }}%
\newcommand\prb{\ref@jnl{Phys.~Rev.~B\ }}%
\newcommand\prc{\ref@jnl{Phys.~Rev.~C\ }}%
\newcommand\prd{\ref@jnl{Phys.~Rev.~D\ }}%
\newcommand\pre{\ref@jnl{Phys.~Rev.~E\ }}%
\newcommand\prl{\ref@jnl{Phys.~Rev.~Lett.\ }}%
\newcommand\pasp{\ref@jnl{PASP\ }}%
\newcommand\pasj{\ref@jnl{PASJ\ }}%
\newcommand\qjras{\ref@jnl{QJRAS\ }}%
\newcommand\skytel{\ref@jnl{S\&T\ }}%
\newcommand\solphys{\ref@jnl{Sol.~Phys.\ }}%
\newcommand\sovast{\ref@jnl{Soviet~Ast.\ }}%
\newcommand\ssr{\ref@jnl{Space~Sci.~Rev.\ }}%
\newcommand\zap{\ref@jnl{ZAp\ }}%
\newcommand\nat{\ref@jnl{Nature\ }}%
\newcommand\iaucirc{\ref@jnl{IAU~Circ.\ }}%
\newcommand\aplett{\ref@jnl{Astrophys.~Lett.\ }}%
\newcommand\apspr{\ref@jnl{Astrophys.~Space~Phys.~Res.\ }}%
\newcommand\bain{\ref@jnl{Bull.~Astron.~Inst.~Netherlands\ }}%
\newcommand\fcp{\ref@jnl{Fund.~Cosmic~Phys.\ }}%
\newcommand\gca{\ref@jnl{Geochim.~Cosmochim.~Acta\ }}%
\newcommand\grl{\ref@jnl{Geophys.~Res.~Lett.\ }}%
\newcommand\jcp{\ref@jnl{J.~Chem.~Phys.\ }}%
\newcommand\jgr{\ref@jnl{J.~Geophys.~Res.\ }}%
\newcommand\jqsrt{\ref@jnl{J.~Quant.~Spec.~Radiat.~Transf.\ }}%
\newcommand\memsai{\ref@jnl{Mem.~Soc.~Astron.~Italiana\ }}%
\newcommand\nphysa{\ref@jnl{Nucl.~Phys.~A\ }}%
\newcommand\physrep{\ref@jnl{Phys.~Rep.\ }}%
\newcommand\physscr{\ref@jnl{Phys.~Scr\ }}%
\newcommand\planss{\ref@jnl{Planet.~Space~Sci.\ }}%
\newcommand\procspie{\ref@jnl{Proc.~SPIE\ }}%

\newcommand\farcm@mss{\mbox{$.\mkern-4mu^\prime$}}%
\let\farcm\farcm@mss 
\newcommand\farcs@mss{\mbox{$.\!\!^{\prime\prime}$}}%
\let\farcs\farcs@mss 
\def\farcm@apj{%
 \mbox{.\kern -0.7ex\raisebox{.9ex}{\scriptsize$\prime$}}%
}%
\def\farcs@apj{%
 \mbox{%
  \kern  0.13ex.%
  \kern -0.95ex\raisebox{.9ex}{\scriptsize$\prime\prime$}%
  \kern -0.1ex%
 }%
}%
\newcommand\ion[2]{#1$\;${\small\rmfamily\@Roman{#2}}\relax}%
\def\load@astro{%
 \dimen@=1\aas@ptsize\p@ 
 \font\astro@font=Astrosym at\dimen@ 
}%
\def\astro#1{\leavevmode\hbox{\astro@font#1}}%
\def\astro@font{%
 \ClassWarning{aastex}{%
  Please use class option `astro', since you are using the astro font.%
 }%
}%
\usepackage{apjfonts}
\usepackage{amsmath}
\DeclareInputText{"0A0}{~}

\renewcommand\tableofcontents{%
  \setlength\itemsep{0.01\itemsep}
  \setlength\parskip{0.01\parskip}
    \setlength\topsep{0.01\topsep}
  \setlength\partopsep{0.01\partopsep}
  \begin{small}
  \leftline {{\bfseries \contentsname\/}}
  \setcounter{secnumdepth}{4}%
  \setcounter{tocdepth}{3}
  \setlength\itemsep{0.01\itemsep}
  \setlength\parskip{0.01\parskip}
 {\@starttoc{toc}}%
\setlength\itemsep{100\itemsep}
\setlength\parskip{100\parskip}
\setlength\partopsep{100\partopsep}
\end{small}
}

\renewcommand*\l@subsection{\@dottedtocline{2}{3.8em}{2.3em}}
\renewcommand*\l@subsubsection{\@dottedtocline{3}{6.1em}{3.2em}}
\newcommand{\bmath}[1]{\mbox{\boldmath\(#1\)}}
\usepackage{fancyhdr}
\numberwithin{equation}{section}
\numberwithin{figure}{section}
\numberwithin{table}{section}
\usepackage[%
backref=page,%
colorlinks=true,%
citecolor=blue,%
hyperindex,%
bookmarks=true,%
bookmarksopen=true,%
bookmarksnumbered=true,%
bookmarksopenlevel=2,%
breaklinks=true%
]{hyperref}
\newcommand{\hlink}[1]{\hyperlink{#1}{\ref*{#1}}} 
\newcommand{\htarget}[1]{\hypertarget{#1}{}}
\def\sep{\unskip, } 
\def\PACS{\hbox {\it PACS:\ }}%
\AtBeginDocument{
  
}

\usepackage{babel}
\makeatother
\begin{document}





\newcommand{\refs}{\mathbf{\texttt{[refs?]}}}
 
\newcommand{\chk}{\mathbf{\texttt{[check!]}}}


\newcommand{\Mo}{M_{\odot}}

\newcommand{\Ro}{R_{\odot}}

\newcommand{\Lo}{L_{\odot}}

\newcommand{\SgrA}{\mathrm{Sgr}\, \mathrm{A}^{\star}}

\newcommand{\Ua}{\widetilde{a}}

\newcommand{\UP}{\widetilde{P}}

\newcommand{\Um}{\widetilde{m}}

\newcommand{\Ut}{\widetilde{t}}

\newcommand{\Urp}{\widetilde{r}_{p}}

\newcommand{\UdE}{\Delta\widetilde{E}}
 
\newcommand{\Urt}{\widetilde{r}_{t}}

\newcommand{\Urh}{\widetilde{r}_{h}}

\newcommand{\UE}{\widetilde{E}}

\newcommand{\UR}{\widetilde{R}}

\newcommand{\HLm}{\widehat{L}_{0}}

\newcommand{\Utm}{\widetilde{t}_{0}}

\newcommand{\UT}{\widehat{T}}

\newcommand{\ULt}{\widetilde{L}_{t}}

\newcommand{\HLt}{\widehat{L}_{t}}

\newcommand{\Utd}{\widetilde{\tau}_{d}}

\newcommand{\Ms}{M_{\star}}

\newcommand{\Rs}{R_{\star}}
 
\newcommand{\Ls}{L_{\star}}

\newcommand{\Ts}{T_{\star}}

\newcommand{\Ns}{N_{\star}}

\newcommand{\Vs}{V_{\star}}

\newcommand{\rs}{r_{\star}}

\newcommand{\ts}{t_{\star}}

\newcommand{\Es}{E_{\star}}

\newcommand{\UWp}{\widetilde{\Omega}_{p}}

\newcommand{\UdJ}{\Delta\widetilde{J}}

\newcommand{\Te}{T_{\mathrm{eff}}}

\newcommand{\Uc}{\widetilde{c}}

\newcommand{\ms}{m_{\bullet}}

\newcommand{\ns}{N_{\bullet}}

\title{{\normalsize {\sffamily\bfseries Physics Reports Review Article}}\\
{\large\bf\dotfill} \\
Stellar Processes Near the\\
Massive Black Hole in the Galactic Center }

\author{Tal Alexander\footnotemark}

\date{\emph{\small Faculty of Physics}\\
\emph{\small The Weizmann Institute of Science}\\
\emph{\small PO Box 26, Rehovot 76100, Israel}}

\maketitle
 \renewcommand{\thefootnote}{\fnsymbol{footnote}}

\footnotetext[1]{Incumbent of the William Z. \& Eda Bess Novick career development chair

{\em Email address: }\url{mailto:tal.alexander@weizmann.ac.il}%
 (Tal Alexander)

{\em URL: }\url{http://www.weizmann.ac.il/~tal/}%
 (Tal Alexander)}

\renewcommand{\thefootnote}{\arabic{footnote}}

 \hrulefill

\begin{abstract}
\begin{singlespace}
\noindent A massive black hole resides in the center of most, perhaps
all galaxies. The one in the center of our home galaxy, the Milky
Way, provides a uniquely accessible laboratory for studying in detail
the connections and interactions between a massive black hole and
the stellar system in which it grows; for investigating the effects
of extreme density, velocity and tidal fields on stars; and for using
stars to probe the central dark mass and to probe post-Newtonian gravity
in the weak- and strong-field limits. \emph{}Recent results, open
questions and future prospects are reviewed in the wider context of
the theoretical framework and physical processes that underlie them.\emph{}\\
\emph{}\\
\emph{keywords:} black hole \sep Milky Way \sep galactic nucleus
\sep stellar dynamics \sep stellar physics \\
\PACS 97.10.-q \sep 98.10.+z\sep98.35.J \sep 98.62.Js  \end{singlespace}

\end{abstract}
 \hrulefill

{\small \tableofcontents{}}\small

\section{Introduction}

\label{s:Intro}

The massive black hole (MBH) in the center of the Milky Way is the
nearest example of a universal phenomenon: central galactic MBHs.
It was first detected as an unusual non-thermal radio source, Sagittarius
A$^{\star}$ ($\SgrA$) (Balick \& Brown \citeyear{Bal74}). Over
the following decades, observations across the electromagnetic spectrum,
together with theoretical arguments, established with ever-growing
confidence that $\SgrA$ is at the dynamical center of the Galaxy
and that it is associated with a very massive and compact dark mass
concentration. This has ultimately led to the nearly inescapable conclusion
that the dark mass is a black hole. At present, observations of $\SgrA$
and its immediate environment offer the strongest empirical evidence
for the existence of MBHs. 

This review deals with one of the many aspects of the MBH phenomenon:
stellar processes near the MBH in the Galactic center (GC). To motivate
this choice of subject matter and to establish the astrophysical context
and the scientific questions of interest, it is necessary first to
address the questions: why study MBHs? why focus on stars near a MBH?
and why in the GC?

\subsection{Astrophysical context}

\label{ss:context}

Observations indicate that most nearby galaxies contain a massive
compact dark object in their center, whose mass lies in the range
$10^{6}\, M_{\odot}\!\lesssim\! m\!\lesssim\mathrm{few\!\times\!10^{9}\,}\Mo$
(Kormendy \& Richstone \citeyear{Kor95}; Magorrian et al. \citeyear{Mag98};
Gebhardt et al. \citeyear{Geb03}; see recent review by Ferrarese
\& Ford \citeyear{Fer05}). It is widely believed that these dark
objects are MBHs, and that they exist in the centers of most, if not
all galaxies. Their number density and mass scale are broadly consistent
with the hypothesis that they are now-dead quasars, which were visible
for a relatively short time in their past as extremely luminous Active
Galactic Nuclei (AGN), powered by the gravitational energy released
by the accretion of gas and stars%
\footnote{It is possible that very low-mass MBHs like the one in the GC have
acquired most of their mass by mergers with other black holes.%
} (Soltan \citeyear{Sol82}; Yu \& Tremaine \citeyear{YuQ02}). Some
present-day galaxies have AGN, although none as bright as quasars.
However, most present-day galactic nuclei are inactive, which implies
that accretion has either almost ceased or switched to a non-luminous
mode. Their inactivity is not due to the lack of gas supply; most
galaxies have more than enough to continue powering an AGN. The {}``dimness
problem'' is one of the key issues of accretion theory, which deals
with the physics of flows into compact objects (e.g. Narayan \citeyear{Nar02}).

The MBH mass $m$ is known to correlate very well with $\sigma_{\mathrm{sph}}$,
the stellar velocity dispersion of the spheroidal component of the
host galaxy (the spheroid is the bulge in disk galaxies, or the entire
galaxy in ellipticals. Ferrarese \& Merritt \citeyear{Fer00}; Gebhardt
et al. \citeyear{Geb00}; Tremaine et al. \citeyear{Tre02}),\begin{equation}
m\simeq1.3\times10^{8}\left(\frac{\sigma_{\mathrm{sph}}}{200\,\mathrm{km\, s^{-1}}}\right)^{\beta_{m/s}}\,\Mo\,,\qquad\beta_{m/\sigma}\sim4-5\,.\label{e:msigma}\end{equation}
 Since the $m/\sigma_{\mathrm{sph}}$ correlation holds on large scales
where the MBH is dynamically unimportant, it suggests a causal connection
between the formation of the MBH and its host galaxy. There are also
claims of correlations between $\sigma_{\mathrm{sph}}$ and the circular
velocity of spiral galaxies on very large scales, which implies a
connection between the MBH mass and the dark matter halo of the galaxy
(Ferrarese \citeyear{Fer02}; Baes et al. \citeyear{Bae03}). Whether
it is the galaxy that somehow influences the growth of the MBH, or
the MBH that somehow regulates the formation of the galaxy, is still
unclear. The nature of the role that MBHs play in the formation and
evolution of galaxies is one of the major unsolved problems of galaxy
formation.

The Galactic MBH is quite normal. Like most MBHs, it is inactive,
and it follows the $m/\sigma_{\mathrm{sph}}$ relation (Valluri et
al. \citeyear{Val05}). With $m\!\sim\!(3$--$4)\!\times\!10^{6}\,\Mo$,
it is one of the least massive MBHs discovered, in keeping with the
relatively small bulge of the Milky Way. What makes it special is
its proximity. At $\sim\!8$ kpc from the Sun, the Galactic black
hole is $\sim\!100$ times closer than the MBH in Andromeda, the nearest
large galaxy, and $\sim\!2000$ times closer than galaxies in Virgo,
the nearest cluster of galaxies. For this reason it is possible to
observe today the stars and gas in the immediate vicinity of the Galactic
MBH at a level of detail that will not be possible for any other galaxy
in the foreseeable future.

In spite of its relative proximity, observations of the GC are challenging
due to strong, spatially variable extinction by interstellar dust,
which is opaque to optical-UV wavelengths. Emission processes related
to the very low-level accretion activity in the GC can be observed
in the radio and mm wavelengths on the long wavelength range of the
spectrum, or in X-rays on the short wavelength range. However stars,
whose spectra are approximately black-body and extend over a limited
wavelength range, emit most of their luminosity in the optical-UV
range, some in the IR and negligibly in longer or shorter wavelengths.
Observations of stars in the GC must be conducted in the infrared,
through the atmospheric transmission {}``windows'' (when observing
from Earth), primarily in the $K$-band%
\footnote{\label{n:Kmag}The $K$-band is centered on $2.2\,\mu\mathrm{m}$
and is $0.6\,\mu\mathrm{m}$ wide. The absolute (intrinsic) $K$-band
magnitude, $M_{K}$, is related to the monochromatic luminosity at
$2.2\,\mu\mathrm{m}$, $L_{K}$, by $M_{K}\!=\!-2.5\log_{10}L_{K}+84.245$,
for $L_{K}$ in $\mathrm{erg\, s^{-1}\,}\mu\mathrm{m^{-1}}$. The
apparent (observed) magnitude is related to the absolute one by $K\!=\! M_{K}\!+\!\mathrm{DM}\!+\! A_{K}$,
where the distance modulus is $\mathrm{DM\!\equiv}\!5\log_{10}(R_{0}/10\,\mathrm{pc})\!=\!14.5$\label{d:DM}
mag for a distance to the GC of $R_{0}\!=\!8\,\mathrm{kpc}$ \label{d:R0}(Eisenhauer
et al. \citeyear{Eis03}) and where $A_{K}\!\sim\!3$ mag is the $K$-band
dust extinction coefficient in the direction of the GC (Rieke et al.
\citeyear{Rie89}). For the Sun, $M_{K\odot}\!=\!3.41$ and $L_{K\odot}\!=\!2.154\!\times\!10^{32}\,\mathrm{erg\, s^{-1}\,\mu m^{-1}}$.%
} ($2.2\,\mu\mathrm{m}$). While the visual extinction along the line
of sight to the GC is $A_{V}\!\sim\!30$ mag , which corresponds to
a transmission ratio of only $1\!:\!10^{12}$ photons, the $K$-band
extinction is only $A_{K}\!\sim\!3$ mag (Rieke et al. \citeyear{Rie89}),
or a transmission ratio of $1\!:\!15$ photons. However, IR astronomy
poses a much more difficult technological challenge than optical astronomy
because of the strong ambient background%
\footnote{History and economics also played a role. Optical astronomy had a
long head-start over IR astronomy, which had to wait for the advent
of modern electronics and cryogenics. In addition, the market for
IR devices is much smaller than that for optical ones. As a result,
the development of efficient IR detectors lagged decades behind optical
detectors. %
}. Rapid advances in IR detectors in the last decade have finally made
it possible to take full advantage of the proximity of the Galactic
MBH. Individual stars can now be tracked as they orbit the MBH in
the very crowded stellar field around it. Their mass, age and evolutionary
stage can be determined from their spectral features, and their radial
(line-of-sight) velocities can be measured from their Doppler-shifted
spectra.

Because of the huge mass ratio between a star and the MBH, stars orbiting
near it are effectively test particles. This is to be contrasted with
the gas in that region, which can be subject to non-gravitational
forces due to thermal, magnetic or radiation pressure. These can complicate
the interpretation of dynamical data and limit its usefulness. The
term {}``near'' is taken here to mean close enough to the MBH so
that the gravitational potential is completely dominated by it, but
far enough so that the stars can survive (that is, beyond the MBH
event horizon, or beyond the radius where stars are torn apart by
the tidal field). In this range, stars directly probe the gravitational
field of the MBH as long as their trajectory does not take them so
close to the MBH that dissipative processes, such as tidal heating,
affect their motion. The event horizon of the MBH in the GC is much
smaller than the tidal radius for most stars, and so effects due to
General Relativity (GR) lead to deviations of only a few percents
from Newtonian motion. To first order, the stellar orbits can be treated
as Keplerian, which substantially simplifies the analysis. However,
with accurate enough astrometric and spectroscopic observations it
may be possible to detect post-Newtonian effects in the orbits and
to probe GR. 

The stars near the MBH are of interest in themselves, in particular
when they can no longer be treated as point particles due to strong
interactions with other stars or with the MBH. The observed stellar
density near the MBH is extremely high, in accordance with theoretical
predictions of the stellar distribution near a MBH. The presence of
so many stars so close to the MBH leads to a variety of interactions
between the MBH and the stars, which may be relevant for feeding the
MBH and can potentially produce an observable signal of the MBH existence.
The high density, large orbital velocities and strong tidal field,
well above those in any other Galactic environment, effectively make
the central $\sim\!0.1\,\mathrm{pc}$ of Galaxy a {}``stellar collider''.
In this extreme environment, otherwise rare dynamical processes that
can affect the inner structure of stars occur relatively frequently.
The formation and evolution of stars in the extreme environment so
close to the MBH challenge theories of stellar dynamics and star formation. 

The Galactic MBH and the stellar environment around it are probably
representative of many similar systems in the universe. It is therefore
likely that insights gained from the study of this uniquely accessible
system can be applied to MBHs in general. In particular, insights
about the role that stars near MBHs play in the feeding, growth and
evolution of the MBH and the galactic nucleus.

\subsection{Science questions}

\label{ss:SciQ}

The study of the stars near the MBH in the GC offers an opportunity
to address many specific questions in the general themes mentioned
above. These broadly fall in four categories.

\begin{description}
\item [The~nature~of~the~dark~mass]The identification of a BH is mainly
done by eliminating all other possibilities, and so an inquiry into
the nature of the dark mass must begin by measuring its primary parameters:
mass, size, position and velocity (\S\ref{s:orbit}). The measured
compactness of the mass distribution can discriminate between a BH
and other alternatives (\S\ref{ss:DM}). The central dark mass may
be composed of several components that have to be disentangled, for
example a binary MBH (\S\ref{ss:2MBH}), or a single MBH surrounded
by a dense cluster of stellar mass compact remnants and a dark matter
cusp of elementary particles (\S\ref{sss:exotic}, \S\ref{sss:distrDM}).
The position and velocity of the dark mass, which lies at the dynamical
center of the Galaxy, are also primary parameters for reconstructing
the Galactic structure and rotation and for calibrating distance indicators
(\S\ref{sss:R0}). 
\item [Post-Newtonian~physics]Stellar orbits very near the MBH should
display post-Newtonian deviations such as periapse angle shifts, Lense-Thirring
precession or gravitational redshift, and thus test various aspects
of GR (\S\ref{ss:GRorbit}). Compact remnants and low-mass main sequence
(MS) stars inspiraling into the MBH could emit detectable gravitational
wave (GW) radiation (\S\ref{sss:GW}). The deflection of light by
the gravitational potential of the MBH may be detected by gravitational
lensing effects (\S\ref{ss:GL}). 
\item [Formation~and~growth~of~MBHs]Observations of stars can provide
clues on the way the MBH forms and grows. Gas can be supplied by stellar
winds from nearby giants or young massive stars, or by stars passing
very near the MBH (\S\ref{ss:1pc}, \S\ref{sss:tscatter}). The
efficiency of tidal disruption and capture can be probed by searching
for stars that are in the process of being captured (\S\ref{sss:squeezar})
or have narrowly avoided such fate (\S\ref{sss:tscatter}). Stars
can interact with a cold remnant accretion disk and reveal its existence
(\S\ref{sss:stardisk}). The MBH may also grow by merging with other
BHs, whether stellar mass BHs (SBHs) (\S\ref{ss:Mseg}) or intermediate
mass black holes (IBHs) (\S\ref{ss:2MBH}, \S\ref{sss:dynfric}). 
\item [Stars~under~extreme~conditions]the unique and extreme conditions
near the MBH (\S\ref{s:dyn}) may be the reason behind the unusual
stellar populations observed there. Observations may reveal evidence
for processes such as extreme mass segregation (\S\ref{ss:Mseg}),
collisional destruction (\S\ref{sss:RGcoll}), tidal spin-up (\S\ref{sss:spinup})
or 3-body exchanges (\S\ref{sss:exchange}) and provide new insights
about unusual modes of star formation near a MBH (\S\ref{sss:insituSF}).
\end{description}

\subsection{Scope and connections to related topics}

\label{ss:scope}

The GC is a complex interacting system. Its different components,
processes and regions are interrelated. Many of the topics discussed
below can be expanded to apply to MBHs in general. Keeping this review
focused on stellar processes near the Galactic MBH inevitably involves
setting somewhat arbitrary limits to its scope.

This review deals with stellar processes in the inner $1$--$2$ parsecs
of the GC, where the MBH dominates the dynamics, and in particular
in the innermost region where stars can still exist. Stellar processes
near a MBH may result in observable effects and thus provide methods
for detecting MBHs. However, some of the effects discussed below are
rather subtle and their usefulness in investigating stellar dynamics
near the MBH is limited to the GC, as it is unlikely that they could
be observed with comparable precision near MBHs in other galaxies. 

In spite of the fact that stars are born of gas and dust and to gas
and dust (and compact remnants) return, gas processes, such as accretion
and outflows, will be discussed only to the extent that they are directly
related to stellar processes near the MBH. Observations of the Milky
Way and a few nearby galaxies show that distinct stellar populations
exist near their MBH (Lauer et al. \citeyear{Lau98}). It is likely
that the environment near a MBH affects the mechanisms of star formation.
However, star formation is poorly understood even in typical galactic
environments, let alone near a MBH, and will be discussed here only
briefly.

The discussion will focus on low mass MBHs ($m\!\lesssim\!10^{8}\,\Mo$),
like the Galactic MBH, where the tidal disruption radius (for solar
type stars) lies outside the event horizon. This allows the possibility
of tidal disruption and limits the role of GR in the dynamics of stars
near the MBH. This situation is not generalizable to massive MBHs
($m\!\gtrsim\!10^{8}\,\Mo$), which are qualitatively different in
these respects. At the other end of the mass spectrum lie IBHs ($\!10^{2}\!\lesssim\! m\!\lesssim\!10^{4}\,\Mo$),
whose existence is still a matter of speculation (e.g. Kaaret et al
\citeyear{Kaa01}; Hopman, Portegies-Zwart \& Alexander \citeyear{Hop04};
Portegies Zwart et al. \citeyear{Por04}; see review by Miller \&
Colbert \citeyear{Col04}). IBHs will also not be discussed here,
except where related directly to the GC. Likewise, although major
(comparable mass) MBH mergers and binary MBHs may play an important
role in the evolution of MBHs in general (e.g. Merritt \& Milosavljevi\'c
\citeyear{Mer04}), it seems unlikely that this is relevant for the
low mass Galactic MBH and the steep stellar cusp around it, and so
this topic will be mentioned only briefly here.

Additional information and discussion of some of the topics outside
the scope of this review can be found in other reviews about the GC:
Genzel \& Townes (\citeyear{Gen87}; early general review on inner
10 pc); Genzel, Hohlenbach \& Townes (\citeyear{Gen94}; comprehensive
review with focus on stellar and gas processes); Morris \& Serabyn
(\citeyear{Mor96}; focus on large spatial scales, gas and star formation);
Melia \& Falcke (\citeyear{Mel01}; focus on accretion and the radio
source $\SgrA$); Yusef-Zadeh, Melia \& Wardle (\citeyear{Yus00};
focus on interactions between stars, gas, dust, clouds and supernovae
remnants) and Alexander (\citeyear{Ale03c}; focus on stellar phenomena).
More observational and theoretical background can be found in Falcke
\& Hehl (\citeyear{Fal03}). 

The acronyms and notations used in this review follow where possible
the usual conventions in the literature. For convenience, the ones
frequently used here are listed in table (\hlink{t:defs}).

\begin{table}

\caption{\label{t:defs}Acronyms and notation frequently used in this review
(by topic, alphabetically) }

{\small \centerline{\htarget{t:defs}}\begin{tabular}{lll|lll}
\multicolumn{6}{l}{}\tabularnewline
\hline 
{\small Term}&
{\small Defined}&
{\small Meaning}&
{\small Term}&
{\small Defined}&
{\small Meaning}\tabularnewline
\hline
\multicolumn{3}{l|}{\textbf{\small Acronyms}}&
\multicolumn{3}{l}{\textbf{\small Stellar dynamics}}\tabularnewline
{\small AGN}&
&
{\small Active Galactic Nucleus}&
{\small $J$}&
{\small p. \pageref{d:Jspec}}&
{\small Specific angular momentum}\tabularnewline
{\small BH}&
&
{\small Black hole}&
{\small $n_{\star}$}&
{\small p. \pageref{d:nstar}}&
{\small Stellar number density}\tabularnewline
{\small DF}&
&
{\small Distribution function}&
{\small $r_{\mathrm{coll}}$}&
{\small p. \pageref{d:rcoll}}&
{\small Collisional radius }\tabularnewline
{\small GC}&
&
{\small Galactic center}&
{\small $t_{c}$}&
{\small p. \pageref{d:tc}}&
{\small Collision time}\tabularnewline
{\small GR}&
&
{\small General Relativity}&
{\small $t_{e}$}&
{\small p. \pageref{d:te}}&
{\small Evaporation time}\tabularnewline
{\small GW}&
&
{\small Gravitational waves}&
{\small $t_{J}$}&
{\small p. \pageref{d:tJ}}&
{\small Angular mom. relaxation time}\tabularnewline
{\small IBH}&
&
{\small Intermediate mass black hole}&
{\small $t_{r}$}&
{\small p. \pageref{d:trel}}&
{\small Relaxation time}\tabularnewline
{\small IMF}&
&
{\small Initial mass function}&
{\small $t_{s}$}&
{\small p. \pageref{d:tseg}}&
{\small Mass segregation time}\tabularnewline
{\small IR}&
&
{\small Infrared}&
{\small $\alpha$}&
{\small p. \pageref{d:alpha}}&
{\small Logarithmic slope of $n_{\star}$}\tabularnewline
{\small MBH}&
&
{\small Massive black hole}&
{\small $\varepsilon$}&
{\small p. \pageref{d:espec}}&
{\small Specific orbital energy}\tabularnewline
{\small MS}&
&
{\small Main sequence}&
{\small $\rho$}&
{\small p. \pageref{d:rho}}&
{\small Mass density}\tabularnewline
{\small NS}&
&
{\small Neutron star}&
{\small $\sigma$}&
{\small p. \pageref{d:sig}}&
{\small 1D stellar velocity dispersion}\tabularnewline
{\small SBH}&
&
{\small Stellar black hole}&
{\small $\psi$}&
{\small p. \pageref{d:relgpot}}&
{\small Relative gravitational potential}\tabularnewline
{\small WD}&
&
{\small White dwarf}&
&
&
\tabularnewline
{\small WR}&
&
{\small Wolf-Rayet star}&
\multicolumn{3}{l}{\textbf{\small Stellar astrometry and orbits}}\tabularnewline
&
&
&
{\small $a$}&
{\small p. \pageref{d:sma}}&
{\small Keplerian semi-major axis}\tabularnewline
\multicolumn{3}{l|}{\textbf{\small Stellar properties}}&
{\small $e$}&
{\small p. \pageref{d:ecc}}&
{\small Keplerian eccentricity}\tabularnewline
{\small $\Es$}&
{\small p. \pageref{d:Eb}}&
{\small Stellar binding energy}&
{\small $i$}&
{\small p. \pageref{d:iincl}}&
{\small Inclination angle}\tabularnewline
{\small $K$}&
{\small p. \pageref{d:R0}}&
{\small $K$-band magnitude}&
{\small $p$}&
{\small p. \pageref{d:rproj}}&
{\small Projected distance from MBH}\tabularnewline
{\small $L_{\star}$}&
{\small p. \pageref{d:Ls}}&
{\small Total stellar luminosity}&
{\small $P$}&
{\small p. \pageref{d:Porb}}&
{\small Keplerian orbital period}\tabularnewline
{\small $\Ms$}&
{\small p. \pageref{d:Ms}}&
{\small Stellar mass}&
{\small $r_{p}$}&
{\small p. \pageref{d:rp}}&
{\small Orbital periapse}\tabularnewline
{\small $\Ns$}&
{\small p. \pageref{d:Ns}}&
{\small Number of stars}&
{\small $r_{a}$}&
{\small p. \pageref{d:ra}}&
{\small Orbital apoapse}\tabularnewline
{\small $\Rs$}&
{\small p. \pageref{d:Rs}}&
{\small Stellar radius}&
{\small $z$}&
{\small p. \pageref{d:rlos}}&
{\small Line of sight distance from MBH}\tabularnewline
{\small $t_{\star}$}&
{\small p.} {\footnotesize \pageref{d:ts}}&
{\small Stellar lifespan}&
{\small $\beta$}&
{\small p. \pageref{d:beta}}&
{\small Velocity in terms of speed of light $c$}\tabularnewline
{\small $T_{\star}$}&
{\small p. \pageref{d:Teff}}&
{\small Effective temperature}&
{\small $\Upsilon$}&
{\small p. \pageref{d:Gparm}}&
{\small Relativistic parameter}\tabularnewline
{\small $V_{e}$}&
{\small p. \pageref{d:Ve}}&
{\small Stellar escape velocity}&
&
&
\tabularnewline
{\small $\Vs$}&
{\small p. \pageref{d:Vc}}&
{\small Stellar circular velocity}&
\multicolumn{3}{l}{\textbf{\small Star--MBH interactions}}\tabularnewline
{\small $\tau_{\star}$}&
{\small p. \pageref{d:td}}&
{\small Stellar dynamical timescale}&
{\small $b^{-1}$}&
{\small p. \pageref{d:bpen}}&
{\small Tidal penetration parameter}\tabularnewline
&
&
&
{\small $r_{t}$}&
{\small p. \pageref{d:rt}}&
{\small Tidal disruption radius}\tabularnewline
\multicolumn{3}{l|}{\textbf{\small MBH properties}}&
{\small $\Gamma_{s}$}&
{\small p. \pageref{d:Gs}}&
{\small Tidal scattering rate}\tabularnewline
{\small $m$}&
{\small p. \pageref{d:m}}&
{\small MBH mass}&
{\small $\Gamma_{t}$}&
{\small p. \pageref{d:Gt}}&
{\small Tidal disruption rate}\tabularnewline
{\small $r_{h}$}&
{\small p. \pageref{d:rh}}&
{\small MBH radius of influence}&
\multicolumn{1}{l}{{\small $\tau_{p}$}}&
{\small p. \pageref{d:tp}}&
{\small Periapse passage timescale}\tabularnewline
{\small $r_{S}$}&
{\small p. \pageref{d:rs}}&
{\small Schwarzschild radius}&
&
&
\tabularnewline
{\small $s$}&
{\small p. \pageref{d:sMBH}}&
{\small Spin parameter}&
\multicolumn{3}{l}{\textbf{\small Gravitational lensing}}\tabularnewline
&
&
&
{\small $A$}&
{\small p. \pageref{d:Amag}}&
{\small Lensing magnification}\tabularnewline
\multicolumn{3}{l|}{\textbf{\small GC properties}}&
{\small $D_{LS}$}&
{\small p. \pageref{d:Dls}}&
{\small Lens--source distance}\tabularnewline
{\small $A_{K}$}&
{\small p. \pageref{d:R0}}&
{\small $K$-band dust extinction}&
{\small $D_{OL}$}&
{\small p. \pageref{d:Dol}}&
{\small Observer--lens distance}\tabularnewline
{\small DM}&
{\small p. \pageref{d:DM}}&
{\small Distance modulus}&
{\small $D_{OS}$}&
{\small p. \pageref{d:Dos}}&
{\small Observer--source distance}\tabularnewline
{\small $R_{0}$}&
{\small p. \pageref{d:R0}}&
{\small Sun--GC distance}&
{\small $x$}&
{\small p. \pageref{d:GLx}}&
{\small Image angular position in terms of $\theta_{E}$}\tabularnewline
{\small $t_{H}$}&
{\small p. \pageref{d:tH}}&
{\small Hubble time ($\sim$age of GC)}&
{\small $y$}&
{\small p. \pageref{d:GLy}}&
{\small Source angular position in terms of $\theta_{E}$}\tabularnewline
{\small $\Theta_{0}$}&
{\small p. \pageref{d:T0}}&
{\small Local Galactic rotation speed}&
{\small $\Gamma_{L}$}&
{\small p. \pageref{d:GLrate}}&
{\small Gravitational lensing rate}\tabularnewline
&
&
&
{\small $\theta_{\mathrm{E}}$}&
{\small p. \pageref{d:ThetaE}}&
{\small Einstein angle}\tabularnewline
&
&
&
&
&
\tabularnewline
\hline
\end{tabular}{\small }}
\end{table}

\section{Observational overview: Stars in the Galactic center}

\label{s:GCstars}

The observational picture of the GC, and in particular that of the
central few parsecs, is a rapidly evolving one. A too detailed focus
on the most recent observations runs the risk of premature obsolescence,
as ongoing observational campaigns continuously yield new data and
insights. This overview will therefore be limited to painting a broad-brush
picture, and defer detailed discussion to the relevant topical sections
below.

It is useful to begin by a brief summary of some stellar properties
that are of particular relevance in this context. The initial stellar
mass sets the stellar lifespan, and together the two determine which
dynamical processes are relevant (for example, a slow dynamical process
that operates on massive objects may be irrelevant because the massive
stars are too short-lived, while the long-lived stars are not massive
enough). The initial mass also sets the evolution of the stellar IR
luminosity, which determines if, and at which evolutionary stage the
star can be observed. Figure (\hlink{f:StellarKt}) shows the stellar
lifespan, the MS $K$-band luminosity and the MS spectral type designation
as function of initial stellar mass, compared to the estimated dynamical
two-body relaxation time in the GC and compared to current and future
$K$-band detection limits (for evolved stars, see Fig. 3.5 in Genzel,
Hollenbach \& Townes \citeyear{Gen94}).

\begin{figure}[!t]
\centering{\htarget{f:StellarKt}\includegraphics[scale=1.2]{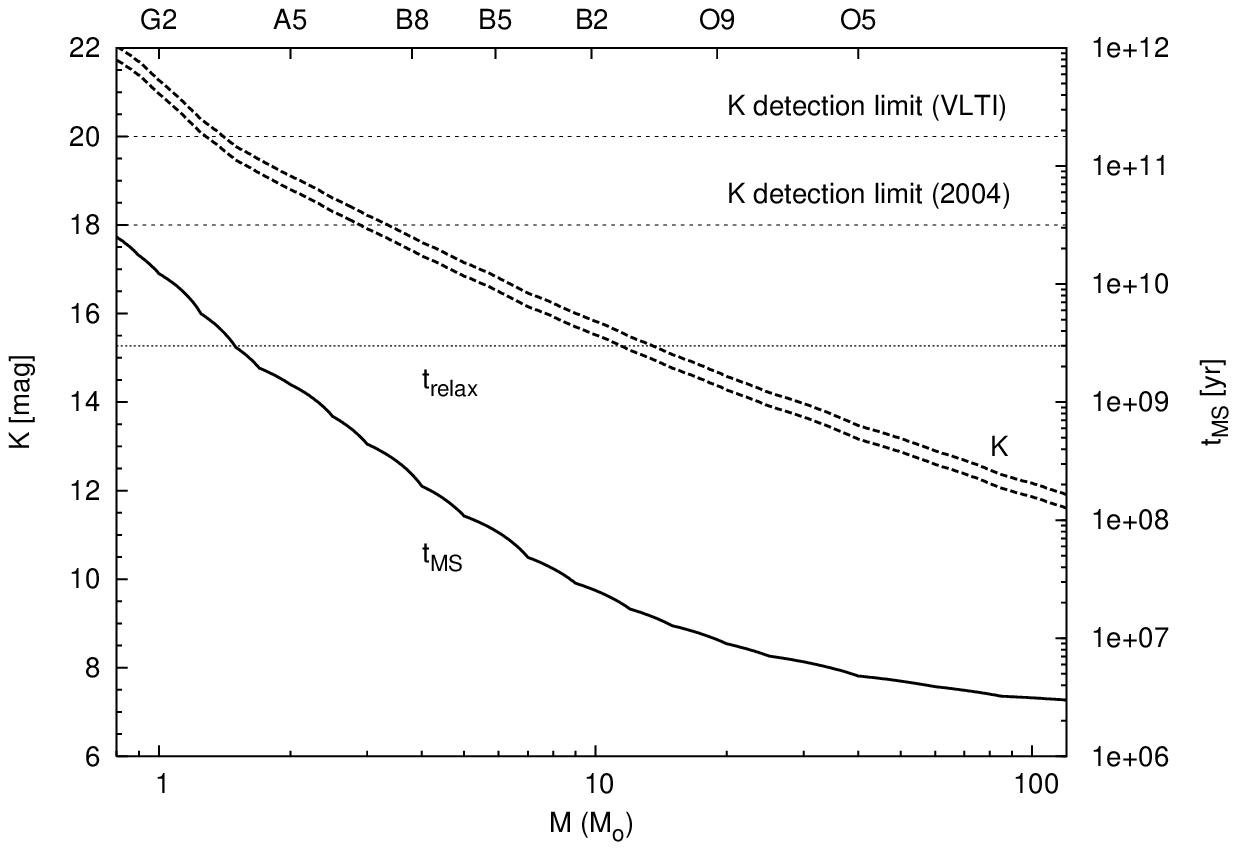}}

\caption{{\small \label{f:StellarKt}Main sequence lifetime and $K$-band
magnitude for zero-age solar metallicity stars (based on the stellar
tracks of Schaller et al. \citeyear{Sch92}). An approximate association
of mass with spectral type is indicated in the upper horizontal axis
(Cox \citeyear{Cox00}). The apparent $K$-band magnitudes are given
for a plausible range of distance and extinction coefficient values:
the top curve for the fiducial distance to the GC of $R_{0}\!=\!8$
kpc (Reid \citeyear{Rei93}; Eisenhauer et al. \citeyear{Eis03})
and $K$-band extinction of $A_{K}\!=\!3$ mag (Rieke et al. \citeyear{Rie89}),
and the bottom curve for the more recent determination of $R_{0}\!=\!7.62$
kpc and $A_{K}\!=\!2.8$ mag. A rough estimate of the two-body relaxation
time in the GC, $t_{r}\!\sim\!3\times\!10^{9}$ yr (Alexander \citeyear{Ale99a}),
the current detection limit in the $K$-band with the VLT ($K\!=\!18$
mag with completeness of 63\%, Genzel et al. \citeyear{Gen03a}) and
the anticipated limit for the VLTI interferometer with adaptive optics
(Paresce et al. \citeyear{Par03}) are also shown. }}
\end{figure}

\subsection{The central 100 parsecs}

\label{ss:100pc}

The stars well outside the dynamical sphere of influence of the Galactic
MBH on the $\lesssim\!100\,\mathrm{pc}$ scale are relevant for understanding
the stellar environment in the inner parsec because they define the
{}``boundary conditions'' for the inner GC; because, given enough
time or under certain dynamical conditions (\S\ref{ss:Mseg}) stars
can migrate from large distances to the center, and because the stellar
population far from the MBH can serve as a control sample for separating
the effects of the MBH from intrinsic stellar phenomena.

The GC lies at the center of the Galactic bulge (typical length-scale
$\sim\!$2 kpc), which is composed of an old ($\sim\!10$ Gyr), passively
evolving stellar population. However, the stellar population in the
central $\sim\!100$ pc, whose density rises toward the center as
$r^{-2}$, appears to have been forming continuously over the lifetime
of the Galaxy. The star formation is probably fed by the large reservoir
of molecular gas (the central molecular zone) that is concentrated
in the inner $\sim\!200$ pc (Serabyn \& Morris \citeyear{Ser96}).
The stellar population in the GC is thus a mixture of the inner extension
of the old bulge population and intermediate-age and young stars from
more recent star formation epochs (Philipp et al. \citeyear{Phi99};
Mezger et al. \citeyear{Mez99}; Figer et al. \citeyear{Fig04}). 

About half of the young stars in the region are found today in three
particularly massive, young ($\lesssim\!5$ Myr) clusters: The Quintuplet
($\sim\!30$ pc from the center in projection, $M\!\sim\!10^{4}\,\Mo$,
$\bar{\rho}\!\sim\!\mathrm{few}\!\times\!10^{3}\,\Mo\,\mathrm{pc^{-3}}$),
The Arches ($\sim\!30$ pc from the center in projection, $M\!\gtrsim\!10^{4}\,\Mo$,
$R\!\sim\!0.2$ pc, $\bar{\rho}\!\sim\!3\!\times\!10^{5}\,\Mo\,\mathrm{pc^{-3}}$),
and the central cluster around the MBH (Figer \citeyear{Fig03}).
The three clusters contain in total hundreds of MS O-stars, tens of
Wolf-Rayet (WR) stars and a few luminous blue variable stars (\S\ref{ss:1pc}),
which are $\sim\!10\%$ of all the massive stars (initial mass$>\!20\,\Mo$)
in the entire Galaxy. They produce $\sim\!10\%$ of the Galactic ionizing
luminosity, and are responsible for the formation of $0.01\,\Mo\,\mathrm{yr^{-1}}$in
stars (a rate per volume of $\sim\!10^{-7}\,\Mo\,\mathrm{yr^{-1}\, pc^{-3}}$,
$250$ times higher than the Galactic mean). This is however only
$\sim\!1\%$ of the total Galactic star formation rate. The disparity
reflects the marked bias in the initial mass function of these clusters
toward the formation of massive hot and luminous stars (Figer et al
. \citeyear{Fig99}; Figer \citeyear{Fig05}). The overall similarity
between the young stellar populations of the Arches and the Quintuplet
and that of the central cluster (\S\ref{ss:1pc}) is noteworthy in
view of the fact that the two off-center clusters show no evidence
of harboring a central BH.

X-Ray observations of the inner 20 pc also reveal $\sim\!2000$ X-ray
sources, which follow the distribution of the stars in the IR. The
hard spectral index of over half of these sources suggests that they
are magnetically accreting white dwarfs (WDs) and X-ray pulsars in
high mass X-ray binaries (where a neutron star (NS) accretes the wind
of a massive companion).

\begin{table}[t]

\caption{\label{t:StellarPop}A mean stellar population model for the inner
GC$^{a}$ }

{\small \centerline{\htarget{t:StellarPop}}\begin{tabular}{lccl}
&
&
&
\tabularnewline
\hline 
{\small Component}&
{\small Mass fraction $^{b}$}&
{\small Number fraction $^{c}$}&
{\small Notes}\tabularnewline
\hline 
{\small Gas consumed}&
{\small $1.8$}&
{\small ---}&
{\small Including reprocessed gas}\tabularnewline
{\small Gas reprocessed}&
{\small $0.8$}&
{\small ---}&
{\small Assumed ejected from GC}\tabularnewline
{\small Live stars ($\Ms\!>\!0.1\,\Mo$)}&
{\small $0.74$}&
{\small $0.85$}&
{\small Mean mass $0.43\,\Mo$}\tabularnewline
{\small Live stars ($\Ms\!>\!0.8\,\Mo$)}&
{\small $0.22$}&
{\small $0.13$}&
{\small Mean mass $0.84\,\Mo$}\tabularnewline
{\small Compact remnants $^{d}$}&
{\small $0.27$}&
{\small $0.15$}&
{\small Mean mass $0.94\,\Mo$}\tabularnewline
{\small WDs $^{e}$ ($0.6\,\Mo$)}&
{\small $0.03$}&
{\small $0.03$}&
{\small Progenitor mass $\textrm{0.8}$--$1.5\,\Mo$}\tabularnewline
{\small WDs $^{e}$ ($0.7\,\Mo$)}&
{\small $0.08$}&
{\small $0.06$}&
{\small Progenitor mass $1.5$--$2.5\,\Mo$}\tabularnewline
{\small WDs $^{e}$ ($1.1\,\Mo$)}&
{\small $0.12$}&
{\small $0.05$}&
{\small Progenitor mass $\textrm{2.5}$--$8.0\,\Mo$}\tabularnewline
{\small NSs $^{e}$ ($1.4\,\Mo$)}&
{\small $0.03$}&
{\small $0.01$}&
{\small Progenitor mass $\textrm{8.0}$--$30\,\Mo$}\tabularnewline
{\small SBHs $^{f}$ ($10\,\Mo$)}&
{\small $0.01$}&
{\small $0.0005$}&
{\small Progenitor mass $>\!30\,\Mo$}\tabularnewline
\hline
\multicolumn{4}{p{5.75in}}{{\small $^{a}$ Continuous star formation over 10 Gyr at a constant
rate with a Miller-Scalo IMF (Miller \& Scalo \citeyear{Mil79}) and
solar metallicity (Alexander \& Sternberg \citeyear{Ale99b}). Calculated
by a stellar population synthesis code (Sternberg, Hoffman \& Pauldrach
\citeyear{Ste03}) using the Geneva stellar evolution tracks (Schaller
et al. \citeyear{Sch92}).}}\tabularnewline
\multicolumn{4}{l}{{\small $^{b}$ Relative to total remaining mass in stars and remnants.}}\tabularnewline
\multicolumn{4}{l}{{\small $^{c}$ Relative to total remaining number of stars and remnants.}}\tabularnewline
\multicolumn{4}{l}{{\small $^{d}$ Average supernovae rate of $1.2\!\times\!10^{-5}\,\mathrm{yr^{-1}}$
per $10^{7}\,\Mo$ of gas consumed. }}\tabularnewline
\multicolumn{4}{l}{{\small $^{e}$ Meylan \& Mayor (\citeyear{Mey91}).}}\tabularnewline
\multicolumn{4}{l}{{\small $^{f}$ Timmes, Woosley \& Weaver (\citeyear{Tim96}).}}\tabularnewline
\hline
\end{tabular}}
\end{table}

It is useful to construct a stellar population model to represent
the continuously star forming population of the central GC. When the
star formation rate is constant, the present-day mass function (number
of stars per stellar mass) at time $t$ is simply related to the initial
mass function (IMF) $\left.\mathrm{d}\Ns/\mathrm{d}\Ms\right|_{0}$by 

\begin{equation}
\left.\frac{\mathrm{d}\Ns}{\mathrm{d}\Ms}\right|_{t}\!\propto\!\left.\frac{\mathrm{d}\Ns}{\mathrm{d}\Ms}\right|_{0}\min[t,\ts(\Ms)]\,,\label{e:PMF}\end{equation}
where $t_{\star}$ is the stellar lifespan\label{d:ts}. The stellar
contents of such a model are listed in table ({\small \hlink{t:StellarPop}}).
This model assumes a 10 Gyr old, continuously star forming population
with a Miller-Scalo%
\footnote{A useful analytic approximation of the normalized Miller-Scalo IMF
({\small \citeyear{Mil79}) is $\ln\left(\left.\mathrm{d}\Ns/\mathrm{d}\Ms\right|_{0}\right)\!=\!-\ln(2\Ms/\Mo)\ln(400\Ms/\Mo)/4-7.11$
for $0.1\!<\!\Ms/\Mo\!<\!125$.}%
} IMF ({\small \citeyear{Mil79}}) and solar metallicity. The $K$-band
luminosity function (number of stars per magnitude) of this model
is well approximated by a power-law down to faint magnitudes ($K\!\lesssim\!20$
mag), $\mathrm{d}\log N_{\star}/\mathrm{d}K\!=\! s$, with $s\!\simeq\!0.35$
(Alexander \& Sternberg \citeyear{Ale99b}). This value broadly agrees
with the observed $K$-band luminosity function in the inner $0.25$
pc (Davidge et al. \citeyear{Dav97}) and that of the Galactic bulge
as observed through Baade's window, $\sim\!0.5$ kpc from the GC (Tiede,
Frogel \& Terndrup \citeyear{Tie95}; Zoccali et al. \citeyear{Zoc03}).

\subsection{The central parsec }

\label{ss:1pc}

The central parsec lies inside the dynamical sphere of influence of
the MBH (\S\ref{sss:scales}). The properties of the stellar population
there are observed to change with decreasing distance from the MBH,
once at $\sim\!0.4$ pc and then again at $\sim\!0.04$ pc (Fig. \hlink{f:VLTfield}). 

\begin{figure}[!t]
\centering{\htarget{f:VLTfield}\includegraphics[%
  clip,
  scale=0.65]{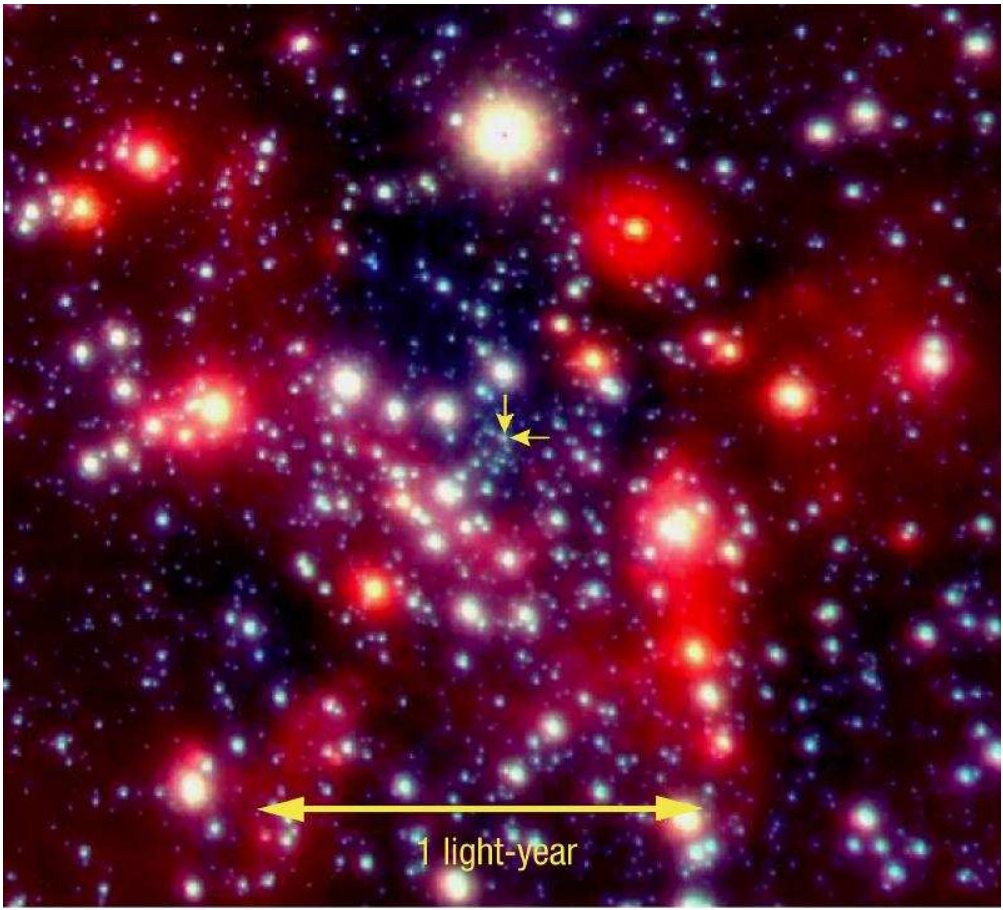}}

\caption{{\small \label{f:VLTfield}A false color composite image of the central
parsec of the GC taken by the Very Large Telescope in the $H$ ($1.65\,\mu\mathrm{m}$),
$K_{s}$ ($2.16\,\mu\mathrm{m}$) and $L^{\prime}$ ($3.76\,\mu\mathrm{m}$)
infrared bands (Genzel et al. \citeyear{Gen03a}). The position of
the MBH is indicated by the two small arrows. A mixture of stellar
types is clearly apparent: Old, intermediate mass cool red giants
and young massive hot blue stars. (Reprinted with permission from
the} \emph{\small Astrophysical Journal}{\small ) }}
\end{figure}

Both the stellar spatial distribution and the stellar population outside
$\sim\!0.4$ pc are a continuation of the large scale $n_{\star}\propto r^{-2}$
mixed distribution of old and young stars in the inner $100$ pc (Genzel
et al. \citeyear{Gen03a}). The old population is dynamically relaxed
and follows the Galactic rotation pattern (McGinn et al. \citeyear{McG89};
Haller et al. \citeyear{Hal96}; Genzel et al. \citeyear{Gen96}). 

Inside $\sim\!0.4$ pc the stellar spatial distribution flattens to
$n_{\star}\!\propto\! r^{-1.4}$ (Genzel et al. \citeyear{Gen03a}).
On approximately the same length-scale (whether or not this is a coincidence
is unclear), there appears a dominant population of massive young
stars (Forrest et al. \citeyear{For87}; Allen, Hyland \& Hillier
\citeyear{All90}; Krabbe et al. \citeyear{Kra95}; Paumard \citeyear{Pau04a}).
This population consists of dynamically unrelaxed blue supergiants
with distinctive helium emission lines in their infrared spectra (Tamblyn
\citeyear{Tam96}; Genzel et al. \citeyear{Gen96}; Genzel et al.
\citeyear{Gen00}). The lack of hydrogen lines and evidence of outflow
in the spectral line shapes ({}``P Cygni profiles'') identify them
as WR stars or Luminous Blue Variables (LBVs) (Najarro et al \citeyear{Naj97}).
WRs are stars with masses of order $\mathrm{few}\!\times\!10\,\Mo$,
lifespans of $\mathrm{few\!\times10^{6}}$ yr undergoing rapid mass
loss by a stellar wind, which removes the hydrogen-rich envelope and
uncovers the helium-rich core. LBVs are WR progenitors that are still
in the process of losing their envelopes. There are tentative detections
of X-ray emission from the shocked winds of a couple of these stars
(Baganoff et al. \citeyear{Bag03}).

The emission line stars co-rotate tangentially in an opposite direction
to the Galactic rotation, in two partially overlapping disk-like structures
(Levin \& Beloborodov \citeyear{Lev03}; Genzel et al. \citeyear{Gen03a};
Fig. \hlink{f:disks}). The two disks are strongly inclined relative
to each other. In projection on the plane of the sky, the inner disk
appears to rotate clock-wise, and the outer disk counter-clockwise.
The inner extent of the disk population is $\sim\!0.04$ pc. Two prominent
concentrations of stars appear among the disk stars. The IRS16 complex
is located on the line-of-nodes (the intersection between the disk
plane and the plane of the sky) of the clockwise disk, and so this
apparent stellar over-density can plausibly be explained as a projection
effect (Genzel et al. 2005, in prep.). The IRS13 complex is part of
the counter-clockwise disk, which is nearly face-on. IRS13 appears
to be a genuine over-concentration of stars with common velocities
(Maillard et al. \citeyear{Mai04}; Sch\"odel et al. \citeyear{Sch05}).

\begin{figure}[!t]
\centerline{\htarget{f:disks}\begin{tabular}{cc}
\includegraphics[%
  scale=0.35]{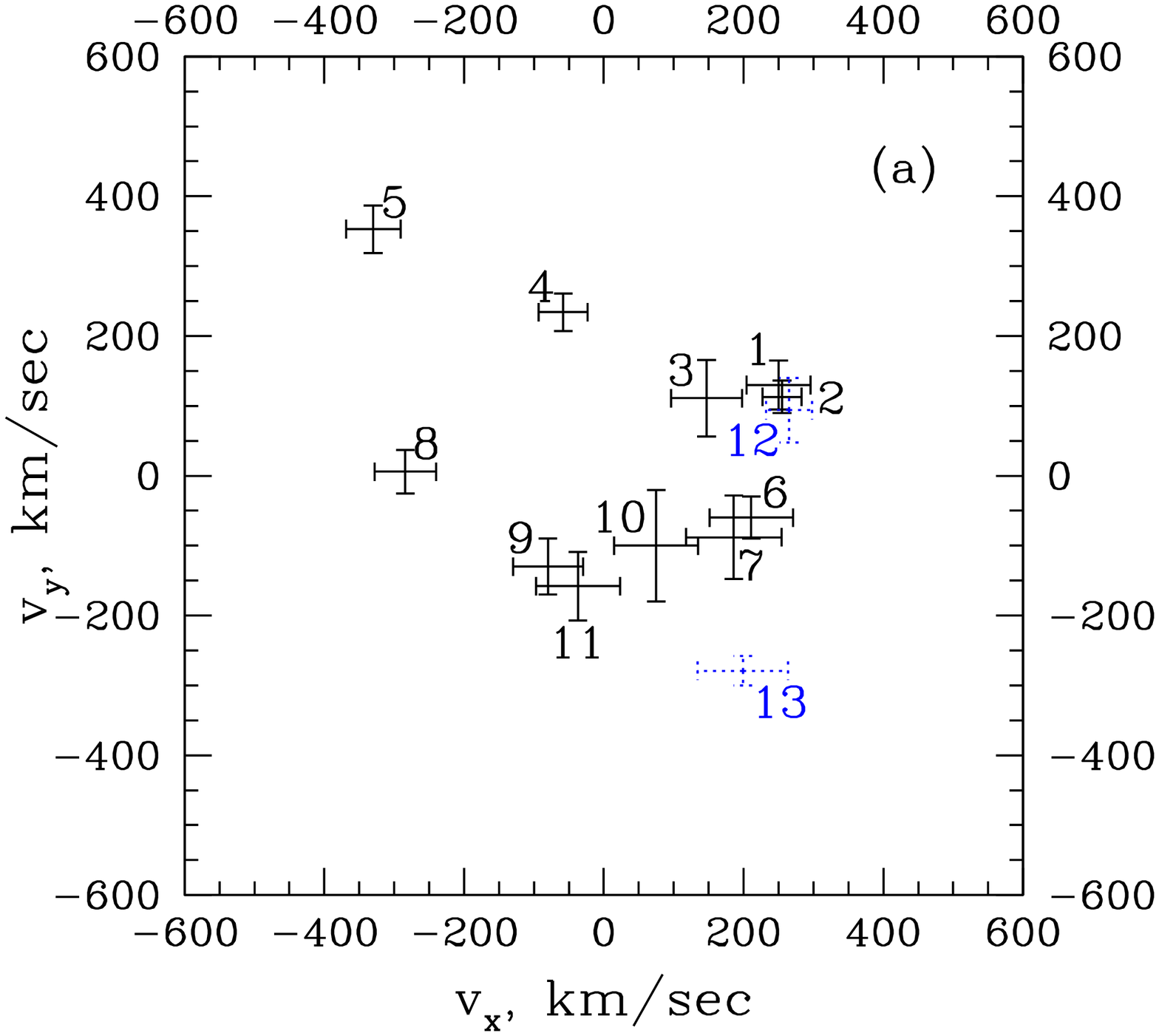}&
\includegraphics[%
  scale=0.35]{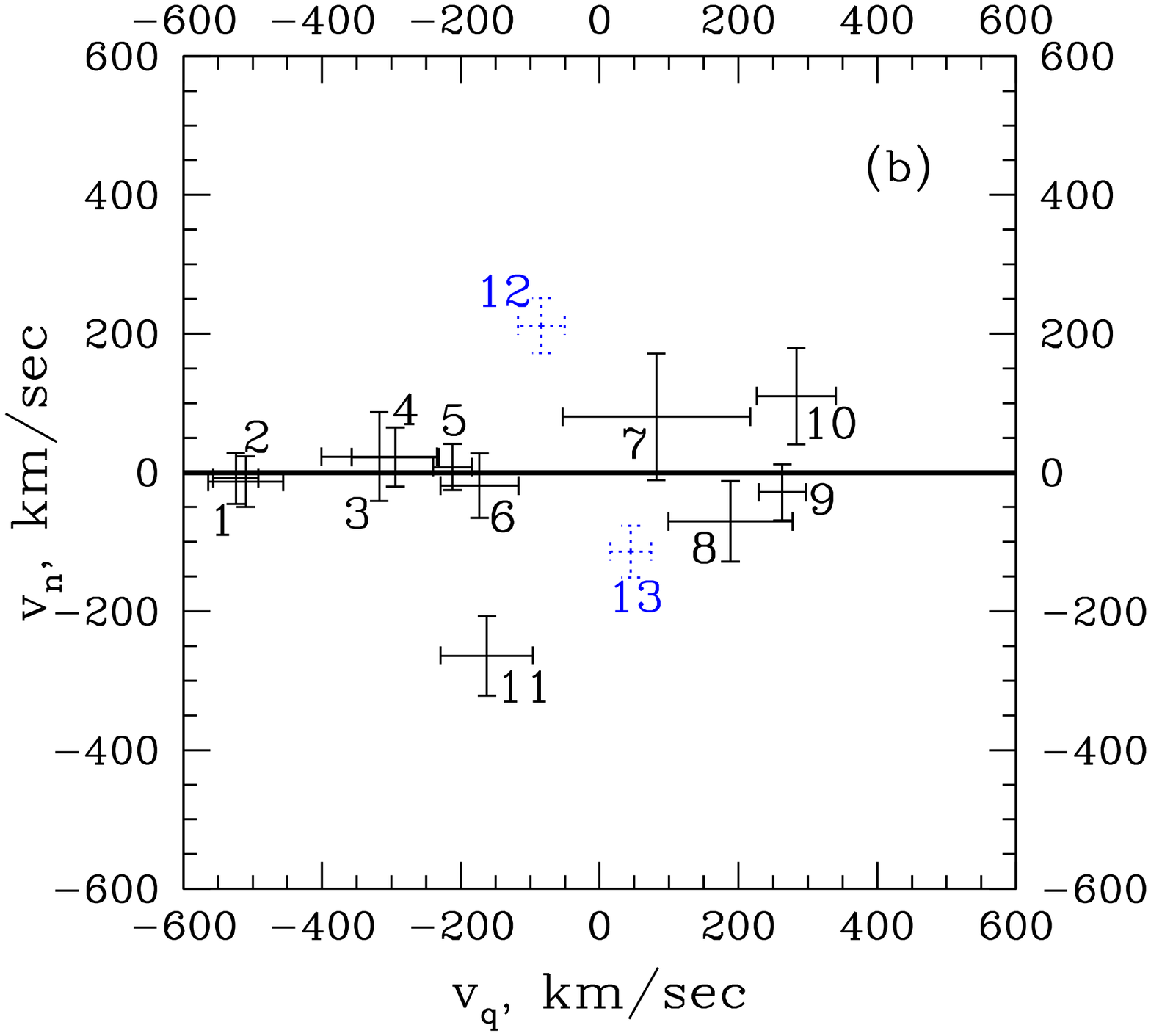}\tabularnewline
\multicolumn{2}{c}{\includegraphics[%
  scale=0.7]{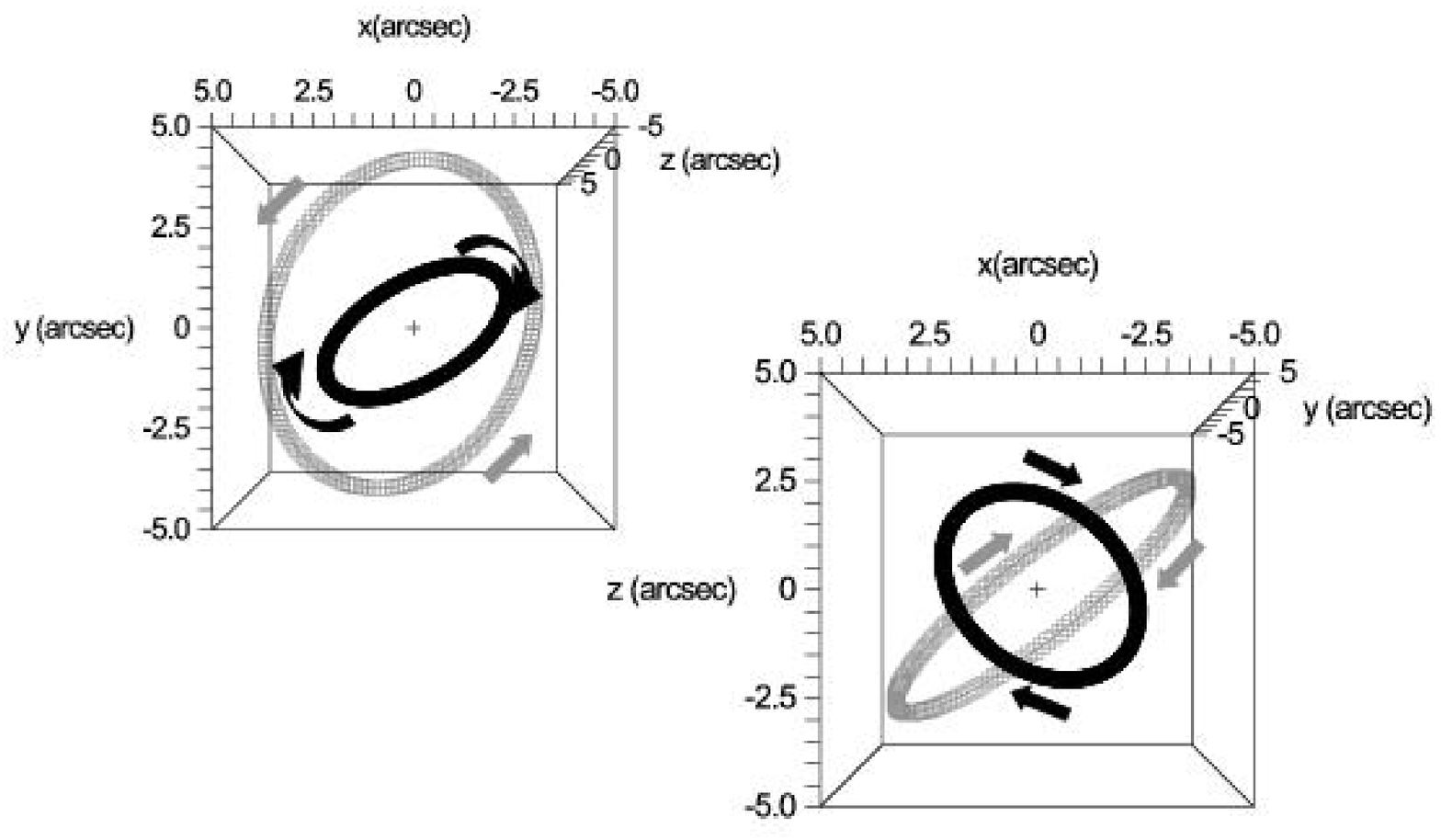}}\tabularnewline
\end{tabular}}

\caption{{\small \label{f:disks}Evidence for disk-like stellar structures
around the Galactic MBH. Top left: Distribution in projected velocity
space of 13 bright emission line stars in the inner $\sim\!5"$ from
a compilation by Genzel et al. (\citeyear{Gen00}). Top right: The
same, after transforming to the disk coordinates. 10 of the 13 stars
lie in a thin disk with opening angle $\lesssim\!10^{\circ}$. (Levin
\& Beloborodov \citeyear{Lev03}. Reprinted with permission from the
Astrophysical Journal). Bottom: Reconstructed orientations of the
the inner clockwise ring and the outer counter-clockwise disk (Genzel
et al. \citeyear{Gen03a}. Reprinted with permission from the} \emph{\small Astrophysical
Journal}{\small ), shown from two vantage points. The coordinates
$x$ and $y$ are in the plane of the sky, and $z$ is along the line
of sight, with the observer at $z\!=\!-\infty$.}}
\end{figure}

Inside $\sim\!0.04$ pc the population changes again. There are no
bright giants, red or blue, and only faint blue stars are observed
there. This population is know as the {}``S-stars'' or {}``S-cluster'',
after their identifying labels%
\footnote{Two labeling conventions are used for the IR sources near the MBH,
the MPE labels (of the form S\#) and the UCLA labels (of the form
S\#-\#, see Ghez et al. \citeyear{Ghe98}). The dual labels of some
stars of special interest very near the MBH are listed here for convenience.

\centerline{\begin{tabular}{p{4em}lllllllll}
MPE: &
S1&
S2&
S4&
S8&
S9&
S11&
S12&
S13&
S14\tabularnewline
UCLA: &
S0-1&
S0-2&
S0-3&
S0-4&
S0-5&
S0-9&
S0-19&
S0-20&
S0-16\tabularnewline
\end{tabular}}%
}. Deep near-IR photometric (Krabbe et al. \citeyear{Kra95}; Genzel
et al. \citeyear{Gen03a}) and spectroscopic (Genzel et al. \citeyear{Gen97};
Eckart, Ott \& Genzel \citeyear{Eck99}; Figer et al. \citeyear{Fig00};
Gezari et al. \citeyear{Gez02}; Ghez et al. \citeyear{Ghe03a}) observations
of that region were all consistent with the identification of these
stars as massive MS stars. It is now spectroscopically confirmed (Ghez
et al \citeyear{Ghe03a}; Eisenhauer et al. \citeyear{Eis05}) that
the S-stars are B0--B9 MS stars (corresponding to masses in the range
$\sim\!3\,\Mo$-- $\sim\!15\!\Mo$). The rotational velocities (stellar
spin) are similar to those of Solar neighborhood stars of the same
spectral type. Thus, there is no indication of anything unusual about
the S-stars, apart for their location very near the MBH. Astrometric
observations (Ghez et al. \citeyear{Ghe05}; Sch\"odel et al. \citeyear{Sch03};
Eisenhauer et al. \citeyear{Eis05}) indicate that, unlike the emission
line stars farther out, the S-stars orbits appear random (isotropic)
and uncorrelated with the orientation of either disk. There are some
marginal statistical indications of a bias toward radial orbits (Genzel
et al. \citeyear{Gen00}; Sch\"odel et al. \citeyear{Sch03}, but
see Ghez et al. \citeyear{Ghe05}). 

The brightest star in the S-cluster, S2/S0-2 ($K\!=\!13.9$ mag),
is a transitional O8--B0 star with a mass of $\sim\!15\,\Mo$, effective
temperature of $\sim30,000$ K, intrinsic bolometric luminosity of
$L\!\sim\!10^{3}\,\Lo$ and a main sequence lifespan of $\sim\!10^{7}$
yr. All the other stars in the cluster are less massive, longer-lived,
cooler and fainter. A B9 MS star has a mass of $\sim\!3\,\Mo$, a
radius of $\sim\!2.1\,\Ro$ and a main sequence lifespan of $\sim\!4\!\times\!10^{8}\,\mathrm{yr}$.
A typical star would correspond to a B2 main sequence star, with a
mass of $\sim\!10\,\Mo$, a radius of $\sim\!4.5\,\Ro$ and main sequence
lifespan of $\sim\!2\!\times\!10^{7}$ yr (Cox \citeyear{Cox00};
Schaller et al. \citeyear{Sch92}). It is not clear what is the relation,
if any, between the population of the He stars on the $0.04$--$0.4$
pc scale and the S-cluster stars inside $0.04$ pc. The nature and
origin of the young stars near the MBH are the subject of ongoing
investigations (\S\ref{s:OBriddle}).

\section{Stellar dynamics at extreme densities}

\label{s:dyn}

The extreme conditions in the central $\sim\!0.1$ pc around the MBH
are unique in the Galaxy. The strong tidal field (total disruption
of a solar type star at a distance ten times the event horizon), high
velocities (up to $\sim\!0.2c$ at disruption) and high stellar density
(up to $\sim\!10^{8}\,\Mo\,\mathrm{pc^{-3}}$, $10^{9}$ higher than
in the vicinity of the Sun) effectively make the environment near
the MBH a {}``stellar collider'' (a total stellar collision rate
of ${\cal O}(10^{-5})\,\mathrm{yr^{-1}}$ in the central $0.02$ pc).
Under such extreme conditions, strong stellar interactions may probe
and modify the stars' internal structure, and otherwise rare dynamical
processes may become frequent enough to have observable consequences.
Before discussing the evidence for the existence of a high density
cusp in the GC (\S\ref{ss:cusp}) and its consequences for stellar
collisions (\S\ref{ss:coll}), it is useful to review briefly the
physical processes and scales that govern the evolution of a stellar
system around a MBH.

\subsection{Physical processes and scales }

\label{ss:process}

\subsubsection{Dynamical processes}

\label{sss:dynproc}

The dynamical processes in a gravitating stellar system can be summarized
by classifying stellar interactions by their distance scale (e.g.
Binney \& Tremaine \citeyear{Bin87}; Heggie \& Hut \citeyear{Heg03}).
In this context the term interaction refers both to remote gravitational
interactions where the stars exchange momentum and energy, as well
as actual physical contact between stars. 

On the largest scale, the stellar orbit is determined by the superposed
interactions with the MBH and all the other stars, which are well
approximated by a smooth gravitational potential. On a shorter scale,
the stellar motions are randomized by two-body scattering when two
stars approach each other so that their mutual interaction dominates
that of the smoothed potential. Over time, scattering leads to the
relaxation of the system and the redistribution of energy and angular
momentum. Massive perturbers, such as stellar clusters, IBHs or molecular
clouds can accelerate the relaxation (Zhao, Haehnelt \& Rees \citeyear{Zha02}).
In addition, very near the MBH, where the orbits are nearly Keplerian,
angular momentum relaxation (but not energy relaxation) can be accelerated
by the process of {}``resonant relaxation'' (Rauch \& Tremaine \citeyear{Rau96}).
In the course of relaxation, the stars, whose mass range spans 2--3
orders of magnitude, are driven toward energy equipartition. However,
equipartition cannot be achieved in a self gravitating system, and
in particular in the steep potential well of a MBH. A two-body interaction
will tend to slow down the heavier star and speed up the lighter one,
but since the typical orbital radius depend only on the velocity (specific
energy), the heavy star will sink to the center, while the lighter
star will drift outward. Over time, this process leads to mass segregation---the
more massive stars are concentrated near the MBH and the lighter stars
are pushed out of the inner region. 

The system continuously loses stars by two-body scattering, either
abruptly, by a single strong encounter that ejects a star, or gradually,
by diffusion to higher energies (evaporation). The lost star takes
away positive energy from the system. The system then becomes more
bound and compact, the collision rate increases, yet more stars are
lost, and a runaway process is launched. This so called {}``gravothermal
catastrophe'' or {}``core collapse'' reflects the fact that a self-gravitating
system has a negative heat capacity---it become hotter when energy
is taken out. Core collapse, if unchecked, will lead to the formation
of an extremely dense stellar core surrounded by a diffuse extended
halo. 

Once the density becomes high enough, very short-range inelastic collisions
are no longer extremely rare, and the fact that the stars are not
point masses, but extended objects with internal degrees of freedom,
becomes significant. In such collisions energy is extracted from the
orbit and invested in the work required to raise stellar tides, or
strip stellar mass. The outcome of inelastic collisions depends critically
on whether a MBH is present or not. If the collision is slow, as it
is in the core of a globular cluster where there is no MBH, then the
typical initial orbit is just barely unbound (parabolic). In this
case, the tidal interaction may extract enough orbital energy for
{}``tidal capture'', and the formation of a binary. However, when
the stars orbit a central MBH, the collisions are fast (The Keplerian
velocity near the MBH exceeds the escape velocity from the star) and
the initial orbits are very unbound (hyperbolic). Even very close
fly-bys cannot take enough energy from the orbit to bind the two stars,
and so they continue on their way separately after having extracted
energy and angular momentum from the orbit. The stars can radiate
away the excess heat on a time scale shorter than the mean time between
collisions, but it is harder to get rid of the excess angular momentum.
It is therefore likely that high rotation is the longest-lasting dynamical
after effect of a close hyperbolic encounter, and that stars in a
high density cusp are spun-up stochastically by repeated collisions
(\S\ref{sss:spinup}). Finally, at zero range, almost head-on stellar
collisions can lead to the stripping of stellar envelopes (\S\ref{ss:cusp}),
the destruction of stars, or to mergers that result in the creation
of {}``exotic stars''. These are stars that cannot be formed in
the course of normal stellar evolution, such as a Thorne-\.Zytkow
object, which is an accreting neutron star (NS) embedded in a giant
envelope (Thorne \& \.Zytkow \citeyear{Tho75}).

\subsubsection{Characteristic scales}

\label{sss:scales}

Several important length-scales and timescales govern the dynamics
of the stellar system around the MBH. These are listed here, with
estimates of their values in the GC. A star of mass $\Ms\!=\!1\,\Mo$\label{d:Ms}
and radius $\Rs\!=\!1\,\Ro$\label{d:Rs} and a MBH mass of $m=3.5\times10^{6}\, M_{\odot}$\label{d:m}
(Sch\"odel et al. \citeyear{Sch03}) are assumed. Physical lengths
are expressed also as angular sizes for a distance to the Galactic
Center of $R_{0}=8\,\mathrm{kpc}$, which corresponds to $1"\!=\!0.039$
pc (Eisenhauer et al. \citeyear{Eis03}). 

\begin{description}
\item [Event~horizon\label{d:rs}]The size of the event horizon of a non-rotating
black hole, the Schwarzschild radius, is \begin{equation}
r_{S}=\frac{2Gm}{c^{2}}=10^{12}\,\mathrm{cm}\sim3\times10^{-7}\,\mathrm{pc}\sim9\,\mu\mathrm{arcsec}\,.\label{e:rsch}\end{equation}

\item [Tidal~radius\label{d:rt}]The tidal radius, $r_{t}$, is the maximal
distance from the MBH where the tidal forces of the MBH can overwhelm
the stellar self-gravity and tear the star apart. Its exact value
depends somewhat on the stellar structure and the energy of the orbit,
but up to a factor of order unity it is \begin{equation}
r_{t}\sim R_{\star}\left(\frac{m}{M_{\star}}\right)^{1/3}=10^{13}\,\mathrm{cm}\sim3\times10^{-6}\,\mathrm{pc}\sim90\,\mu\mathrm{arcsec}\,.\label{e:rt}\end{equation}

\item [Collision~radius\label{d:rcoll}]The collision radius, $r_{\mathrm{coll}}$,
is the minimal distance from the MBH where large angle deflections
by close gravitational encounters are possible. Closer to the MBH
the relative velocity between the interacting stars, $v$, exceeds
the escape velocity from the stellar surface, $V_{e}$\label{d:Ve},
and the distance of closest approach required for a large angle deflection
becomes smaller than the stellar radius (i.e. when $V_{e}^{2}\!\sim\! G\Ms/\Rs\!<\! Gm/r\!\sim\! v^{2}$).
Stars undergoing such a physical collision are destroyed since the
orbital energy exceeds the stellar binding energy. At distances smaller
than\begin{equation}
r_{\mathrm{coll}}\!\sim\!\left(\frac{m}{\Ms}\right)\Rs\!\sim\!0.08\,\mathrm{pc}\sim2\,\mathrm{arcsec\,,}\label{e:rc}\end{equation}
 two-body relaxation gradually becomes inefficient. 
\item [MBH~radius~of~influence\label{d:rh}]The radius of influence,
$r_{h}$, is the region where the MBH dominates the dynamics. It is
customary to define $r_{h}\!=\! Gm/\sigma^{2}$, where $\sigma$\label{d:sig}
is the 1D stellar velocity dispersion away from the MBH. However,
in most cases this is not a well defined quantity because $\sigma$
depends on radius ($\sigma$ is constant only in an isothermal density
distribution, which is not the case in the GC). Kinematically, the
MBH potential dominates that of the stars out to a distance where
the enclosed stellar mass is $\Ms(<\! r)\!\sim\!{\cal {O}(}m)$. For
the isothermal distribution $\Ms(<\! r_{h})\!=\!2m$. The measured
enclosed mass in the GC indicates that $\Ms(\lesssim\!2\,\mathrm{pc})\!\simeq\! m$
and $\Ms(\lesssim\!4\,\mathrm{pc})\!\simeq\!2m$ (Sch\"odel et al.
\citeyear{Sch03}), and so the MBH radius of influence in the GC can
be estimated at\begin{equation}
r_{h}\sim10^{19}\,\mathrm{cm}\sim3\,\mathrm{pc}\sim90\,\mathrm{arcsec}\,.\label{e:rh}\end{equation}

\item [Age~of~GC]It is assumed that the age of the GC, being the center
of the old spheroid component of the Galaxy, the bulge, is of the
same order as the age of the universe (roughly, the Hubble time)\label{d:tH},\begin{equation}
t_{H}\sim10\,\mathrm{Gyr}\,.\end{equation}

\item [Dynamical~timescale]The dynamical, or orbital time $t_{d}$, is
the time it takes a star to cross the system \begin{equation}
t_{d}\sim\frac{r}{v}\sim2\pi\sqrt{\frac{\mathrm{r}^{3}}{GM(<\! r)}}\sim2\times10^{5}\,\mathrm{yr}\,(\mathrm{at}\,3\,\mathrm{pc})\sim300\,\mathrm{yr}\,(\mathrm{at}\,0.03\,\mathrm{pc})\,,\label{e:tdyn}\end{equation}
where $r$ is the typical size of the system and $M(<\! r)$ is the
total enclosed mass (MBH and stars).
\item [Two-body~relaxation~timescale\label{d:trel}]The two-body relaxation
time, $t_{r}$, is related to $\sigma$, the mean stellar mass $\left\langle M_{\star}\right\rangle $
and the stellar number density $n_{\star}$\label{d:nstar} by (e.g.
Binney \& Tremaine \citeyear{Bin87})\begin{equation}
t_{r}\sim\frac{0.34\sigma^{3}}{G^{2}\left\langle M_{\star}\right\rangle ^{2}n_{\star}\ln\Lambda}\sim{\cal {O}(}10^{9}\,\mathrm{yr)}\,,\label{e:trel}\end{equation}
 where a Maxwellian velocity distribution is assumed and $\log\Lambda\!\sim\!{\cal {O}(}10)$
is the Coulomb logarithm. $\Lambda$ is the ratio between the largest
and smallest impact parameters possible in the system for elastic
collisions%
\footnote{The relaxation time can be derived approximately from the rate for
large deflection angle collisions, $t_{r}^{-1}\!\sim\! n_{\star}\sigma\Sigma$,
where the cross-section is given by $\Sigma\!\sim\!\pi\left(G\left\langle \Ms\right\rangle /\sigma^{2}\right)^{2}$.
The actual rate is larger by a factor $\log\Lambda$ due to the additional
contribution of the many weaker collisions at large impact parameters.%
}. Since near the MBH $\sigma^{3}\!\propto r^{-3/2}$, the relaxation
time is not a strong function of $r$ in a stellar cusp where roughly
$n_{\star}\!\propto r^{-3/2}$, as is the case in the GC (\S\ref{ss:cusp}).
The old stars near the MBH are expected to be dynamically relaxed
since since $t_{r}\!<\! t_{H}$. The relaxation time can also be approximately
expressed in terms of $\Ns$, the number of stars enclosed within
radius $r$ ,\begin{equation}
t_{r}(r)\sim\left(\frac{m}{\left\langle \Ms\right\rangle }\right)^{2}\frac{P(r)}{2\pi\log(0.4\Ns)N_{\star}}\,,\label{e:trelNP}\end{equation}
where $P$ is the circular orbital period. 
\item [Segregation~timescale\label{d:tseg}]The mass segregation timescale
for stars of mass $\Ms$ is of the same order as the relaxation timescale,
and scales with $\Ms$ as (\S\ref{ss:Mseg})\begin{equation}
t_{s}\sim t_{r}\frac{\left\langle \Ms\right\rangle }{\Ms}\qquad(\Ms>\left\langle \Ms\right\rangle )\,.\label{e:tseg}\end{equation}
 
\item [Evaporation~time\label{d:te}]The evaporation time scale is \begin{equation}
t_{e}\sim300t_{r}\,.\label{e:tevap}\end{equation}
 Evaporation is unimportant in the GC since $t_{e}\!\gg\! t_{H}$
(and also because the GC is not an isolated system, but is embedded
in the bulge, so evaporated stars can be replaced).
\item [Collision~timescale\label{d:tc}]The rate, per star, of grazing
collisions between two stars of mass and radius $M_{\star}^{a}$,
$R_{\star}^{a}$ and $M_{\star}^{b}$, $R_{\star}^{a}$, each, is
\begin{equation}
t_{c}^{-1}=4\sqrt{\pi}n_{\star}\sigma\left(R_{\star}^{a}+R_{\star}^{b}\right)^{2}\left[1+\frac{G\left(M_{\star}^{a}+M_{\star}^{b}\right)}{2\sigma^{2}\left(R_{\star}^{a}+R_{\star}^{b}\right)}\right]\sim10^{-9}\,\mathrm{yr}^{-1}\,(\mathrm{at}\,0.02\,\mathrm{pc})\,,\label{e:tcoll}\end{equation}
where it is assumed that the stars follow a mass independent Maxwellian
velocity distribution with 1D velocity dispersion $\sigma$ (this
is a good approximation near the MBH, see \S\ref{sss:relax}). The
two terms contributing to the total rate are the geometric cross-section
and the gravitational focusing term due to the mutual attraction of
the colliding stars, which dominates when the typical stellar velocities
are much smaller than the escape velocity from the stars. For $M_{\star}^{a}\!=\! M_{\star}^{b}$,
this corresponds to $\sigma^{2}<GM_{\star}/2R_{\star}=V_{\mathrm{e}}^{2}/4$.
The collision rate diverges for $\sigma/V_{e}\!\rightarrow\!0$ or
$\sigma/V_{e}\!\rightarrow\!\infty$ and is lowest when $\sigma/V_{e}\!=\!1/2$.
For the high velocities near the MBH, gravitational focusing is only
important for collisions with compact remnants where $\Rs\!\ll\!\Ro$. 
\end{description}

\subsubsection{Keplerian power-law stellar cusps}

\label{sss:relax}

Theoretical studies of the co-evolution of a MBH and the stellar system
around it indicate that a wide range of dynamical scenarios all lead
to the formation of a stellar density cusp, a region of diverging
density around the MBH (e.g. Bahcall \& Wolf \citeyear{Bah76,Bah77};
Young \citeyear{You80} Lee \& Goodman \citeyear{Lee89}; Quinlan,
Hernquist \& Sigurdsson \citeyear{Qui95}). The dynamical evolution
of the stars in the radius of influence depends on the ratio between
the relaxation time and the age of the system. If $t_{r}\!>\! t_{H}$,
two-body relaxation can be neglected, the MBH grows adiabatically
and the resulting stellar distribution depends on the initial conditions.
For example, an initially isothermal distribution will give rise to
a $n_{\star}\!\propto\! r^{-3/2}$ cusp (Young \citeyear{You80}).
Other initial conditions will result in a variety of stellar distributions,
some falling as steeply as $n_{\star}\!\propto\! r^{-5/2}$ (Lee \&
Goodman \citeyear{Lee89}; Quinlan, Hernquist \& Sigurdsson \citeyear{Qui95}).
However, when $t_{r}\!<\! t_{H}$, two-body relaxation erases the
initial conditions, and the final configuration will depend only on
the boundary conditions and the stellar mass function.

The formation history of the Galactic MBH is not known. Estimates
of the stellar density and velocity indicate that $t_{r}\!<\! t_{H}$
near the MBH (\S\ref{sss:scales}), and so the old stars there are
expected to have relaxed dynamically by two-body interactions. This
is consistent with their observed isotropic velocity field and spatial
distribution (Genzel et al \citeyear{Gen96}; Ghez et al \citeyear{Ghe98};
Genzel et al \citeyear{Gen00}; \S\ref{s:GCstars}). The DF of a
spherically symmetric system with an isotropic velocity field is a
function of $\varepsilon$ only. Bahcall \& Wolf (\citeyear{Bah77})
show that the distribution function (DF) $f$ (stars per $\mathrm{d}\mathbf{r}\mathrm{d}\mathbf{v}$
interval) of such a system that is relaxed and consists of stellar
masses in the range $M_{1}<M_{\star}<M_{2}$ and a moderately varying
mass function $\mathrm{d}\Ns/\mathrm{d}\Ms$\label{d:Ns}, can be
approximated near the MBH by a power-law with a mass-dependent index\begin{equation}
f(\varepsilon;\Ms)\propto\epsilon^{p_{M}}\,,\qquad p_{M}\equiv\frac{\Ms}{4M_{2}}\,,\label{e:BWsol}\end{equation}
where $\varepsilon\!=\!-v^{2}/2+\psi(r)$ is the specific energy of
the star\label{d:espec}, $\psi\!=\!-\phi\!>\!0$\label{d:relgpot}
is the relative gravitational potential, $\phi$ is the gravitational
potential, and $f\!=\!0$ for $\varepsilon<0$. The power-law index
varies in the range $0\!<\! p_{M}\!\le\!1/4$ across the mass spectrum.
The corresponding stellar number density (stars per $\mathrm{d}\mathbf{r}$)
is (see Eq. \ref{e:nstar})\begin{equation}
n_{\star}(r;\Ms)\propto r^{-3/2-p_{M}}\,.\label{e:BWcusp}\end{equation}
 Thus the most massive component of a relaxed multi-mass system is
expected to have an $r^{-7/4}$ density profile. and the lightest
one an $r^{-3/2}$ density profile. In the simple case of a single
mass population ($M_{1}\!=\! M_{2}$), $p_{M}\!=\!1/4$ and $n_{\star}\!\propto\! r^{-7/4}$
(Bahcall \& Wolf \citeyear{Bah76}; see also Binney \& Tremaine \citeyear{Bin87},
p. 547 for a simple derivation by dimensional analysis). Note that
these results apply also to the Galactic dark matter halo elementary
particles in the GC (Gnedin \& Primack \citeyear{Gne04}), which are
expected to interact gravitationally with the cusp stars and settle
into a $r^{-3/2}$ cusp near the MBH (\S\ref{sss:distrDM}).

Dynamical simulations (e.g. Murphy Cohn \& Durisen \citeyear{Mur91};
Baumgardt, Makino \& Ebisuzaki \citeyear{Bau04}) confirm that when
the population consists of a single stellar mass, or when the most
massive stars come to dominate the central stellar mass density due
to mass segregation, their DF has indeed a power-law index $p_{M}\!\sim\!0.25$.
The simulations also indicate that the central distribution of the
less massive stars is flatter, with $p_{M}\!\lesssim\!0$. Note that
Eq. (\ref{e:BWsol}) was derived under various assumptions about the
boundary conditions far from the MBH and about the stellar mass function,
and its general applicability has yet to be investigated in full by
numeric simulations. Nevertheless, it will be assumed here to hold
generally, since it is broadly consistent with the numerical results
and is analytically convenient. Note that the applicability of the
Bahcall-Wolf solution to the GC is based on the assumption that the
GC has evolved in isolation for at least a relaxation time before
the present epoch. In particular, it is assumed that it has not experienced
a strong perturbation such as associated, for example, with a merger
of two comparable mass galaxies and their MBHs (a {}``major merger'').
This assumption is consistent with the low mass of the MBH, and with
the steep stellar cusp observed around it (\S\ref{ss:cusp}), since
major mergers are thought to result in flattened stellar cores due
to ejection of stars by the two merging MBHs (Milosavljevi\'c et
al \citeyear{Mil02}).

The Bahcall-Wolf solution applies to point particles. This assumption
no longer holds very near the MBH, where the collision rate is high
because of the very high stellar density and velocity. Stars on tight
orbits around the MBH cannot survive for long, and so eventually most
of the population there will consist of stars that are on very wide,
marginally bound (parabolic) orbits, which spend only a small fraction
of their time in the collisionally dominated region. These marginally
bound stars have a flatter spatial distribution, of the form (e.g.
Binney \& Tremaine \citeyear{Bin87}, p. 551)

\begin{equation}
n_{\star}\propto r^{-1/2}\,.\end{equation}

Near the MBH, at $r\!\ll\! r_{h}$ the potential is Keplerian to a
good approximation, $\psi(r)\!=\! Gm/r$, and the orbits can be characterized
by their semi-major axis $a$\label{d:sma} (or their period $P$\label{d:Porb})
and their eccentricity $e$\label{d:ecc} ($0\!\le\! e\!\le\!1$ for
bound orbits),

\begin{equation}
a=\frac{Gm}{2\varepsilon}\,,\qquad e^{2}=1-\frac{J^{2}}{Gma}\,,\qquad P=2\pi\sqrt{\frac{a^{3}}{Gm}}\,,\qquad(m\!\gg\!\Ms),\label{e:Kepler}\end{equation}
where $J$ is the magnitude of the specific angular momentum\label{d:Jspec},
such that the orbital energy is given by \begin{equation}
\varepsilon=-\frac{1}{2}v^{2}+\frac{Gm}{r}=-\frac{1}{2}v_{r}^{2}-\frac{J^{2}}{2r^{2}}+\frac{Gm}{r}\,,\label{e:Ekepler}\end{equation}
 where $v_{r}$ is the radial velocity. The stellar densities per
$\mathrm{d}\mathbf{r}$, $\mathrm{d}\varepsilon$ and $\mathrm{d}a$
that correspond to a DF of the form $f(\varepsilon)\!=\! A\varepsilon^{p}$
(Eq. \ref{e:BWsol}) are, respectively, (e.g. Sch\"odel et al \citeyear{Sch03},
Eqs. A1--A8)

\begin{eqnarray}
n_{\star}(r) & = & (2\pi)^{3/2}\frac{\Gamma(1+p)}{\Gamma(5/2+p)}(Gm)^{3/2+p}Ar{}^{-3/2-p}\,,\nonumber \\
n_{\star}(\varepsilon) & = & \sqrt{2}\pi^{3}(Gm)^{3}A\varepsilon^{p-5/2}\,,\nonumber \\
n_{\star}(a) & = & 2^{2-p}\pi^{3}(Gm)^{3/2+p}Aa^{1/2-p}\,.\label{e:nstar}\end{eqnarray}

In the general spatially spherically symmetric case (anisotropic velocity
field), the DF $f(\varepsilon,J)$ depends both on $\varepsilon$
and $J$. The distribution of specific energy and angular momentum,
$n(\varepsilon,J)$ (stars per $\mathrm{d}\varepsilon\mathrm{d}J$)
is \begin{equation}
n_{\star}(\varepsilon,J)=8\pi^{2}Jf(\varepsilon,J)P_{r}(\varepsilon,J)\,,\label{e:NeL}\end{equation}
 where $P_{r}$ is the radial period. For Keplerian orbits $P_{r}\!=\! P$
and the distribution of specific energy and eccentricity (stars per
$\mathrm{d}\varepsilon\mathrm{d}e$) is (cf Cohn \& Kulsrud \citeyear{Coh78})

\begin{equation}
n_{\star}(\varepsilon,e)=\left[2\sqrt{2}\pi^{3}(GM)^{3}f(\varepsilon,J)\varepsilon^{-5/2}\right]e\,,\end{equation}
It then follows that for Keplerian orbits with \emph{isotropic} velocities
the normalized distribution of eccentricities $n(e)$ (stars per $\mathrm{d}e$
interval) is simply (cf Binney \& Tremaine \citeyear{Bin87})

\begin{equation}
n_{\star}(e)=2e\,.\label{e:ne}\end{equation}
Thus, isotropically distributed stars on Keplerian orbits tend to
have high eccentricities, with a fraction $1-e^{2}$ having eccentricities
$>\! e$ (for example, 10\% have $e\!>\!0.95$). 

The velocity distribution is tied to the spatial density distribution
through the Jeans Equation (the continuity equation of the stellar
orbits in phase space in terms of the stellar density and velocity
dispersion)

\begin{equation}
\frac{v_{c}^{2}}{\sigma^{2}}=-\frac{d\ln n_{\star}}{d\ln r}-\frac{d\ln\sigma^{2}}{d\ln r}\,,\label{e:Jeans}\end{equation}
where $v_{c}\!=\!\sqrt{GM(<\! r)/r}$ is the circular velocity and
a steady state, isotropic, non-rotating system is assumed. The assumption
of steady-state is justified because the dynamical timescale is much
shorter than the relaxation timescale. The assumptions of approximate
isotropy and non-rotation are observationally justified for the old
stellar population and for the stars in the central $1"$ (\S\ref{ss:1pc}).
Since very near the MBH the stellar mass is negligible, $M(<\! r)\!\simeq\! m$,
the velocity dispersion is Keplerian, $\sigma^{2}\propto r^{-1}$,
and the density is a power-law, $n_{\star}\propto r^{-\alpha}$ (here
$\alpha\!=\!3/2+p$),\label{d:alpha} it follows from the Jeans equation
that

\begin{equation}
\sigma_{M}^{2}=\left(\frac{1}{5/2+p_{M}}\right)\frac{Gm}{r}\,.\label{e:sigM}\end{equation}
 This result justifies the approximation that the velocity dispersion
in a relaxed stellar system around a MBH is mass-independent, since
$\sigma_{M}^{2}$ changes by less than 10\% over the entire mass range,
in marked contrast to the $\sigma_{M}^{2}\propto M_{\star}^{-1}$
dependence of equipartition.

Theoretical arguments indicate that the velocity field of a relaxed
system very near a MBH should be approximately Maxwellian and isotropic
locally (Quinlan, Hernquist \& Sigurdsson \citeyear{Qui95}). This
is indeed observed to be the case in the GC. The measured radial velocity
distribution is well fitted by a Gaussian distribution (Genzel et
al. \citeyear{Gen96}) and the measured velocities in the plane of
the sky and along the line-of-sight are consistent with an isotropic
velocity dispersion (Ghez et al \citeyear{Ghe98}; Genzel et al \citeyear{Gen00}).
The Jeans equation (Eq. \ref{e:Jeans}) indicates that the steeper
the cusp (larger $p_{M}$), the larger $v_{c}/\sigma$, and so the
fraction $f_{u}$ of unbound stars (those with $v\!>\! v_{e}\!=\!\sqrt{2}v_{c}$)
is smaller%
\footnote{The stars are bound to the GC as a whole (MBH and stars), but when
they approach the MBH they move with $v\!>\! v_{e}$ in the two-body
MBH-star system.%
}, \begin{equation}
f_{u}(p_{M})=\sqrt{\frac{2}{\pi}}\int_{\sqrt{5+2p_{M}}}^{\infty}u^{2}e^{-u^{2}/2}\mathrm{d}u\,.\end{equation}
The fraction of unbound stars varies greatly across the range of interest
because the integration over the exponential tail of the Maxwellian
distribution depends sensitively on the lower limit. For example,
in a flat core ($n_{\star}\!\sim\!\mathrm{const.}$), $f_{u}(-3/2)\!\simeq\!0.6$,
whereas in a steep cusp (e.g. $n_{\star}\!\propto\! r^{-2.5}$), $f_{u}(1)\!\simeq\!0.07$.
Because unbound stars have wide orbits and spend most of their time
far away from the MBH, the stellar population in a shallow cusp is
well mixed and representative of the average population over a large
volume. In contrast, the stellar population in a steep cusp is localized
and can therefore develop and maintain properties that differ from
those of the general population.

\subsection{The stellar cusp in the Galactic center}

\label{ss:cusp} 

The basic premise of the attempt to quantify the global properties
of the stellar distribution in the inner GC is that there is an underlying
quasi-steady state distribution that includes most of the mass and
most of the stars, and which is traced by the old faint stars. In
contrast, the young stars and bright giants, which dominate the light,
are less reliable tracers because they could be far from steady-state,
or subject to statistical fluctuations due to their small numbers,
or susceptible to environmental effects such as stellar collisions
or photoionization (Sellgren et al. \citeyear{Sel90}). 

The proximity of the GC allows, in principle, a direct test of the
theoretical predictions about the distribution of stars around a MBH.
However, even in the GC the derivation of the 3D mass and number density
distributions from the observations is hindered by several substantial
obstacles. 

\begin{enumerate}
\item The star counts are inevitably incomplete. Only stars brighter than
the detection threshold can be observed. To observe fainter stars,
the instrumental photometric sensitivity must be improved, and with
it the angular resolution in order to resolve individual stars in
the crowded field near the MBH and avoid the source confusion problem.
The dynamic range of the detectors must also be increased, otherwise
the bright stars in the frame will saturate the detectors and out-shine
their faint neighbors. These instrumental requirements can rarely
all be satisfied simultaneously. The degree of incompleteness can
be estimated and compensated for in the statistical sense, by simulating
the detection procedure with mock data.
\item Obscuration ({}``extinction'') by a non-uniform distribution of
interstellar dust can bias the derived stellar distribution. To compensate
for such effects, it is necessary to map the dust across the field,
for example by comparing the observed stellar colors with their assumed
intrinsic ones. The analysis of the extinction gradient across the
central parsec suggests that any bias due to dust extinction is unlikely
to be large (Alexander \citeyear{Ale99a}; but see strong gradients
in extinction maps by Scoville et al. \citeyear{Sco03}).
\item It is crucial to separate the old and young stars in order make a
meaningful comparison to stellar dynamical models, since the two populations
are dynamically distinct (\S\ref{ss:1pc}). The phase space distribution
of the young stars reflects the initial conditions at birth and their
still unknown formation mechanism (\S\ref{s:OBriddle}), whereas
the distribution of the old stars has had time to adjust to the presence
of the MBH and is probably dynamically relaxed. The stellar type can
be determined, at least for the brighter stars, by their spectra or
infrared colors. 
\item The observed 2D star counts have to be translated to 3D densities.
This can be done by first making assumptions about the symmetries
of the DF (e.g. spherical symmetry and isotropic velocity; Genzel
et al \citeyear{Gen96}), deprojecting the observed surface density
and line of sight velocity dispersion (Eq. \ref{e:Abel}) and relating
the deprojected quantities to the enclosed mass via the Jeans equation
(Eq. \ref{e:Jeans}). Alternatively, the star counts can be fitted
to a projected 3D model distribution with a small number of free parameters
(e.g. a broken power-law cusp), for example by maximal likelihood
criteria (Alexander \citeyear{Ale99a}; Genzel et al \citeyear{Gen03a};
see also Merritt \& Tremblay \citeyear{Mer94} for a more sophisticated
approach, yet to be applied to GC data). This method does not take
into account the velocity information, which in any case is not available
for all the stars. 
\item Only the luminous stars are directly observed. However, most of the
mass is expected to be in the very faint low mass stars, compact remnants
or other forms of dark mass, if such exist (\S\ref{ss:DM}). In order
to obtain the mass density $\rho(r)$\label{d:rho}, it is necessary
either to measure the diffuse mass dynamically, which is very difficult
to do in the center, where the potential is dominated by the MBH (\S\ref{sss:distrDM}),
or else assume a stellar number-to-mass ratio, $\Ns/M$, in order
to translate the stellar number density $n_{\star}(r)$ to mass density.
Such a procedure is very uncertain, since $\Ns/M$ probably depends
strongly on the distance from the MBH due to mass segregation.
\end{enumerate}
\begin{figure}[!t]
\centering{\htarget{f:cusp}\includegraphics[%
  scale=0.65]{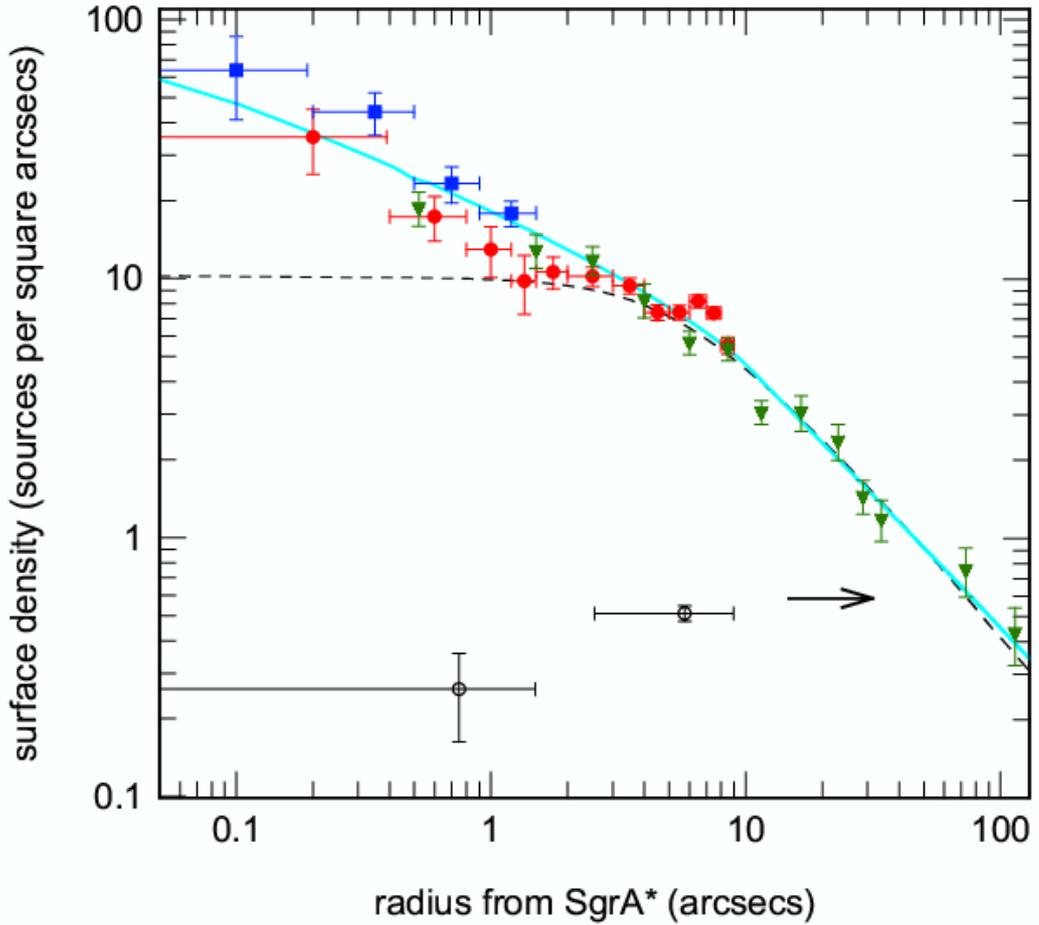}}

\caption{{\small \label{f:cusp}Measurements of the surface density distribution
of the stellar counts around the Galactic MBH (Genzel et al \citeyear{Gen03a}).
The observed surface density density corresponds to a space number
density of $n_{\star}\!\propto\! r^{-\alpha}$ with $\alpha\!=\!1.0\!\pm\!0.1$
inside 0.4 pc, and $\alpha\!=\!2.0\!\pm\!0.1$ outside. (Reprinted
with permission from the} \emph{\small Astrophysical Journal}{\small ).}}
\end{figure}

The light distribution in the GC approximately follows a $\sim\! r^{-2}$
power-law from as far away as $\sim\!100$ pc, where it cannot be
directly related to the central super-massive BH, to $\sim\!1$ pc
from $\SgrA$ (Serabyn \& Morris \citeyear{Ser96}; \S\ref{s:GCstars}).
Early attempts to determine the stellar distribution inside the central
parsec concentrated on identifying the core radius $r_{c}$, where
the surface density falls to half its central value, by measuring
the surface density of the brightest stars (the only ones observable
at the time) or of the total surface brightness. These studies yielded
conflicting results, from $r_{c}\!\sim\!0.05$ pc (Allen \citeyear{All83};
Scoville \citeyear{Sco04}) to $\sim1$ pc (Rieke \& Rieke \citeyear{Rie88}).
It appears that the differences between the various estimates depend
on whether the light distribution or the number counts are used and
whether or not the bright stars are included in the fits. A consensus
has yet to emerge on this question. 

Generally, star counts follow the mass more faithfully than the total
light distribution, which is completely dominated by the light of
giants, even though they are only a small fraction of the population
both by number and mass. Analyses of the stellar distribution based
on star counts (Alexander \citeyear{Ale99a}; Genzel et al \citeyear{Gen03a})
indicate that a stellar cusp rises monotonously all the way into the
central $\sim\!0.1"$ ($0.004$ pc) (Fig. \hlink{f:cusp}). The number
density was determined by fitting the star-counts to a 3-parameter
broken power-law (break radius and the inner and outer exponents)
using maximal likelihood and taking into account completeness corrections.
The number density was then converted to a mass density by assuming
a constant $\Ns/M$ ratio and normalizing to the enclosed stellar
mass $\Ms(<\! r)\!=\! M(<\! r)\!-\! m$ far from the MBH. The total
enclosed mass $M(<\! r$) was measured dynamically on the $\lesssim\!10$
pc scale, and the MBH mass $m$ was measured dynamically close to
the center (\S\ref{ss:mass}). The resulting mass model is (Genzel
et al \citeyear{Gen03a}; assuming $R_{0}\!=\!8$ kpc)\begin{equation}
\rho(r)\!=\!\rho_{0}\left(\frac{r}{r_{0}}\right)^{-\alpha},\quad\rho_{0}\!=\!1.2\!\times\!10^{6}\,\Mo\,\mathrm{pc^{-3}},\quad r_{0}\!=\!0.4\,\mathrm{pc},\quad\alpha\!=\!\left\{ \begin{array}{cc}
1.4\pm0.1 & r\!<\! r_{0}\\
2.0\pm0.1 & r\!\ge\! r_{0}\end{array}\right.\,.\label{e:cusp}\end{equation}
 This translates to an enclosed stellar mass in the inner cusp ($r<r_{0}$)
of \begin{equation}
M_{\star}(<\! r)\!=\!6\!\times\!10^{5}(r/r_{0})^{1.6}\, M_{\odot}\,,\label{e:Mencl}\end{equation}
and extremely high densities near the MBH, $\rho(0.04\,\mathrm{pc})=3\!\times\!10^{7}\, M_{\odot}\mathrm{\, pc^{-3}}$
and $\rho(0.004\,\mathrm{pc})=8\!\times\!10^{8}\, M_{\odot}\mathrm{\, pc^{-3}}$.
Note however that the total mass enclosed in the high density regions
is very small, only $\sim\!8000\,\Mo$ inside $0.04$ pc and $\sim\!200\,\Mo$
inside $0.004$ pc. These estimates of extreme densities are consistent
with the observed gradual depletion of bright giants with decreasing
distance to the center, which is naturally explained by a combination
of projection effects and collisional destruction in a steep density
cusp (\S \ref{ss:coll}).

The observed power-law is consistent with the Bahcall-Wolf solution
for the less massive stars in a relaxed population. This may hint
that the mass density inside $r_{0}$ is dominated by dark objects
more massive than the $\Ms\!<\!10\,\Mo$ stars which constitute the
majority of the observed population. If this is indeed the case, as
is further explored in \S\ref{ss:Mseg}, then $\Ns/M$ must rise
toward the center, contrary to what was assumed, and so the mass model
(Eq. \ref{e:cusp}) should be interpreted as representing the run
of the density in the luminous stars only. The total mass density
distribution may have a steeper slope than the observed number density,
and the mass density close to the center may be yet higher than the
estimates above%
\footnote{For a given, dynamically determined enclosed mass $M_{0}$ within
$r_{0}$, a power-law density $\rho(r)\!=\!\left[4\pi M_{0}\left/(3\!-\!\alpha)r_{0}^{3}\right.\right](r/r_{0})^{-\alpha}$
increases with increasing $\alpha$ inside $r_{\alpha}\!=\! r_{0}\exp[1/(\alpha-3)]$
and decreases outside $r_{\alpha}$.%
}.

\subsection{Mass segregation}

\label{ss:Mseg}

It is quite likely that the stars observed near the MBH constitute
only a small fraction of the total extended mass there. Mass segregation
is expected to drive a flow of NSs and SBHs to the center since these
compact remnants are more massive than the mean stellar mass in the
unsegregated parent population (table {\small \hlink{t:StellarPop}})
and since, unlike the short-lived massive MS stars, the remnants are
effectively eternal and have time to segregate significantly (Morris
\citeyear{Mor93}; Miralda-Escud\'e \citeyear{Mir00}). SBHs, which
are the most massive stellar remnants, are expected to be the most
centrally concentrated.

The farther from the center a compact object is born, the longer it
takes it to sink in. The process can be described in terms of dynamical
friction, the drag experienced by a massive object $M_{\bullet}$
as it moves through a background of lighter masses (the drag is due
to the work done by the massive object in focusing the lighter masses
across and behind it). When the velocity DF is Maxwellian, and in
the limit $M_{\bullet}\!\gg\!\left\langle \Ms\right\rangle $, the
deceleration due to dynamical friction, $\dot{v}_{\mathrm{df}}$,
depends only on the total mass density of the background stellar population
and not on the individual masses (Chandrasekhar \citeyear{Cha43};
Binney \& Tremaine \citeyear{Bin87}, Eqs. 7--17, 7--23),

\begin{equation}
\dot{v}_{\mathrm{df}}=-\frac{4\pi\log\Lambda G^{2}M_{\bullet}\rho(r)}{v^{2}}\left[\mathrm{erf}(X)-\frac{2X}{\sqrt{\pi}}\exp(-X^{2})\right]\,,\end{equation}
where $X\!\equiv\! v/\sqrt{2}$$\sigma$ (Note that the dynamical
friction timescale for $v\!\sim\!\sigma$, $t_{\mathrm{df}}\!\equiv\!\left|v/\dot{v}_{\mathrm{df}}\right|\!\sim\! t_{r}\left\langle \Ms\right\rangle /M_{\bullet}$,
is the mass segregation timescale, Eq. \ref{e:tseg}). To estimate
the size of the {}``collection basin'' from which compact objects
can reach the center over the lifetime of the Galaxy $t_{H}\!\sim\!10$
Gyr, consider the orbital evolution of a compact remnant of mass $M_{\bullet}$
that is initially on a circular orbit with velocity $v_{c}$ at some
radius $r$ from the center. The dynamical friction drag exerts a
torque $\dot{L}\!=\!\dot{v}_{\mathrm{df}}r$ that causes the orbit
to decay and the mass to spiral inward. It is convenient to approximate
here the central density distribution as an isothermal cusp, $\rho(r)\!=\! v_{c}^{2}/4\pi Gr^{2}$,
whose circular velocity $v_{c}\!=\!\sqrt{2}\sigma$ is independent
of the radius, so that $\dot{L}=v_{c}\dot{r}=\dot{v}_{\mathrm{df}}r\,.$
This assumption does not introduce a large error since the stellar
density distribution outside the central regions is approximately
isothermal (\S\ref{ss:cusp}), and most of the time spent by the
inspiraling object is at large radii. The solution of the torque equation
relates the time available for reaching the center, $t_{\mathrm{df}}$,
to the maximal initial distance $\max r_{\mathrm{df}}$,\begin{equation}
\max r_{\mathrm{df}}=\left(\frac{\log\Lambda}{1.17}\frac{GM_{\bullet}t_{\mathrm{df}}}{v_{c}}\right)^{1/2}\sim5\,\mathrm{pc}\,,\label{e:rdf}\end{equation}
for $t_{\mathrm{df}}\!=\! t_{H}$, an SBH of mass $M_{\bullet}\!=\!10\,\Mo$,
and for $\log\Lambda\sim\!10$ and $v_{c}\!\sim\!150\,\mathrm{km\, s^{-1}}$
on the $\lesssim10$ pc scale (Kent \citeyear{Ken92}). This is in
very good agreement with estimates based on more realistic density
profiles for the GC (Morris \citeyear{Mor93}; Miralda-Escud\'e \citeyear{Mir00}). 

It is difficult to estimate reliably the number of SBHs, $N_{\bullet}$,
that have sunk to the center in the course of the Galaxy's lifetime,
or the characteristic spatial extent $r_{\bullet}$ of the central
SBH cluster. Such an estimate depends on various uncertain quantities
and properties: the IMF, the star formation history of the GC, the
relation between the initial stellar mass and the final SBH mass (the
SBH mass function) and the dynamical conditions in the GC. In addition,
the magnitude of the SBH natal {}``kick'' velocity could affect
the segregation efficiency, since faster SBHs sink more slowly (Eq.
\ref{e:rdf}). Estimates lie in the range $N_{\bullet}\!=\!2.4\!\times\!10^{4}$
within $r_{\bullet}\!=\!0.7$ pc (for $M_{\bullet}\!=\!7\,\Mo$ SBHs;
Miralda-Escud\'e \& Gould \citeyear{Mir00}) to between $N_{\bullet}\!=\!3.2\!\times\!10^{4}$
and $N_{\bullet}\!=\!7.4\!\times\!10^{5}$ within $r_{\bullet}\!=\!0.8$
pc (for $M_{\bullet}\!=\!10\,\Mo$ SBHs; Morris \citeyear{Mor93}).
However, values of $N_{\bullet}\!\gtrsim\!10^{5}$ are inconsistent
with the current dynamical measurements of the enclosed stellar mass
on these scales (\S\ref{ss:cusp}). Inside $r_{\bullet}$ the SBHs
are expected to be distributed in a $r^{-7/4}$ cusp (\S\ref{sss:relax}).

Irrespective of the uncertainty in $N_{\bullet}$, this is an extreme
degree of concentration. The inner parsec of the Galaxy may contain
up to $\sim\!10^{-2}$ of all Galactic SBHs, in $\sim\!10^{-10}$
of the Galactic volume. Such a degree of mass segregation will significantly
affect the run of the $\Ns/M$ ratio with radius, which may fall by
an order of magnitude from $r\!\sim\!10$ pc to $r\!\sim\!0.01$ pc
(M. Freitag, private. comm.).

It is possible to obtain a rough upper limit on the central concentration
of SBHs from purely dynamical considerations. The more concentrated
the SBH cluster, the higher the rate at which two-body scattering
will deflect SBHs into event horizon-crossing orbits. The {}``drain
limit'' is set by considering how many SBHs can be packed inside
a radius $r_{\bullet}$, in steady state, so that the number scattered
into the MBH over the age of the Galaxy equals the number enclosed
(Alexander \& Livio \citeyear{Ale04}; \S\ref{sss:tdrate})\begin{equation}
\frac{1}{N_{\bullet}}\frac{\mathrm{d}N_{\bullet}}{\mathrm{d}t}\sim\frac{1}{\log(2\sqrt{r_{\bullet}/r_{S}})t_{r}}<\frac{1}{t_{H}}\,.\label{e:qdrain}\end{equation}
 Equation (\ref{e:qdrain}) translates to the upper bound (solved
numerically),\begin{equation}
\max N_{\bullet}(<r_{\bullet})\!\sim\!\frac{2\log(2\sqrt{r_{\bullet}/r_{S}})}{3\log(0.4\max N_{\bullet})}\left(\frac{m}{M_{\bullet}}\right)^{2}\frac{P(r_{\bullet})}{t_{H}}\,.\label{e:drainlim}\end{equation}
 This limit is particularly useful close to the MBH, where the enclosed
stellar mass estimates become unreliable due to the uncertainties
in the determination of the MBH mass and of $\Ns/M$. The drain limit
for $10\,\Mo$ SBHs is $\sim\!10^{4}$ inside $r_{\bullet}\!=\!0.1$
pc.

The central cluster of massive compact remnants is not easy to detect.
Three possible approaches are discussed in the literature: (1) direct
detection of the compact objects; (2) detection of mass segregation
by the central depletion of light {}``test particles''; (3) indirect
detection of the compact objects by their effects on the environment.
The evidence for mass segregation around the Galactic MBH is at present
only circumstantial. 

Direct dynamical detection of the smoothed potential of the extended
mass by deviations from Keplerian motion requires very high astrometric
precision over multiple orbital periods. The current upper limit is
still $\sim\!30$ higher than predicted by the drain limit (Mouawad
et al \citeyear{Mou05}; \S\ref{sss:distrDM}). Orbital perturbations
by occasional close interactions with individual compact remnants
may be detectable with future very large telescopes ($\gg\!10\,\mathrm{m}$)
(Weinberg, Milosavljevi\'c \& Ghez \citeyear{Wei04}). The gravitational
lensing properties of the MBH with respect to background stars will
be modified by the SBHs orbiting it, in particular when the light
rays of a lensed image pass near one of the SBHs (Gould \& Loeb \citeyear{Gou92};
Miralda-Escud\'e \& Gould \citeyear{Mir00}; \S\ref{ss:GL}). However,
the detection of background stars lensed by an MBH--SBH {}``binary''
requires a photometric sensitivity $\sim\!100$ times larger ($K\!\sim\!23$
mag) than is presently available and even then the events are predicted
to be rare ($\sim\!0.06\,\mathrm{yr^{-1}}$) (Chanam\'e, Gould \&
Miralda-Escud\'e \citeyear{Cha01}). Accretion of the dilute interstellar
gas by fast moving compact remnants in the inner $\sim\!0.1$ pc will
probably not lead to significant X-ray emission (Haller et al \citeyear{Hal96};
Pessah \& Melia \citeyear{Pes03}). On the larger, $1$ pc scale there
is an observed over-abundance of transient X-ray sources (4 sources
observed compared to 0.2 expected on average). This is interpreted
as evidence for a high central concentration of NS and SBHs, of which
some are accreting from a binary companion acquired in the course
of a 3-body exchange encounter with a stellar binary (Muno et al \citeyear{Mun04}).

Chanam\'e \& Gould (\citeyear{Cha02}) propose using the angular
distribution of millisecond pulsars around the MBH for detecting mass
segregation. Millisecond pulsars, being NSs, are lighter than the
average stellar mass in the SBH-dominated center. Since they are older
than the relaxation time (their ages are estimated from the spin-down
timescale, $P/2\dot{P}\!\sim\!\mathrm{few\!\times\!10^{9}}$ yr),
they should be depleted in the center due to mass segregation. However,
the radio pulses are hard to detect because they are strongly scattered
by the interstellar plasma in the GC. Lazio et al (\citeyear{Laz03})
report finding only 10 candidate pulsars in the central $\sim200$
pc of the GC. Deeper surveys at higher frequencies will be required
to avoid the scattering and to significantly improve these statistics.
Another class of potential light test particles are the long-lived,
low mass MS progenitors ($\Ms\!\lesssim\!2\,\Mo$) of the He burning
{}``horizontal branch / red clump'' giants (Alexander et al., in
preparation). While the low mass progenitors are themselves too faint
to be observed by current instruments, their giant progeny are luminous
enough ($K\!\sim\!16$) and are readily detected. Observations show
that these giants gradually disappear from the population with decreasing
projected distance from the MBH (Genzel et al \citeyear{Gen03a}).
One possible interpretation is that this is due to mass segregation
operating on the progenitors. However detailed calculations are still
lacking, and other explanations, such as collisional destruction (\S\ref{ss:coll}),
have yet to be ruled out.

Finally, the compact remnants could be revealed by their effect on
the relatively few burning stars in the center. Collisional destruction
by SBHs, perhaps accompanied by the creation of exotic collision products
(e.g. Thorne-\.Zytkow objects) is a possibility that has not yet
been explored in detail. Strong, non-destructive interactions between
SBHs and MS stars of similar mass may be responsible for the capture
of the young B-stars observed there (Alexander \& Livio \citeyear{Ale04};
\S\ref{sss:exchange}).

\subsection{Stellar Collisions}

\label{ss:coll}

With central densities above $10^{8}\,\Mo\,\mathrm{pc^{-3}}$, the
inner cusp is the densest environment in the Galaxy, exceeding even
the core density of massive globular clusters by up to a factor of
$100$. In this extreme density the mean time between physical stellar
collisions, $t_{\mathrm{coll}}$ (Eq. \ref{e:tcoll}) is shorter than
the age of system $\sim\! t_{H}$, which is also roughly the lifespan
of a $\sim\!1\,\Mo$ star. Since the probability to avoid a collision
is $\exp(-t_{H}/t_{\mathrm{coll}})\!\ll\!1$, the dynamics and stellar
population in the high density inner cusp are completely dominated
by collisions. The GC is essentially a naturally occurring {}``stellar
collider''.

\subsubsection{Collisional destruction of giants}

\label{sss:RGcoll}

Images of the inner arcseconds of the GC clearly reveal a central
cavity in the distribution of luminous late-type red giants (Fig.
\hlink{f:VLTfield}). The depletion is gradual (Fig. \hlink{f:SGdepletion}).
Inside $\sim\!5"$ (0.2 pc projected distance) there are no $K\!<\!10$
mag late-type giants,  and inside $\sim\!1.5"$ (0.06 pc projected
distance) there are even no $K\!<\!12$ mag late-type giants. Apart
for the luminous, short-lived He stars (\S\ref{ss:1pc}), the maximal
stellar luminosity gradually decreases in the inner $2"$. 

A possible explanation for this trend, selective obscuration of the
center by a centrally concentrated distribution of dust, is not supported
by an analysis of the measured extinction across the inner parsec
(Alexander \citeyear{Ale99a}). Alternatively, stellar collisions
could be responsible for the depletion of the luminous late-type giants
(Lacy, Townes \& Hollenbach\citeyear{Lac82}; Phinney \citeyear{Phi89};
Genzel et al. \citeyear{Gen96}) or the disappearance of CO absorption---a
distinctive spectral signature of red giants---in the integrated light
(Sellgren \citeyear{Sel90}; Figer et al \citeyear{Fig00}). Luminous
red giants have very large extended envelopes, and therefore a large
cross-section for collisions with other stars. A strong collision
may strip the giant's envelope leaving behind an almost bare burning
core. This will drastically lower its IR luminosity because the IR
spectral range lies in the Raleigh-Jeans part of the stellar blackbody
spectrum (for effective temperature\label{d:Teff} $T_{\star}\!\gtrsim\!4000\,\mathrm{K}$),
and so the IR luminosity scales as $L_{\mathrm{IR}}\propto R_{\star}^{2}T_{\star}$
whereas the total stellar luminosity scales as $L_{\star}\propto R_{\star}^{2}T_{\star}^{4}$\label{d:Ls}.
Consider a collision that disperses the envelope of a $\sim\!100\, R_{\odot}$
red supergiant and leaves a $\sim\!1\, R_{\odot}$ burning core. Studies
of rapid mass transfer in close binaries (e.g. Podsiadlowski, Rappaport
\& Pfahl \citeyear{Pod02}) indicate that a stripped core retains
its nuclear luminosity. The factor $10^{4}$ decrease in surface area
must then be compensated by a factor 10 increase in the effective
temperature, which translate to a factor $10^{3}$ (7.5 mag) decrease
in the IR luminosity.

Despite its low surface gravity, a red giant is quite robust against
collisions. The dynamical timescale in the envelope is of the order
of a year and the thermal timescale is of the order of several decades,
so a collision in which a star {}``punches a hole'' in the giant's
envelope will probably not have a lasting effect. However, hydrodynamical
studies of collisions between dwarf stars and giants indicate that
the envelope can be completely disrupted when the impact parameter
is significantly smaller than the stellar radius. This happens either
by the ejection the giant's core from the extended envelope, which
then rapidly disperses, or by the formation of a common envelope binary,
which stirs the envelope and leads to its evaporation on a time scale
much shorter than the giant phase (Livne \& Tuchman \citeyear{Liv88};
Davies \& Benz \citeyear{Dav91}; Rasio \& Shapiro \citeyear{Ras90},\citeyear{Ras91};
Bailey \& Davies \citeyear{Bai99}). 3-body collisions between a binary
and a giant are even more efficient in disrupting the envelope (Davies
et al. \citeyear{Dav98}). However, it is hard to estimate the overall
contribution of 3-body collisions because the binary fraction near
the MBH is uncertain. Note that collisional destruction is less likely
to occur in early-type blue supergiants, which do not have a well-separated
compact core / extended envelope structure. 

Figure (\hlink{f:SGdepletion}) compares the observed distribution
of stars in the plane of projected distance $p$ and magnitude $K$
with theoretical contours of the expected mean number of surviving
giants in a collisionally dominated cusp. The stellar spectral type
is indicated, where such information is available. The early-type
stars, which are too young to be dynamically relaxed, clearly stand
out in the $p$-$K$ plane. They are too short-lived to be affected
by stellar collisions. In contrast, the old late-type giants are relaxed
and susceptible to collisions. The theoretical calculations estimate
the collisions rate (Eq. \ref{e:tcoll}) and survival probability
by assuming solar mass projectiles, use cross-sections for collisional
destruction calibrated by hydrodynamical simulations, and take into
account the stellar DF (Eqs. \ref{e:cusp}, \ref{e:sigM}) and mass
function (Eq. \ref{e:PMF}) of the GC and the evolution of the stellar
radius and $K$-band luminosity along the stellar tracks (Alexander
\citeyear{Ale99a}). The possible effects of mass segregation in enhancing
the collision rate and efficiency are not taken into account. 

The results show that two distinct effects can cause the observed
depletion. One is purely geometrical. The number of stars enclosed
in an annulus of projected radius $p$ and width $\mathrm{d}p$ scales
as $p^{2-\alpha}\mathrm{d}p$ in a $n_{\star}\!\propto\! r^{-\alpha}$
cusp. For cusps shallower than $r^{-2}$ the number of stars decreases
with $p$, and so does the probability of observing the rarer, luminous
giants. The second effect is dynamical. Once the stellar density rises
above $\sim\!5\!\times\!10^{7}\,\Mo\,\mathrm{pc^{-3}}$, collisions
become effective. The additional collisional depletion, above that
due to projection, improves the fit of the model to the data in the
inner $\sim\!1.5"$. Thus, the observed gradual depletion of bright
late-type giants toward the center cannot, in itself, be taken as
proof of a high density cusp. However, taken together with the observed
rise in the surface density (\S\ref{ss:cusp}), both properties indicate
that a steep, high density cusp provides a self-consistent interpretation
of the data.

\begin{figure}[!t]
\begin{center}\centering{\htarget{f:SGdepletion}\includegraphics[%
  width=0.9\textwidth,
  keepaspectratio]{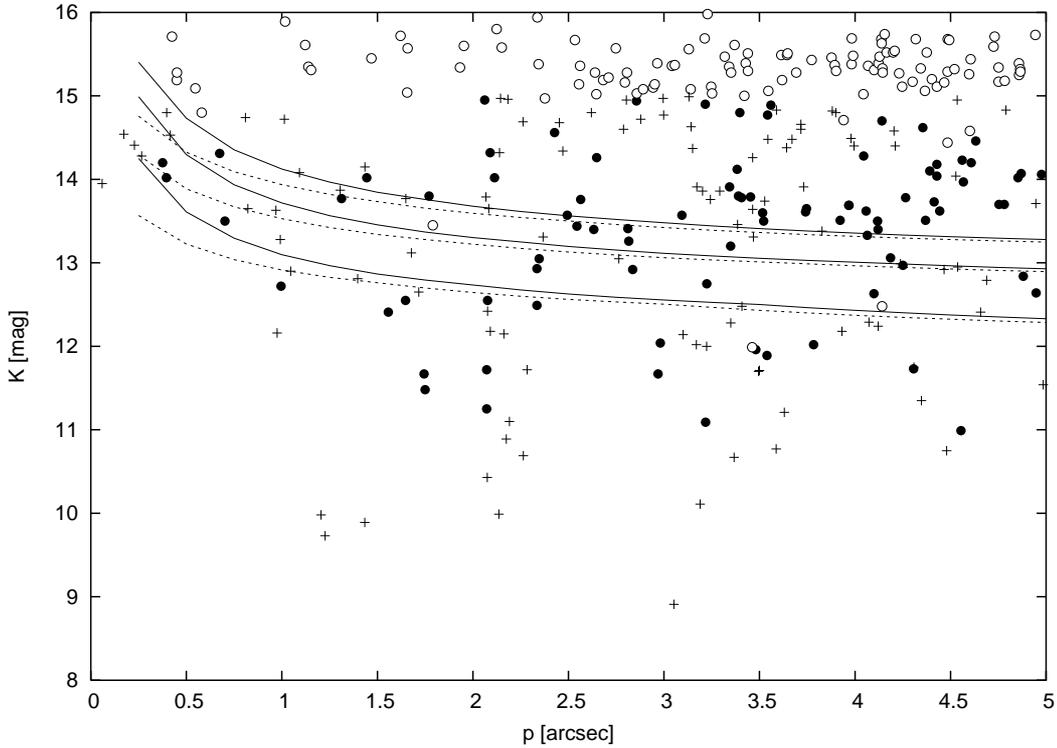}}\end{center}

\caption{{\small \label{f:SGdepletion}Evidence for the collisional destruction
of the envelopes of bright red giants in a high density stellar cusp
around the Galactic MBH (updated version of figure 9 in Alexander
\citeyear{Ale99a}). The apparent stellar $K$-band magnitude of stars
in the inner $5"$ is plotted against the projected angular distance
from the black hole, $p$ (VLT CONICA/NAOS data from Trippe, Ott et
al. 2005, in prep.). The stars are sorted into spectral types (filled
circles: late-type, crosses: early-type, open circles: unknown) according
to their spectral features (when available) or according to their
CO-band color $m(\mathrm{CO})\!\equiv\! m(2.29\mu\mathrm{m})\!-\! m(2.26\mu\mathrm{m})$
(from the Gemini science demonstration dataset), which is sensitive
to the characteristic CO absorption feature in late-type giants (Genzel
et al. \citeyear{Gen03a}). Most of the low luminosity stars without
spectral identifications are probably also old late-type stars. The
three dashed contour lines represent model predictions for the decrease
in surface density of bright stars due to projection. The three solid
lines include also the effects of collisional destruction in a high
density $n_{\star}\!=\!1.2\!\times\!10^{6}(p/10")^{-1.4}$ stellar
cusp (\S\ref{ss:cusp}). The model predicts, on average, 1.5 (top
contour), 1.0 (central contour), and 0.5 (bottom contour) dynamically
relaxed (old) stars per 0.25 arcsecond bin that are brighter than
the contour level. }}
\end{figure}

\subsubsection{Tidal spin-up }

\label{sss:spinup}

The Keplerian velocity in the inner parsec around the MBH is a substantial
fraction of the escape velocity from a typical star, and even exceeds
it at $r\!<\! r_{\mathrm{coll}}$ (\S\ref{sss:scales}). Stars moving
rapidly in the dense stellar cusp near the MBH will suffer over their
lifetimes numerous high-velocity, close tidal encounters with other
stars. Although such encounters transfer some energy and angular momentum
from the hyperbolic orbit ($\varepsilon\!<\!0$ relative to the other
star) to the colliding stars, they rarely remove enough energy for
tidal capture. This is in marked contrast to the situation in the
high density cores of globular clusters, where the colliding stars
are on nearly zero-energy orbits and close collisions can lead to
the formation of tight binaries. The effects of hyperbolic encounters
on the stars are mostly transient. The dynamical and thermal relaxation
timescales are short compared to the stellar lifespan, and thus apart
from some mass-loss in very close collisions, the star is largely
unaffected. It is however more difficult for the star to shed the
excess angular momentum since magnetic breaking (the torque applied
to a star when the stellar wind resists being swept by the rotating
stellar magnetic field) typically operates on timescales of the order
of the stellar lifespan (Gray \citeyear{Gra92}). Fast rotation is
therefore the longest lasting dynamical after-effect of a close encounter.
Over time, the contributions from many, randomly oriented tidal encounters
will lead to a {}``random walk'' buildup of the stellar spin (Alexander
\& Kumar \citeyear{Ale01a}).

The possibility that stars in MBH cusps are rapid rotators may have
interesting implications for their evolution and the interpretation
of their observed properties (see review by Maeder \& Meynet \citeyear{Mae00}).
For example, rotationally induced mixing may reveal itself in the
spectral line ratios, and rotation may be directly observed in the
spectral line profiles (but high rotation will be suppressed by expansion
in the giant phase). Detection of such signatures in the spectra of
the observed giants can provide additional evidence for the existence
of an underlying dense cusp of low-mass MS stars, which at present
cannot be directly observed.

Consider a a star of mass $M_{\star}$ and radius $R_{\star}$ that
undergoes a single tidal encounter with a perturbing mass $m$ (here
a star or a compact object). When the tidal deformations in the star
are small, the energy taken from the orbit and invested in raising
the tides can be described by a linear multipole expansion in the
periapse distance (distance of closest approach) $r_{p}\!=\! a(1-e)$
\label{d:rp} (Press \& Teukolsky \citeyear{Pre77}),

\begin{equation}
\frac{\Delta E_{t}}{\Es}=\left(\frac{m}{\Ms}\right)^{2}\sum_{l=2}^{\infty}\frac{T_{l}\left(\eta,e\right)}{\left(r_{p}/\Rs\right)^{2l+2}}\sim\frac{T_{2}}{\left(r_{p}/r_{t}\right)^{6}}\,,\label{e:dEt}\end{equation}
where $\Es\!\equiv\! G\Ms^{2}/\Rs$\label{d:Eb} is the stellar binding
energy (up to a factor of order unity depending on the stellar structure),
$T_{l}$ are the dimensionless tidal coupling coefficients, which
depend on the eccentricity $e$ ($e\!>\!1$ for an unbound orbit,
Eq. \ref{e:Kepler}) and on the dimensionless periapse crossing timescale
$\eta\!\equiv\!\tau_{p}/\tau_{\star}$ where $\tau_{p}\!=\!\sqrt{r_{p}^{3}/G(\Ms+m)}$
\label{d:tp} is the periapse crossing timescale and $\tau_{\star}\!=\!\sqrt{\Rs^{3}/G\Ms}$\label{d:td}
is the stellar dynamical timescale. The tidal coefficients $T_{l}$
can be calculated numerically for any given stellar model and orbit
(e.g. Alexander \& Kumar \citeyear{Ale01a}). For many applications
it suffices to use the lowest order multipole, $T_{2}$. The angular
velocity at periapse is\begin{equation}
\Omega_{p}^{2}=\frac{v_{\infty}^{2}}{r_{p}^{2}}+\frac{2G(\Ms+m)}{r_{p}^{3}}\,,\end{equation}
 where $v_{\infty}$ is the relative velocity at infinity. Generally,
$T_{2}$ peaks when $\Omega_{p}$ matches the frequency of the fundamental
mode of the star, $\Omega_{p}\!\sim\!{\cal {O}(}\tau_{\star}^{-1})$.
The higher the velocity of the encounter (the larger $e$, Eq. \ref{e:Kepler}),
the larger the value of $r_{p}$ (larger $\eta$) for resonance with
the fundamental mode. The distance dependence of $\Delta E_{t}$ (to
lowest order) can be understood by noting that the energy in the tidal
oscillations is proportional to the square of the amplitude of the
tidal elongation, $\Delta\Rs/\Rs\!\sim\!\left(\Rs/r_{p}\right)^{3}\left(m/\Ms\right)$.

The orbital energy, $\Delta E_{t}$, and angular momentum, $\Delta J_{t}$,
that are transferred to the star in an impulsive tidal encounter (such
as a hyperbolic one) are related by (e.g. Kumar \& Quataert \citeyear{Kum98})

\begin{equation}
\Delta E_{t}=\Delta J_{t}\Omega_{p}\,.\label{e:dEimp}\end{equation}
 For rigid body rotation%
\footnote{The timescale for angular momentum re-distribution due to convective
transport in a red giant is $\sim\!1$ yr (Zahn \citeyear{Zah89})
and so rigid rotation is achieved on a timescale similar to that of
the collision itself. The timescale for angular momentum re-distribution
in radiative MS stars is not well known, although it is likely to
be shorter than the stellar lifespan. %
}, the change in the angular velocity of the star due to the tidal
interaction is $\Delta\Omega_{t}=\Delta J_{t}/I\,,$where $I$ is
the star's moment of inertia (assumed here constant), so that

\begin{equation}
\frac{\Delta\Omega_{t}}{\Omega_{b}}=\frac{\sqrt{G\Ms^{3}\Rs}}{I\Omega_{p}}\left(\frac{m}{\Ms}\right)^{2}\sum_{l=2}^{\infty}\frac{T_{l}\left(\eta,e\right)}{\left(r_{p}/\Rs\right)^{2l+2}}\,,\label{e:dWt}\end{equation}
 where $\Omega_{b}\!\equiv\!\sqrt{G\Ms/\Rs^{3}}$ is the centrifugal
breakup angular velocity, where the star sheds mass from its equator. 

Equations (\ref{e:dEt}) and (\ref{e:dWt}) apply only for soft encounters
($r_{p}\gtrsim2\Rs$), where the linear expansion holds. Hydrodynamic
simulation can then be used to extend Eq. (\ref{e:dWt}) to strong
encounters. Simulations show that as $r_{p}$ decreases, $\Delta\Omega_{t}$
first grows faster then predicted by linear theory, but then as $r_{p}$
approaches $\Rs$, the formal divergence of $\Delta\Omega_{t}$ is
truncated when mass is ejected from the tidally disturbed star and
carries away the excess angular momentum (Alexander \& Kumar \citeyear{Ale01a}).
Figure (\hlink{f:spinup}) shows the predicted mean rotation of solar
mass stars in the GC cusp after 10 Gyr of stochastic tidal spin-up,
assuming inefficient magnetic breaking. The stellar spin reaches an
r.m.s value of $\Omega/\Omega_{b}\!\sim\!0.3$ at the center (rotational
velocity of $0.3V_{\star}\sim\!130\,\mathrm{km\, s^{-1}}$), about
60 times higher than is observed in solar type stars in normal environments.
The spin-up effect falls only slowly with distance from the MBH ($\Omega/\Omega_{b}\!\sim\!0.1$
at 0.3 pc) because the higher tidal coupling in slower collisions
compensates for the lower collision rate there. Thus atypically high
spin could be a characteristic of long-lived, low mass stars over
a substantial fraction of the MBH's sphere of influence. The spin-up,
being a random walk process, grows with time as $t^{1/2}$ and become
significant only for long-lived stars. It is not relevant for short-lived
massive MS stars or for the relatively short post-MS giant phase.
For example, a $5\,\Mo$ star with $V_{\star}\!\sim\!600\,\mathrm{km\, s^{-1}}$
and a MS lifespan of $t_{\star}\!\sim\!10^{8}\,\mathrm{yr}$ will
be spun up close to the MBH to only $0.3V_{\star}\sqrt{t_{\star}/t_{H}}\!\sim\!20\,\mathrm{km\, s^{-1}}$
(scaling from the calculated spin-up for a $1\,\Mo$ star and assuming
a similar spin-up efficiency). This is much less than the typical
rotational velocity observed in such stars in normal environments,
whose origin is related to their formation and not to tidal spin-up.

\begin{figure}[!t]
\centering{\htarget{f:spinup}\includegraphics[%
  scale=0.5,
  angle=270]{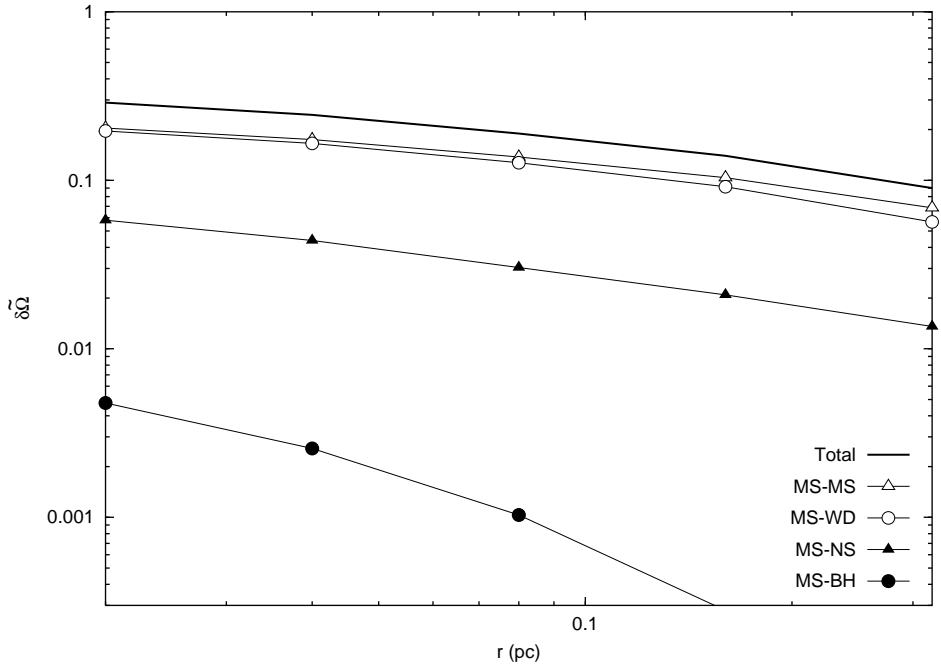}}

\caption{{\small \label{f:spinup}The average spin-up of a solar type star
by star-star tidal interactions over 10 Gyr as function of distance
from the Galactic MBH (Alexander \& Kumar \citeyear{Ale01a}). An
$\alpha\!=\!1.5$ density cusp is assumed. The rotation grows over
time in a random walk fashion by repeated close passages. The angular
rotation frequency $\delta\widetilde{\Omega}\!\equiv\!\delta\Omega/\Omega_{b}\!=\!1$
corresponds to rotation at the centrifugal break-up velocity. In addition
to the total spin-up, the separate contributions from collisions with
main sequence stars (MS), white dwarfs (WD), neutron stars (NS) and
stellar black holes (BH) are shown. (Reprinted with permission from
the} \emph{\small Astrophysical Journal}{\small ).}}
\end{figure}

\subsubsection{Stellar mergers and exotic objects}

\label{sss:mergers}

A possible outcome of a collision between two stars is a merger, where
a new object, possibly more massive than either of its building blocks,
is created. The high rate of stellar collisions near the MBH raises
the question whether it is possible to create massive stars from lower
mass stars by successive mergers, or exotic stars by the capture of
a compact object inside a normal star (Morris \citeyear{Mor93}).
This is of interest in view of the puzzling presence of young massive
stars near the MBH, where normal modes of star formation are thought
to be impossible (\S\ref{s:OBriddle}). The hypothesis that the very
massive emission line stars in the central $0.4$ pc are merger products
is made less compelling by the fact that such stars are also found
in the young massive stellar clusters in the central $\sim\!50\,\mathrm{pc}$,
where the central densities are much smaller (\S\ref{ss:100pc}).
In addition, the collision rate is much too low at the relatively
large distance of the emission line stars (Eq. \ref{e:tcoll}). However,
it is still relevant to consider the merger scenario for the S-stars,
which are both much lighter and closer to the MBH.

A successful chain of mergers requires that the mean time between
mergers be shorter than the lifespan of the merger product (usually
assumed to be similar to the MS lifespan of a normal star of the same
mass, and therefore becoming shorter as the merger chain progresses).
The time constraint does not appear to be an insurmountable obstacle,
if glancing collisions can lead to a merger (Genzel et al. \citeyear{Gen03a}).
However, it is very unlikely that the merging efficiency can be so
high for such high velocity collisions. 

The first requirement for an efficient merger is that the impactor
star (the smaller of the two with mass $M_{i}$) can be stopped in
the target star (the larger of the two with mass $M_{t}$). This translates
to the requirement that the target mass contained in the cylinder
{}``punched out'' by the impactor equals or exceeds the impactor
mass, that is approximately $M_{i}\!\lesssim\! R_{i}^{2}R_{t}(M_{t}/R_{t}^{3})$,
where $R_{i}$ and $R_{t}$ are the radii of the impactor and target.
Thus, the stopping condition can be formulated as the requirement
that the column density of the impactor be less or equal to that of
the target,\begin{equation}
\frac{M_{i}}{R_{i}^{2}}\lesssim\frac{M_{t}}{R_{t}^{2}}\,.\label{e:Ncol}\end{equation}

Effective stopping does not in itself imply an efficient merger, since
when $v_{\infty}>V_{e}$, the specific kinetic energy of the impactor
is enough to unbind a unit mass of the target star. The total amount
of mass that can be ejected from the target depends then also on the
mass ratio between the impactor and the target. The relative velocity
at impact is $\ge\! V_{e}$ even when $v_{\infty}\!=\!0$, which is
of the order of the sound speed in the star. Therefore such collisions
will be supersonic, and the kinetic energy can be efficiently converted
to heat by the shocks and drive the mass loss. The outcome of a penetrating
encounter depends therefore not only on the mass ratio and the relative
velocity, but also on the assumed stellar $\Ms$-to-$\Rs$ relation,
and on the impact parameter (since the column density is a strongly
varying function of position in the star). 

The actual situation revealed by numerical simulations is more complex
than implied by the simplified discussion above. For example, an impactor
can be destroyed by heating even when it is not stopped by the target,
and mergers can occur in slow enough collisions even without initial
physical contact between the two stars if they are tidally captured
and form a decaying binary. The simulations generally show that the
merger efficiency decreases rapidly when $v_{\infty}\!\gtrsim\! V_{e}$,
but the exact details depend strongly on the assumed stellar structure
model. Lai et al. (\citeyear{Lai93}) find that mergers can occur
even when the relative velocity is as high as $v_{\infty}\!\sim\!1.5V_{e}$,
provided that the collision is very close to head-on. However, Freitag
\& Benz (\citeyear{Fre05}), using more realistic stellar models,
find that mergers are much less efficient. Freitag, G\"urkan \& Rasio
(\citeyear{Fre04b}) find no mergers in Monte-Carlo simulations of
stellar collisions in a galactic nucleus with a MBH. Thus, massive
star buildup by mergers very near the MBH does not appear likely.

A related process is the capture of a compact object (NS of SBH) inside
a star, and the possible formation of an exotic, accretion powered
''star'' (Thorne-\.Zytkow object). The typical relative velocity
at impact between the compact object and the star is $v\!\sim\!{\cal {O}}(Gm/r)$,
which is much smaller than the escape velocity from the compact object,
$V_{e}\!\sim\!{\cal {O}}(c)$. The collisional cross-section of the
compact object is then dominated by the gravitational focusing term
(see Eq. \ref{d:tc}), which is much larger than its physical size.
Therefore, the effective size of the impactor is the Bondi accretion
radius, $R_{i}\sim{\cal O}(GM_{i}/v^{2})$ (Bondi \citeyear{Bon52}),
and the column density ratio between the compact impactor and the
target star is $\sim\!(R_{t}/r)^{2}\left.m^{2}\right/(M_{i}M_{t})$.
This ratio can be of order unity if the collision is atypically slow
or occurs at a large enough distance (e.g. $r\!\sim\!0.1\,\mathrm{pc}$
for $m\!=\!3.5\!\times\!10^{6}\,\Mo$ and a collision between a $M_{i}\!=\!1.4\,\Mo$
NS and a $M_{t}\!=\!10\,\Mo$, $R_{t}\!=\!4.5\,\Rs$ star). A star
powered by accretion on a NS may appear as a red giant or supergiant
with peculiar photospheric abundances (Eich et al. \citeyear{Eic89};
Biehle \citeyear{Bie91},\citeyear{Bie94}), whereas one accreting
on a SBH may look like a WR blue giant (Morris \citeyear{Mor93}).
The evolution of such configurations in uncertain.

\section{Probing the dark mass with stellar dynamics}

\label{s:orbit}

Stars near the MBH are effectively test particles since their mass
is negligible compared to that of the MBH, $\Ms/m\!\sim\!10^{-7}$--$10^{-5}$.
As long as their trajectories do not take them too close to the MBH,
where dissipative processes can affect their motion (\S\ref{s:inter}),
stars directly probe the gravitational potential near the MBH. This
is in contrast to gas dynamics, where the possible influence of non-gravitational
forces due to thermal pressure, radiation pressure or magnetic fields
can substantially complicate the interpretation of dynamical data
and limit its usefulness. The analysis of stellar orbits near the
MBH, well within the radius of influence, is further simplified by
the fact that the orbits are Keplerian to a good approximation (Eq.
\ref{e:Kepler}). 

The type of information about the dark mass that can be extracted
from the data, the level of analysis possible, and the methods used,
all depend on the nature of the observations (stellar position, proper
motion, radial velocity, magnitude, colors, spectral classification),
on their quality, on the properties of stars observed (proximity to
the MBH, velocity, luminosity, age), and on the total duration of
the monitoring. Two approaches can be used to weigh and locate the
dark mass (\S\ref{ss:mass}). One is statistical, which typically
requires less data for each star (and so was the first to be applied),
but needs a meaningfully large sample and is limited by various necessary
assumptions about the parent population. The second involves the derivation
of partial or full orbital solutions for individual stars. In principle,
a single well sampled stellar orbit is all that is needed to obtain
the mass and location of the dark mass (assuming the dark mass is
fully enclosed in the orbit; \S\ref{ss:DM}). However, this is only
feasible near the MBH, where the periods are short enough for accelerations
to be measured. Farther away from the MBH ($r\!\gtrsim\!0.1\,\mathrm{pc}$,
$P\!\gtrsim\!1000\,\mathrm{yr}$) there is no substitute for statistical
mass estimators to measure the total enclosed mass (stars and MBH).

Following initial attempts to probe the potential of the central dark
mass by gas kinematics (Wollman et al. \citeyear{Wol76}; Lacy et
al. \citeyear{Lac80}), the use of stellar kinematics progressed to
include radial velocities (McGinn et al. \citeyear{McG89}; Sellgren
et al. \citeyear{Sel90}; Krabbe et al. \citeyear{Kra95}; Haller
et al. \citeyear{Hal96}; Genzel et al. \citeyear{Gen97}), stellar
proper motions (Eckart \& Genzel \citeyear{Eck96}; Ghez et al. \citeyear{Ghe98}),
and stellar accelerations (Ghez et al. \citeyear{Ghe00}; Eckart et
al. \citeyear{Eck02}). Each of these subsets of the full orbital
data were used to place some constraints on the dark mass (\S\ref{sss:orbsol}).
Finally, it is now possible, using high-precision infrared astrometry
and spectroscopy, to follow the trajectories and radial velocity curves
of individual stars as they orbit the MBH (e.g. Sch\"odel et al.
\citeyear{Sch02}; Ghez et al. \citeyear{Ghe03a}; Eisenhauer et al.
\citeyear{Eis03},\citeyear{Eis05}). The combined input from stellar
orbits and from high precision radio measurements of the position
and proper motion of $\SgrA$ have all but ruled out non-BH alternatives
for the dark matter in the center of the GC (\ref{ss:DM}). This has
also made it possible to directly measure the distance to the GC (\S\ref{sss:R0})
and strongly constrain the presence of a secondary BH in the GC (\S\ref{ss:2MBH}).
A very different class of stars that may provide information on the
central dark mass are high-velocity stars, well on their way out of
the Galaxy (\S\ref{ss:hivstar}). The discovery of such stars can
prove that the dark mass must be extremely compact to eject them so
energetically, and is therefore likely a MBH.

\begin{figure}[!t]
\centering{\htarget{f:S2spectrum}\begin{tabular}{c}
\includegraphics[%
  scale=0.52]{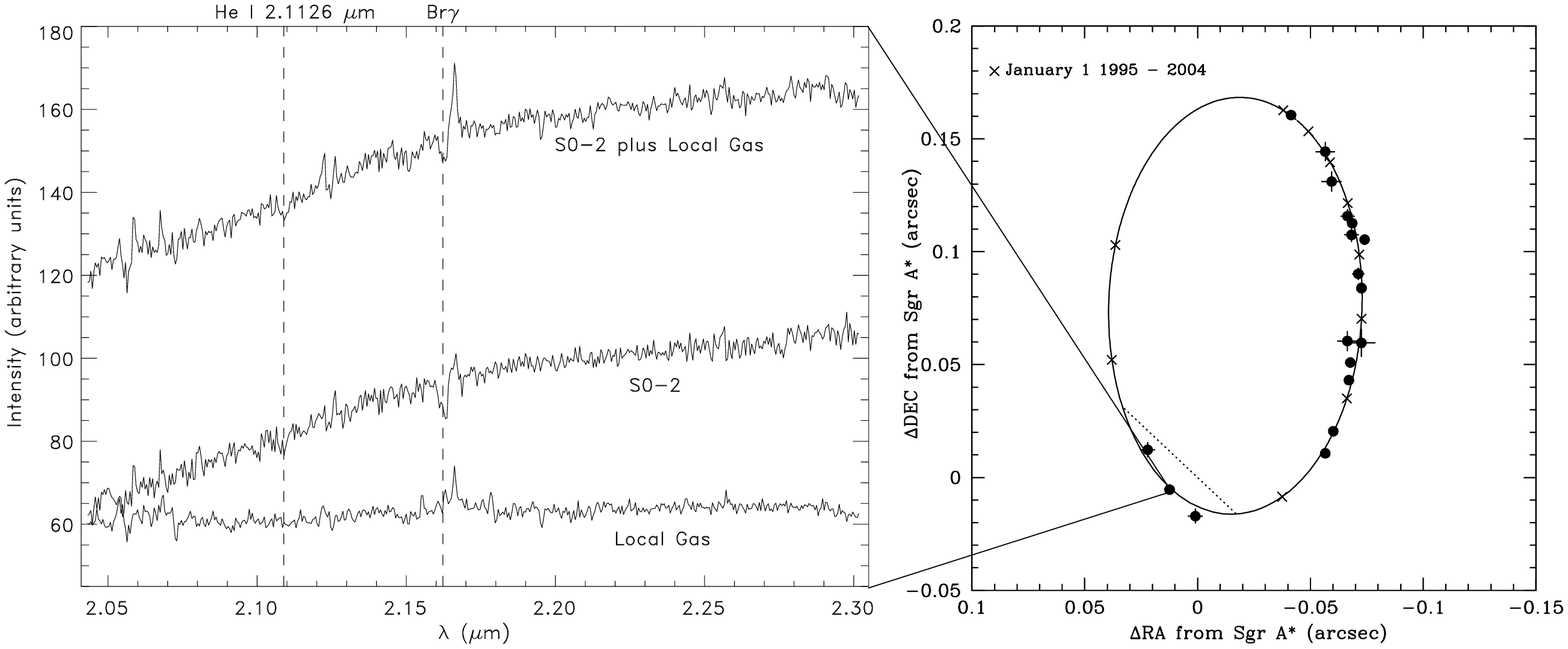}\tabularnewline
\includegraphics[%
  scale=0.6]{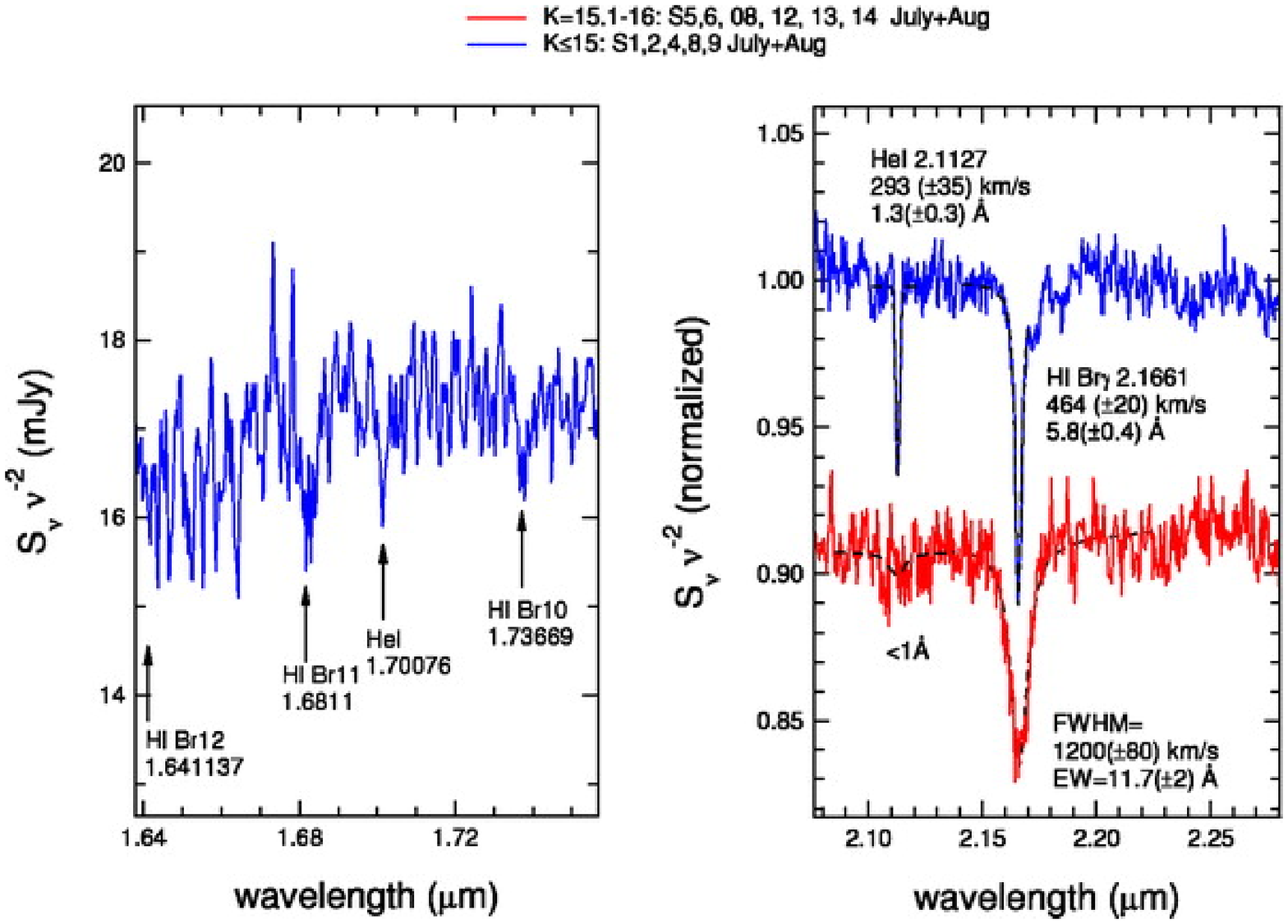}\tabularnewline
\end{tabular}}

\caption{{\small \label{f:S2spectrum}Infrared spectra of the early-type (hot)
S-cluster stars. Top: Infrared spectrum and orbit reconstruction of
S2/S0-2 (Ghez et al. \citeyear{Ghe03a}. Reprinted with permission
from the} \emph{\small Astrophysical Journal}{\small ). Bottom: Composite
spectra of 5 bright stars ($K\!<\!15$ mag, top spectra left and right)
in the $H$ and $K$-bands and 6 faint stars ($15<\! K\!<\!16$ mag,
bottom spectrum left) in the $K$-band (Eisenhauer et al. \citeyear{Eis05}.
Reprinted with permission from the} \emph{\small Astrophysical Journal}{\small ).}}
\end{figure}

\subsection{Weighing and pinpointing the dark mass}

\label{ss:mass}

The primary parameters of the dark mass are its mass, size, location
and velocity. An empirical determination of their values has many
implications. The mass and size fix the mean density, which determines
whether the dark mass is indeed a MBH (\S\ref{ss:DM}). The formal
density associated with a BH (mass over volume of a Euclidean sphere
whose radius is the event horizon) is \begin{equation}
\rho_{\mathrm{BH}}\sim\frac{3}{4\pi}\frac{m}{r_{S}^{3}}=\frac{3(c^{2}/2G)^{3}}{4\pi m^{2}}=\left(\frac{m}{4\!\times\!10^{6}\,\Mo}\right)^{-2}\times\left\{ \begin{array}{l}
1.1\!\times\!10^{3}\,\mathrm{g\, cm^{-3}}\\
1.7\!\times\!10^{25}\,\Mo\,\mathrm{pc^{-3}}\end{array}\right.\,.\label{e:rhoBH}\end{equation}
The higher the lower bound on $\rho/\rho_{\mathrm{BH}}$, the more
plausible is the assumption that it is a MBH, and the less likely
are alternative non-singular configurations. The question of the position
the dark mass can be decomposed into three separate issues. (1) Where
is the dynamical center (the center of attraction) in the GC? (2)
Where is the dynamical center relative to the center of mass of the
GC? (3) Where is the dynamical center relative to the radio source
$\SgrA$? The three positions need not be the same. 

The dark mass will execute Brownian-like motion relative to the stellar
cluster around it due to small gravitational {}``kicks'' arising
from Poisson fluctuations in the stellar number density (e.g. Chandrasekhar
\citeyear{Cha44}). The amplitude of the Brownian excursions away
from the cluster's center of mass will depend on how massive the dark
mass is (the more massive it is, the less it will move) and on the
properties of the stellar system around it (\S\ref{sss:distrDM}).
It should be emphasized that due to the long range nature of gravitational
interactions, Brownian motion in a self-gravitating system is fundamentally
different from that in a gas. The statistical mechanics of the gravitational
Brownian motion of a massive object, which itself dominates the potential,
are still a subject of active investigations. Neither is it obvious
that $\SgrA$ must lie at the dynamical center. This is of course
expected if the source is an accreting MBH, since the emission should
arise inside $\sim\!10r_{S}\!\sim\!10^{13}\,\mathrm{cm}$. However,
the extremely low luminosity of $\SgrA$, now also detected in the
X-ray (Baganoff et al. \citeyear{Bag01}) and IR (Genzel et al. \citeyear{Gen03b}),
does not require a MBH. It could easily be powered by accretion on
a stellar mass compact object. In that extreme case, $\SgrA$ would
merely be a test particle in orbit around the dark mass. To explore
this possibility, it is necessary first to pinpoint the location of
the radio source $\SgrA$ in the IR frame, where the stars are observed
and where the dynamical center is identified%
\footnote{The IR reference frame is defined by the average angular position
$\mathbf{\left\langle p\right\rangle \!=\!\Sigma_{\mathnormal{i\!=\!1}}^{\mathnormal{\Ns}}p_{\mathnormal{i}}}/\Ns$
of a statistically large sample of stars around $\SgrA$. While each
star is moving, $\mathbf{\left\langle p\right\rangle }$ is assumed
to be nearly stationary and the relative motion between $\left\langle \mathbf{p}\right\rangle $
and the central dark mass is assumed to be negligibly small (Eisenhauer
et al. \citeyear{Eis03}).%
}. This was achieved by the discovery of 7 red giants and supergiants
in the inner $15"$ whose extended mass-loss envelopes ($R\!\sim\!5\Rs$)
emit radio maser radiation (coherent light emission analogous to laser
light) from a molecular transition of SiO (Menten et al. \citeyear{Men97};
Reid et al. \citeyear{Rei03}). Since these giants are also bright
IR sources, they can be observed in both radio and IR and used to
align the two reference frames and correct for various distortions.
This enabled the determination of the position of $\SgrA$ relative
to the IR stellar observations to within 10 mas ($1\sigma$) and established
that $\SgrA$ indeed coincides with the dynamical center as determined
by the orbits (\S\ref{sss:orbsol}). The radio/IR alignment also
played a crucial role in the detection of the low level IR accretion
emission from $\SgrA$ (Genzel et al. \citeyear{Gen03b}; Ghez et
al. \citeyear{Ghe04}). 

Finally, a precise measurement of the position of the MBH along the
line of sight, the distance to the GC, is important for reconstructing
the structure and dynamics of the Galaxy (\S\ref{sss:R0}).

\subsubsection{Statistical estimators}

\label{sss:stat}

Astrometric observations of stars can provide the 2D projected angular
positions $\mathbf{p}(t_{i})$ \label{d:rproj} at discrete times
$\{ t_{i}\}$, which by differencing yield the angular velocity (proper
motion) $\dot{\mathnormal{\mathbf{p}}}(t)$ and angular acceleration
(proper acceleration) $\ddot{\mathbf{p}}(t)$. Spectroscopic observations
can measure the Doppler shift of spectral features and provide the
line of sight velocity $\dot{z}(t_{i})$, \label{d:rlos} and the
line of sight acceleration $\ddot{z}(t)$. The position $z$ along
the line of sight can not be measured directly. In practice, not all
this information is available for all the stars. Spectroscopy requires
more photons than photometry, and so faint stars may have proper motion
information, but not radial velocities. In addition, a reliable measurement
of the Doppler shift requires distinct spectral features, which are
not always available in the IR range of the spectrum. Proper motion
can be measured only when the astrometric errors are smaller than
the proper displacement, and so slow stars far from the center may
have radial velocity information, but not proper motion. The astrometry
of stars near the dense center can be unreliable because of source
confusion and because they can be {}``lost'' among the other stars
due to their high velocity and large displacement between consecutive
observation times.

The available stellar data can be described statistically in terms
of the mean surface number density $\Sigma(\mathbf{p})$ and the projected
velocity dispersion of the $i$'th velocity component $\sigma_{p}^{2}(\mathbf{p})\!=\!\left\langle [v_{i}(\mathbf{p})-\bar{v}_{i}(\mathbf{p})]^{2}\right\rangle $
(assuming a value of $R_{0}$ for converting proper motion to physical
velocity). The means are taken over binned projected areas. It is
usually assumed, as is indicated by the observed $\Sigma$, that the
density distribution is spherically symmetric. In that case the bins
are annuli of projected radius $p$ and width $\Delta p$.

The Jeans equation is the most direct way of using the mean quantities
to estimate the enclosed mass (e.g. Binney \& Tremaine \citeyear{Bin87}
\S4.2.1d). It is assumed here for simplicity that the velocity dispersion
is isotropic (see analysis of the non-isotropic case by Genzel et
al. \citeyear{Gen00}). The projected quantities can be inverted by
the Abel integrals,\begin{equation}
n(r)=-\frac{1}{\pi R_{0}^{2}}\int_{r}^{\infty}\frac{\mathrm{d}\Sigma}{\mathrm{d}p}\frac{\mathrm{d}p}{\sqrt{(R_{0}p)^{2}-r^{2}}}\,;\qquad n(r)\sigma^{2}(r)=-\frac{1}{\pi R_{0}^{2}}\int_{r}^{\infty}\frac{\mathrm{d}(\Sigma\sigma_{p}^{2})}{\mathrm{d}p}\frac{\mathrm{d}p}{\sqrt{(R_{0}p)^{2}-r^{2}}}\,.\label{e:Abel}\end{equation}
The derived functions%
\footnote{In practice, the discrete $p$-bins that are used for estimating $\sigma_{p}^{2}$
and $\Sigma$ are often too coarse for a stable solution of Eq. (\ref{e:Abel}).
Instead, some simple parametrized models are assumed for $n(r)$ and
$\sigma^{2}(r)$, and the best fit parameters are obtained by the
inverse equations $\Sigma(p)\!=\!2R_{0}^{2}\int_{p}^{\infty}nr(r^{2}-R_{0}^{2}p^{2})^{-1/2}\mathrm{d}r$
and $\Sigma(p)\sigma_{p}^{2}(p)\!=\!2R_{0}^{2}\int_{p}^{\infty}n\sigma^{2}r(r^{2}-R_{0}^{2}p^{2})^{-1/2}\mathrm{d}r$.%
} $n(r)$ and $\sigma^{2}(r)$ are then used to estimate the total
enclosed mass $M(r)$ from the Jeans equation (Eq. \ref{e:Jeans}).
If there is a compact mass in the center, the enclosed mass curve
will converge to $m$ with decreasing radius, otherwise it should
decrease to zero (Fig. \hlink{f:Mencl}). 

\begin{figure}[H]
\centerline{\htarget{f:Mencl}\includegraphics[%
  scale=1.2]{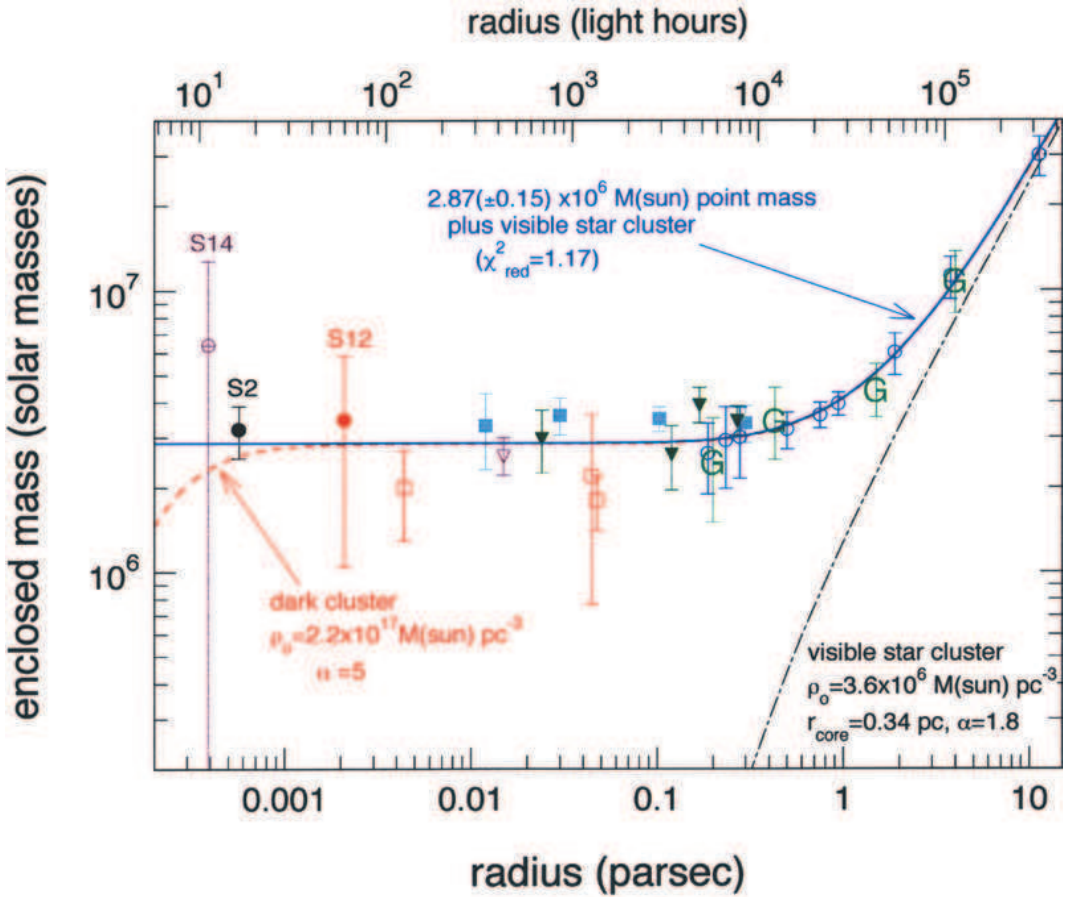}}

\caption{{\small \label{f:Mencl}Enclosed mass distribution in the GC (Sch\"odel
et al. \citeyear{Sch03}) for $R_{0}\!=\!8$ kpc (Reid \citeyear{Rei93}).
Filled circles: Orbital solutions. Filled triangles: Leonard-Merritt
projected mass estimator (data from Sch\"odel et al. \citeyear{Sch03}
and Ott et al. \citeyear{Ott03}). Open triangle: Bahcall-Tremaine
mass estimator (Ghez et al. \citeyear{Ghe98}). Filled rectangles:
Jeans eq. solution for a parameterized anisotropic model (Genzel et
al. \citeyear{Gen00}). Open circles: Jeans eq. solution for a parameterized
isotropic model using radial velocities of late-type stars (Genzel
et al. \citeyear{Gen96}). Open rectangles: Non-parametric mass estimator
for an isotropic model (Chakrabarty \& Saha \citeyear{Cha01b}). {}``G'':
Mass estimates from gas Doppler shifts (Genzel \& Townes \citeyear{Gen87}).
Solid line: Best fit to all data for a model of a central point mass
with a stellar cluster. Dash-dot line: Mass in the stellar cluster
alone---this model clearly does not fit the data. Dashed line: Visible
cluster with a Plummer model of a hypothetical very compact dark cluster---this
model marginally fits the data. (Reprinted with permission from the}
\emph{\small Astrophysical Journal}{\small ) }}
\end{figure}

When the data is too sparse to allow meaningful estimates of $\Sigma$
and $\sigma^{2}$, more elementary statistical estimators can be used
that do not require binning or inversion from projected to 3D quantities.
These can be obtained by taking moments of the Jeans equation, $\left\langle r^{k}M(<\! r)\right\rangle $
for integer $k$, and averaging over the stellar density (the virial
theorem is also derived this way, see Leonard \& Merritt \citeyear{Leo89}).
Bahcall \& Tremaine (\citeyear{Bah81}) show that the virial theorem
mass estimator \begin{equation}
M_{\mathrm{VT}}\!=\!\frac{3\pi}{2G}\left\langle \dot{z}^{2}\right\rangle \left/\left\langle 1/R_{0}p\right\rangle \right.\,,\end{equation}
where $\left\langle \ldots\right\rangle $ denotes the sample average,
is very biased and has other undesirable statistical properties. They
propose instead a better behaved {}``projected mass estimator''
for spherically symmetric and isotropic systems of test particles
orbiting a dominant point mass, which uses line of sight motions,\begin{equation}
M_{\mathrm{BT}}=\frac{16}{\pi G}R_{0}\left\langle \dot{z}^{2}p\right\rangle \,.\end{equation}

Leonard \& Merritt (\citeyear{Leo89}) propose a related mass estimator
for a spherically symmetric non-rotating system, which uses proper
motions and is anisotropy-independent, \begin{equation}
M_{\mathrm{LM}}=\frac{16}{\pi G}R_{0}^{3}\left\langle (\frac{2}{3}\dot{p}_{\parallel}^{2}+\frac{1}{3}\dot{p}_{\perp}^{2})p\right\rangle \,,\end{equation}
where $\dot{p}_{\parallel}$ and $\dot{p}_{\perp}$ are the proper
motions parallel and transverse to $\mathbf{p}$. This estimator is
unbiased only when the sample covers the entire radial extent of the
cluster and for cluster density profiles that have a finite mass.
However, the stellar data in the GC do not satisfy either of these
requirements (the observed power-law density, Eq. \ref{e:cusp}, can
not be extrapolated to infinity as the total mass does not converge).
The bias in $M_{\mathrm{LM}}$ then depends both on the anisotropy
and the steepness of the cusp. This can be partially compensated for
by correction factors (Genzel et al. \citeyear{Gen00}). 

Chakrabarty \& Saha (\citeyear{Cha01b}) propose a non-parametric
mass estimation algorithm for a spherical and isotropic system that
can use both line of sight velocities and proper motions. The method
uses maximum likelihood to find the energy-binned DF and radius-binned
potential that best reproduce the observed data.

A different approach for estimating the mass, the {}``Orbital Roulette''
method, was proposed by Beloborodov \& Levin (\citeyear{Bel04}).
The method assumes (as does the virial theorem) that at any given
time the stars are at randomly distributed phases in their orbits.
For any trial mass distribution, the orbits can be calculated and
the phases tested for randomness. If the trial mass is too low, the
stars will tend to have high velocities relative to the local escape
velocity of the trial potential, and therefore will appear to be atypically
close to periapse. Conversely, if the trial mass is too high, the
stars will appear to be atypically close to apoapse ($r_{a}\!=\! a(1+e)$,
\label{d:ra}the maximal distance along the orbit). The best-fit mass
distribution and its confidence interval can be identified by the
location and shape of the phase-randomness maximum. This method cannot
be applied directly to the stellar data in the GC, since it requires
all 6 phase space coordinates of the stars, whereas only 5 are available
at best. However, in the special case of stars in the inner thin disk,
$z$ can be derived from the constraint on the angular momentum $\mathbf{J}$.
Initial results indicate that the Roulette method mass estimate is
consistent with the higher values obtained by orbital mass estimates
(\S\ref{sss:orbsol}) (A. Beloborodov and Y. Levin, private comm.).

\begin{table}

\caption{\label{t:Mstat}Statistical mass estimates$^{a}$ of the central
dark mass (in $10^{6}\,\Mo$)}

\begin{center}\centerline{{\small \htarget{t:Mstat}}\begin{tabular}{lllllll}
&
&
&
&
&
&
\tabularnewline
\hline 
{\small JE $^{b}$}&
{\small VT $^{b}$}&
{\small BT $^{b}$}&
{\small LM $^{b}$ }&
{\small NP $^{b}$}&
{\small Distance}&
{\small Refs $^{c}$}\tabularnewline
\hline
{\small $2.5$--$3.2$ z}&
{\small $1.7\!\pm\!0.6$ z}&
{\small $1.9\!\pm\!0.6$ z}&
---&
---&
{\small $<\!0.2$ pc}&
{\small {[}1{]}}\tabularnewline
{\small $2.5\!\pm\!0.4$ b}&
{\small $2.6\!\pm\!0.8$ p}&
{\small $2.6\!\pm\!0.8$ p}&
---&
---&
{\small $<\!0.08$ pc}&
{\small {[}2{]}}\tabularnewline
---&
{\small $2.5\!\pm\!0.2$ p}&
{\small $2.6\!\pm\!0.2$ p}&
---&
---&
{\small $<\!0.2$ pc}&
{\small {[}3{]}}\tabularnewline
{\small $3.3$ b}&
{\small $2.5$--$2.6$ p}&
{\small $3.1\!\pm\!0.3$ p}&
{\small $2.9\!\pm\!0.4$ p}&
---&
{\small $<\!0.2$pc }&
{\small {[}4{]}}\tabularnewline
---&
---&
---&
---&
{\small $2.0\!\pm\!0.7$ p}&
{\small $<\!0.004$ pc}&
{\small {[}5{]}}\tabularnewline
---&
---&
---&
---&
{\small $2.2_{-1.0}^{+1.6}$ z}&
{\small $<\!0.005$ pc}&
{\small {[}5{]}}\tabularnewline
---&
---&
---&
---&
{\small $1.8_{-0.3}^{+0.4}$ b}&
{\small $<\!0.05$ pc}&
{\small {[}5{]}}\tabularnewline
\hline
\multicolumn{7}{l}{{\small $^{a}$ Proper motion data is designated by (p), radial velocities
by (z), both by (b).}}\tabularnewline
\multicolumn{7}{l}{{\small $^{b}$ JE: Jeans Equation, VT: Virial theorem, BT: Bahcall-Tremaine}}\tabularnewline
\multicolumn{7}{l}{{\small $^{b}$ LM: Leonard-Merritt, NP: Non-parametric}}\tabularnewline
\multicolumn{7}{l}{{\small $^{c}$ {[}1{]} Genzel et al. (\citeyear{Gen96}), {[}2{]}
Eckart \& Genzel (\citeyear{Eck97}), {[}3{]} Ghez et al. (\citeyear{Ghe98})}}\tabularnewline
\multicolumn{7}{l}{{\small $^{c}$ {[}4{]} Genzel et al. (\citeyear{Gen00}), {[}5{]}
Chakrabarty \& Saha (\citeyear{Cha01b})}}\tabularnewline
\hline
\end{tabular}}\end{center}
\end{table}

Table ({\small \hlink{t:Mstat}}) compares the statistical mass estimates
for the central dark mass that were obtained by different authors
using various methods and different data sets. The results all lie
in the range $m\!\sim\!(2$--$3)\!\times\!10^{6}\,\Mo$.

\subsubsection{Orbital estimators}

\label{sss:orbsol}

The first step toward orbital reconstruction is to detect accelerations
(curvature) in the stellar motion. Measurements of acceleration can
provide quite strong constraints on the properties of the dark mass
even without full orbital information. The mean enclosed density of
the dark mass can be directly obtained from projected quantities through
Newton's law of gravity (for an assumed angular position of the MBH),\begin{equation}
\bar{\rho}_{2\mathrm{D}}=\frac{3}{4\pi}\frac{m}{r^{3}}=\frac{3}{4\pi G}\frac{\ddot{r}}{r}=\frac{3}{4\pi G}\frac{\ddot{p}}{p}\,,\label{e:rhoBH2D}\end{equation}
where $\bmath{r}$ and $\bmath{\ddot{{r}}}$ are the true 3D position
and acceleration. The last equality holds because $\bmath{r}$ and
$\bmath{\ddot{{r}}}$ are parallel, and so are decreased in projection
by the same factor. For example, the position and acceleration of
the star S2 in 1997/1998 ($p\!=\!150\,\mathrm{mas}$, $\ddot{p}\!=\!5.4\,\mathrm{mas\, yr}^{-2}$,
Ghez et al. \citeyear{Ghe00}; Eckart et al. \citeyear{Eck02}) translate
to a lower bound on the density of $\bar{\rho}_{2\mathrm{D}}\!=\!2\!\times\!10^{12}\,\Mo\,\mathrm{pc^{-3}}$
or $\rho/\rho_{\mathrm{BH}}\!=\!3\!\times10^{-14}$ for $m\!=\!4\!\times\!10^{6}\,\Mo$.
Furthermore, since the projected acceleration vectors point to the
projected position of the center of acceleration, they can locate
the MBH in the IR frame (Ghez et al. \citeyear{Ghe00}; Eckart et
al. \citeyear{Eck02}) (Fig. {\small \hlink{f:acc}}).

\begin{figure}[!t]
\centerline{{\small \htarget{f:acc}}\includegraphics[%
  scale=0.5]{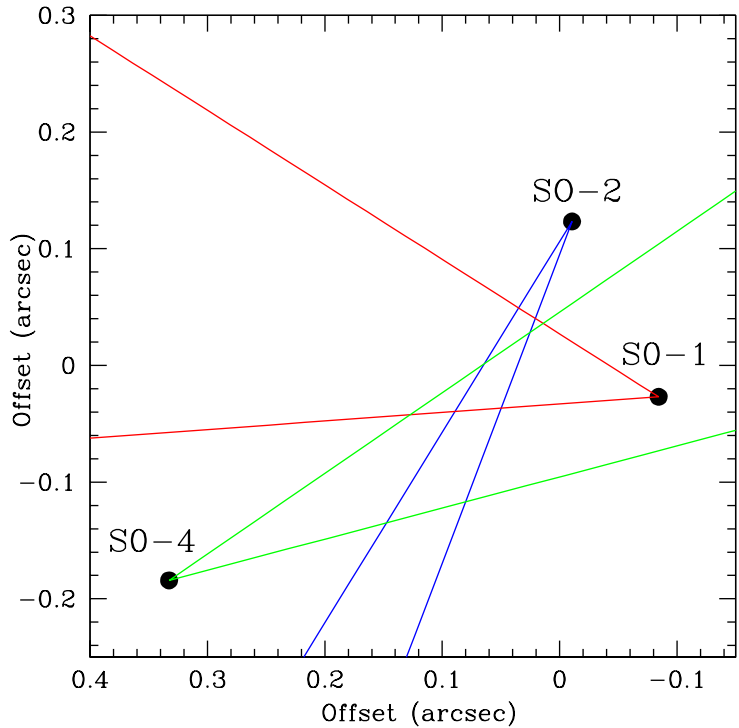}}

\caption{{\small \label{f:acc}Pinpointing the position of the MBH in the
IR frame by projected acceleration vectors (Ghez et al. \citeyear{Ghe00}).
The $1\sigma$ error wedges on the acceleration of 3 stars overlap
in a region $\sim\!0.05''$ from the nominal position of $\SgrA$
(at $0,0$) as was initially derived from the alignment of the radio
and IR grids by maser giants (Menten et al. \citeyear{Men97}) (Reprinted
with permission from} \emph{\small Nature}{\small ).}}
\end{figure}

When the inclination of the orbit is not known, the acceleration provides
a lower limit on the MBH mass (for an assumed value of $R_{0}$),
\begin{equation}
m>R_{0}^{3}\ddot{p}p^{2}/G=m\cos^{3}i\,,\end{equation}
where $i$ \label{d:iincl}is the inclination angle (see footnote
\ref{n:orbparms}). The measured accelerations of a few stars in the
inner $\sim\!0.5"$ give a lower bound of $m\gtrsim10^{6}\,\Mo$ (Ghez
et al \citeyear{Ghe00}; Eckart et al. \citeyear{Eck02}; Sch\"{o}del
et al. \citeyear{Sch03}). 

There is a close similarity between the S-stars orbiting the MBH and
a planetary system (for example, $\Ms/m\!\sim\! M_{\mathrm{Earth}}/\Mo$
and $P_{\mathrm{Jupiter}}\!\lesssim\! P_{\star}\!\lesssim\! P_{\mathrm{Pluto}}$).
The potential of the MBH completely dominates the dynamics, and perturbations
by other stars can be neglected over the typical time span of the
monitoring campaigns, $\sim\!{\cal {O}}(10\,\mathrm{yr})$. Classical
astronomical techniques for deriving the parameters of binary star
systems from astrometric and spectroscopic observations can be directly
applied, with only a few modifications (Salim \& Gould \citeyear{Sal99}). 

When considered separately, each of the stars forms a binary system
with the MBH. A binary orbit is described by 14 parameters: $6$ phase
space parameters for the star and the MBH each, and the two masses%
\footnote{\label{n:orbparms}It is customary to describe a binary orbit in the
reduced mass frame by 6 parameters (apart for the reduced mass): 2
parameters that determine the shape of the Keplerian ellipse, the
semi-major axis $a$ and the eccentricity $e$ (Eq. \ref{e:Kepler}),
1 parameter that determines the phase of the orbit, the time of periapse
passage $t_{0}$, 2 angles that determine the orientation of the orbital
plane in space, the inclination angle $i$ (the angle between the
line of sight and the normal to the orbital plane) and the longitude
of the ascending node $\Omega$ (the position angle of the intersection
between the plane of the sky and the orbital plane, the line of nodes),
and 1 angle that determines the orientation of the ellipse in the
orbital plane, the argument of pericenter $\omega$ (the angle between
the line of nodes and the semi-major axis). %
}. The mass of the star and the reflex motion of the MBH can be neglected
because of the very large mass ratio. However, any constant velocity
the MBH may have must be taken into account. This is in contrast with
the case of visual stellar binaries (binaries where both stars are
resolved), where the proper motion of the binary's center of mass
is irrelevant since the \emph{relative} position between the two stars
is directly observed. Until recently the MBH was not detected in the
IR, where the stars are observed, and so the stellar orbits were measured
against the IR grid, and \emph{not} relative to the position of the
MBH itself. Now that $\SgrA$ is also observed in the IR%
\footnote{$\SgrA$ is today easily detected in its flaring state, less so in
its quiescent state, where it is only visible in part of the exposures.
It is quite likely that with some improvements in the photometric
sensitivity, $\SgrA$ will always be observable.%
} (Genzel et al. \citeyear{Gen03b}; Ghez et al. \citeyear{Ghe04}),
the problem of solving the orbit will become easier. The results discussed
here are still for the case of an invisible MBH. 

In principle, one well sampled stellar orbit with precise astrometric
measurements is enough to solve all the parameters of the system apart
for two degeneracies: (1) Only the combination $m/R_{0}^{3}$ is fixed
by the solution, but not $m$ and $R_{0}$ separately (cf Eq. \ref{e:Kepler}).
(2) There remains a twofold degeneracy in the sign of the radial velocity.
These degeneracies can be resolved with radial velocity information,
in which case the MBH mass and the distance to the Galactic center
can be determined (\S\ref{sss:R0}). The visual binary method for
deriving distances utilizes the fact that the star's radial velocity
$\dot{z}$ is measured via the Doppler shift of the stellar spectral
features in terms of an absolute velocity, whereas the proper motion
$\mathbf{\dot{p}}$ is measured in terms of an angular velocity. The
two are tied together by the orbital solution, $\dot{\mathbf{r}}_{\mathrm{sol}}(t)\!=\!(R_{0}\mathbf{\dot{p}},\dot{z})$,
thereby yielding the distance to the binary.

At present, the available orbital data for any single star is not
sufficiently accurate to fully constrain the solution. Tighter constraints
can be obtained by solving simultaneously several orbits (Salim \&
Gould \citeyear{Sal99}). Since all the stars are orbiting a common
point, the MBH, the ratio of data points to free parameters is much
improved and this can significantly reduce the errors on the derived
parameters (Fig. \hlink{f:6orbits}). The best fit orbital parameters
and their errors are then obtained by a non-linear least squares fit
to the data. 

\begin{figure}[!t]
\centerline{\htarget{f:6orbit}\includegraphics[%
  clip,
  scale=0.75]{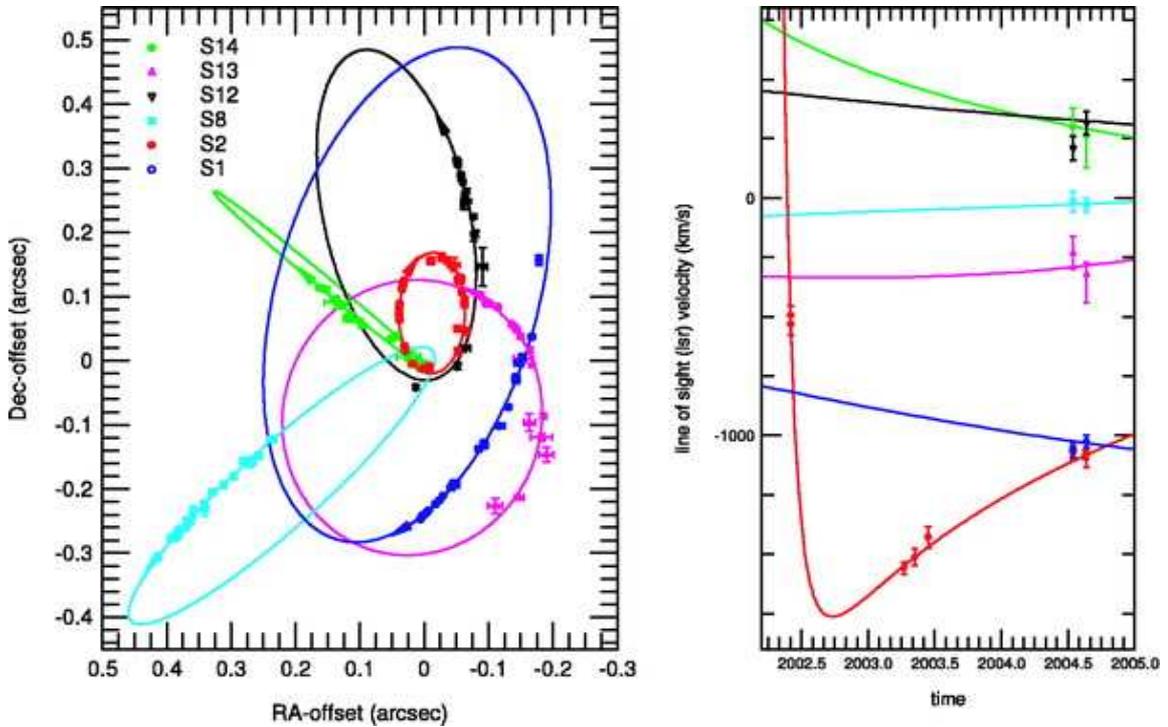}}

\caption{{\small \label{f:6orbits}The projected orbits (left) and the radial
velocity curve (right) of the 6 S-stars used by Eisenhauer et al.
(\citeyear{Eis05}) to solve for the mass, position and distance of
the MBH. The points with error bars are the observed astrometric and
spectroscopic data and the solid curves are the best fit multi-orbit
solutions. (Reprinted with permission from the} \emph{\small Astrophysical
Journal}{\small ).}}
\end{figure}

Full 3D orbital solutions provide yet tighter constraints on the density
of the central dark mass than Eq. (\ref{e:rhoBH2D}). If the mass
is fully contained in the orbit, as suggested by the fact that the
orbit is consistent with Keplerian motion (\S\ref{ss:DM}), the mean
density is

\begin{equation}
\bar{\rho}_{3\mathrm{D}}\!=\!\frac{3}{4\pi}\frac{m}{r_{p}^{3}}\,.\label{e:rhoBH3D}\end{equation}
The smallest periapse discovered to date, $r_{p}\!=\!0.22\,\mathrm{mpc}$
for the star S0-16, raises the lower bound on the mean density to
$\bar{\rho}_{3\mathrm{D}}\!=\!8\!\times\!10^{16}\,\Mo\,\mathrm{pc^{-3}}$,
or $\rho/\rho_{\mathrm{BH}}\!=\!10^{-9}$ for $m\!=\!4\!\times\!10^{6}\,\Mo$
(Ghez et al. \citeyear{Ghe05}). However, the tightest constraints
on the compactness of the dark mass, $\rho\!>\!7\!\times\!10^{21}\,\Mo\,\mathrm{pc^{-3}},$
or $\rho/\rho_{\mathrm{BH}}\!=\!10^{-4}$ come not from stellar orbits,
but from radio observations (Reid \& Brunthaler \citeyear{Rei04}).
These yield a lower limit on the mass associated with $\SgrA$ ($\gtrsim\!4\!\times\!10^{5}\,\Mo$)
based on the small peculiar velocity ($<\!1\,\mathrm{km\, s^{-1}}$
perpendicular to the Galactic plane), and an upper limit on the extent
of the dark mass ($\lesssim\!10^{13}\,\mathrm{cm}$) based on the
assumption that it is contained within the millimeter radio emitting
region (Doeleman et al. \citeyear{Doe01}; Bower et al. \citeyear{Bow04}).

Table ({\small \hlink{t:Morb}}) compares the orbital estimates for
the mass, position and velocity of the dark mass that were obtained
by different authors. The mass estimates all lie in the range $m\!\sim\!(3$--$4)\!\times\!10^{6}\,\Mo$,
which is systematically higher than those derived from the statistical
estimators $m\!\sim\!(2$--$3)\!\times\!10^{6}\,\Mo$, (\S\ref{sss:stat},
table {\small \hlink{t:Mstat}}). The discrepancy with the orbital
mass estimates is most marked for the low value from the non-parametric
method (Chakrabarty \& Saha \citeyear{Cha01b}). This is puzzling,
since non-parametric estimators are expected be robust, but so are
orbital mass estimators, which unlike the statistical estimators do
not require assumptions about the properties of the stellar cluster.
Orbital mass estimates are expected, if anything, to be slightly lower
than the statistical mass estimates because they probe the potential
closest to the MBH ($r\!\lesssim\!0.02$ pc), where the mass fraction
of any extended mass (e.g. stars, compact objects) should be negligible.
The systematic differences cannot be attributed to uncertainties in
$R_{0}$, since all methods assume $R_{0}\!=\!8$ kpc. This discrepancy
has not been studied in detail yet, and the reasons for it remain
unclear. 

\begin{table}

\caption{\label{t:Morb}Orbital mass, position and velocity estimates for
the central dark mass}

\begin{center}\centerline{{\small \htarget{t:Morb}}\begin{tabular}{cccccll}
&
&
&
&
&
&
\tabularnewline
\hline 
{\small $m$ $^{a}$}&
{\small $\Ns$ $^{b}$}&
{\small $R_{0}$}&
{\small $\Delta p$ $^{c}$}&
{\small $\dot{p}$}&
{\small Fixed parameters}&
{\small Refs $^{e}$}\tabularnewline
{\small $(10^{6}\,\Mo)$}&
&
{\small (kpc)}&
{\small (mas)}&
{\small (mas yr$^{-1}$)}&
{\small for MBH $^{d}$}&
\tabularnewline
\hline
{\small $3.7\!\pm\!1.5$ p}&
{\small 1}&
{\small ---}&
{\small ---}&
{\small ---}&
{\small $\dot{\mathbf{p}}\!=\!\dot{z}\!=\!0$, $\mathbf{p}$, $R_{0}$}&
{\small {[}1{]}}\tabularnewline
{\small $3.6\!\pm\!0.4$ p}&
{\small 6}&
{\small ---}&
{\small 1.5}&
{\small ---}&
 {\small $\dot{\mathbf{p}}\!=\!\dot{z}\!=\!0$, $R_{0}$}&
{\small {[}2{]}}\tabularnewline
{\small $4.1\!\pm\!0.6$ b}&
{\small 1}&
{\small ---}&
{\small 2.4}&
{\small ---}&
 {\small $\dot{\mathbf{p}}\!=\!\dot{z}\!=\!0$, $R_{0}$ }&
{\small {[}3{]}}\tabularnewline
{\small $3.59\!\pm\!0.59$ b}&
{\small 1}&
{\small $7.94\!\pm\!0.42$}&
{\small 1.7}&
{\small ---}&
{\small $\dot{\mathbf{p}}\!=\!\dot{z}\!=\!0$}&
{\small {[}4{]}}\tabularnewline
{\small $3.7\!\pm\!0.2$ p}&
{\small 7}&
{\small ---}&
{\small 1.3}&
{\small $1.5\!\pm\!0.5$}&
 {\small $\dot{z}\!=\!0$, $R_{0}$ }&
{\small {[}5{]}}\tabularnewline
{\small $3.61\!\pm\!0.32$ b}&
{\small 6}&
{\small $7.62\!\pm\!0.32$}&
{\small 1.7}&
{\small ---}&
{\small $\dot{\mathbf{p}}\!=\!\dot{z}\!=\!0$}&
{\small {[}6{]}}\tabularnewline
\hline
\multicolumn{7}{l}{{\small $^{a}$ Proper motion data is designated by (p), both proper
motion and radial velocities by (b).}}\tabularnewline
\multicolumn{7}{l}{{\small $^{b}$ Number of stars in orbital fit. }}\tabularnewline
\multicolumn{7}{l}{{\small $^{c}$ The uncertainty in the position of the center of acceleration
in the IR grid, $\Delta p\!=\!(\Delta p_{x}^{2}\!+\!\Delta p_{y}^{2})^{1/2}$.}}\tabularnewline
\multicolumn{7}{l}{{\small $^{d}$ When fixed, $R_{0}$ is assumed to be $8\,\mathrm{kpc}$.}}\tabularnewline
\multicolumn{7}{l}{{\small $^{e}$ {[}1{]} Sch\"odel et al. (\citeyear{Sch02}), {[}2{]}
Ghez et al. (\citeyear{Ghe03b}), {[}3{]} Ghez et al. (\citeyear{Ghe03a})}}\tabularnewline
\multicolumn{7}{l}{{\small $^{e}$ {[}4{]} Eisenhauer et al. (\citeyear{Eis03}), {[}5{]}
Ghez et al. (\citeyear{Ghe05}), {[}6{]} Eisenhauer et al. (\citeyear{Eis05})}}\tabularnewline
\hline
\end{tabular}}\end{center}
\end{table}

\subsubsection{Distance to the Galactic Center}

\label{sss:R0}

The Sun-GC distance, $R_{0}$, is a key parameter in the reconstruction
of Galactic structure and dynamics as observed from our vantage point
at the Solar system. For example, the current uncertainties in the
values $R_{0}$ and the local Galactic rotation speed $\Theta_{0}$\label{d:T0}
preclude the derivation of useful limits on the spheroidal axes ratios
of the dark matter halo around the Galaxy, which could provide information
on the nature of the dark matter (Olling \& Merrifield \citeyear{Oll01};
see discussion in Weinberg, Milosavljevi\'c \& Ghez \citeyear{Wei04}).
The distance to the GC is also an important (though not the only)
lower rung in the extragalactic distance scale, useful for calibrating
standard candles such as RR Lyrae stars, Cepheids and red clump giants.
Until recently, the adopted distance to the GC, $R_{0}\!=\!8.0\!\pm\!0.5\,\mathrm{kpc}$
(Reid \citeyear{Rei93}) was based on the average of various primary
(geometric), secondary (standard candles) and tertiary (theoretical
models) distance indicators (see brief summary in Eisenhauer et al.
\citeyear{Eis03}). Geometric distance indicators are in principle
the most reliable, \emph{if} the assumptions about physical properties
of the system used are well founded.

An early attempt to use stellar observations in the GC to determine
the distance to the GC was based on a general relation that holds
in a spherical system and relates the line of sight velocity to the
projected velocity through $R_{0}$ (e.g. Leonard \& Merritt \citeyear{Leo89}),\begin{equation}
\left\langle \dot{z}^{2}p\right\rangle =R_{0}^{2}\left\langle (\frac{2}{3}\dot{p}_{\parallel}^{2}+\frac{1}{3}\dot{p}_{\perp}^{2})p\right\rangle \,.\end{equation}
 The combined radial and proper motion data yield $R_{0}\!=\!8.2\!\pm\!0.9$
kpc (Genzel et al. \citeyear{Gen00}).

The visual binary method (Salim \& Gould \citeyear{Sal99}; Jaroszynski
\citeyear{Jar99}) uses precision measurements of proper motions and
radial velocities of stars orbiting the MBH to provide a direct geometric
determination of the Sun-MBH distance (\S\ref{sss:orbsol}). The
only assumption is that the orbit is Keplerian. By virtue of its simplicity,
this method is essentially free of systematic uncertainties due to
uncertainties in the astrophysical modeling.

The first application of the visual binary method was based on the
orbit of the star S2 (Eisenhauer et al. \citeyear{Eis03}). Since
the data was not restrictive enough to constrain the 13 free parameters
(\S\ref{sss:orbsol}), it was assumed that the 3 velocity components
of the MBH can be taken as zero, in keeping with the low peculiar
velocity measured in the radio ($\sim\!25\,\mathrm{km\, s^{-1}}$,
Backer \& Sramek \citeyear{Bac99}; Reid et al. \citeyear{Rei99},
\citeyear{Rei03a}) and the small error in the estimated velocity
of the Solar system toward the GC. This analysis yielded $R_{0}\!=\!7.94\!\pm\!0.42$
kpc, $\Theta_{0}\!=\!220.7\!\pm\!12.7\,\mathrm{km\, s^{-1}}$ and
$m\!=\!(3.6\!\pm\!0.6)\!\times\!10^{6}\,\Mo$. A subsequent analysis
with more data (still assuming zero velocity for the MBH) yielded
compatible values, $R_{0}\!=\!7.62\!\pm\!0.32$ kpc and $m\!=\!(3.6\!\pm\!0.3)\!\times\!10^{6}\,\Mo$
(Eisenhauer et al. \citeyear{Eis05}).

\subsection{Constraints on non-BH dark mass alternatives}

\label{ss:DM}

A black hole, by its nature, is an object whose existence is hard
to prove. Other than demonstrating the absence of a hard surface and
a diverging gravitational redshift (the signature of an event horizon
as seen by a stationary distant observer), one has to make do with
placing ever tighter constraints on the volume enclosing the dark
mass. Such limits can then be used to exclude non-BH dark mass models
(that is, mass distributions that extend beyond their event horizon),
such as a cluster of stellar-mass dark objects (\S\ref{sss:compact})
or {}``balls'' of exotic particles (\S\ref{sss:exotic}). Even
if the existence of a MBH can be established, it is still possible,
indeed quite likely, that it is surrounded by an extended distribution
of dark mass, for example a dense cluster of massive compact stellar
remnants that have sunk to the bottom of the potential well by the
dynamical process of mass segregation (\S\ref{sss:distrDM}, \S\ref{ss:OBsols}).
The stellar orbits tracked around the central dark mass probe the
gravitational potential there at distances as small as $\sim\!10$
light hours ($\sim\!10^{15}$ cm). The information provided by these
orbits offers the strongest case yet for the existence of a MBH in
a galactic nucleus. Gravitational lensing may also be used to probe
the distribution of the dark mass (\S\ref{sss:GLGC}, \S\ref{sss:extGL}).
Figure (\hlink{f:DMdensity}) summarizes the constraints on the size
and density of the dark mass that are discussed below.

\begin{figure}[!t]
\centering{\htarget{f:DMdensity}\includegraphics[%
  clip,
  scale=0.75]{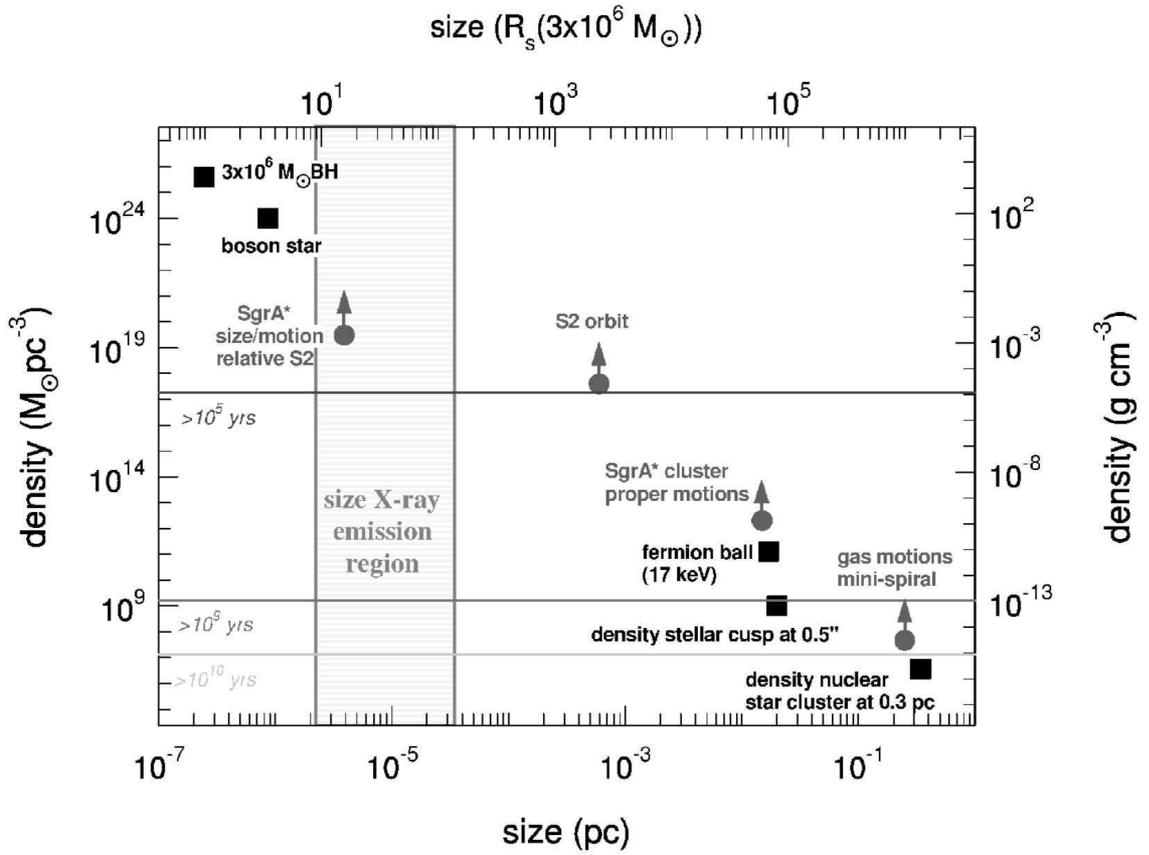}}

\caption{{\small \label{f:DMdensity}Compilation of the constraints on the
nature of the dark mass in the GC (Sch\"odel et al. \citeyear{Sch03}).
Filled circles: Lower limits on the size and density of the dark mass.
Grey area: Constraints on the size of the variable X-ray emitting
region (Baganoff et al. \citeyear{Bag01}). Filled squares: Different
dark mass candidates, including the visible star cluster and its central
cusp, the heavy fermion ball (Tsiklauri \& Viollier \citeyear{Tsi98}),
the boson star (Torres et al. \citeyear{Tor00}) and a $3\!\times\!10^{6}\,\Mo$
MBH. Horizontal lines: Lifetimes of hypothetical dark clusters of
astrophysical objects, neutron stars, white dwarfs, or stellar black
holes (Maoz \citeyear{Mao98}). The tightest constraint on the mass
density, $\rho\!>\!{\cal {O}}(0.1\,\mathrm{g\, cm^{-3})}$ (not shown)
come from the radio proper motion of $\SgrA$ compared to the velocities
in the surrounding star cluster} (Reid \& Brunthaler \citeyear{Rei04})
{\small under the assumption that the dark mass is enclosed in the
millimeter radio source (Doeleman et al. \citeyear{Doe01}). All configurations
are excluded except a black hole and a boson star. (Reprinted with
permission from the} \emph{\small Astrophysical Journal}{\small ) }}
\end{figure}

\subsubsection{Cluster of dark stellar-mass objects}

\label{sss:compact}

The most conservative dark mass alternative to a MBH is a very dense
cluster of dark, or low-luminosity stellar mass objects. Possible
candidates include very low mass stars ($\Ms\!\gtrsim\!0.09\,\Mo$),
brown dwarfs ($0.01\,\Mo\!\lesssim\!\Ms\!\lesssim\!0.09\,\Mo$, failed
stars that are not massive enough to ignite hydrogen), planets ($\Ms\!\lesssim\!0.01\,\Mo$,
where gravity is negligible compared to the electrostatic forces)
or compact stellar remnants: white dwarfs (WDs) ($\Ms\!\lesssim\!1.4\,\Mo$),
NSs ($1.4\,\Mo\!\lesssim\!\Ms\!\lesssim\!3\,\Mo$) and stellar mass
black holes ($\Ms\!\gtrsim\!3\,\Mo$). It is not at all obvious that
stellar evolution and stellar dynamics can form a cluster of dark
objects that is massive and concentrated enough to play the role of
the central dark mass in the GC. However, irrespective of the formation
history of such a hypothetical cluster, it is possible to invoke stability
arguments to obtain robust upper limits on the cluster lifespan against
the dynamical processes of evaporation and run-away collisions (Maoz
\citeyear{Mao98}; \S\ref{sss:scales}). The lower this upper limit
is compared to the age of the GC, the less likely is the dark cluster
hypothesis. 

Maoz (\citeyear{Mao98}) obtained such limits for a sample of galaxies,
including the Milky Way, where a dark mass $m$ was detected inside
a radius $r_{m}$. To obtain the most conservative limit, it was assumed
that (1) the hypothetical cluster has the least centrally concentrated
observed density profile (a Plummer model, $\rho\!=\!\rho_{0}/\left[1\!+\!\left(r/r_{0}\right)^{2}\right]^{5/2}$),
so as to minimize the collision rate; (2) it is composed of equal-mass
objects, since otherwise mass segregation would accelerate its dynamical
evolution; (3) the objects are at zero temperature, so as to minimize
their radius for a given mass and thus minimize the collision rate
(Eq. \ref{e:tcoll}). A maximal lifespan $t_{m}$ was obtained for
each galaxy in the sample, based on the measured values of $m$ and
$r_{m}$, and by choosing the dark object that maximizes $t_{m}$
($0.6\,\Mo$ WDs in the case of the GC). The dark mass in the GC was
estimated to have $t_{m}\!<\!10^{8}\,\mathrm{yr}\!\ll\! t_{H}$. This
argues against the possibility that the dark mass is a dark cluster.

\subsubsection{Exotic elementary particles}

\label{sss:exotic}

One interesting MBH alternative is that the dark masses in the centers
of galaxies are the high density peaks of the cold dark matter distribution
that is dynamically observed to exist on galactic and intergalactic
scales (Tsiklauri \& Viollier \citeyear{Tsi98}; Munyaneza \& Viollier
\citeyear{Mun02}). This idea's appeal lies in its economical use
of a single component to explain dark mass on all scales. If dark
matter is composed of fermionic elementary particles, then a massive
compact {}``Fermion Ball'' could support itself against gravity
by degeneracy pressure, without having a hard surface (thereby avoiding
tell-tale emission from matter impacting on the surface) and without
being relativistic. Dynamically, such a configuration would be indistinguishable
from a MBH for any process that occurs outside the ball. Deviations
from Keplerian motion will become apparent only for orbits that cross
through the ball. Also, a Fermi ball, unlike a low-mass MBH, will
not necessarily induce tidal disruption since its size can exceed
the tidal disruption radius (\S\ref{ss:tide}).

The maximal Fermi ball mass that can be supported by the degeneracy
pressure of a fermion of mass $m_{f}$ is set by the Oppenheimer-Volkoff
limit (Oppenheimer \& Volkoff \citeyear{Opp39}), the GR analog of
the Chandrasekhar mass for WDs, $\max m\!\sim\! g_{f}^{-1/2}m_{\mathrm{P}}^{3}/m_{f}^{2}$,
where $g_{f}$ is the spin degeneracy factor of the fermion and $m_{\mathrm{P}}\!\equiv\!\sqrt{\hbar c/G}$
is the Planck mass. By choosing a small enough $m_{f}$ it is then
possible to explain the entire range of galactic central dark masses
as Fermi balls.  The maximal measured central dark mass ($\max m\lesssim3\!\times\!10^{9}\,\Mo$)
then implies an upper limit on the fermion mass of

\begin{equation}
\max m_{f}=0.62\left(g_{f}/2\right)^{-1/4}\sqrt{\frac{m_{\mathrm{p}}}{\max m}}m_{\mathrm{p}}\sim\left(g_{f}/2\right)^{-1/4}15\,\mathrm{keV}\, c^{-2}\,.\end{equation}

The orbit of S0-2 is consistent, up to the measurement errors, with
a Keplerian orbit around a point mass $m$. This one orbit can be
used to rule out the universal Fermi ball alternative to MBHs (Munyaneza,
Tsiklauri \& Viollier \citeyear{Mun98}; Sch\"odel et al. \citeyear{Sch02}).
The lack of deviations from a Kepler orbit near the orbital periapse
implies that the Fermi ball must be wholly contained within the periapse
$r_{p}$. The velocity of S2 at periapse is $\lesssim0.02c$, and
so a Fermi ball that fills the enclosed volume can be treated in the
non-relativistic limit. The polytrope mass--radius relation (Chandrasekhar
\citeyear{Cha39}) for a cold, non-relativistic degenerate Fermi ball
results in a lower limit on the fermion mass \begin{equation}
\min m_{f}=1.76\left(g_{f}/2\right)^{-1/4}\left(\frac{\ell_{\mathrm{P}}}{r_{p}}\right)^{3/8}\left(\frac{m_{\mathrm{P}}}{m}\right)^{1/8}m_{\mathrm{P}}\,\sim\left(g_{f}/2\right)^{-1/4}70\,\mathrm{keV}\, c^{-2}\,,\end{equation}
where $\ell_{\mathrm{P}}\!\equiv\!\sqrt{\hbar G/c^{3}}$ is the Planck
length. Thus, there is no fermion mass that can simultaneously accommodate
the orbital constraints of S2 on the extent of the dark mass in the
GC and support the largest measured central dark masses. This rules
out the possibility that the same component singly explains both the
diffuse dark matter on galactic and inter-galactic scales and the
compact dark matter in the centers of galaxies. 

Another proposed exotic particle alternative to a MBH, which cannot
be ruled out by the stellar orbits, is the {}``Boson Star''. Such
an object is composed of a scalar field and is supported against collapse
by the Heisenberg uncertainty principle (Torres, Capozziello \& Lambiase
\citeyear{Tor00}). A wide range of boson star masses can be realized,
including MBH masses, depending on the assumptions about the boson
particle masses and their self-interactions. A boson star does not
extend much beyond its event horizon, and is therefore strongly relativistic,
but it does not possess a singularity, an horizon or a hard surface.
Because both are so compact, stellar orbits cannot empirically distinguish
between a MBH and a boson star. However, there are theoretical arguments
against the dark mass being a boson star, since it is not clear that
it can avoid accreting some matter and collapsing to a MBH. A boson
star, unlike a MBH, has no capture orbits and so high velocity matter,
such as that released in tidal disruption events, will not accumulate
in the center. However, this may not be enough to suppress the accretion
of compact remnants on tightly bound orbits (\S\ref{sss:distrDM})
or gas flowing from the inner edge of an accretion disk.

\subsubsection{Extended dark mass around the MBH }

\label{sss:distrDM}

Even if a MBH does exist in the GC, it is still possible that there
are other dark components around it, a cluster of massive compact
objects (\S\ref{ss:Mseg}), a cusp of exotic particles, or a combination
of the two. It should be noted that irrespective of the presence of
the MBH, the Galactic dark matter halo is expected to peak at the
GC (e.g. Navarro, Frenk \& White \citeyear{Nav96}). The possibility
that a steep, high density $\rho\!\sim\!\rho_{h}(r/r_{h})^{-\alpha}$
cusp of self-interacting dark matter particles could form under the
dynamical influence of the MBH (\S\ref{sss:relax}) was investigated
in the hope that the enhanced particle annihilation rate ($\propto\!\rho^{2}$)
could produce a detectable signal (Gondolo \& Silk \citeyear{Gon99};
but see opposing conclusions by Merritt \citeyear{Mer04b}). The detection
of strong TeV emission from the vicinity of $\SgrA$ (Kosack et al.
\citeyear{Kos04}; Aharonian et al. \citeyear{Aha04}), further adds
to the interest in this possibility. However, with a (very uncertain)
normalization of only $\rho_{h}\!\sim\!100\,\Mo\,\mathrm{pc^{-3}}$
(Gnedin \& Primack \citeyear{Gne04}), the mass in the dark matter
particles is too small relative to the stellar cusp (Eq. \ref{e:cusp})
to be probed by the stellar orbits.

The smoothed potential of an extended dark mass component can be probed
by deviations from an exact Keplerian orbit that will be displayed
by a star that passes through \emph{}the mass distribution. Such orbits
will display a retrograde shift of the angle of periapse (a receding
rosette-like orbit), because the stellar acceleration at distance
$r$ from the center is proportional to the enclosed mass $M(<\! r$)
(spherical symmetry assumed), and it decreases as the star approaches
periapse. This shift is in the opposite sense of the effect expected
for a GR orbit around a point mass (an advancing rosette-like orbit,
\S\ref{ss:GRorbit}). The retrograde periapse shift per orbit $\Delta\omega$
due to a small spherical potential perturbation $\delta\phi$ on top
of the potential of the MBH, $\phi\!=\!\phi_{\mathrm{MBH}}\!+\!\delta\phi$,
is given by (e.g. Landau \& Lifshitz \citeyear{Lan69})\begin{equation}
\Delta\omega(a,e)=\frac{\partial}{\partial J}\left(\frac{2}{J}\int_{0}^{\pi}r^{2}\delta\phi\mathrm{d}\omega\right)\,.\end{equation}
For a powerlaw extended mass distribution, parametrized as $M_{e}(<\! r)\!=\! m(r/r_{m})^{3/2-p}$
(cf Eq. \ref{e:nstar}), the periapse shift per orbit can be evaluated
numerically by (see also Ivanov, Polnarev \& Saha \citeyear{Iva05})

\begin{equation}
\Delta\omega(a,e)=\left(\frac{a}{r_{m}}\right)^{3/2-p}\frac{2(1-e^{2})^{3/2-p}}{e(1-2p)}\int_{0}^{\pi}\frac{4(2\!-\! p)e+\left[4(2\!-\! p)\!-\!(3\!-\!2p)(1-e^{2})\right]\cos\omega}{(1+e\cos\omega)^{7/2-p}}\mathrm{d}\omega\,.\end{equation}

Mouawad et al. (\citeyear{Mou05}) show that the lack of detectable
deviations from Keplerian motion in the orbit of S2 places an upper
limit of $M_{e}(<\! r_{a})/m\!<\!0.05$ on the extended mass, where
$r_{a}\!\simeq\!0.01\,\mathrm{pc}$ is the apoapse of S2. This upper
limit still exceeds by a factor of $\sim\!30$ the dynamical upper
limit on the population of SBHs that can exist there (\S\ref{ss:Mseg}).
Occasional orbital perturbations due to close encounters with individual
compact objects can distinguish a dark matter component consisting
of compact objects from a smooth particle dark matter distribution.
Such perturbations may be detected by future very large telescopes
($\gg\!10\,\mathrm{m}$) (Weinberg, Milosavljevi\'c \& Ghez \citeyear{Wei04}). 

The orbits are not sensitive to the mass distribution inside the smallest
periapse (provided it is spherical). An upper limit on the fraction
of the extended mass on these small scales ($\lesssim\!\mathrm{few\!\times\! mpc}$)
can be obtained from the Brownian motion of the mass that is associated
with the radio source $\SgrA$ due to gravitational interactions with
stars around it. The amplitude of the Brownian motion is strongly
constrained by the radio limits on the peculiar motion of $\SgrA$
(motion in excess of the reflex motion induced by the Sun's orbit
around the Galaxy). Brownian motion in a gravitating system is different
from that in a gas, where the inter-molecular forces are short range
forces. Equipartition arguments can not be directly applied. Inside
$r_{h}$ the potential is dominated by the MBH and the stars are {}``carried
away'' with it as it moves; conservation of momentum rather than
equipartition dictates the relation between the MBH velocity and mass
and those of the stars. Neither is it possible to assign a well-defined
temperature to the stellar system; the MBH itself affects the velocity
field ({}``temperature'') of the stars near it (Merritt \citeyear{Mer04a}),
and even outside $r_{h}$ the stellar distribution is generally not
isothermal (Chatterjee, Hernquist \& Loeb \citeyear{Cha02a}, \citeyear{Cha02b}).
Reid \& Brunthaler (\citeyear{Rei04}) use the radio observations
to first show that if the mass associated with $\SgrA$ is some fraction
of the total central mass, then it must lie very close to the center
of the region delimited by the stellar orbits, otherwise it would
have orbited rapidly around the dark mass, in contradiction with the
upper bound on the peculiar motion of $\SgrA$ ($\sim\!1\,\mathrm{km\, s^{-1}}$
perpendicular to the galactic disk). They then proceed to assume that
$\SgrA$ does lie in the center, and use N-body simulations of the
Brownian motion to translate the upper bound on the peculiar motion
to a lower bound of $\gtrsim\!0.4\!\times\!10^{6}\,\Mo$ on the mass
associated with $\SgrA$ ($\sim\!0.1$ of the dynamical mass measured
by the orbits). This limit is consistent with the results of a more
detailed formal analysis of Brownian motion in gravitating systems
(Laun \& Merritt \citeyear{Lau05}). It is unlikely that the large
mass fraction ($\sim\!0.9$) that remains unconstrained is in fact
non-BH dark matter. However, an order of magnitude improvement in
the velocity resolution will be needed to confirm that the MBH indeed
dominates the dark mass.

\subsection{Limits on MBH binarity}

\label{ss:2MBH}

Galaxies as observed today are thought to have formed by a chain of
successive mergers of smaller galaxies. High redshift quasars and
the large number of AGN found at smaller redshifts suggest that these
mergers must also involve the formation of a binary MBH in the center
of the newly merged galaxy, after the two MBHs rapidly sink in due
to dynamical friction (\S\ref{ss:Mseg}). The binary subsequently
decays, initially by dynamical processes, and finally by the emission
of GW, and the two MBHs merge (see review by Merritt \& Milosavljevi\'c
\citeyear{Mer04}). In a {}``major merger'' the two MBHs have comparable
masses, whereas in a minor merger the mass ratio is large. The formation
history of the Galactic MBH is not known, but there are several arguments
that suggest that the Galactic MBH did not undergo a major merger,
at least since it reached its present mass scale. Its small mass can
be easily supplied, over time, by the stellar winds near it (\S\ref{ss:Windfeed})
and so there is no compelling reason to assume mergers. Another argument
against a major merger is the observed stellar distribution around
the MBH (\S\ref{ss:cusp}). Initially the binary's orbit decays dynamically,
by ejecting stars that pass near it out of the galactic core by 3-body
interactions. A total stellar mass of the same order as the binary
mass is ejected during a major merger, leaving behind a galaxy with
a shallow central density distribution (Milosavljevi\'c et al. \citeyear{Mil02}).
The GC, however, has a steep inner cusp. Furthermore, a binary MBH
is expected to induce a tangential anisotropy in the stellar orbits
by ejecting or destroying stars on radial orbits that approach it
(Gebhardt \citeyear{Geb03}). This is not observed in the GC (Sch\"odel
et al. \citeyear{Sch03}).

Minor mergers, on the other hand, could well have occurred. The presence
of a $\sim\!10^{3}\,\Mo$ IBH is in fact suggested by the apparently
self-bound stellar cluster IRS13, $\sim\!0.15$ pc (projected) from
the MBH (Maillard et al. \citeyear{Mai04}; but see opposing view
by Sch\"odel et al. \citeyear{Sch05}). It is also suggested by the
{}``sinking cluster'' model, which seeks to explain the young population
in the center as originating in a dense, star forming stellar cluster
with an IBH, which sank to the center by dynamical friction (Hansen
\& Milosavljevi\'c \citeyear{Han03}; \S\ref{sss:dynfric}).

The binary MBH hypothesis can be constrained directly by the stellar
orbits. The good fit of S2's motion to a Keplerian orbit suggests
that a hypothetical binary MBH must be contained in a region much
smaller than the semi-major axis of S2, $\sim\!5$ mpc (see Jaroszynski
\citeyear{Jar00} for a quantitative analysis). However, it is very
unlikely that such a tight binary (of moderate mass ratio) exists
now in the GC, given its very short lifespan against orbital decay
and coalescence by the emission of GW radiation. The orbit of circular
binary MBH (masses $m_{1}$ and $m_{2}$ and semi-major axis $a_{12}$)
would shrink to a point by the emission of GW radiation in time 

\begin{equation}
t_{\mathrm{GW}}=\frac{5}{256}\frac{c^{5}a_{12}^{4}}{G^{3}(m_{1}+m_{2})m_{1}m_{2}}\textrm{\,,}\end{equation}
 (Peters \citeyear{Pet64}; Shapiro \& Teukolsky \citeyear{Sha83}
Eq. 16.4.10). For example, if $m_{1}\!=\! m_{2}\!=\! m/2\!=\!1.5\!\times\!10^{6}\,\Mo$
and $a_{12}\!=\!0.1$ mpc, then the lifespan of the binary is only
$t_{\mathrm{GW}}\!=\!10^{4}$ yr. MBH Binaries with smaller mass ratios
are, however, much longer lived, and cannot be ruled out by this argument.

The MBH motion in the plane of the sky, $v\sim\!2\,\mathrm{mas\, yr^{-1}\!=\!}76(R_{0}/8\,\mathrm{kpc})\,\mathrm{km\, s^{-1}}$,
is obtained from the full orbital solutions (Ghez et al. \citeyear{Ghe05}).
A second BH orbiting the MBH would induce a reflex motion of the MBH
around their common center of mass. This can be translated to an upper
bound on the mass and distance of such a BH (assuming it lies outside
the solved stellar orbits) of \begin{equation}
m_{\mathrm{2}}\lesssim v\sqrt{\frac{m_{1}r}{G}}\simeq5\times10^{5}\sqrt{\frac{r}{0.08\,\mathrm{pc}}}\,\Mo\,,\label{e:Mibh}\end{equation}
 assuming a circular orbit in the plane of the sky and $m_{\mathrm{2}}\!\ll\! m_{1}\!=\!3\!\times\!10^{6}\,\Mo$
(see Hansen \& Milosavljevi\'c \citeyear{Han03} for a a more detailed
analysis of the astrometric constraints on $m_{2}$ and $a$) 

Tighter upper bounds can be obtained from the limits on the peculiar
motion of $\SgrA$ in the radio of $18\!\pm\!7\,\mathrm{km\, s^{-1}}$
in the plane of the galaxy and $0.4\!\pm\!0.9\,\mathrm{km\, s^{-1}}$
perpendicular to it (Reid \& Brunthaler \citeyear{Rei04}). However,
these are \emph{not} motions measured directly relative to the central
cluster, but rather measured relative to extragalactic radio sources,
and so include also the reflex motion, $\Theta_{0}/R_{0}$, which
can not be subtracted exactly, and the possible Brownian motion of
the central cluster relative to the Galactic center of mass, which
is unknown. The upper limit on the perpendicular motion of $\sim\!1\,\mathrm{km\, s^{-1}}$
translates through Eq. (\ref{e:Mibh}) to an upper limit of $m_{2}\!\lesssim\!10^{4}\sqrt{r/0.005\,\mathrm{pc}}\,\Mo$
(for $r\!>\!0.005\,\mathrm{pc}$). This upper bound is still consistent
with a $10^{3}$--$10^{4}\,\Mo$ IBH in IRS13.

\subsection{High-velocity runaway stars}

\label{ss:hivstar}

A possible stellar signature of the existence of a MBH, or possibly
a binary MBH, would be the discovery of very high velocity stars tens
of kpc away \emph{}from the GC moving with $v\!\sim\!{\cal {O}}(1000\,\mathrm{km\, s^{-1}})$
on nearly radial orbits relative to the GC (the local escape velocity
from the Galaxy at the Sun is $\sim600\,\mathrm{km\, s^{-1}}$, e.g.
Wilkinson \& Evans \citeyear{Wil99}). The stars can acquire such
high velocities by exchange interactions deep in the potential of
the MBH, either by the break-up of a binary on a nearly radial orbit
(Hills \citeyear{Hil88}), by an encounter with another single star,
or an encounter with a binary MBH (Yu \& Tremaine \citeyear{YuQ03}).
A slow-moving star with velocity $v_{\infty}$ far from the MBH will
acquire near periapse a very high velocity, $v^{2}\!\sim\!2Gm/r_{p}\!\gg\! v_{\infty}^{2}$.
A small velocity perturbation by a gravitational interaction with
a second mass, $\delta v$, can add $v\delta v\!\gg\! v_{\infty}^{2}/2$
to the orbital energy of the star, and take away the same amount from
the other mass (a similar mechanism forcefully expels tidal ejecta,
see \S\ref{sss:tdafter}). Yu \& Tremaine (\citeyear{YuQ03}) find
that stellar binary exchanges or encounters with a binary MBH can
populate the Galaxy with high velocity stars at a substantial rate
($\gtrsim10^{-5}\,\mathrm{yr^{-1}}$), resulting in up to $\sim\!10^{3}$
high velocity stars ($v_{\infty}\!>\!1000\,\mathrm{km\, s^{-1}}$)
inside the Solar circle.

Intriguingly, Brown et al. (\citeyear{Bro05}) report the discovery
of the highest velocity star ever observed in the Galactic halo, moving
with a radial velocity of $710\,\mathrm{km\, s^{-1}}$ at a distance
of $\sim\!55\,\mathrm{kpc}$. Its direction of motion could be consistent
with radial motion away from the GC, if the (unknown) transverse velocity
components are small. The stellar spectrum does not identify its spectral
type unambiguously, but one of the two possibilities is a B9 MS star,
such as those found in the S-cluster (the other possibility is a blue
horizontal branch giant).

\section{Probing post-Newtonian gravity near the MBH}

\label{s:GRprobe}

General Relativity (GR) is the least tested of the theories of the
four fundamental forces of nature. An important goal of the study
of stars near the Galactic MBH is to detect post-Newtonian effects
and probe GR in the weak and strong field limits near a super-massive
object. Three classes of effects are discussed in this context. The
first is relativistic celestial mechanics. Stars are observed moving
with velocities of up to $v\!\sim\mathrm{few\!\times\!0.01}c$ at
distances of $\lesssim\!10^{3}r_{S}$ from a mass of $\sim\!4\!\times\!10^{6}\,\Mo$.
This regime of GR dynamics is virtually unexplored%
\footnote{Gas processes probe the dynamics much closer to the MBH, but are also
much harder to interpret both observationally and theoretically. Emission
from accreting gas is observed from the strong field near the event
horizon of MBHs. There is tentative evidence of gravitational redshift
in the Fe $K\alpha$ line profiles in AGN (e.g. Fabian et al. \citeyear{Fab95},
but see e.g. Page et al. \citeyear{Pag04}). The emission from Galactic
SBHs and AGN sometimes display quasi-periodic oscillations that can
be interpreted as coming from the last stable orbit around a spinning
BH (see Genzel et al. \citeyear{Gen03b}; Aschenbach et al. \citeyear{Asc04}
for such evidence from the Galactic MBH).%
} (\S\ref{ss:GRorbit}). For comparison, the high precision confirmation
of the predictions of GR in the Hulse-Taylor binary pulsar (PSR 1913+16)
were for masses of $\sim\!1.4\, M_{\odot}$ with $v\!\sim\!0.003c$
at $r_{p}\!\sim\!2\!\times\!10^{5}r_{S}$ (Taylor \& Weisberg \citeyear{Tay89}).
  The second class of weak field phenomena that may be relevant
in the GC is gravitational lensing (\S\ref{ss:GL}). The lensed stars
are not necessarily related to the stars near the MBH, since it is
only their projected angular distance to the MBH that has to be small,
but their true (3D) position can be anywhere behind the MBH. The third
class of GR effects that may be relevant in the GC is the emission
of GW from very low-mass stars that spiral into the MBH. This dissipative
process is discussed separately in (\S\ref{sss:GW}).

\subsection{Relativistic orbital effects}

\label{ss:GRorbit}

At present, all the available orbital data can be adequately modeled,
within the measurement errors, in terms of Newtonian motion. With
improved resolution, higher precision and a longer baseline, deviations
from Keplerian orbits may be detectable. This will likely be a challenging
task. The chances of success will be much improved if future observations
detect fainter stars on even tighter and more eccentric orbits than
the relatively bright ones observed today. This is because post-Newtonian
effects are larger the tighter the orbit, while the uncertainty due
to the dynamical influence of the unknown amount of extended mass
around the MBH (\S\ref{sss:distrDM}) is minimized. However, it is
not clear how many faint, low-mass stars exist in the inner few mpc,
since mass segregation and possibly also collisional destruction by
the massive remnants there are expected to suppress their numbers
there. 

The highest stellar velocity recorded to date in the GC was that of
the star S0-16 during periapse passage (Ghez et al. \citeyear{Ghe05}),
which reached $\sim\!12000\,\mathrm{km\, s^{-1}}$ ($\beta\!\sim\!0.04$,
$\beta\!\equiv\! v/c$ \label{d:beta}) at a distance of $r_{p}\!\sim\!600r_{S}$
from the MBH. This corresponds to a relativistic parameter $\Upsilon\!\equiv\! r_{S}/r$
\label{d:Gparm} at periapse of $\Upsilon(r_{p})\!\sim\!1.6\!\times\!10^{-3}$.
For comparison, the relativistic parameter on the surface of a WD
is a $\mathrm{few\!\times\!10^{-4}}$ and on a NS a $\mathrm{few}\!\times\!10^{-1}$.
The relativistic parameter changes along the orbit. When the orbit
is only mildly relativistic, it obeys a nearly Keplerian energy equation
(Eqs. \ref{e:Kepler}, \ref{e:Ekepler}) and $\Upsilon$ can be expressed
as \begin{equation}
\Upsilon=\beta^{2}+\frac{Gm}{c^{2}a}\equiv\beta^{2}+\Upsilon_{0}\,.\label{e:Gparm}\end{equation}
When $e\!\gg\!0$ and the star is near periapse, $\beta^{2}\!\gg\!\Upsilon_{0}$
and $\Upsilon\!\sim\!\beta^{2}$. The deviations from Newtonian mechanics
become more pronounced as $\Upsilon$ increases. For $\Upsilon\!\ll\!1$
(as is the case here) it is useful to expand the post-Newtonian deviations
in $\beta$, and classify each effect by the order of its $\beta$-dependence
(equivalently, by the order of its $\sqrt{r_{S}/r_{p}}$ or $\sqrt{r_{S}/a}$
dependence).

\subsubsection{Gravitational redshift}

\label{sss:GRz} 

The measurement of stellar proper motions and accelerations in the
inner arcsecond of the GC (Eckart \& Genzel \citeyear{Eck96}; Ghez
et al. \citeyear{Ghe98}) preceded by more than a decade the measurement
of radial velocities for these stars (Ghez et al. \citeyear{Ghe03a};
Eisenhauer et al. \citeyear{Eis03}). For this reason, most theoretical
studies of the detection of post-Newtonian orbits in the GC focused
exclusively or mainly on proper motions and effects related to them,
such as periapse shift (Jaroszynski \citeyear{Jar98b}; Fragile \&
Mathews \citeyear{Fra00}; Rubilar \& Eckart \citeyear{Rub01}; see
also general treatise on relativistic celestial mechanics by Brumberg
\citeyear{Bru91}). However, with the advent of adaptive optics assisted
IR imaging spectroscopy, it now appears that it is the \emph{}radial
velocities\emph{,} rather than the proper motions, that may provide
the tightest constraints on deviations from Newtonian orbits. The
current quality of spectroscopic observations allows the determination
of radial velocities from the Doppler shift of stellar spectra to
within $\delta v\!\sim\!25\,\mathrm{km\, s^{-1}}$, or $\delta\lambda/\lambda\!\sim\!10^{-4}$
(Eisenhauer et al. \citeyear{Eis05}). This limit is partly instrumental
and partly due to the fact that the stars very close to the MBH have
hot ({}``early type'') stellar spectra (\S\ref{ss:1pc}, \S\ref{s:OBriddle})
with only a few strong IR lines, which are typically broad with $\left\langle v_{\mathrm{rot}}\sin i\right\rangle \!\sim\!150\,\mathrm{km\, s^{-1}}$
(Ghez et al. \citeyear{Ghe03a}; Eisenhauer et al. \citeyear{Eis05};
Fig. \hlink{f:S2spectrum}). 

The observed radial velocity $\beta_{r}$ can be expanded in terms
of the true magnitude of the 3D stellar velocity $\beta$ as 

\begin{equation}
\beta_{r}=\Delta\lambda/\lambda=B_{0}+B_{1}\beta+B_{2}\beta^{2}+{\cal {O}}(\beta^{3})\,.\label{e:br}\end{equation}
Several Newtonian and post-Newtonian terms contribute to the three
leading orders. The relativistic Doppler shift \begin{equation}
\beta_{D}\!=\!\frac{1+\beta\cos\vartheta}{\sqrt{1-\beta^{2}}}-1\!\simeq\!\beta\cos\vartheta+\frac{1}{2}\beta^{2},\end{equation}
can be decomposed to the first order classical Doppler effect ($\vartheta$
is the angle between the velocity vector $\bmath{\beta}$ and the
line of sight) and to the second order transverse relativistic Doppler
redshift. The gravitational redshift of a source at distance $r$
from the MBH as measured by an observer at infinity (Eq. \ref{e:Gparm})
is\begin{equation}
\beta_{z}=\frac{\Delta\lambda}{\lambda}=\left(1-\frac{r_{S}}{r}\right)^{-1/2}-1\simeq\frac{1}{2}\Upsilon(r)=\frac{1}{2}\left[\Upsilon_{0}(a)+\beta^{2}\right]\,.\label{e:GRz}\end{equation}
The gravitational redshift contributes a constant term, which depends
on the orbital semi-major axis, or equivalently, the energy, and a
second order term of the same magnitude as the transverse Doppler
redshift. 

The Roemer (or light travel time) delay is a Newtonian effect caused
by the fact that the light travel time from the star to the observer
changes with the orbital phase when the orbital plane is not observed
face-on (see also Loeb \citeyear{Loe03}). The phase-dependent mismatch
between the signal's emission and arrival times leads to apparent
deviations from the Newtonian orbit. The exact value of the effect
depends on the orbital parameters. Its magnitude and $\beta$-dependence
can be estimated by considering a circular edge-on orbit, in which
case the maximal delay is the light crossing time of the diameter
of the orbit $\delta t\!\sim\!2a/c$ and the maximal \emph{}apparent
velocity shift is $\beta_{R}\!\sim\!|\dot{\beta}_{r}|\delta t\!\sim\!(Gm/ca^{2})\delta t=2\beta^{2}$.
The Roemer time delay enters radial velocity measurements as a ${\cal {O}}(\beta^{2})$
correction, with a pre-factor $B_{R}$ that is typically larger than
that of either the transverse Doppler shift or the gravitational redshift.
Note that the Roemer delay expresses itself in the proper motions
as a \emph{relative} shift $\delta r/a$ of order ${\cal {O}}(\beta$),
since $\delta r/a\!\sim\!(2a/a)(\delta t/P)\!\sim\!2\beta$.

The observed radial velocity (Eq. \ref{e:br}) can therefore be written
as \begin{equation}
\beta_{r}=\left[\beta_{\odot}+\beta_{z,\mathrm{gal}}+\beta_{z,\star}+\frac{1}{2}\Upsilon_{0}\right]+[\cos\vartheta]\beta+[1+B_{R}]\beta^{2}+{\cal {O}}(\beta^{3})\,,\label{e:br2}\end{equation}
where the constant coefficient $B_{0}$ is expressed in terms of (1)
the constant velocity shift $\beta_{\odot}$ due to the compound motion
of the Sun and the Earth relative to the GC as well as the blueshift
due to the potential wells of the Sun, Earth and planets ; (2) the
constant velocity shift $\beta_{z,\mathrm{gal}}$ due to redshift
by the potential of the Galaxy; (3) the gravitational redshift $\beta_{z,\star}$
due to the star's potential%
\footnote{The constant terms $\beta_{\odot}$, $\beta_{z,\mathrm{gal}}$ and
$\beta_{z,\star}$ are very small compared to $B_{2}\beta^{2}$ ($cB_{2}\beta^{2}\!\sim\!{\cal {O}}(100\,\mathrm{km\, s^{-1}})$
at periapse for S2/S0-2). The peculiar radial motion of the sun relative
to the GC is estimated to be $<\!10\,\mathrm{km\, s^{-1}}$ (Gould
\& Popowski \citeyear{Gou98}) and the transverse Doppler redshift
is $\Theta_{0}{}^{2}/2c\!\simeq\!0.08\,\mathrm{km\, s^{-1}}$. The
constant gravitational redshift due the the Galactic potential (in
addition to that of the BH, which is taken into account by Eq. \ref{e:GRz})
is $c\beta_{z,\mathrm{gal}}\!\sim\!\Theta_{0}{}^{2}/c\!\simeq\!0.16\,\mathrm{km\, s^{-1}}$.
The gravitational blueshift due to the proximity of the Sun is only
$-v_{\oplus}^{2}/c\!=\!-0.003\,\mathrm{km\, s^{-1}}$, where $v_{\oplus}\!\simeq\!30\,\mathrm{km\, s^{-1}}$
is the Earth's orbital velocity. Likewise, the redshift due to the
star's potential and the blueshift due to the Earth's potential are
only $c\beta_{z,\star}\sim\!1\,\mathrm{km\, s^{-1}}$ and $-GM_{\oplus}/cR_{\oplus}\!\simeq\!-0.0002\,\mathrm{km\, s^{-1}}$
respectively. %
}, and (4) The constant contribution $\Upsilon_{0}/2$ from the gravitational
redshift of the star in the potential well of the MBH.

Since stars are observed with $\beta\!\sim\!\mathrm{few}\!\times\!0.01$
at periapse and since $B_{2}\!=\!1+B_{R}\!\sim\!{\cal {O}}(1)$, it
follows that second order effects are detectable with existing instruments,
$B_{2}\beta^{2}\!\sim\!10^{-3}\!>\!\delta\lambda/\lambda\!\sim\!10^{-4}$.
Third order GR effects, such as frame dragging (\S\ref{sss:LTeffect}),
Gravitational lensing and the Shapiro delay (relevant only for a nearly
edge-on orbit and even then very small, Kopeikin \& Ozernoy \citeyear{Kop99}),
or GR periapse shift (\S\ref{sss:GRperishift}), which is only a
$\Delta\beta\!\sim\!\Delta\omega r_{p}/\tau_{p}\!\sim\!{\cal {O}}(\beta^{3})$
(Eq. \ref{e:GRprecess}) effect in the radial velocities (although
it is a ${\cal {O}}(\beta^{2})$ effect in the proper motions) will
not be detectable with the available radial velocity data. It should
be noted that second order effects involving proper motion, such as
the relativistic periapse shift, will be very hard to detect with
the present astrometric precision. 

In practice, the ability to measure the post-Newtonian deviations
from a Newtonian orbit requires an accurate determination of the Keplerian
parameters, and so reliable proper motion data are still needed. In
particular, radial velocity data for a single star on a ${\cal {O}}(\beta^{2})$
post-Newtonian orbit can always be expressed as a Newtonian orbit
(including the classical Roemer effect) with suitably modified orbital
parameters (the degeneracy between $\Upsilon_{0}$ and $\beta_{\odot}$
in Eq. \ref{e:br2} is an expression of this fact). Multiple orbits
and proper motions are needed to break this degeneracy (Zucker \&
Alexander 2005, in preparation). Post-Newtonian orbit solving simulations
with proper motions and radial velocities, such as those conducted
by Weinberg, Milosavljevi\'c \& Ghez (\citeyear{Wei04}) for Newtonian
orbits, are needed to assess the practical observational requirements
for detecting these effects.

A similar approach to detecting post-Newtonian effects with radio
pulsars, which may exist around the MBH, was studied by Pfahl \& Loeb
(\citeyear{Pfa04}). The problems are that it is unknown whether such
a population exists, and that radio observations at the required sensitivity
will be extremely challenging (cf \S\ref{ss:Mseg}).

\subsubsection{Periapse shift}

\label{sss:GRperishift}

The deviation of the GR potential from the Newtonian $1/r$ point
mass potential leads to rosette-like orbits with a prograde shift
in the argument of pericenter $\omega$ (see footnote \ref{n:orbparms})
of $\Delta\omega$ per orbit, which for a non-rotating black hole
is (Weinberg \citeyear{Wei72}, Eq. 8.6.11)\begin{equation}
\Delta\omega=\frac{3\pi}{\left(1-e^{2}\right)}\frac{r_{S}}{a}\,.\label{e:GRprecess}\end{equation}
 The periapse advance is a ${O}(\beta^{2})$ effect. The effect of
the periapse advance on the orbital shape is maximal at the apoapse.
The change in the angular position of the apoapse, $\mathbf{p}_{a}$,
per orbit (neglecting corrections for the orbital orientation) depends
only on the orbital eccentricity, \begin{equation}
\Delta p_{a}\sim\Delta\omega\frac{a(1+e)}{R_{0}}=\frac{3\pi r_{S}}{(1-e)R_{0}}\,.\end{equation}
 For S2/S0-2, the shortest period star detected to date ($P\!\sim\!15\,\mathrm{yr}$,
$a\!\sim\!4.5\,\mathrm{mpc}$, $e\!\sim\!0.88$ Eisenhauer et al.
\citeyear{Eis05}; Ghez et al. \citeyear{Ghe05}), the shift is $\Delta p_{a}\!\sim\!1$
mas per period. A similar shift is predicted for the star S14/S0-16
with the most eccentric orbit solved to date ($P\!\sim\!38\,\mathrm{yr}$,
$e\!\sim\!0.94$). A more efficient use of the orbital data is to
fit an orbital model including the periapse shift (i.e. with a variable
argument of periastron $\omega$) to all the data points. Either way,
unless a very eccentric, very short period star is discovered, the
detection of the periapse advance will require very long monitoring
(see detailed discussions of the observability of the effect by Jaroszynski
\citeyear{Jar98b}; Fragile \& Mathews \citeyear{Fra00}; Rubilar
\& Eckart \citeyear{Rub01}; Weinberg, Milosavljevi\'c \& Ghez \citeyear{Wei04}). 

The prospects of discovering the GR periapse advance are further complicated
by the likely existence of a distribution of faint stars and dark
stellar remnants around the MBH, which introduces a retrogade periapse
shift (\S\ref{sss:distrDM}, \S\ref{ss:Mseg}). Measurements of
the periapse shifts in several stellar orbits will be needed to disentangle
the two countervailing effects (Rubilar \& Eckart \citeyear{Rub01}).

\subsubsection{Frame dragging}

\label{sss:LTeffect}

There is some evidence from the variability of the IR and X-ray accretion
flares that the Galactic MBH is spinning in at least half its maximal
allowed rate (Genzel et al. \citeyear{Gen03b}; Aschenbach et al.
\citeyear{Asc04}). A spinning mass with specific spin $S_{\bullet}$
induces secular changes in the trajectory of a test particle orbiting
 it through gravitomagnetic GR effects known as {}``frame dragging''
or Lense-Thirring precession%
\footnote{A related but distinct gravitomagnetic effect, not discussed here,
is the Lense-Thirring precession of the spin of the test particle
itself.%
}. The $S_{\bullet}$-induced change in the longitude of the ascending
node $\Omega$ (see footnote \ref{n:orbparms}) corresponds to the
precession of the orbital angular momentum vector $\mathbf{J}$ around
the MBH spin vector $\mathbf{S}_{\bullet}$. The change in the argument
of pericenter, $\omega$, corresponds to a periapse shift, in addition
to the $S_{\bullet}$-independent shift (Eq. \ref{e:GRprecess}). 

In the weak field limit, the change per orbital period in $\Omega$
is (Thirring \& Lense \citeyear{Thi18})\begin{equation}
\Delta\Omega_{\mathrm{LT}}=\sqrt{2}\pi\frac{s}{(1-e^{2})^{3/2}}\left(\frac{r_{S}}{a}\right)^{3/2}\,,\label{e:dWLT}\end{equation}
and the period of the precession is\begin{equation}
P_{\Omega_{\mathrm{LT}}}=4\pi\frac{r_{S}}{c}\frac{(1-e^{2})^{3/2}}{s}\left(\frac{a}{r_{S}}\right)^{3}\,,\end{equation}
where $s\!\equiv\! S_{\bullet}/(Gm/c)$\label{d:sMBH} is the dimensionless
spin parameter ($0\!\le\! s\!\le\!1$). The change per orbital period
in the argument of pericenter is \begin{equation}
\Delta\omega_{\mathrm{LT}}=-3\Delta\Omega_{\mathrm{LT}}\cos i\,,\label{e:dwLT}\end{equation}
where $\cos i\equiv\mathbf{J}\cdot\mathbf{S}_{\bullet}/JS_{\bullet}$.
The frame dragging shifts are ${\cal {O}}(\beta^{3}$) effects.

The predicted magnitude of the effect in the GC is small (Jaroszynski
\citeyear{Jar98b}), and it is unlikely that it will be measured directly
from the orbits even with the proposed Thirty Meter Telescope (Weinberg,
Milosavljevi\'c \& Ghez \citeyear{Wei04}). However, over the lifespan
of one of the young S-stars near the MBH, $t_{\star}\!\sim\!10$ Myr,
the accumulated precession can be substantial if the orbit is eccentric
enough. Levin \& Beloborodov (\citeyear{Lev03}) note that for the
eccentric, short-period star S2/S0-2 ($P_{\star}\!=\!15.2$ yr, $e\!=\!0.87$,
Sch\"odel et al. \citeyear{Sch02}) the Lense-Thirring period is
$P_{\Omega_{\mathrm{LT}}}\!\sim\!{\cal {O}}(10^{7}\,\mathrm{yr})$.
They propose that S2 was initially a member of the thin star disk
(\S\ref{ss:1pc}), orbiting in the plane of the disk but on a particularly
eccentric orbit, and that it was gradually dragged out of the plane
by the Lense-Thirring effect, to the point where its orbit today appears
unrelated to the disk. Such effect, if measured in two stars could
be used to determine the spin axis. They further propose that this
could be the origin of the other S-stars as well. However, the orbital
solutions of several other S-stars do not support this idea, since
for a large fraction of the S-stars $P_{\Omega_{\mathrm{LT}}}\!\gg\! t_{\star}$
(Eisenhauer et al. \citeyear{Eis05}).

\subsection{Gravitational lensing}

\label{ss:GL}

Gravitational lensing of stars by the Galactic MBH differs from the
other stellar processes discussed here in that it is a process that
affects the light emitted by the stars, rather than the stars themselves,
and in that the lensed stars are typically well outside the dynamical
region of influence of the MBH. However, the lensed images can masquerade
as stars near the MBH or as accretion flares from the MBH and complicate
the interpretation of the observations. On the other hand, gravitational
lensing may be used to probe the dark mass and the stars around it,
and to locate the MBH on the IR grid, where the stars are observed.

There is yet no clear evidence for gravitational lensing by the MBH.
This is consistent with various estimates that suggest that there
are not enough luminous sources behind the MBH for gravitational lensing
to be important for present-day observations, although future, deep
observations may detect lensing events (Wardle \& Yusef-Zadeh \citeyear{War92};
Jaroszynski \citeyear{Jar98a}; Alexander \& Sternberg \citeyear{Ale99a};
Alexander \& Loeb \citeyear{Ale01c}). Nevertheless, it is worthwhile
to consider the possible roles of gravitational lensing in the observations
and study of the Galactic Center. This is important not only in anticipation
of future observations, but also because the estimates of the lensing
probability are quite uncertain (they involve models of the unobserved
far side of the Galaxy) and may under-estimate the true rate of such
events.

\subsubsection{Point mass gravitational lens}

\label{sss:pointGL}

The MBH dominates the central potential and so gravitational lensing
in the GC can be described to first approximation as lensing by a
point mass. A light ray arriving from a background source to the observer
is bent by an angle $\alpha_{L}$ as it passes near a point mass lens
(Fig. \hlink{f:GL}). In the small angle approximation, which is valid
as long the impact parameter $b_{L}$ is much larger than the event
horizon ($b_{L}\!\gg\! r_{S}$), the bending angle is given by 

\begin{equation}
\alpha_{L}=2r_{S}/b_{L}\,,\label{e:bend}\end{equation}
 (see Schneider, Ehlers \& Falco \citeyear{SEF92} for a comprehensive
treatment of gravitational lensing). The bending angle of a point
mass gravitational lens diverges towards the center ($b_{L}\!=\!0$)
and decreases as the impact parameter grows. This is in the opposite
sense to a spherical glass lens, where the bending angle is zero when
the ray goes through the lens center and increases with the impact
parameter. Thus, unlike a spherical lens, a point mass gravitational
lens does not produce a faithful image of the lensed source, but rather
breaks, warps and/or flips the image. A point mass lens creates two
images of the source, one on either side of the lens. The two images
are always in focus at the observer, regardless of the distance of
the source behind the lens. By symmetry, the two images, the lens
and the (unobserved) source all lie on one line (Fig. {\small \hlink{f:GL2im}}).
The typical angular cross-section of the lens is given by the Einstein
angle,\label{d:ThetaE}

\begin{equation}
\theta_{E}=\sqrt{2r_{S}\frac{D_{LS}}{D_{OS}D_{OL}}}\,,\label{e:ThetaE}\end{equation}
 where $D_{OL}$\label{d:Dol} is the observer-lens distance ($R_{0}$
for the Galactic MBH), $D_{LS}$\label{d:Dls} is the lens-source
distance, and $D_{OS}$\label{d:Dos} is the observer-source distance%
\footnote{In a flat background spacetime, which is the relevant background for
lensing on Galactic scales, $D_{OS}\!=\! D_{OL}\!+\! D_{LS}$. In
curved spacetime, which is the relevant background for lensing on
cosmological scales, the distances are the angular diameter distances,
and this simple sum no longer holds. %
}. The physical size of the Einstein angle at the source plane is $R_{E}\!=\! D_{OS}\theta_{E}$\label{d:RE}.
At a fixed observer-lens distance, $\theta_{E}$ reaches a maximal
value for a source at infinity, \begin{equation}
\theta_{E}\!=\!\theta_{\infty}\sqrt{\frac{D_{LS}}{D_{OS}}}\,,\qquad\theta_{\infty}=\sqrt{\frac{4Gm}{c^{2}D_{OL}}}\sim2"\,\,\mathrm{in\, the}\,\mathrm{GC}\,.\label{e:ThetaEinf}\end{equation}

\begin{figure}[!t]
\centering{{\small \htarget{f:GL}}\includegraphics[%
  scale=0.9]{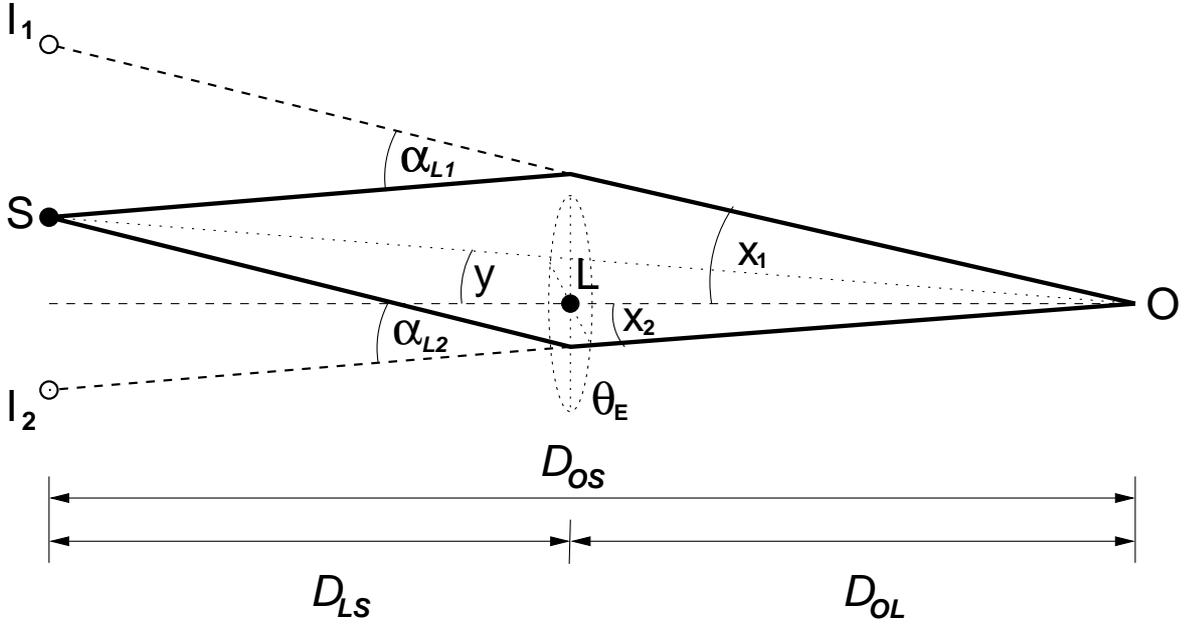}}

\caption{{\small \label{f:GL}Schematics of gravitational lensing by a point
mass in the small bending angle limit. An observer (O) sees light
rays from a background source (S), which is located at a distance
$D_{OS}$ away, deflected by angles $\alpha_{L1}$, $\alpha_{L2}$
in the potential of a massive lens (L), which is located at a distance
$D_{OL}$ from the observer and $D_{LS}$ from the source. The lens
mass and the distances define an Einstein ring of angular opening
$\theta_{\mathrm{E}}$ (Eq. \ref{e:ThetaE}), perpendicular to the
optical axis (O--L). The source (S), which is not observed directly,
is located at an angle $y\theta_{\mathrm{E}}$ relative to the optical
axis. It appears as two lensed images ($\mathrm{I}{}_{1}$, $\mathrm{I}_{2}$)
at angular positions $x_{1}\theta_{\mathrm{E}}$ and $x_{2}\theta_{\mathrm{E}}$,
one inside the Einstein ring and one outside it. }}
\end{figure}

The relation between $y$\label{d:GLy}, the angular position of the
source relative to the observer-lens axis (the optical axis) and $x_{1,2}$\label{d:GLx},
the angular positions of the images, can be derived from the geometry
of the light paths in the small bending angle limit (Fig. \hlink{f:GL}),

\begin{equation}
y=x_{1,2}-1/x_{1,2}\,,\qquad\mathrm{or}\qquad x_{1,2}=\frac{1}{2}\left(y\pm\sqrt{4+y^{2}}\right)\label{e:yx}\end{equation}
 where $x_{1,2}$ and $y$ are measured in terms of $\theta_{E}$
and $x_{2}\!<\!0$ by definition. The angular separation between the
two images is ${\cal {O}}(\theta_{E})$. Gravitational lensing conserves
surface brightness, and so the magnification $A_{1,2}$ \label{d:Amag}in
the flux of each image relative to that of the unlensed source is
proportional to the apparent change in the angular area of the source,

\begin{equation}
A_{1,2}=\left|\partial\mathbf{y}\left/\partial\mathbf{x}_{1,2}\right.\right|^{-1}=\left|1-x_{1,2}^{-4}\right|^{-1}\,.\label{e:GLA}\end{equation}
 The primary image at $x_{1}$ is always magnified. The secondary
image at $x_{2}$ can be demagnified to zero. The two magnifications
obey the relations 

\begin{equation}
A_{1}=A_{2}+1\geq1\,,\qquad A\equiv A_{1}+A_{2}=\frac{y^{2}+2}{y\sqrt{y^{2}+4}}\,.\label{e:A}\end{equation}
 When $y\!=\!0$ the magnification formally diverges and the image
appears as a ring of angular size $\theta_{E}$, the Einstein ring.
This divergence is avoided in practice by the finite size of the source
(e.g. a star). Finite sized sources are also sheared tangentially
around the Einstein ring as the magnification increases. In the limit
of high magnification, or small source angle $(y\!\ll\!1)$, $A\!\sim\!1/y$.
It is customary to define $A(y\!=\!1)\!=\!1.34$ as the threshold
of a microlensing event. 

Relative motion between the source, lens and observer will manifest
itself as two moving images with time-variable magnifications (Fig.
{\small \hlink{f:GL2im}}). The term {}``macrolensing'' is used
to describe the situation where the images can be resolved, and {}``microlensing''
the situation where the images are not separately resolved, and only
their joint variable magnified flux is observable. When the relative
angular velocity $\dot{y}$ between the source and lens is constant,
the microlensing light-curve is given by substituting $y(t)\!=\![y_{0}^{2}+\dot{y}^{2}(t-t_{0})^{2}]^{1/2}$
in Eq. (\ref{e:A}), where $y_{0}$ is the angular impact parameter
of the source trajectory relative to the lens and $t_{0}$ is the
time of maximal magnification when $y\!=\! y_{0}$ (Fig. {\small \hlink{f:GL2im}}).
Microlensing light-curves are distinguished from other astrophysical
flares by their characteristic time-symmetric functional form and
by their wavelength-independence, which is a consequence of the equivalence
principle.

\begin{figure}[!t]
\centering{{\small \htarget{f:GL2im}}\begin{tabular}{cc}
\includegraphics[%
  width=0.45\textwidth,
  keepaspectratio,
  angle=270]{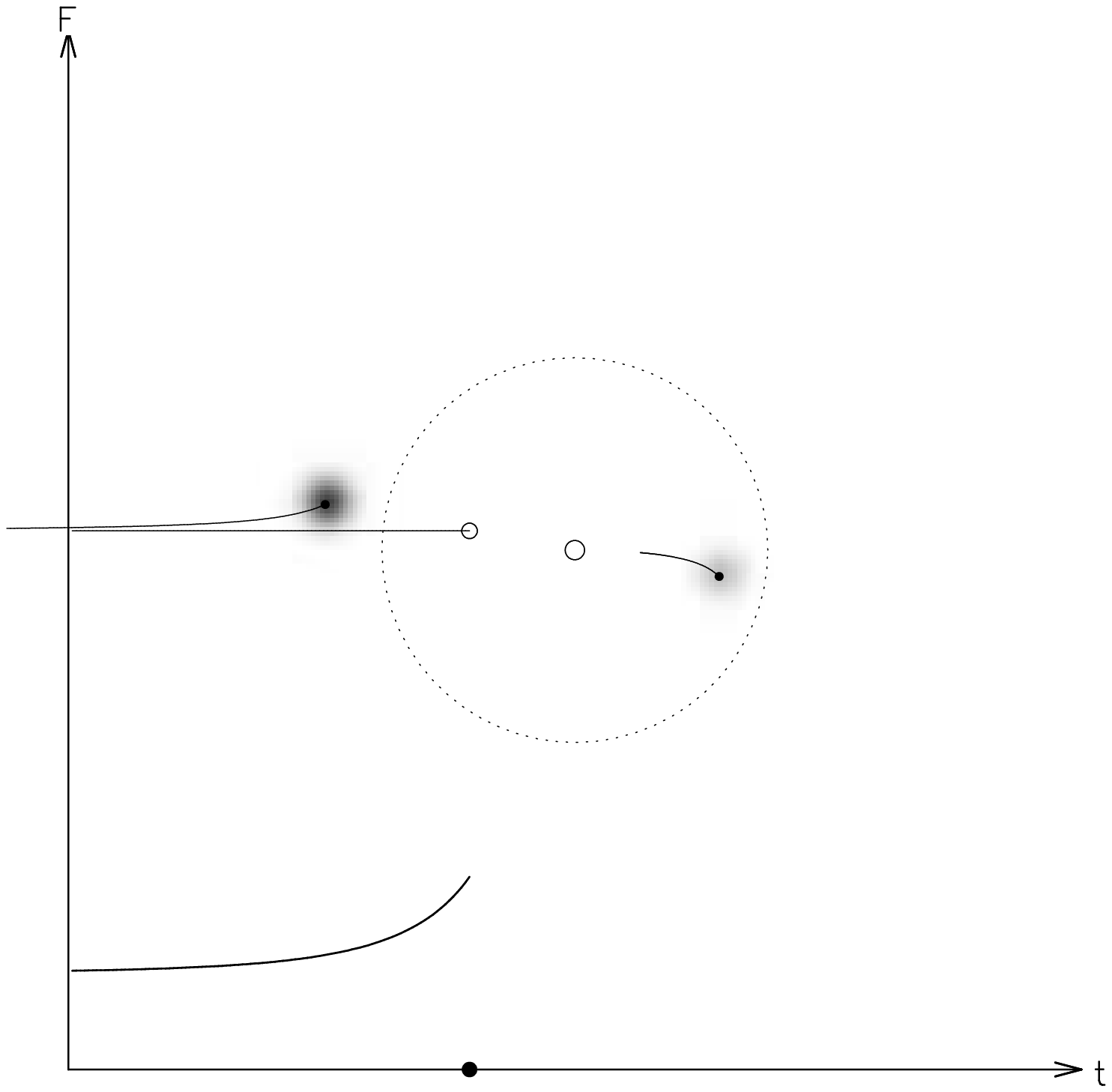}&
\includegraphics[%
  width=0.45\textwidth,
  keepaspectratio,
  angle=270]{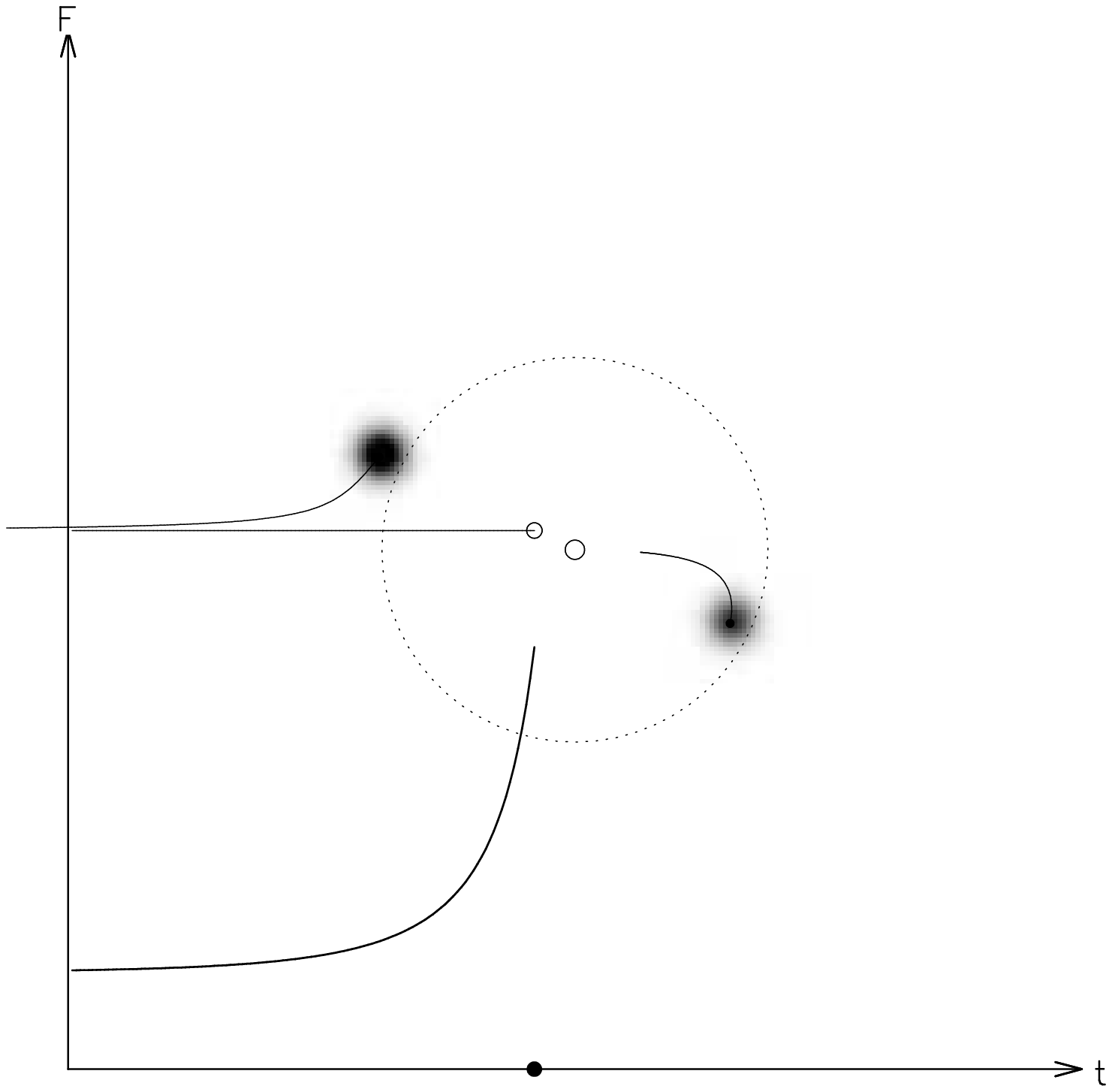}\tabularnewline
\includegraphics[%
  width=0.45\textwidth,
  keepaspectratio,
  angle=270]{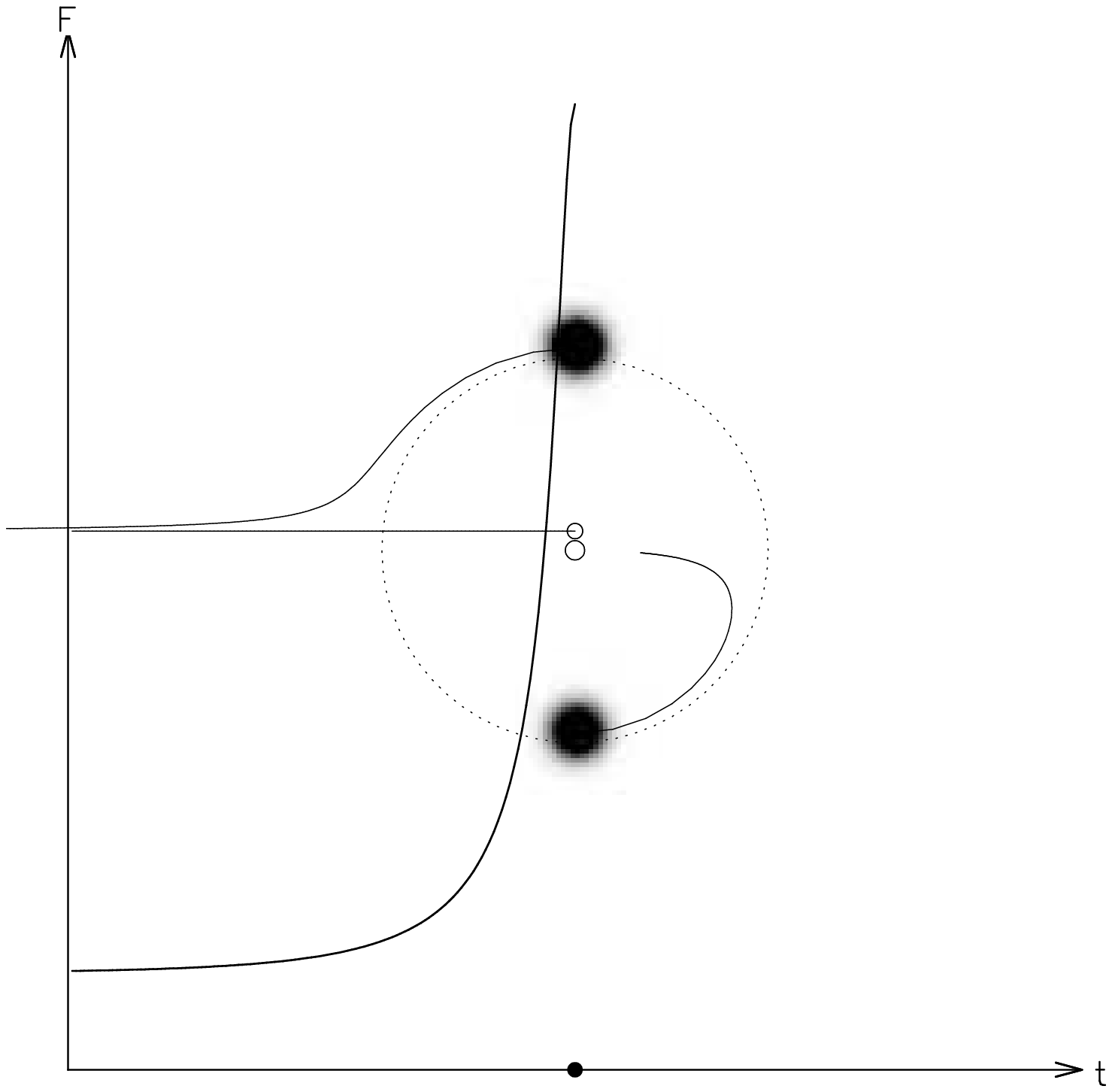}&
\includegraphics[%
  width=0.45\textwidth,
  keepaspectratio,
  angle=270]{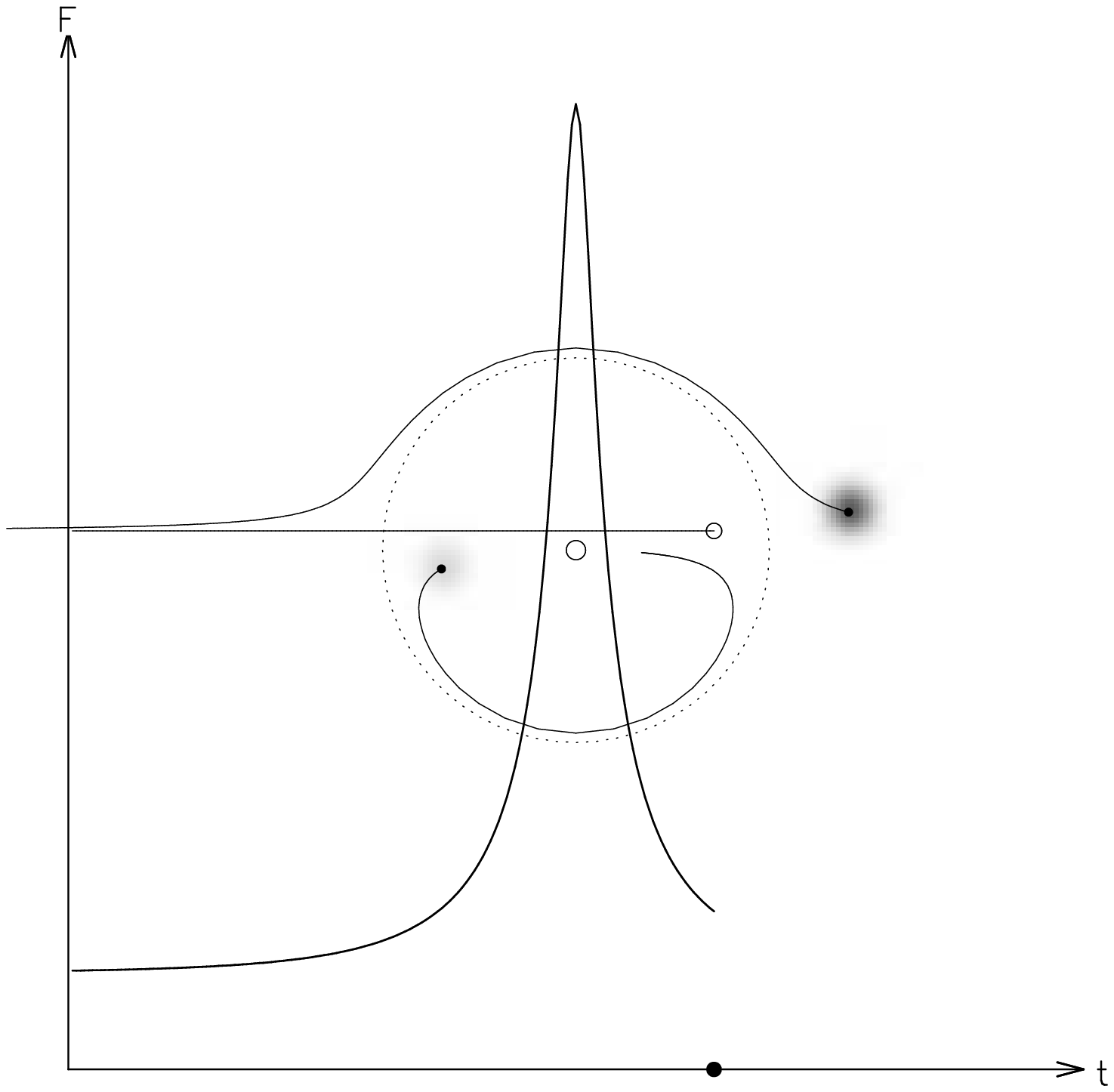}\tabularnewline
\end{tabular}}

\caption{{\small \label{f:GL2im}A time sequence (left to right, top to bottom)
of a gravitational lensing event of a point source by a point mass.
The two resolved images and their trajectories (fuzzy circles, whose
size and shade are proportional to the flux, with curved lines tracking
the image trajectories) are superimposed on a plot of the microlensing
light-curve (total flux $F$ from both images as function of time
$t$ in arbitrary units). Note that the tangential shear of the images
is not shown here. The background source, which is not observed directly,
(open circle with straight line tracking the source trajectory) moves
in projection from left to right behind the MBH (open circle at center)
with an impact parameter of $0.1\theta_{E}$. The two images move
in tandem clockwise about the Einstein ring (large dotted circle).
The strongly magnified image (top) is always outside the Einstein
ring and is always brighter than the source. The weakly magnified
image (bottom) is always inside the Einstein ring and can be strongly
demagnified. }}
\end{figure}

\subsubsection{Gravitational lensing by the Galactic MBH}

\label{sss:GLGC}

In order to plan the observational strategy for detecting gravitational
lensing, or to estimate how likely it is that an observed flare is
due to lensing, it is necessary to evaluate the detection probability.
Two quantities are commonly used for this purpose, the optical depth
and the lensing rate. The probability $P_{L}\!=\!1-\exp(-\tau_{L})$
for finding at any given time a star lensed by the Galactic MBH is
defined in terms of an optical depth $\tau_{L}=\int_{D_{OL}}^{\infty}n_{\star}\pi R_{E}^{2}dD_{OS}$,
which expresses the mean number of background sources within an angle
$\theta_{E}$ of the line of sight to the MBH. Rough estimates predict
$\tau_{L}\!\sim\!{\cal {O}}(1)$ (Alexander \citeyear{Ale01d}; Alexander
\& Loeb \citeyear{Ale01c}) for lensing by the MBH (assuming no limits
on the photometric sensitivity). 

The optical depth is a useful quantity in situations such as Galactic
microlensing searches ($\tau_{L}\!\sim\!{\cal {O}}(10^{-6})$, e.g.
Afonso et al. \citeyear{Afo03}), when there are many possible lensed
lines of sights, and $\sim\!\tau_{L}^{-1}\!\sim\!10^{6}$ background
stars must be monitored simultaneously to find the rare one that is
lensed by an intervening star. The observational situation for gravitational
lensing by the MBH in the Galactic Center is different because the
position of the lens is known and fixed, and so there is only one
line of sight to monitor. The optical depth does not take into account
the relative motions of the observer, lens and source, which reshuffle
their random alignment and introduces a timescale to the problem.
A more relevant quantity in this situation is the lensing rate $\Gamma_{L}$\label{d:GLrate},
which indicates how long it is necessary to monitor the GC on average
to detect a lensing event with peak flux above a flux detection threshold
$F_{0}$,

\begin{equation}
\Gamma_{L}(>F_{0})\simeq\int_{D_{OL}}^{\infty}n_{\star}v_{\perp}\frac{2R_{E}}{A}dD_{OS}\,,\qquad\mathrm{for}\: A\geq\frac{F_{0}}{\left.L_{\star}\right/4\pi D_{OS}^{2}}\,,\label{eq:GLrate}\end{equation}
where $v_{\perp}$ is the velocity of the sources perpendicular to
the line of sight relative to the optical axis, including the contribution
from the reflex motion due to the observer, $\Ls$ is the source luminosity
and $A\!\gg\!1$ is assumed. For practical applications, equation
(\ref{eq:GLrate}) has to be modified to take into account the range
of stellar luminosities and velocities, dust extinction, the total
duration of the observations $T$ and the sampling rate $\Delta T$
(the mean duration of events amplified by more than $A$ is $\overline{t}_{L}\!=\!\pi R_{E}\left/2Av_{\perp}\right.$;
only events with $\Delta T\!<\!\overline{t}_{L}\!<\! T$ can be detected)
(Alexander \& Sternberg \citeyear{Ale99a}).

The observational limitations, $F_{0}$, $T$ and $\Delta T$ and
the telescope's angular resolution $\delta\phi$, place restrictions
on $D_{LS}$ and the impact parameter for which sources can be detected,
and affect the appearance of the lensing event, the typical timescales
and the peak magnifications that are likely to be observed. The angular
distance between the two images close to peak magnification is $\sim\!2\theta_{E}$
(Eq. \ref{e:yx}). A lensed star will appear as a microlensing event
if it is closer than $D_{\mu}$ behind the MBH, 

\begin{equation}
D_{\mu}=\frac{R_{0}}{\left(\theta_{\infty}/\delta\phi\right)^{2}-1}\sim0.2\,\mathrm{pc}\,\left(\frac{R_{0}}{8\,\mathrm{kpc}}\right)\left(\frac{\delta\phi/10\,\mathrm{mas}}{\theta_{\infty}/2"}\right)^{2}.\end{equation}
 Stars farther away will appear as two resolved images. The Einstein
angle grows from zero to $\theta_{\infty}$with source distance behind
the MBH (Eq. \ref{e:ThetaEinf}), and so the lensing cross-section
and the lensing duration (for a given $\max A$ and for constant $v_{\perp}$)
also increase. In addition, high magnification events typically have
longer time scales because the source trajectory must have a smaller
impact parameter and so spends more time inside the Einstein radius.
Therefore, observations with limited temporal sampling will tend to
pick out high magnification events. 

A couple of early detections of transient flaring events very close
to $\SgrA$ (Genzel et al. \citeyear{Gen97}; Ghez et al. \citeyear{Ghe98})
were considered as lensing event candidates. For one of these a light
curve was recorded, but as it was under-sampled only estimates of
a timescale ($\sim\!1$ yr) and a typical magnification ($\sim\! A>5$)
could be derived from it. The a-posteriori probability of detecting
a lensing event (for the relevant observational limitations) was estimated
to be only $0.5\%$, but on the other hand, the observed timescale
and magnification were close to the expected median value (Alexander
\& Sternberg \citeyear{Ale99b}). The interpretation of this event
remains inconclusive. The likely alternative to lensing is that these
were faint stars near the MBH that were observable only intermittently
when they were well separated from the more luminous stars nearby.
It should be noted that the $K$-band IR flares observed from $\SgrA$
have a typical timescale of ${\cal O}(1\,\mathrm{hr})$ (although
year-long variability is observed in longer IR wavelengths, Cl\'enet
et al. \citeyear{Cle04b}). They are too frequent to be microlensing
of near background stars and furthermore, their irregular light-curves
and quasi-periodic oscillations are inconsistent with the time-symmetric
microlensing light-curves (Genzel et al. \citeyear{Gen03b}).

Macrolensed image pairs, if hidden among the many stars observed around
the MBH, can be used to pinpoint the MBH on the IR grid directly from
the astrometric data, independently of other methods (\S\ref{ss:mass})
and with no assumptions about $m$, $R_{0}$ or the distribution of
stars behind the MBH (Alexander \citeyear{Ale01d}). The MBH lies
on the line connecting the two images of any background source it
gravitationally lenses (Fig. {\small \hlink{f:GL2im}}), and so the
intersection of these lines fixes its position. The point-mass lensing
equations (Eqs. \ref{e:yx}, \ref{e:A}) imply simple relations between
the fluxes of the two images, and between their angular positions
and velocities and as measured relative to the projected position
of the MBH,

\begin{equation}
-\theta_{1}\left/\theta_{2}\right.=-\dot{\theta}_{r,1}\left/\dot{\theta}_{r,2}\right.=\dot{\theta}_{t,1}\left/\dot{\theta}_{t,2}\right.=\sqrt{F_{1}\left/F_{2}\right.}\,,\label{e:pinpoint}\end{equation}
where $\dot{\theta}_{r,i}$ is the radial component of the projected
angular velocity of image $i$ ($i\!=\!1,2$), $\dot{\theta}_{t,i}$
is the tangential component and $F_{i}$ the flux of the image. The
simultaneous search for the position of the MBH and for lensed image
pairs proceeds by enumerating over a grid of possible MBH positions.
For each trial position, scores are assigned to all possible pairs
of stars according to how well they satisfy Eq. (\ref{e:pinpoint}).
The MBH position is then chosen as the one maximizing the total lensing
score, and the likeliest lensed image pair candidates can be recorded
for spectroscopic validation (lensed image pairs should have exactly
the same spectra, up to small deviations due to differences in the
dust extinction along the two light paths). Such a search on an early
astrometric data set produced a statistical detection of the MBH inside
the error region on the center of acceleration (Fig. {\small \hlink{f:acc}}),
with a chance detection probability of only $5\!\times\!10^{-4}$,
where most of the signal came from the intersection of two pairs of
lensed image candidates, tentatively identified as far background
blue and red supergiants (simple models of the distribution of stars
and dust in the far side of the Galaxy predict none within $\theta_{E}$
of the MBH). However, a subsequent recalibration of the data, and
additional color information on the lensed image candidates (Cl\'enet
et al. \citeyear{Cle04a}), cast strong doubt on this result. This
method has yet to be applied to the newer, more comprehensive and
accurate astrometric data sets. 

The MBH can modify the observed stellar surface density inside $\sim\!\theta_{\infty}$
(Wardle \& Yusef-Zadeh \citeyear{War92}). A lens magnifies by enlarging
the angular size of the unlensed sky behind it. Since surface brightness
is conserved, the fluxes of sources are magnified by the same amount.
If all the stars can be detected even without magnification, then
the effect of lensing is to decrease the surface density of sources.
However, if the magnification reveals faint stars, which could not
be observed unlensed, then the observed surface density can be larger
than the unlensed density if enough faint sources are magnified above
the detection threshold to over-compensate for the decrease in surface
density ({}``positive magnification bias''), or it could be smaller
if the converse is true ({}``negative magnification bias''). The
lensed luminosity function (stellar number surface density $\Sigma$
per flux interval) is related to the unlensed one by

\begin{equation}
\left.\left(\frac{\mathrm{d}\Sigma}{\mathrm{d}F}\right)_{\mathrm{lensed}}\right|_{F}=A^{-2}\left.\left(\frac{\mathrm{d}\Sigma}{\mathrm{d}F}\right)_{\mathrm{unlensed}}\right|_{F/A}\,.\label{e:GLbias}\end{equation}
 In many cases the luminosity function is well approximated by a power-law,
$\mathrm{d}\Sigma\left/\mathrm{d}F\right.\propto F^{-s}$ (\S\ref{ss:100pc}).
For $s\!=\!2$ the decrease in the total surface density is exactly
balanced by the magnification of faint stars above the detection threshold.
The chances for the detection of this effect in the Galactic Center
appear small. A statistically meaningful detection requires a very
high surface density that probably exceeds even that around the MBH
(Wardle \& Yusef-Zadeh \citeyear{War92}), and furthermore, models
of the stellar luminosity function in the inner Galactic Center suggest
that $s\sim2$ (Alexander \& Sternberg \citeyear{Ale99a}).

Gravitational lensing will also affect the astrometric measurements
of stars in the near vicinity of the MBH (Jaroszynski \citeyear{Jar98a}).
The bright lensed image of a star on nearly edge-on orbits (within
about $\pm\!5^{\circ}$of the line-of-sight to the MBH, as measured
in a coordinate system centered on the MBH) will exhibit a small astrometric
deviation of ${\cal {O}}(0.1\,\mathrm{mas)}$ in the Keplerian orbit
as it passes behind the MBH (Nusser \& Broadhurst \citeyear{Nus04}).
The astrometric precision required to detect such a tiny effect is
well beyond present-day capabilities. If measured by future telescopes,
the form of the orbital deviation can distinguish between different
dark mass distributions (\S\ref{ss:DM}) and be used to break the
$m/R_{0}^{3}$ degeneracy of astrometric Keplerian orbits (\S\ref{sss:orbsol}).

\subsubsection{Beyond the point lens approximation}

\label{sss:extGL}

The point mass lens model and the small deflection angle assumption
provide a very useful approximation for describing gravitational lensing
by the Galactic MBH. Nevertheless, there are several reasons to explore
more complicated mass distributions and large angle lensing. First,
it would be useful if gravitational lensing could be used to dispel
any remaining doubts that the dark compact mass in the Galactic Center
is indeed a MBH, and not some other extended distribution of matter
(e.g. Capozziello \& Iovane \citeyear{Cap99}; Nusser \& Broadhurst
\citeyear{Nus04}; \S\ref{ss:DM}). Unfortunately, it can be shown
that he behavior of high-magnification light curves near peak magnification
is universal and independent of the details of the lens (Schneider,
Ehlers \& Falco \citeyear{Sch92}, Eq. 11.21b). For spherically symmetric
mass distributions this implies that the light curves differ only
in the low magnification tails, which are much harder to observe.
The astrometric effects of lensing may provide a more promising way
to probe the dark matter distribution (\S\ref{ss:DM}). Second, large
angle lensing probes GR in the strong field limit. GR corrections
to the small angle approximation result in a variety of interesting
effects (Virbhadra \& Ellis \citeyear{Vir00}; Petters \citeyear{Pet03};
Bozza \& Mancini \citeyear{Boz04}), such as {}``retrolensing'',
where the Galactic MBH can redirect by $\theta\!\sim\!\pi$ the light
from a close luminous \emph{foreground} source (for example the star
S2) back at the observer (De Paolis et al. \citeyear{DeP03}). However,
these effects are minute and will be extremely hard to observe even
by the next generation of instruments. Third, the MBH is surrounded
by a massive stellar cluster. Because the stellar mass is not smoothly
distributed but is composed of discrete point masses, its effect on
the lensing properties of the MBH is much larger than one may naively
estimate by adding the stellar mass to that of the MBH (Alexander
\& Loeb \citeyear{Ale01c}). Enhanced lensing by the presence of discrete
stellar mass objects may be used to discover SBHs that have accumulated
near the MBH due to mass segregation (Miralda-Escud\'e \& Gould \citeyear{Mir00};
Chanam\'e, Gould \& Miralda-Escud\'e \citeyear{Cha01}; \S\ref{ss:Mseg}).

The effect of stars on lensing by the MBH is similar to that of planets
on microlensing by a star, a phenomenon that was studied extensively
for the purpose of planet detection (e.g. Gould \& Loeb \citeyear{Gou92}).
The lensing cross-section of an isolated star is $\theta_{E}^{2}(M_{\star})/\theta_{E}^{2}(M_{\bullet})=M_{\star}/M_{\bullet}\lesssim10^{-6}$
times smaller than that of the MBH (Eq. \ref{e:ThetaE}). However,
when the star lies near $\theta_{E}(M_{\bullet})$, the shear of the
MBH distorts its lensing cross-section, which develops a complex topology,
becomes radially elongated and is increased by up to an order of magnitude.
As stars orbit the MBH, their elongated cross-sections scan the lens
plane. If these happen to intersect one of the images of a background
source that is lensed by the MBH, the image will be split into 2 or
4 sub-images whose angular separation will be of order $\theta_{E}(M_{\star})$,
and so the sub-images will not be individually resolved. However,
their combined flux will be significantly magnified. This will increase
the probability of high magnification events over what is expected
for lensing by the MBH alone. The light curves of such events will
no longer be symmetric as they are for a point mass, but will exhibit
a complex temporal structure, and their typical variability timescales
will increase sharply for images that lie near $\theta_{E}(M_{\bullet})$
because of the increased stellar cross-section for lensing. Enhanced
lensing by stars in the Galactic Center is estimated to increase the
probability of $A>5$ lensing events by $\sim\!2$ and of $A>50$
events by $\sim\!3$ (Alexander \& Loeb \citeyear{Ale01c}).

\section{Strong star--MBH interactions }

\label{s:inter}

Some of the most spectacular stellar phenomena that can occur in the
GC involve strong interactions with the MBH. In this context {}``strong''
is defined as any interaction that cannot be described by Newtonian
gravity operating on point masses. This could be because the stars'
internal degrees of freedom can no longer be ignored, or because dissipative
interactions have to be included, or because GR effects must be taken
into account. 

Early interest in such processes focused on the consumption of stars
by the MBH, either directly, when a whole star crosses the event horizon,
or indirectly, when the gaseous debris left after the tidal disruption
of a star is accreted by the MBH (Hills \citeyear{Hil75}; Frank \&
Rees \citeyear{Fra76}; Lightman \& Shapiro \citeyear{Lig77}; Frank
\citeyear{Fra78}; Rees \citeyear{Ree88}). The interest was driven
by the implications for the growth of MBHs, for feeding luminous accretion
in AGN and for detecting MBHs in quiescent galaxies by tidal disruption
flares. However, tidal disruption is only one of several possible
modes of interaction with the MBH. A useful analogy can be made between
the MBH and the stars around it and an atomic nucleus and its surrounding
electrons. In spite of the fundamental differences between a macroscopic
classical system and a microscopic quantum one, the analogy to atomic
processes suggests a classification scheme for the various modes of
strong star--MBH interactions. This not only provides the {}``bookkeeping''
needed to keep track of all the permutations, but also provides some
analytical tools for calculating cross-sections, rates and branching
ratios. The following atomic-like processes, and their possible observational
signatures, are discussed here.

\begin{description}
\item [Annihilation]Direct infall preceded by tidal disruption, where a
substantial fraction of the stellar mass-energy is released (\S\ref{ss:tide}).
\item [Absorption]Direct infall where the stellar mass-energy is absorbed
by the MBH (\S\ref{ss:tide}).
\item [Deep~inelastic~scattering]Tidal scattering, where the star is
strongly perturbed by the MBH tidal field during a close {}``fly
by'', but is not destroyed (\S\ref{sss:tscatter}).
\item [Metastable~decay~vs~collisional~ionization~/~de-excitation]Gradual
orbital inspiral due to a dissipational force, such as gravitational
wave emission or tidal heating, in the presence of orbital perturbations
that can throw the star into the MBH or deflect it to a wide orbit
(\S\ref{sss:inspiral}).
\item [Charge~exchange]Three-body exchange where an incoming star knocks
out one that is on a bound orbit to the MBH (the 3rd body) and replaces
it (\S\ref{sss:exchange}).
\item [Ionization]Three-body encounters where an incoming star eject a
star tightly bound to the MBH into a wide, unbound orbit (\S\ref{sss:exchange}). 
\end{description}

\subsection{Tidal disruption}

\label{ss:tide}

The tidal interaction between a star and a black hole is characterized
by three length-scales: the stellar radius $\Rs$, the Schwarzschild
radius $r_{S}$ and the tidal radius $r_{t}$. The ratios of these
length-scales can be expressed in terms of the ratios of the masses
and escape velocities, \begin{eqnarray}
\frac{r_{t}}{\Rs} & \sim & \left(\frac{m}{\Ms}\right)^{1/3}\,,\nonumber \\
\frac{r_{t}}{r_{S}} & \sim & \left(\frac{c}{V_{e}}\right)^{2}\left(\frac{m}{\Ms}\right)^{-2/3}\,,\nonumber \\
\frac{r_{S}}{\Rs} & = & \left(\frac{c}{V_{e}}\right)^{-2}\left(\frac{m}{\Ms}\right)\,,\label{e:rtide}\end{eqnarray}
where $c$, the speed of light, is the escape velocity from the BH
and $V_{e}^{2}\!=\!2G\Ms/\Rs$ is the escape velocity from the stellar
surface (for a solar type star $V_{e}\!\sim\!600\,\mathrm{km\, s^{-1}}$
and $(c/V_{e})^{2}\!\sim\!2.5\!\times\!10^{5}$). 

The nature and outcome of a {}``grazing collision'' between a black
hole and a star ($r_{p}\!\sim\! r_{S}\!+\!\Rs$) depends on the relative
magnitude of the black hole's tidal field, the stellar self-gravity
and its gas pressure and whether the disruption occurs in the Newtonian
or GR limit (see discussion by Gomboc \& \v{C}ade\v{z} \citeyear{Gom05}).
By definition, the tidal energy extracted from the orbit and deposited
in the star as it crosses the tidal disruption radius is of order
of its binding energy, $\Es$. It also follows from the definition
of $r_{t}$ that the periapse crossing time at the tidal radius, $\tau_{t}$,
is of the order of the star's dynamical timescale, since $\tau_{t}\!\sim\!\sqrt{r_{t}^{3}/Gm}\!=\!\sqrt{\Rs^{3}/G\Ms}\!=\!\tau_{\star}$
($\tau_{\star}\!\sim\!1600\,\mathrm{s}$ for a solar type star). Therefore,
\emph{complete} tidal disruption occurs as fast or faster than the
time it takes for the stellar gravity or gas pressure to react. These
can therefore be neglected during the disruption and the particles
composing the star approximately follow free-fall trajectories in
the gravitational field of the black hole (Laguna et al. \citeyear{Lag93};
Kochanek \citeyear{Koc94}). This approximation holds until the stream
of disrupted gas completes an orbit around the black hole and intersects
itself. At that point strong shocks likely occur and the gas hydrodynamics
have to be taken into account. When $r_{p}\!>\! r_{t}$, the tidal
interaction is weaker and slower, lasting $\tau_{p}\!\sim\!\tau_{t}b^{3/2}$,
where $b^{-1}\!\equiv\! r_{t}/r_{p}$ is the {}``penetration parameter''
\label{d:bpen}. It may then become longer then the stellar dynamical
time scale to the extent that the stellar gravity and hydrodynamics
do have to be taken into account. The ratio $r_{t}/r_{S}$ determines
whether the disruption occurs in the Newtonian limit ($r_{t}/r_{S}\!\gtrsim\!10$)
or in the GR limit ($r_{t}/r_{S}\!\lesssim\!\mathrm{few}$). 

The different regimes of tidal disruption can be classified by the
mass ratio $m/\Ms$ and Eq. (\ref{e:rtide}). In order of increasing
$m/\Ms$, these are 

\begin{enumerate}
\item $m/\Ms\!\ll\!1$, which corresponds to $r_{S}\ll\! r_{t}\!\ll\!\Rs$.
In this case the tidal interaction is overall weak and Newtonian (only
a small part of the star overlaps with the tidal disruption zone),
and the process is dominated by the star's gravity and pressure. The
star is not disrupted. If the black hole passes through the star it
may accrete a small fraction of its mass. A more significant effect
may occur if the black hole is captured inside the star and sinks
to the center. This tidal regime could apply to the formation exotic
stars that cannot be formed in the course of normal stellar evolution,
such as Thorne-\.Zytkow objects (Thorne \& \.Zytkow \citeyear{Tho75}),
which are giant envelopes with an accreting compact object in their
center.
\item $m/\Ms\!\sim\!1$, which corresponds to $r_{S}\!\ll\! r_{t}\!\sim\!\Rs$.
In this case a strong Newtonian interaction will occur, which would
result in significant mass loss and possible disruption, depending
on the central concentration of the stellar density profile. The process
will be accompanied by strong shocks. This tidal regime applies to
close interaction between a SBH and a massive star (e.g. the 3-body
exchange scenario, \S\ref{sss:exchange}).
\item $m/\Ms\!\sim\!(c/V_{e})^{2}$, which corresponds to $\Rs\!\sim\! r_{S}\!<r_{t}$.
In this case the disruption is complete and occurs in the Newtonian
limit (the subsequent flow of the gaseous debris into the black hole
is relativistic). This regime applies to disruption by intermediate
mass black holes ($m\!\sim\!10^{3}$--$10^{4}\,\Mo$), whose existence
is still a matter of speculation (\S\ref{ss:scope}).
\item $(c/V_{e})^{2}\!<\! m/\Ms\!<\!(c/V_{e})^{3}$, which corresponds to
$\Rs\!\ll\! r_{S}\!<\! r_{t}$. Tidal disruption by the MBH in the
GC falls in this regime ($r_{t}\sim\!9r_{S}\sim\!160\!\Rs$ for a
solar type star). The disruption can be treated as Newtonian to a
good approximation.
\item $m/\Ms\!>\!(c/V_{e})^{3}$, which corresponds to $\Rs\!\ll r_{t}\!\ll r_{S}$.
This is the case for a MBH with mass $\gtrsim\!10^{8}\,\Mo$ (for
solar type stars). The event horizon lies outside the (formal) tidal
radius and the star falls into the MBH as a point particle on a General
Relativistic trajectory without being significantly perturbed. 
\end{enumerate}
The various processes and phenomena associated with the tidal disruption
of stars are reviewed by Rees (\citeyear{Ree88}).

\subsubsection{Tidal disruption rate}

\label{sss:tdrate}

A star is tidally disrupted if the MBH mass is small enough so that
$r_{t}\!>\! r_{S}$ and if its orbital angular momentum $J$ is small
enough so that $r_{p}\!\le\! r_{t}$. Such orbits are called {}``loss-cone''
orbits%
\footnote{The term {}``loss-cone'' originates in plasma physics. For motion
on a straight line (no gravity), the volume in velocity space that
corresponds to disruption orbits originating from a point is a cone.
With gravity's effects included, it is actually a cylinder (Miralda-Escud\'e
\& Gould \citeyear{Mir00}). %
} and in a spherical potential have\begin{equation}
J^{2}\!\le\! J_{lc}^{2}(\varepsilon)\!\equiv\!2r_{t}^{2}\left(\psi(r_{t})-\varepsilon\right)\!\simeq\!2Gmr_{t}\,,\label{e:Jlc}\end{equation}
 where the last approximate equality assumes that $r_{t}\!\ll\! r_{h}$
(the potential is dominated by the MBH) and $\varepsilon\!\ll\!\psi(r_{t})$
(the star is consumed from a weakly bound orbit). In terms of Keplerian
orbits (Eq. \ref{e:Kepler}), which are a valid approximation inside
$r_{h}$, tidal disruption requires that $r_{p}\!=a(1-e)\!\le\! r_{t}$,
that is either a very tightly bound orbit (small $a$, high $\varepsilon$)
or a very eccentric one (high $e$, low $J$). The solid angle subtended
by the loss cone is usually very small. The loss-cone's opening angle
for parabolic orbits is easily derived from angular momentum conservation
(Frank \& Rees \citeyear{Fra76}) 

\begin{equation}
\vartheta_{lc}^{2}\sim\frac{r_{t}}{a}=(1-e)\simeq\frac{1}{2}\left(\frac{J_{lc}}{J_{c}}\right)^{2},\label{e:ThetaLC}\end{equation}
where the typical radius is taken as $r\!\sim\! a$ and where 

\begin{equation}
J_{c}^{2}=Gma=\left.(Gm)^{2}\right/2\varepsilon\,,\label{e:Jc}\end{equation}
 is the maximal (circular orbit) angular momentum for specific energy
$\varepsilon$. For example, $\vartheta_{lc}\!\sim\!10^{-3}$ for
a solar mass star on a $a\!=\!1\,\mathrm{pc}$ orbit around a $3\!\times\!10^{6}\,\Mo$
MBH. 

Loss-cone theory deals with the processes by which stars enter loss-cone
orbits, the rate at which this occurs, and how these depend on the
parameters of the system. This problem has been studied extensively
over the years (Hills \citeyear{Hil75}; Frank \& Rees \citeyear{Fra76};
Lightman \& Shapiro \citeyear{Lig77}; Cohn \& Kulsrud \citeyear{Coh78};
Magorrian \& Tremaine \citeyear{Mag99}; Syer \& Ulmer \citeyear{Sye99};
Alexander \& Hopman \citeyear{Ale03b}; Wang \& Merritt \citeyear{Wan04}).
While different approaches and assumptions lead to somewhat different
quantitative results, the process is well understood qualitatively.

Any stars that are initially on loss-cone orbits are promptly destroyed
on an orbital timescale once they reach periapse. From that time on,
the tidal disruption rate, $\Gamma_{t}$,\label{d:Gt} is set by the
rate at which these orbits can be replenished by the relaxation processes
that redistribute stars in phase space to high-$\varepsilon$ or low-$J$
orbits. Diffusion in $\varepsilon$-space occurs on the relaxation
timescale, which roughly corresponds to the timescale for $\varepsilon$
to change by order unity, \begin{equation}
t_{r}\!\sim\!\varepsilon/\dot{\varepsilon}\,.\end{equation}
The relaxation time is approximately independent of energy in the
GC (Eq. \ref{e:trel}), and is assumed here to be also independent
of $J$. Since $t_{r}$ is also the timescale for $J^{2}$ to change
by order $J_{c}^{2}$ (Eq. \ref{e:Jc}), the angular momentum relaxation
timescale, which roughly corresponds to the timescale for $J$ to
change by order unity, is \label{d:tJ}\begin{equation}
t_{J}\sim J^{2}/\dot{\left(J^{2}\right)}\sim[J/J_{c}(\varepsilon)]^{2}t_{r}\,,\label{e:tJ}\end{equation}
 where the square root dependence of $J$ on $t_{J}$ reflects the
random walk (diffusive) nature of the process%
\footnote{In fact, it is the velocity component perpendicular to the velocity
vector, $\Delta v_{\perp}$, that executes a symmetric random walk,
and not directly $J$ itself. This introduces a drift term ($\propto\! t$)
to the evolution of $J$, which is small compared to the diffusive
term ($\propto\!\sqrt{t}$) as long as $t\!\lesssim\! t_{J}$ and
can be neglected (Hopman \& Alexander \citeyear{Hop05}).%
}. Typically, $J_{lc}\!\ll\! J_{c}$. In principle, stars can enter
the loss-cone, $J\!<\! J_{lc}(\varepsilon)$ either by a decrease
in $J$, or by an increase in $\varepsilon$ (up to the last stable
orbit). In practice, diffusion in $J$-space is much faster: although
$t_{J}\!\sim\! t_{r}$ when $J\!\sim\! J_{c}$, once $J$ becomes
smaller than $J_{c}$ by order unity, $t_{J}\!\ll\! t_{r}$. Because
$J$-diffusion is so much more efficient than $\varepsilon$-diffusion,
only $J$-diffusion need be considered.

The way a star enters the loss-cone depends on $\Delta J(\varepsilon)$,
the r.m.s change in $J(\varepsilon)$ over an orbital (dynamical)
time $P(\varepsilon)$, \begin{equation}
\Delta J(\varepsilon)\sim\sqrt{\frac{P(\varepsilon)}{t_{r}(\varepsilon)}}J_{c}(\varepsilon)\,.\end{equation}
 The ratio $\Delta J(\varepsilon)/J_{lc}$ defines two dynamical regimes
of loss-cone re-population (Lightman \& Shapiro \citeyear{Lig77}).
In the {}``Diffusive regime'' of stars with large $\varepsilon$
(tight short-period orbits), $\Delta J(\varepsilon)\!\ll\! J_{lc}$
and so the stars slowly diffuse in $J$-space. The loss-cone remains
nearly empty at all times since any star inside it is promptly destroyed.
At $J\!\gg\! J_{lc}$ the DF is nearly unperturbed by the existence
of the loss cone. In particular, it can remain very nearly spherically
symmetric and depend only on $\varepsilon$ (\S\ref{ss:cusp}), but
it falls logarithmically to zero at $J\gtrsim J_{lc}$. In the {}``full
loss-cone regime'' (sometimes also called the {}``pinhole'' or
{}``kick'' regime) of stars with small $\varepsilon$ (wide long-period
orbits), $\Delta J(\varepsilon)\!\gg\! J_{lc}$ and so the stars can
enter and exit the loss-cone many times before reaching periapse.
As a result, the DF is nearly isotropic at all $J\!\gtrsim\! J_{lc}$. 

The tidal disruption flux density (stars disrupted per $\mathrm{d}t\mathrm{d\varepsilon}$)
in the diffusive regime, $F_{\mathrm{dif}}$, is derived from the
orbital diffusion coefficients for two-body scattering via the Fokker-Planck
equation, with the boundary conditions that the stellar density fall
to zero for $J\!\rightarrow\! J_{lc}$ and approach the isotropic
distribution $n_{\mathrm{iso}}(\varepsilon)$ for $J\!\rightarrow\! J_{c}$
(Lightman \& Shapiro \citeyear{Lig77}), \begin{equation}
F_{\mathrm{dif}}(\varepsilon;J_{lc})\sim\frac{n_{\mathrm{iso}}(\varepsilon)}{\ln[J_{c}(\varepsilon)/J_{lc}(\varepsilon)]t_{r}(\varepsilon)}\,.\label{e:qdif}\end{equation}
The diffusion flux is regulated mainly by the relaxation time and
is only weakly dependent on the size of the loss-cone (typically $5\!\lesssim\!\ln(J_{c}/J_{lc})\!\lesssim\!10$)%
\footnote{The diffusion timescale is longer than $t_{r}$ by the logarithmic
factor because the stellar phase-space density approaches zero near
the loss-cone.%
}. The logarithmic term reflects the fact that the diffusion occurs
in a 2D space of solid angle (Lightman \& Shapiro \citeyear{Lig77}).
It can also be understood qualitatively by analogy to heat diffusion
on the surface of a sphere from the equator ({}``$J_{c}$''), to
a small hole at the pole ({}``$J_{lc}$'') (Frank \& Rees \citeyear{Fra76}). 

In the full loss-cone regime far from the MBH, where the DF is nearly
isotropic, the tidal disruption flux density is simply $\propto\!\pi\vartheta_{lc}^{2}$.
Since the loss-cone is refilled every orbital time, the flux density
is\begin{equation}
F_{\mathrm{pin}}(\varepsilon;J_{lc})\sim\frac{J_{lc}^{2}(\varepsilon)}{J_{c}^{2}(\varepsilon)}\frac{n_{\mathrm{iso}}(\varepsilon)}{P(\varepsilon)}\,.\label{e:qpin}\end{equation}

The transition between the two regimes occurs at a critical energy
$\varepsilon_{c}$ (or equivalently a critical typical orbital radius
$r_{c}$), where $F_{\mathrm{dif}}(\varepsilon_{c})\!=\! F_{\mathrm{pin}}(\varepsilon_{c})$,
or approximately where $\Delta J(\varepsilon)\!\sim\! J_{lc}$. The
total tidal disruption rate is then given by \begin{equation}
\Gamma_{t}\sim\int_{0}^{\varepsilon_{c}}F_{\mathrm{pin}}(\varepsilon;J_{lc})\mathrm{d\varepsilon}+\int_{\varepsilon_{c}}^{\infty}F_{\mathrm{dif}}(\varepsilon;J_{lc})\mathrm{d\varepsilon}\,.\label{e:Gt}\end{equation}
 An analysis of the behavior of the integrands in Eq. (\ref{e:Gt})
for typical observed values of stellar cusp slopes, reveals that most
of the contribution comes from the diffusive regime at values near
$\varepsilon_{\max}\!\sim\!\max(\varepsilon_{c},\varepsilon_{h})$,
where $\varepsilon_{h}\!\sim\!\psi(r_{h}),$ or equivalently, near
$r_{\max}\!=\!\min(r_{c},r_{h})$ (Lightman \& Shapiro ; Syer \& Ulmer
\citeyear{Sye99}; for numeric examples see Magorrian \& Tremaine
\citeyear{Mag99}). In most galaxies $r_{c}\!\lesssim\! r_{h}$ (Wang
\& Merritt \citeyear{Wan04}). Making use of this property, the tidal
disruption rate can be approximated to within an order of magnitude
by 

\begin{equation}
\Gamma_{t}\sim\frac{\Ns(<r_{h})}{\ln(\sqrt{r_{h}/r_{t}})t_{r}(r_{h})}\,.\label{e:Gtrh}\end{equation}
Taking for the GC $r_{h}\!\sim\!2$ pc, $r_{t}\!\sim\!10^{13}\,\mathrm{cm}$,
$\Ns(<\! r_{h})\!\sim\! m/\left\langle \Ms\right\rangle \sim\!3.5\!\times\!10^{6}$
and $t_{r}(r_{h})\!\sim\!4\!\times\!10^{9}\,\mathrm{yr}$ (Eq. \ref{e:trelNP}),
the estimated rate is $\Gamma_{t}\!\sim\!10^{-4}\,\mathrm{yr^{-1}}$,
in agreement with more detailed calculations ($\Gamma_{t}\!\sim\!5\!\times\!10^{-5}\,\mathrm{yr^{-1}}$,
Syer \& Ulmer \citeyear{Sye99}).

It should be noted that there are dynamical mechanisms that may increase
the tidal disruption rate beyond that predicted by two-body scattering
in a spherical system. The possibilities include the effects of massive
perturbers, for example giant molecular clouds (Zhao, Haehnelt \&
Rees \citeyear{Zha02}), which indeed exist in the GC on scales of
1--2 pc from the MBH; enhanced rates due to resonant scattering (factor
of $\sim\!2$ rate increase; Rauch \& Tremaine \citeyear{Rau96};
Rauch \& Ingalls \citeyear{Rau98}); deviations from spherical symmetry
(factor of $\sim\!2$ rate increase; Magorrian \& Tremaine \citeyear{Mag99});
or chaotic orbits in triaxial potentials (factor of $10$--$100$
rate increase; Poon \& Merritt \citeyear{Poo02}; Merritt \& Poon
\citeyear{Mer04c}).

\subsubsection{Tidal disruption and its aftermath}

\label{sss:tdafter}

The observations of tidal disruption in the GC are realistically limited
to those after-effects that persist for at least $\Gamma_{t}^{-1}\!\sim\!10^{4}\,\mathrm{yr}$
(the mean time between events). As described below, there are two
main channels by which energy is released in a tidal disruption event:
one is a luminous accretion flare lasting a few years, and the other,
which is much less luminous but longer-lasting, is a shock wave, analogous
to a supernova blast, that sweeps through the interstellar medium
and can energize it substantially for $\sim\!10^{5}\,\mathrm{yr}$
(e.g. Khokhlov \& Melia \citeyear{Kho96}).  Observationally, the
two effects complement each other. While tidal disruption flares may
be, and perhaps were already observed in cosmological galaxy surveys
(e.g. Komossa \citeyear{Kom02}), the supernova-like shock wave may
be observed in the GC. 

The hydrodynamics, radiative properties, early and late evolution
of tidal disruption in the Newtonian and relativistic limits were
studied extensively by numerous authors, both analytically (Rees \citeyear{Ree88};
Phinney \citeyear{Phi89}; Cannizzo, Lee \& Goodman \citeyear{Can90};
Kochanek \citeyear{Koc94}; Loeb \& Ulmer \citeyear{Loe97}; Ulmer,
Paczy\'nski \& Goodman \citeyear{Ulm98}; Ulmer \citeyear{Ulm99};
Menou \& Quataert \citeyear{Men01}; Ivanov \& Novikov \citeyear{Iva01};
Ivanov, Chernyakova \& Novikov \citeyear{Iva02}) and by simulations
(Nolthenius \& Katz \citeyear{Nol82}; Evans \& Kochanek \citeyear{Eva89};
Khokhlov, Novikov \& Pethick \citeyear{Kho93b}; Frolov et al. \citeyear{Fro94};
Laguna et al. \citeyear{Lag93}; Diener et al. \citeyear{Die97};
Ayal, Piran \& Livio \citeyear{Aya00}; Bogdanovi\'c et al. \citeyear{Bog04};
Gomboc \& \v{C}ade\v{z} \citeyear{Gom05}). While there is considerable
uncertainty about details, a general picture has emerged.

It is useful to begin by considering the energy budget of a tidal
disruption event. The star starts on a nearly parabolic orbit ($E_{\mathrm{orb}}\!\simeq\!0)$
and so the total energy of the system (ignoring for now its rest-mass
energy) is the stellar binding energy, $\Es\!<\!0$. The outcome of
tidal disruption is that some energy is extracted out of the orbit
to unbind the star and accelerate the debris. Initially about $M_{in}\!\sim\!\Ms/2$
of the stellar mass becomes tightly bound to the MBH with energy $E_{in}$,
($-E_{\mathrm{in}}\!\gg\!\left|\Es\right|)$, while the remainder
$M_{\mathrm{out}}\!\sim\!\Ms/2$ of the stellar mass is forcefully
ejected with energy $E_{out}$, ($+E_{\mathrm{out}}\!\gg\!\left|\Es\right|$),
so that $E_{\mathrm{in}}\!+\! E_{\mathrm{out}}\!=\!\Es$. 

There are 3 distinct energy scales in the process, each associated
with a different time scale. The lowest is the stellar binding energy
$\Es$. By virtue of the virial theorem, the internal heat of the
star is of order $U\!\sim\!-\Es/2$. During the disruption, on the
periapse passage timescale $\tau_{p}\!\sim\!{\cal {O}}(1\,\mathrm{hr})$,
the inner parts of the star are exposed as it is stretched and the
heat can be radiated away. This may lead to a short flare immediately
after disruption (Gomboc \& \v{C}ade\v{z} \citeyear{Gom05}). 

Next in scale is the kinetic energy of the ejecta. While the \emph{mean}
specific energy of the debris is $-\Es/\Ms\!\sim\!\Vs^{2}$ ($\Vs$
is the circular velocity on the stellar surface\label{d:Vc}), the
spread about this value can be orders of magnitude larger. As the
star is disrupted, tidal torquing spins it up in the direction of
the orbit, imparting an excess velocity $\gtrsim\! V_{e}\!=\!\sqrt{2}V_{\star}$
above the mean orbital velocity $v_{p}\!\gg\! V_{\star}$ to the stellar
hemisphere on the far side from the MBH, and a comparable velocity
deficit on the near side. The difference in velocities leads to a
spread in the specific kinetic energies of about $(v_{p}\!+\! V_{e})^{2}/2\!-\!(v_{p}\!-\! V_{e})^{2}/2\!=\!2^{3/2}v_{p}V_{\star}\!\gg\!\Vs^{2}$
(Rees \citeyear{Ree88}). The spread in energy can also be expressed
in terms of the work done on the stellar debris by the tidal force
(Lacy, Townes \& Hollenbach \citeyear{Lac82}),\begin{equation}
W\sim\Ms\Delta\phi_{\mathrm{MBH}}(r_{p})\sim\frac{Gm\Ms}{r_{p}^{2}}\Rs\sim-b^{-2}\left(\frac{m}{\Ms}\right)^{1/3}\Es\gg-\Es\,,\label{e:Wtd}\end{equation}
where $\Delta\phi_{\mathrm{MBH}}(r_{p})$ is the MBH's gravitational
potential difference across the star at periapse. In a typical disruption
event in the GC, $W\!\gtrsim\!100\Es$ (for $b\!\sim\!1$) and thus
the debris is released with specific energies in the range $(\Es\!\pm W)/\Ms$.
Note that while $W\!\gg\!-\Es$, it is still much smaller than the
orbital kinetic energy of the star at disruption, $-b^{-1}(m/\Ms)^{2/3}\Es$.
The kinetic energy carried by the ejected debris can significantly
exceed that released by a normal supernova ($\sim\!10^{51}\,\mathrm{erg}$)
if the orbit is highly penetrating, \begin{equation}
E_{\mathrm{out}}\!\sim\! W\!\sim\!4\!\times10^{52}\,\mathrm{erg}\left(\frac{\Ms}{\Mo}\right)^{2}\left(\frac{\Rs}{\Ro}\right)^{-1}\left(\frac{m/\Ms}{10^{6}}\right)^{1/3}\left(\frac{b}{0.1}\right)^{-2}\,.\end{equation}
Unlike a spherical supernova explosion, the mass will be ejected mostly
in one half of the orbital plane (Khokhlov \& Melia \citeyear{Kho96}). 

The highest energy scale is that due to accretion onto the MBH. The
bound debris will eventually circularize around the MBH after the
stream of orbiting gas will cross itself, collide and shock. In order
to fall into the MBH, the gas has to lose its angular momentum. If
a mechanism exists that can exert torque on the rotating gas (for
example shear due to effective viscosity mediated by magneto-hydrodynamic
turbulence), it will gradually spiral inward on a viscous dissipation
timescale. The gravitational binding energy of the gas to the MBH
will be converted to heat, until it reaches the last stable circular
orbit around the MBH ($3r_{S}$ for a Schwarzschild BH), where it
will promptly fall into the event horizon on a dynamical timescale.
If the gas can dissipate the heat by the emission of radiation, or
acceleration of particles, then between $\eta\!=\!0.057$ (for a non-rotating
BH) to $\eta\!=\!0.42$ (for a maximally spinning BH) of its rest
mass energy can be extracted by the time it reaches the last stable
orbit (e.g. Shapiro \& Teukolsky \citeyear{Sha83}). If, on the other
hand, energy dissipation is inefficient, the heat may be carried by
the flow (advected) into the MBH and add to its rest mass (advection
dominated accretion flow {[}ADAF{]} solution; Rees et al. \citeyear{Ree82},
Abramowicz et al. \citeyear{Abr95}; Narayan \& Yi \citeyear{Nar94},
\citeyear{Nar95a}; Narayan, Yi \& Mahadevan \citeyear{Nar95b}).
Alternatively, the gas will not fall into the MBH but will escape
as a wind (advection dominated inflow-outflow solutions {[}ADIOS{]},
Blandford \& Begelman \citeyear{Bla99}) or by convection outward
(convection dominated accretion flow {[}CDAF{]}, Quataert \& Gruzinov
\citeyear{Qua00}; Narayan, Igumenshchev \& Abramowicz \citeyear{nar00}).
In the case of high radiative efficiency, \begin{equation}
E_{\mathrm{acc}}\sim\eta(\Ms/2)c^{2}\sim\eta b^{2}\left(\frac{\Ms}{m}\right)^{1/3}\left(\frac{c}{V_{e}}\right)^{2}E_{\mathrm{out}}>E_{\mathrm{out}}\,.\end{equation}

Immediately after the disruption and acceleration at periapse, the
debris' self gravity and gas pressure are dynamically insignificant
compared to its kinetic energy, and the gas streams follow ballistic
(free falling) Keplerian orbits. The orbits are very eccentric $1\!-\!\left\langle e\right\rangle \!=\! r_{p}/a\!=2b(\Ms/m)^{2/3}$
with a large spread in orbital period and apoapse, extending from
$P\!\rightarrow\!\infty$ ($r_{a}\!\rightarrow\!\infty$) of the barely
bound gas, down to a minimal period $P_{\mathrm{min}}$, ($r_{a}\!\simeq\! r_{t}(m/\Ms)^{1/3}$),
which corresponds to that of the most tightly bound gas with specific
binding energy $b^{-2}(m/\Ms)^{1/3}\Vs^{2}$ (Eqs. \ref{e:Wtd}, \ref{e:Kepler};
Rees \citeyear{Ree88}; Evans \& Kochanek \citeyear{Eva89}). The
minimal period is\begin{equation}
P_{\mathrm{min}}=\frac{\pi}{\sqrt{2}}\sqrt{\frac{m}{\Ms}}b^{3}\tau_{\star}\sim0.1\,\mathrm{yr\,}\left(\frac{m/\Ms}{10^{6}}\right)^{1/2}\left(\frac{\Rs}{\Ro}\right)^{3/2}\left(\frac{\Ms}{\Mo}\right)^{-1/2}b^{3}\,.\end{equation}
After an initial delay of $P_{\mathrm{min}}$ the debris will start
returning to the point of disruption at a rate $\dot{M}\!=\!\left(\mathrm{d}M/\mathrm{d}E\right)\left(\mathrm{d}E/\mathrm{d}P\right)$.
Simulations show that the spread in energy of the bound debris is
nearly uniform, $\mathrm{d}M/\mathrm{d}E\simeq\mathrm{const}$ (e.g.
Evans \& Kochanek \citeyear{Eva89}; Ayal, Piran \& Livio \citeyear{Aya00}),
and so (Eq. \ref{e:Kepler})\begin{equation}
\dot{M}_{in}(t)\simeq\frac{\Ms}{3P_{\mathrm{min}}}\left(\frac{t}{P_{\mathrm{min}}}\right)^{-5/3}\,,\qquad(t>P_{\mathrm{min}}).\label{e:dMindt}\end{equation}
When $r_{p}\!\gtrsim\! r_{S}$, GR correction have to be taken into
account. These lead to a stronger tidal interaction, and consequently
a higher spread in energies (Eq. \ref{e:Wtd}), a shorter $P_{\mathrm{min}}$
and larger peak debris return rate (Ulmer \citeyear{Ulm99}). 

Initially, the mass return rate exceeds the Eddington-limited accretion
rate%
\footnote{The luminosity released by matter accreting on a mass $m$ exerts
an outward radiation pressure on the infalling mass. The radiation
pressure balances gravity and halts the accretion when the luminosity
reaches the Eddington luminosity, $L_{\mathrm{Edd}}\!=\!1.5\!\times\!10^{38}(m/\Mo)\,\mathrm{erg\, s^{-1}}$
(for isotropic luminosity and pressure due to Thomson scattering).
The Eddington accretion rate $\dot{M}_{\mathrm{Edd}}\!\equiv\! L_{\mathrm{Edd}}/c^{2}\!=\!3\!\times\!10^{-9}(m/\Mo)\,\Mo\,\mathrm{yr^{-1}}$
is defined as the maximal possible for unit efficiency. The actual
Eddington-limited accretion rate is $\max\dot{M}\!=\!\dot{M}_{\mathrm{Edd}}/\eta$. %
}\begin{equation}
\max\dot{M}\!=\!\frac{\dot{M}_{\mathrm{Edd}}}{\eta}\!\simeq\!0.03\left(\frac{m}{10^{6}\,\Mo}\right)\left(\frac{\eta}{0.1}\right)^{-1}\,\Mo\,\mathrm{yr^{-1}}\,.\label{e:Medd}\end{equation}
 In the GC, $\max\dot{M}\!\sim\!0.1\,\Mo\,\mathrm{yr^{-1}}$. The
return rate remains higher than the Eddington-limited rate for no
more than a few years, 

\begin{equation}
t_{\mathrm{in}}=\left(\frac{\pi}{3\sqrt{6}}\right)^{2/5}\eta^{3/5}b^{6/5}\left(\frac{\Ms}{m}\right)^{2/5}\left(\frac{\tau_{\star}}{\tau_{\mathrm{Edd}}}\right)^{2/5}\tau_{\mathrm{Edd}}=4\left(\frac{\eta}{0.1}\right)^{3/5}b^{6/5}\left(\frac{m/\Ms}{10^{6}}\right)^{-2/5}\left(\frac{\tau_{\star}}{10^{4}\,\mathrm{s}}\right)^{2/5}\,\mathrm{yr\,,}\end{equation}
where $\tau_{\mathrm{Edd}}\!\equiv\! m/\dot{M}_{\mathrm{Edd}}$ is
the $e$-folding timescale of Eddington accretion with unit efficiency.
A similar estimate is obtained by considering the time for accretion
of all the debris at the Eddington-limited rate if all the mass returned
at once, $t_{\mathrm{Edd}}\!\sim\! M_{in}/\max\dot{M}\!\sim\!5\,\mathrm{yr}$.
The high initial mass return rate does not necessarily imply an early
phase of Eddington-limited accretion. In order for the disruption
event to produce a luminous flare, both the time for circularization,
which sets the flow pattern required for dissipation, and the timescale
for viscous dissipation, which extracts the angular momentum, have
to be short enough so as not to be an impediment to the accretion.
This indeed appears to be the case.

Circularization is expected to be efficient. The evolution of the
bound debris is driven by shocks between streams of gas that converge
to their common point of origin at periapse. Relativistic precession
(Eq. \ref{e:GRprecess}) will also lead to shocks between incoming
and outgoing streams. These shocks will redistribute angular momentum
and energy between the gas streams. Some fraction of the initially
bound debris will become unbound after undergoing strong shocks and
heating on the second periapse crossing, leaving perhaps $\sim\!\Ms/4$
of the gas bound to the MBH (Ayal, Piran \& Livio \citeyear{Aya00}).
While the exact details are uncertain, it is expected that circularization
will occur rapidly, on a timescale of $t_{\mathrm{circ}}\!\sim\!\mathrm{few\!\times\! P_{min}}$
(Ulmer \citeyear{Ulm99}). 

The viscous accretion timescale at the tidal radius is $t_{\mathrm{acc}}\sim\!2\pi\tau_{t}/\alpha\pi h^{2}\!=\!3\!\times\!10^{-4}(\tau_{\star}/\tau_{\star,\odot})/\alpha\pi h^{2}\,\mathrm{yr}$,
where $\alpha$ is the dimensionless viscosity parameter (the proportionality
factor between the viscous stress tensor and the pressure) and $h$
is the ratio of the thickness of the accretion disk to its radius
($h\!\sim\!1$ for a thick disk). When $\dot{M}\!\gtrsim\!\dot{M}_{\mathrm{Edd}}$,
the radiation pressure puffs up the disk ($h\!\rightarrow\!1$) and
decreases $t_{\mathrm{acc}}$. The value of $\alpha$ is uncertain,
but $10^{-3}\!\lesssim\!\alpha\!\lesssim\!10^{-1}$ seems plausible%
\footnote{Alternatively, if $10^{-5}\!\lesssim\!\alpha\!\lesssim\!10^{-3}$
and $\eta\!\gtrsim\mathrm{few\!\times\!0.01}$, the configuration
may be analogous to a Thorne-\.Zytkow object, in the form of a large
($\sim\!100r_{t}$) optically thick star-like object with $\Ts\!\sim\!10^{4}\,\mathrm{K},$
powered by accretion. Such an object will emit at the Eddington limit
mainly in the optical (Loeb \& Ulmer \citeyear{Loe97}), unlike the
UV/X-ray flares discussed here.%
} (Ulmer \citeyear{Ulm99}), so that probably $t_{\mathrm{acc}}\!\lesssim\!$0.01
yr. It then follows that $t_{\mathrm{acc}}\!<\! t_{\mathrm{circ}}\!<\! t_{\mathrm{Edd}}$,
which means that for the first few years after a disruption event
the accretion will probably proceed at a super-Eddington rate ($L\!\sim\! L_{\mathrm{Edd}}$,
small $\eta$, $\dot{M}\!>\!\dot{M}_{\mathrm{Edd}}$) with a radiatively
inefficient accretion flow, such as ADAF, ADIOS and/or CDAF (Rees
\citeyear{Ree88}; Menou \& Quataert \citeyear{Men01}).

Most of the work on the observational signature of tidal disruption
concentrated on {}``tidal flares'', which correspond to this initial
luminous phase, lasting months to years. Tidal flares are expected
to peak in the UV/X-ray range, as can be estimated by assuming that
the Eddington luminosity is emitted as blackbody radiation from a
surface whose typical size probably lies between the last stable orbit
and the tidal radius. The blackbody temperature is then $T_{\mathrm{bb}}\!=\!(L_{\mathrm{Edd}}/4\pi r^{2}\sigma_{\mathrm{SB}})^{1/4}$,
and 

\begin{equation}
T_{\mathrm{bb}}(r_{t})\sim3\!\times\!10^{5}\left(\frac{m}{10^{6}\,\Mo}\right)^{1/12}\mathrm{K}\,,\qquad\mathrm{or}\qquad T_{\mathrm{bb}}(3r_{S})\sim7\!\times\!10^{5}\left(\frac{m}{10^{6}\,\Mo}\right)^{-1/4}\mathrm{K}\,,\end{equation}
(Assuming a solar-type star for $r_{t}$), almost independently of
the MBH mass. This temperature corresponds to a blackbody peak at
$\sim\!0.1$ keV. 

There are a few tentative detections of tidal flares in other galaxies
(see review by Komossa \citeyear{Kom02}). However, the short duration
of the flare and the mean time between tidal disruption events in
the GC (\S\ref{sss:tdrate}) translate to a very low duty cycle of
$\sim\!10^{-4}$. The probability of directly observing a tidal flare
in the Galaxy is small.

A few years after the tidal disruption event, when the return rate
drops below the Eddington rate, the gas will cool efficiently and
accrete via a radiatively efficient, geometrically thin accretion
disk, at a rate set by the viscous timescale. The accretion rate in
the thin disk is expected to fall off slowly with time (Cannizzo et
al. \citeyear{Can90}). However, this result is at odds with observations.
The predicted average luminosity due to accretion of debris from a
past tidal disruption event is 2--4 orders of magnitude higher than
that observed in the GC, or in the centers of nearby galaxies M31
and M32 (Menou \& Quataert \citeyear{Men01}; \S\ref{sss:stardisk}).
This discrepancy has been interpreted as indicating that most of the
debris was blown away during the early super-Eddington accretion phase.
Following that, the luminous accretion phase shuts off early, either
because the low-mass accretion disk spontaneously switches over to
a low radiative efficiency solution when $\dot{M}\!\lesssim\!10^{-2}\dot{M}_{\mathrm{Edd}}$,
or else because increasingly larger parts of the cold outer disk become
neutral, unable to sustain magneto-hydrodynamic turbulence and therefore
effectively inviscid (Menou \& Quataert \citeyear{Men01}).

\subsubsection{Signatures of tidal disruption}

\label{sss:tdsign}

While there is as yet no clear evidence of tidal disruption in the
GC, there are several ideas about the form such evidence could take.

The ejection of tidal debris with energies that far exceed those released
in normal supernovae could provide a long-term signature of tidal
disruption, in the form of unusual supernova remnant-like structures.
This was suggested as the origin of the Sgr A East radio source (Khokhlov
\& Melia \citeyear{Kho96}). Sgr A East is a prominent non-thermal
radio continuum source with an elongated shell-like structure extending
over 10--20 pc, offset from $\SgrA$, but engulfing it at the edge
of the structure (e.g. Yusef-Zadeh \& Morris \citeyear{Yus87}). The
shell structure and the non-thermal spectrum are reminiscent of a
supernova remnant. Early estimates of the energy required to evacuate
the radio-emitting cavity inside the dense surrounding molecular clouds
yielded an extreme lower bound of $E\!>\!4\!\times\!10^{52}$ erg
(Mezger et al. \citeyear{Mez89}). The high energy requirement, large
size and non-spherical shape are in line with what is expected from
a deep tidal disruption event. However, more recent estimates of the
energy from X-ray observations (Maeda at al. \citeyear{Mae02}; Sakano
et al. \citeyear{Sak04}) yield a much lower value, which together
with the metal-rich abundances observed in Sgr A East, are consistent
with the energy and elemental output of a single normal core collapse
supernova (but see \S\ref{sss:detonate} about the possibility of
nucleosynthesis during a tidal detonation of a low mass star). The
single supernova interpretation is also supported by observations
of molecular transitions, which allow an estimate of the kinetic energy
in molecular gas that is being swept by the expansion of Sgr A East
(Herrnstein \& Ho \citeyear{Her05}).

A tidal disruption event is expected to produce a short luminous UV/X-ray
accretion flare (\S\ref{sss:tdafter}), which will likely also have
a hard X-ray component, as is seen from accreting MBHs in AGN. These
primary X-ray photons will encounter gas clouds as they propagate
from the MBH, and will be Thomson-scattered into the line-of-sight
or absorbed and re-emitted as fluorescence lines, primarily in the
6.4 keV K$\alpha$ line from neutral Iron ({}``X-ray reflection nebula'').
The reflected photons arriving at the observer at any given time were
scattered (or re-emitted) from a parabolic surface around the MBH.
Reflected X-ray emission is not an exclusive signature of tidal disruption.
There could well be other causes for episodic accretion flares or
short high-state epochs. However, the spatial distribution and temporal
changes in the scattered X-ray continuum photons and the emission
lines can at least provide information on the position of the source
and the time and duration of the flare, as the paraboloid of equal
arrival time sweeps the volume around the MBH at the speed of light
(Sunyaev \& Churazov \citeyear{Sun98}). The X-ray reflection effect
could effectively lengthen by $\sim\!2$ orders of magnitude the time
available for detecting an accretion flare. However, even if the distribution
of gas clouds extends far from the center, it is unlikely that the
fluorescence could be detected as long as $\Gamma_{t}^{-1}\!\sim\!10^{4}\,\mathrm{yr}$
years after the tidal disruption event ($\lesssim\!1$ kpc from the
center) due the geometrical dilution of the primary X-ray flux with
distance. 

The present observed level of diffuse X-ray emission from the central
$\sim\!700$ lyr of the GC was used to argue that $\SgrA$ has not
radiated at the Eddington level for even a day in the last 400 yr
(Sunyaev, Markevitch \& Pavlinsky \citeyear{Sun93}). The detected
6.4 keV line emission from several sites near the GC (Koyama et al.
\citeyear{Koy96}; Murakami et al. \citeyear{Mur00}, \citeyear{Mur01})
was found to be well reproduced by an X-ray reflection nebula model.
The luminosity required to reproduce the 6.4 keV line exceeds by 3
orders of magnitude any observed X-ray point source in the region,
and by 6 orders of magnitude the present average X-ray luminosity
of $\SgrA$ (Baganoff et al. \citeyear{Bag01}; Goldwurm et al. \citeyear{Gol03})
. This was taken as evidence that $\SgrA$ was active $>\!300$ yr
ago. The detection of high energy X-rays ($\le\!200$~keV) from a
giant molecular cloud further supports this interpretation (Revnivtsev
et al. \citeyear{Rev04}). The reconstructed X-ray luminosity and
hard X-ray spectrum slope are characteristic of low-luminosity AGN,
and suggest that $\SgrA$ was active for at least 10 yr some 300--400
years ago. However, recent X-ray observations with high spectral resolution
(Predehl et al. \citeyear{Pre03}) do not support this interpretation,
due to the absence of a strong 7.1 keV absorption edge that is expected
to accompany the 6.4 keV emission line in X-ray reflection nebulae,
and due to the lack of the expected inverse correlation between the
6.4 keV line surface brightness and distance from the center.

Another possible long-term after-effect of a strong tidal encounter
is tidal stripping, the partial disruption of evolved giants (Davies
\& King \citeyear{Dav05}). Evolved giant stars have an extended,
cool, low density envelope and a hot, massive compact core. Because
of its low surface gravity, the envelope can be stripped off the core
even at large periapse distances where denser main-sequence stars
are safe. The bare core, while no longer burning, will likely remain
hot ($T\!>\!10^{5}\,\mathrm{K}$) and luminous ($L\!>\!10^{2}\,\Lo$)
for $10^{3}$--$10^{6}$ yr after the tidal stripping event. The hot
core will appear as a {}``supersoft X-ray source'', which can be
detected in the GC and in the centers of other nearby galaxies (Di
Stefano et al. \citeyear{Dis01}). The challenge lies in distinguishing
tidally stripped cores from normal young hot white dwarfs and the
He cores of massive stars that lost their envelopes by strong winds
or by interactions with a binary companion. In addition, Collisional
stripping (\S\ref{sss:RGcoll}) is expected to produce a similar
effect.

\subsubsection{Tidal detonation}

\label{sss:detonate}

When a disrupted star crosses deep inside the tidal radius, $b^{-1}\!\gtrsim\!5$,
it experiences a very strong and rapid tidal compression during periapse
passage. Figure (\hlink{f:disrupt}) shows a qualitative demonstration
of the effect by a simple (low-resolution Newtonian) Smoothed Particle
Hydrodynamics (SPH) simulation%
\footnote{SPH is an algorithm for simulating the hydrodynamics of 3D self-gravitating
fluids, which is commonly used in the study of stellar collisions
(Lucy \citeyear{Luc77}; Gingold \& Monaghan \citeyear{Gin77}; see
review by Monaghan \citeyear{Mon92}). The star is represented by
discrete spherical mass elements, each with an internal density distribution
that peaks in the center and falls to zero at the edge. The total
density at a point is the sum of densities in all overlapping spheres
that include the point. The resulting density field is continuous
and differentiable, and so its thermodynamic properties can be evaluated
everywhere once an equation of state is specified. Every time step,
the positions of the mass elements are updated according to the gravitational
force and the pressure gradient, and the sphere sizes are readjusted
to reflect the changes in the local space density of elements.%
}. The compression occurs perpendicular to the orbital plane and it
more than compensates for the stretching of the star along the orbit,
resulting in a rapid increase in the central density. This happens
despite the fact that the tidal force tensor is traceless (i.e. the
force lines do not converge)---Both the inward- and outward-directed
tidally induced velocities relative to the star's center, $V_{t}\!\sim\!\Rs/\tau_{p}$,
are of the same order of magnitude. However, expansion along one dimension
leads to a relative increase in that dimension of only order unity,
whereas collapse in another dimension leads to an arbitrarily large
relative decrease in that dimension, so the net effect is a vanishing
volume with diverging central density (i.e., on the periapse passage
timescale $\tau_{p}$ the volume shrinks roughly as $\Rs^{3}\rightarrow\Rs(\Rs+V_{t}\tau_{p})(\Rs-V_{t}\tau_{p})\rightarrow0$).
An order of magnitude estimate indicates that the maximal central
compression ratio can be as high as $\max\rho_{c}/\rho_{c}$$\sim\! b^{-3}$,
where $\rho_{c}$ is the initial stellar central density. This upper
limit is derived by assuming the full conversion of the kinetic energy
of the adiabatically collapsing stellar ellipsoids into internal heat
(Carter \& Luminet \citeyear{Car83}). The actual compression ratio
that can be achieved, and its ultimate consequences depend on the
uncertain details of the response of the gas hydrodynamics and geometry
to the sudden compression via increased nuclear energy production
and shocks. Simulations indicate that the maximal compression ratio
is significantly smaller than the kinematic upper limit (Luminet \&
Marck \citeyear{Lum85}; Khokhlov, Novikov \& Pethick \citeyear{Kho93b}),
but still very high, $\max\rho_{c}/\rho_{c}$$\sim\! b^{-1.5}$--
$b^{-2}$ (Bicknell \& Gingold \citeyear{Bic83}; Laguna et al. \citeyear{Lag93}). 

It has been suggested that such a strong tidal compression could lead
to a ignition of nuclear reactions and the sudden release of energy,
thereby producing a supernova-like explosion and an energetic outflow
of metal-enriched stellar debris (Carter \& Luminet \citeyear{Car82,Car83};
Carter \citeyear{Car92}). The nuclear energy produced during the
compression is $E_{\mathrm{nuc}}\!\sim\! Q\tau_{p}$, where $Q$ is
the nuclear power. The nuclear power from a tidally compressed burning
core scales very roughly as $Q\!\sim\!\Ls\left(\rho_{c}/\rho_{\star}\right)^{k}\left(T_{c}/\Ts\right)^{n}$,
where $\Ls$, $\rho_{\star}$ and $\Ts$ are the initial equilibrium
luminosity, central stellar density and temperature. For hydrogen
burning, $k\!=\!1$, $n\!\sim\!4$ for the $p$-$p$ chain and $n\!\sim\!16$
for the CNO cycle; for helium burning (triple-$\alpha$), $k\!=\!2$
and $n\!\sim\!40$. As a crude estimate, consider a simplified scenario
where the compressed stellar core continues burning by the same nuclear
process, but with scaled-up power. Assuming adiabatic contraction
of an ideal gas (i.e. neglecting nuclear burning during the contraction),
with $P\!\propto\rho^{\gamma}$ and $T\!\propto\!\rho^{\gamma-1}$
for $\gamma\!=\!5/3$, including the scaling of the periapse passage
time, $\tau_{p}\!\sim\!\tau_{\star}b^{3/2}$, and parameterizing $\max\rho_{c}/\rho_{c}$$\sim\! b^{-u}$
($u\!=\!3/2$ will be assumed below), the nuclear energy produced
varies with $b$ as 

\begin{equation}
E_{\mathrm{nuc}}\!\sim\!\tau_{\star}\Ls b^{-s}\,,\qquad s\equiv u[k+n(\gamma-1)]-3/2\,.\end{equation}
For a solar type star ($p$-$p$ burning, $\Es\!\sim\!4\!\times\!10^{48}\,\mathrm{erg}$
and $\tau_{\star}\!=\!1600$ sec, $s\!=\!10$), the energy produced
is negligible, $E_{\mathrm{nuc}}/\Es\!\sim\!2\!\times\!10^{-12}b^{-10}\!\ll1$,
as long as $b^{-1}\!<\!13$. The $p$-$p$ chain is too slow to be
an important source of energy during the brief tidal compression.
However, for a $10\,\Mo$ main sequence star, such as are observed
near the Galactic MBH (CNO cycle burning, $\Es\!\sim\!8\times10^{49}\,\mathrm{erg}$,
$\tau_{\star}\!\sim\!4800$ sec and $\Ls\!\sim\!6\!\times\!10^{3}\,\Lo$,
$s\!=\!16$), $E/\Es\!\sim\!10^{-9}b^{-16}\!\sim\!1$ even for a moderate
penetration parameter of $b^{-1}\!\sim\!3.6$. CNO cycle burning accelerated
by the tidal compression will cause the star to explode while it is
being disrupted. A helium burning giant, where $E_{\mathrm{nuc}}\!\propto\! b^{-41.5}$,
will be even more susceptible to tidal detonation.

Since the cross-section for a close encounter with the MBH with periapse
$\le r_{p}$ scales as $r_{p}$ (\S\ref{ss:dissipation}), tidal
detonation events are expected to occur at an even smaller rate than
that of tidal disruption events, $\Gamma_{d}\!\simeq\!\Gamma_{t}b\!\lesssim\!10^{-5}\,\mathrm{yr}$
(\S\ref{sss:tdrate}), and so it is not likely that such an event
will be observed in the GC in real time. Moreover, the tidal detonation
is energetically negligible compared to the kinetic energy that is
transferred to the debris by the tidal field from the stellar orbital
energy (\S\ref{sss:tdafter}). However, tidal detonation can affect
the aftermath of a tidal disruption event more subtly. By increasing
the velocity at which debris is expelled relative to its center of
mass along the orbital plane, it can modify the properties of the
subsequent accretion on the MBH by making the bound matter more tightly
bound and strongly relativistic. An additional effect of {}``pancake
detonation'' is to modify the chemical composition of the interstellar
medium around the MBH. Pichon (\citeyear{Pic85}) finds that nucleosynthesis
by pancake detonation leads to the synthesis of proton-enriched isotopes,
in contrast with core-collapse supernovae, which synthesize neutron-enriched
isotopes. Unusual nucleosynthesis may also be associated with a related
process, the tidal detonation of a WD caused by a combination of tidal
compression and enhanced self-gravity due to GR terms. The compression
can lead to pyconuclear reactions and a thermonuclear runaway under
highly degenerate conditions, even when the WD mass is well below
the Chandrasekhar mass (Wilson \& Mathews \citeyear{Wil04}).

\begin{figure}[!t]
\centerline{\htarget{f:disrupt}\begin{tabular}{cc}
\includegraphics[%
  scale=0.34,
  angle=270]{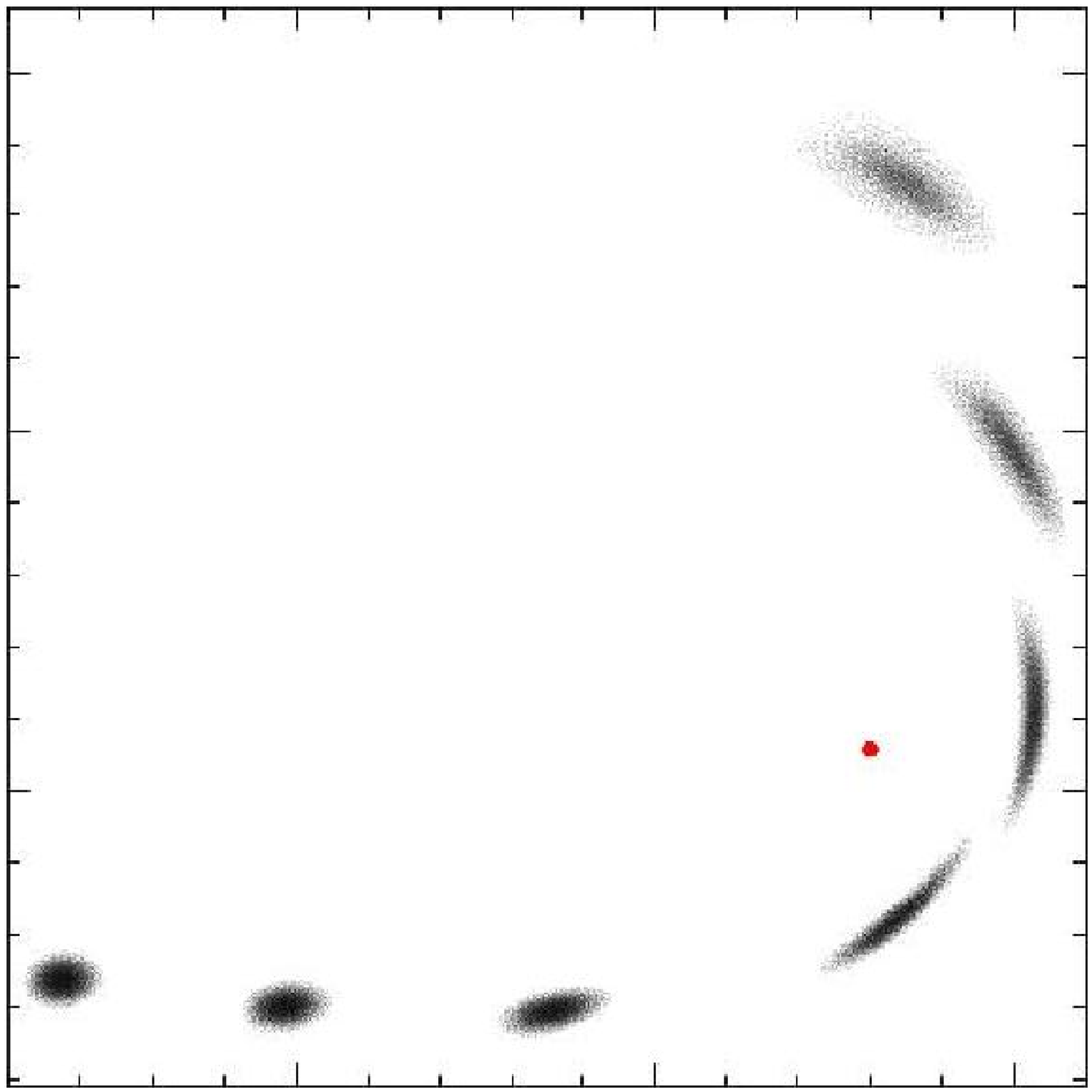}&
\includegraphics[%
  scale=0.34,
  angle=270]{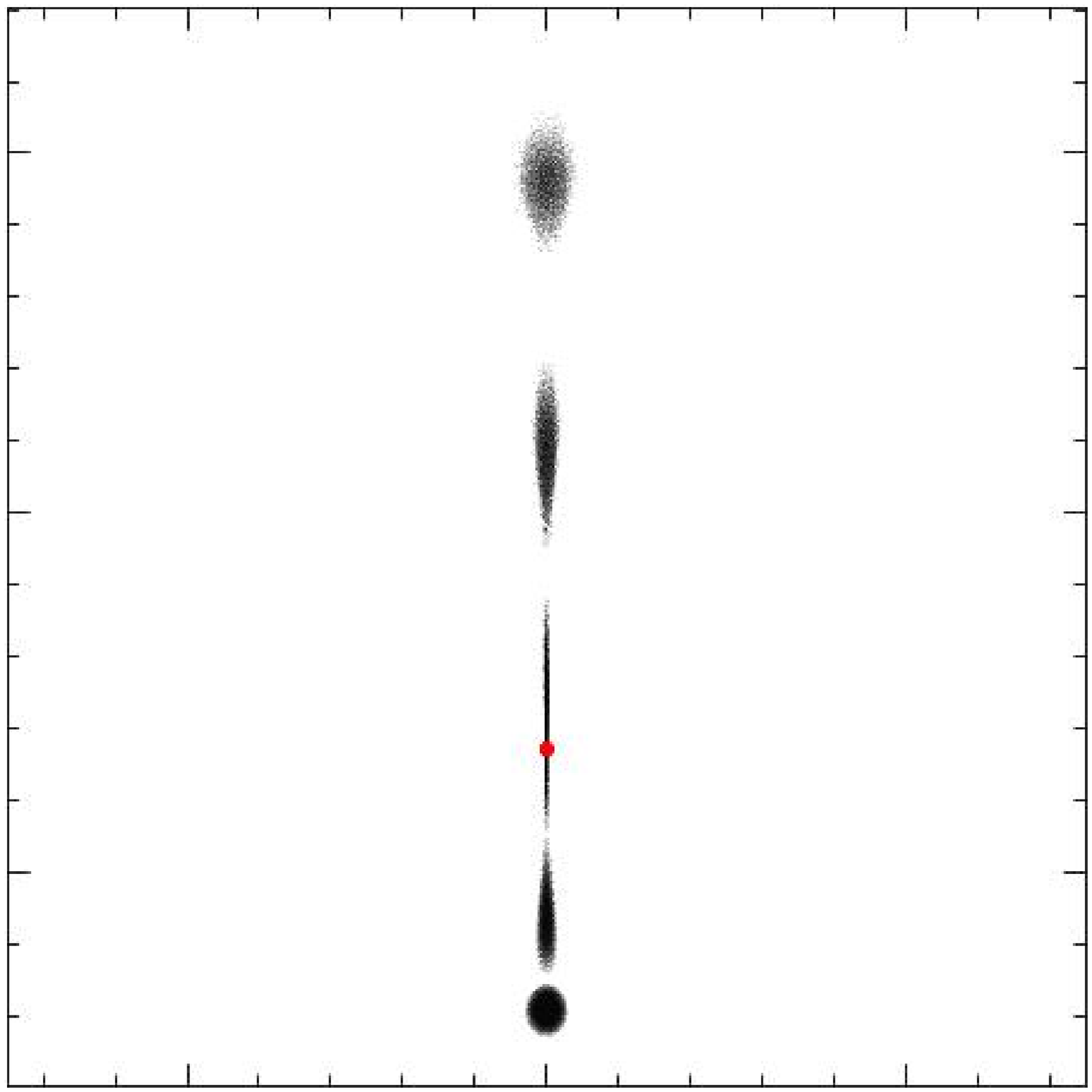}\tabularnewline
\multicolumn{2}{c}{\includegraphics[%
  scale=0.4,
  angle=270]{rhoc_evol.ps}}\tabularnewline
\end{tabular}}

\caption{{\small \label{f:disrupt}A sequence of snapshots from an SPH simulation
of a $\Ms\!=\!10\,\Mo$, $\Rs\!=\!4.5\,\Ro$ star (represented by
an $n=1.5$ ideal gas polytrope) being disrupted by a very close passage
to a $3\!\times\!10^{6}\,\Mo$ MBH on a parabolic orbit with $r_{p}\!=\!4\,\Rs\!=\!0.06r_{t}\!=\!1.4r_{S}$.
The simulation assumes Newtonian gravity and neglects nuclear burning,
and so while the results are qualitatively correct, they are quantitatively
inaccurate: GR effects analogous to periapse shift, omitted here,
will actually cause the expanding debris in penetrating encounters
($b^{-1}\!\gg\!1$) to fan out in a thin crescent-like shape centred
on the MBH (Laguna et al. \citeyear{Lag93}). Top left: face-on view
of the star as it orbits (counter clockwise) the MBH (dot at bottom
right corner). Top right: edge-on view of the same orbit (the two
left-most snapshots shown in the face-on view are omitted from the
edge-on view, for clarity). Bottom: time evolution of the central
density. Time is measured in terms of the star's dynamical time $\tau_{\star}\!=\!4840\,\mathrm{s}$.
Note the strong and rapid tidal compression at periapse (a density
increase by a factor of 11 on a timescale of $\sim\!100\,\mathrm{s}$)
.}}
\end{figure}

\subsection{Dissipative interactions with the MBH}

\label{ss:dissipation}

A star whose orbit passes just outside the tidal disruption radius,
or the event horizon if $r_{S}\!>\! r_{t}$, will survive the close
encounter with the MBH. Such ''near misses'' often involve some
orbital energy loss, $\Delta E$, for example by the emission of GW
(when $r_{S}\!<\! r_{p}\!\lesssim\!\mathrm{few\times\!}r_{S}$) or
by the work invested in raising stellar tides and exciting stellar
oscillations (when $r_{t}\!<\! r_{p}\!\lesssim\!\mathrm{few\times\!}r_{t}$).
If the star maintains its nearly radial trajectory for many orbital
periods, the dissipational energy losses will gradually accumulate
and cause the orbit to shrink and the star to spiral in. This will
continue until either the dissipation shuts off, or the star is destroyed
by the MBH, or is disrupted by the dissipational energy. An initial
close encounter with the MBH is in itself not enough to guarantee
that the star will reach the final stages of inspiral. The inspiral
timescale is usually orders of magnitude longer than the orbital timescale.
During that time the star is vulnerable to the same scattering process
that deflected it into the low-$J$ orbit in the first place. It can
be scattered again to an orbit with a larger periapse, where dissipation
is inefficient, or to a loss-cone orbit where it is promptly destroyed.
Thus the probability for completing inspiral is small, and the rate
of inspiral events is much smaller than that of direct infall events
(\S\ref{sss:inspiral}). In the GC $r_{t}\!>\! r_{S}$, and most
stars that undergo a non-disruptive close tidal encounter with the
MBH are subsequently scattered to a wide orbit and survive after experiencing
an extreme tidal distortion, spin-up, mixing and mass-loss that may
affect their evolution and appearance. Such {}``tidally scattered''
stars eventually comprise a few percents of the stellar population
within the MBH radius of influence (\S\ref{sss:tscatter}).

If the inspiraling star avoids being scattered to a wider orbit or
falling directly into the MBH, its orbital period $P$ will decrease
every peri-passage and the mean dissipated power, $\sim\!\Delta E/P$,
will increase. If the emitted power grows high enough, it may have
observed in the form of GW emission (\S\ref{sss:GW}), or high tidal
luminosity (\S\ref{sss:squeezar}). A different class of mechanisms
for orbital energy dissipation (not relevant for the present-day GC)
is by drag against a massive accretion disk, if one is present (\S\ref{sss:stardisk}).
This may play a role in feeding stars to MBHs.

\subsubsection{Orbital inspiral into the MBH}

\label{sss:inspiral}

Inspiral can be treated as an extension of loss-cone theory. The inspiral
problem was studied in the context of GW emission from compact objects
orbiting a MBH (Hils \& Bender \citeyear{Hil95}; Freitag \citeyear{Fre01};
Sigurdsson \& Rees \citeyear{Sig97b}; Sigurdsson \citeyear{Sig97a};
Ivanov \citeyear{Iva02}; see review by Sigurdsson \citeyear{Sig03};
Hopman \& Alexander \citeyear{Hop05}), and was generalized to arbitrary
dissipation mechanisms and applied to tidal capture by Alexander \&
Hopman (\citeyear{Ale03b}) and Hopman, Portegies-Zwart \& Alexander
(\citeyear{Hop04}). 

Inspiral processes are characterized by an inspiral timescale, $t_{0}(\varepsilon_{0},J_{0})$,
which is the time needed for a star starting at an orbit with initial
energy $\varepsilon_{0}$ and angular momentum $J_{0}$ to complete
the inspiral ($\varepsilon\!\rightarrow\!\infty$, $P\!\rightarrow\!0$).
The dissipated specific energy per orbit, $\Delta\varepsilon$, typically
increases sharply as $r_{p}$ is decreased and correspondingly $t_{0}$
decreases sharply with $r_{p}$ (Eq. \ref{e:t0GW}). Over the range
of interest the inspiral time satisfies $P\!\ll\! t_{0}\!\ll\! t_{r}$
(when $P\!\lesssim\! t_{0}$, $r_{p}$ is so small that the inspiral
is indistinguishable from direct infall. Conversely, when $t_{0}\!\lesssim\! t_{r}$,
$r_{p}$ is so large that dissipation is negligible). The condition
$t_{0}\!\ll\! t_{r}$ also implies that $\varepsilon$-diffusion by
scattering is negligible during the inspiral. The same is not generally
true for $J$-diffusion. However, when the initial inspiral time is
also much shorter than the angular momentum relaxation timescale (Eq.
\ref{e:tJ}), 

\begin{equation}
t_{0}(\varepsilon_{0},J_{0})\ll t_{J}(\varepsilon_{0},J_{0})\,,\label{e:t0tp}\end{equation}
then the inspiral is completed with $J\!\sim\! J_{0}$ nearly constant.
When Eq. (\ref{e:t0tp}) is satisfied the star decouples from the
random perturbations that governed its dynamics in the initial scattering-dominated
phase, and it enters the final dissipation-dominated phase where its
orbit is deterministic and governed by the MBH potential and the dissipation
mechanism only.

It is not necessary to specify $J_{0}$ to obtain a \emph{statistical}
description of $t_{0}$. The mean inspiral time $\bar{t}_{0}$ of
stars that complete the inspiral is effectively a function of $\varepsilon_{0}$
only because as long as $t_{J}\!<\! t_{0}\!<t_{r}$, $J$-diffusion
by scattering will continue until such time when $J$ satisfies $t_{0}\!\ll\! t_{J}$.
Thus, the requirement of successful inspiral, together with the dynamics
of random walk in $J$-space in the presence of dissipation, define
a mean value of angular momentum in the inspiral phase, $\bar{J}_{0}(\varepsilon_{0})$.
The value of $\bar{J}_{0}$ and the distribution of $J_{0}$ around
it can be calculated once the specific properties of the stellar system
and the dissipation mechanism are given (Hopman \& Alexander \citeyear{Hop05}).

In analogy to the case of direct disruption, Eq. (\ref{e:t0tp}) sets
a lower limit, $\varepsilon_{c}$, on the initial energy (or an upper
limit, $r_{c}$, on the initial orbital radius) that is required for
completing inspiral. Stars with $\varepsilon_{0}\!>\!\varepsilon_{c}$
(the diffusive regime) will typically complete inspiral, whereas stars
with $\varepsilon_{0}\!<\!\varepsilon_{c}$ will be typically scattered
to high-$J$ or loss-cone orbits and therefore have a vanishingly
small chance of completing inspiral. Because $t_{0}\!\gg\! P$, the
phase-space volume available for the diffusive regime of inspiral
is much smaller than that of direct infall ($\varepsilon_{c}^{\mathrm{inspiral}}\!\gg\!\varepsilon_{c}^{\mathrm{infall}}$,
$r_{c}^{\mathrm{inspiral}}\!\ll\! r_{c}^{\mathrm{infall}}$) and consequently
the inspiral rate is generally much smaller than that of direct infall.
Inspiral can be likened to a race between scattering and dissipation,
in which dissipation can win only if the track is short.

It is sometimes the case that the energy dissipated per orbit, $\Delta\varepsilon$,
is constant over the course of the inspiral to a good approximation
(for example when $\Delta\varepsilon$ depends mainly on the periapse
distance, which remains nearly fixed when $t_{0}\!\ll\! t_{J}$ and
when the dissipated angular momentum $\Delta J$ is small). In that
case, the inspiral timescale in a Keplerian potential, $\hat{t}_{0}$
(calculated relative to a given time based on the values of $\Delta\varepsilon$,
$\varepsilon_{0}$ and $P_{0}$ at that time), is \begin{equation}
\hat{t}_{0}=\frac{1}{\Delta\varepsilon}\int_{\varepsilon_{0}}^{\infty}P(\varepsilon)\mathrm{d}\varepsilon=2\frac{\varepsilon_{0}P_{0}}{\Delta\varepsilon}\propto\frac{1}{\varepsilon_{0}^{1/2}\Delta\varepsilon}\,,\label{e:t0constE}\end{equation}
and the period and the semi-major axis evolve simply as\begin{equation}
P=P_{0}(1-t/\hat{t}_{0})^{3}\,,\qquad a=a_{0}(1-t/\hat{t}_{0})^{2}\,.\label{e:t0constPa}\end{equation}
The number of inspiral orbits by time $t\!<\!\hat{t}_{0}$ is then
given by\begin{equation}
N_{\mathrm{orb}}=\frac{\hat{t}_{0}}{2P_{0}}\left[(1-\frac{t}{\hat{t}_{0}})^{-2}-1\right]\,.\label{e:t0constNorb}\end{equation}

\subsubsection{Gravitational wave emission}

\label{sss:GW}

A compact enough object that can survive tidal disruption at a distance
of $r_{p}\!\sim\!\mathrm{few\!\times\!}r_{S}$, will dissipate its
orbital energy and angular momentum by the emission of GW radiation.
As the orbit decays, both the emitted power and frequency of the GW
increase. The specific orbital energy and angular momentum (per unit
stellar mass) lost each orbit by the emission of GW are ($r_{p}\!>\!3r_{S}$,
$m\!\gg\!\Ms$) (Peters \citeyear{Pet64})

\begin{equation}
\Delta\varepsilon_{\mathrm{GW}}=-\frac{8\pi}{5\sqrt{2}}f(e)\frac{\Ms c^{2}}{m}\left(\frac{r_{p}}{r_{S}}\right)^{-7/2}\,,\qquad f(e)=\frac{1+\frac{73}{24}e^{2}+\frac{37}{96}e^{4}}{(1+e)^{7/2}}\,.\label{e:dEgw}\end{equation}

\begin{equation}
\Delta J_{\mathrm{GW}}=-\frac{16\pi}{5}g(e)\frac{G\Ms}{c}\left(\frac{r_{p}}{r_{S}}\right)^{-2}\,,\qquad g(e)=\frac{1+\frac{7}{8}e^{2}}{(1+e)^{2}}\,.\label{e:dJgw}\end{equation}

An orbit in the Schwarzschild metric ends in the MBH if its angular
momentum is smaller than some threshold, which for bound orbits is
almost independent of $\varepsilon$. The effective loss cone is therefore
defined as the threshold angular momentum for a zero-energy orbit
 (e.g. Shapiro \& Teukolsky \citeyear{Sha83}) 

\begin{equation}
J_{lc}=\frac{4Gm}{c}\,.\end{equation}
 Stars begin their inspiral on extremely eccentric orbits. The interplay
between dissipation and scattering in the presence of a mass sink
conserves the trend toward very high eccentricity until the very last
stages of the inspiral (Hopman \& Alexander \citeyear{Hop05}). For
$e\!\sim\!1$ the periapse can be expressed as $r_{p}/r_{S}=4(J/J_{lc})^{2}$
and 

\begin{equation}
\Delta\varepsilon_{\mathrm{GW}}=\varepsilon_{1}\left(\frac{J}{J_{lc}}\right)^{-7}\,,\qquad\varepsilon_{1}\equiv\frac{85\pi}{3\!\times\!2^{13}}\frac{\Ms c^{2}}{m}\,.\label{e:DEgwJ}\end{equation}
The GW inspiral time increases rapidly with $J$,\begin{equation}
\hat{t}_{0}\simeq2P_{0}\frac{\varepsilon_{0}}{\varepsilon_{1}}\left(\frac{J}{J_{lc}}\right)^{7}\,.\label{e:t0GW}\end{equation}

GW inspiral in the GC is of interest in anticipation of the planned
\emph{Laser Interferometer Space Antenna} (LISA) mission. LISA is
expected to detect GW emitted by compact objects spiraling into MBHs%
\footnote{LISA is sensitive to GW in the frequency range $10^{-4}\!\lesssim\!\nu\!\lesssim\!10^{-1}\,\mathrm{Hz}$
and is optimized for the range $10^{-3}\!\lesssim\!\nu\!\lesssim\!10^{-2}\,\mathrm{Hz}$,
which corresponds to the frequency at the last stable circular orbit
(LSCO) for MBHs of up to $\lesssim\!10^{7}\,\Mo$ ($\nu_{\mathrm{LSCO}}\sim\sqrt{Gm/(3r_{S})^{3}}/2\pi=c^{3}/(12\sqrt{6}\pi Gm)\sim2\!\times\!10^{-3}(m/10^{6}\,\Mo)^{-1}\,\mathrm{Hz}$).
Extreme mass-ratio GW sources are of special interest because they
are the cleanest probes of spacetime near a MBH, albeit with a weak
signal (Eq. \ref{e:dEgw}; see review by Glampedakis \citeyear{Gla05}).
This is to be contrasted with the yet intractable GR physics of 2
merging MBHs or NSs. The expected GW signal from inspiralling stellar
mass objects at any given time is below the detector noise, but over
the mission's lifetime of several years, $\sim\!10^{5}$ orbital periods
can be observed from an active source. It should then be possible
to extract the signal from the noise by utilizing the properties of
the GW waveforms, which depend on the orbital parameters, in particular
the eccentricity. %
} at cosmological distances (Barack \& Cutler \citeyear{Bar04}; Gair
et al \citeyear{Gai04}). The estimated rates are $10^{-9}$--$10^{-8}\,\mathrm{WD\,\, yr^{-1}}$
per galaxy (Hils \& Bender \citeyear{Hil95}; Sigurdsson \& Rees \citeyear{Sig97b};
Freitag \citeyear{Fre01} ; Ivanov \citeyear{Iva02}; Hopman \& Alexander
\citeyear{Hop05}) and $10^{-8}$--$10^{-4}\,\mathrm{SBH\, yr^{-1}}$
per galaxy (Miralda-Escud\'e \& Gould \citeyear{Mir00}; see review
article on rate estimates by Sigurdsson \citeyear{Sig03}). For typical
galactic density profiles, the rate depends only weakly on the relaxation
time, because of the near cancellation of the $t_{r}$-dependence
of the scattering rate into the {}``inspiral cone'' and that of
the volume of the diffusive regime (\S\ref{sss:inspiral}). Furthermore,
the rate depends only weakly on $m$ because of the $m/\sigma$ relation
(Eq. \ref{e:msigma}) (Hopman \& Alexander \citeyear{Hop05}). Thus,
the chances of detecting GW sources in any given galaxy are low, irrespective
of its dynamical properties. 

The one exception is the Milky Way, where the proximity of the MBH
makes it possible to detect even very low-mass objects with a weak
GW signal ($\Delta E_{\mathrm{GW}}\!\propto\!\Ms^{2}$, Eq. \ref{e:dEgw})
(Freitag \citeyear{Fre03}). Very low-mass MS stars dominate the population
and their numbers far exceed those of compact objects (table \ref{t:StellarPop}).
The mean density of a star increases with decreasing stellar mass
and reaches a maximum at $\Ms\!<\!0.1\,\Mo$, near the transition
to brown dwarfs. Such stars can therefore survive the tidal field
of a $\sim\!10^{6}\,\Mo$ MBH and become GW sources. The mean number
of active GW sources in the GC at any given time is $N_{GW}\!\sim\!\Gamma_{i}\hat{t}_{0}$,
where $\hat{t}_{0}$ is the inspiral time from an initial orbit with
$P_{0}\!\sim\!10^{4}\,\mathrm{s}$, the lowest frequency detectable
by LISA. Freitag (\citeyear{Fre03}) finds a capture rate of $\Gamma_{i}\sim\!10^{-6}\,\mathrm{yr^{-1}}$
for low mass MS stars in the GC (but see Alexander \& Hopman \citeyear{Ale03b}
who find a significantly smaller rate). This correspond to a few active
GW sources at any given time in the GC, $N_{GW}\!\sim\!\Gamma_{i}\hat{t}_{0}\!>\!1$
for $\hat{t}_{0}\!=\!2\!\times\!10^{6}\,\mathrm{yr}$ (Eqs. \ref{e:DEgwJ},
\ref{e:t0GW} with $m\!=\!3\!\times\!10^{6}\,\Mo$, $\Ms\!=\!0.05\,\Mo$,
$r_{p}/r_{S}=10$ and $P_{0}\!=\!10^{4}\,\mathrm{s}$). It is not
likely that LISA will detect GW from compact objects in the GC. The
rates quoted above translate to $N_{GW}\!\ll\!1$ for the inspiral
of SBHs and NSs and to $N_{GW}\!<\!1$ for WDs (Freitag \citeyear{Fre03}).

\subsubsection{Tidal heating, squeezars, and tidal capture}

\label{sss:squeezar}

Dissipative interactions of MS stars with the Galactic MBH proceed
primarily via tidal heating, because the tidal disruption radius for
a solar mass star lies well outside the event horizon, at $r_{t}\!\sim\!10r_{S}$
. At that distance the tidal interaction dominates over the energy
released by GW emission and the interaction can be treated to good
approximation in the Newtonian limit.

Inspiral by GW emission differs from inspiral by tidal heating in
that the GW radiation is not localized in the star (the typical wavelength
is $\lambda_{\mathrm{GW}}\!\sim\! r_{p}\!>\! r_{S}\!\gg\!\Rs$). In
contrast, the tidal energy extracted from the orbit is localized and
thermalized \emph{inside} the star. A star that is initially on a
zero energy (parabolic) orbit has to dissipate many orders of magnitude
its own binding energy in order to circularize, \begin{equation}
E_{c}=\frac{Gm\Ms}{2r_{p}}\sim\frac{1}{2}\left(\frac{m}{\Ms}\right)^{2/3}\frac{\Es}{b}\sim10^{4}\Es\label{e:Ec}\end{equation}
where it is assumed that the circularization occurs at the initial
periapse radius%
\footnote{If the impulsive relation holds between the tidal energy and angular
momentum (Eq. \ref{e:dEimp}), then $r_{p}\!=\!\mathrm{const}$ and
the initial $r_{p}$ is also the circularization radius. Conversely,
if orbital angular momentum is conserved, then the circularization
radius is $2r_{p}$.%
}. It is therefore unlikely that the star will survive tidal inspiral
around a MBH. It will ultimately be destroyed, either by expanding
beyond its tidally limited maximal size and breaking up ($r_{t}$
increases with $\Rs$ until it overtakes $r_{p}$; Eq. \ref{e:rt}),
or by exceeding its own Eddington luminosity and evaporating. Before
that happens, the star will exist in a transient phase as a {}``squeezar'',
a star whose atypically high luminosity is powered by tidal interactions
with the MBH rather than by its own nuclear burning (Alexander \&
Morris \citeyear{Ale03a}). It should be noted that tidal capture
is possible around an IBH because in that case $E_{c}/\Es\!\sim{\cal {O}}(10)$
(Eq. \ref{e:Ec}), and the tidal heat can be radiated away in ${\cal O}(10^{4})$
yr (Hopman, Portegies Zwart \& Alexander \citeyear{Hop04}).

The energy deposition in the star in one periapse passage can be parametrized
as (cf \S\ref{sss:spinup}, Eq. \ref{e:dEt})\begin{equation}
\frac{\Delta E_{t}}{\Es}=\frac{T_{2}(b^{3/2})}{b^{6}}\left[\frac{\Rs(t)}{\Rs(0)}\right]^{5}\,,\label{e:dEsq}\end{equation}
 where possible stellar expansion is taken into account by the last
term. As energy and angular momentum are exchanged between the orbit
and the star, the star's mechanical and thermal properties change,
and with them its response to the tidal force. Because of the feedback
between the tidal interaction and the stellar properties, it is difficult
to predict how the star and its orbit evolve in the course of tidal
inspiral. A similar problem is encountered in studies of the circularization
of tidal capture stellar binaries in clusters. The results of such
studies suggest two useful limits that likely bracket the true response:
surface heating with radiative cooling (McMillan, McDermott \& Taam
\citeyear{McM87}) in {}``hot squeezars'', and bulk heating with
adiabatic expansion (Podsdiadlowski \citeyear{Pod96}) in {}``cold
squeezars''. Hot squeezars dissipate the tidal heat in a very thin
surface layer that expands moderately and radiates at a significantly
increased effective temperature. This may apply to stars with large
convective envelopes, where most of the energy in the oscillations
is carried by eigenmodes that attain their maximal amplitude at the
surface and dissipate there. Cold squeezars dissipate the tidal heat
in their bulk, due to possible non-linear coupling between the high-amplitude,
low-order oscillations that are directly excited by the tidal force,
and an infinitude of high order modes which effectively dissipate
instantaneously. Bulk heating leads to quasi-adiabatic, self-similar
expansion at roughly constant effective temperature that doesn't rise
much above its original value.

With these and a few additional simplifying assumptions (motion in
a Keplerian potential with constant $r_{p}$) it is possible to calculate
the evolution of the stellar properties and the orbit. The orbital
evolution is simple when the approximation $\Delta E_{t}\!=\!\mathrm{const}$
is valid (Eqs. \ref{e:t0constE}, \ref{e:t0constPa}, \ref{e:t0constNorb}).
This is assumed for hot squeezars, but is not applicable to cold squeezars
where $\Delta E_{t}$ is not constant due to the adiabatic expansion
(Eq. \ref{e:dEsq}). In that case the evolution has to be calculated
numerically. 

Numeric modeling of squeezar evolution (Alexander \& Morris \citeyear{Ale03a})
show that cold squeezars can brighten by up to $\Delta K\!\sim\!3$
mag. The temperature of hot squeezars can rise up to $\lesssim\!5$
time the original effective temperature, and the star can brighten
by up to $\Delta K\!\sim\!4$ mag (Fig. \hlink{f:squeezar}). The
luminosity and temperature increase could place low mass MS squeezars
($\Ms\!\gtrsim\!2\,\Mo$) for the last $\lesssim\!10^{4}$ yr of their
life in the range of the S-stars. However, none of the S-stars for
which an orbital solution is known approach the MBH close enough for
tidal heating to be significant. Neither does the short squeezar cooling
time allow for the possibility that these stars were tidally heated
in the past and then scattered to wider orbits. Estimates of the inspiral
rate for various tidal dissipation laws and their corresponding squeezar
lifespans suggest that the mean number of squeezars in the GC at any
given time is 0.1--1. Thus, the chances of detecting squeezars in
the GC appear to be low.

\begin{figure}[!t]
\centering{\htarget{f:squeezar}\includegraphics[scale=1.1]{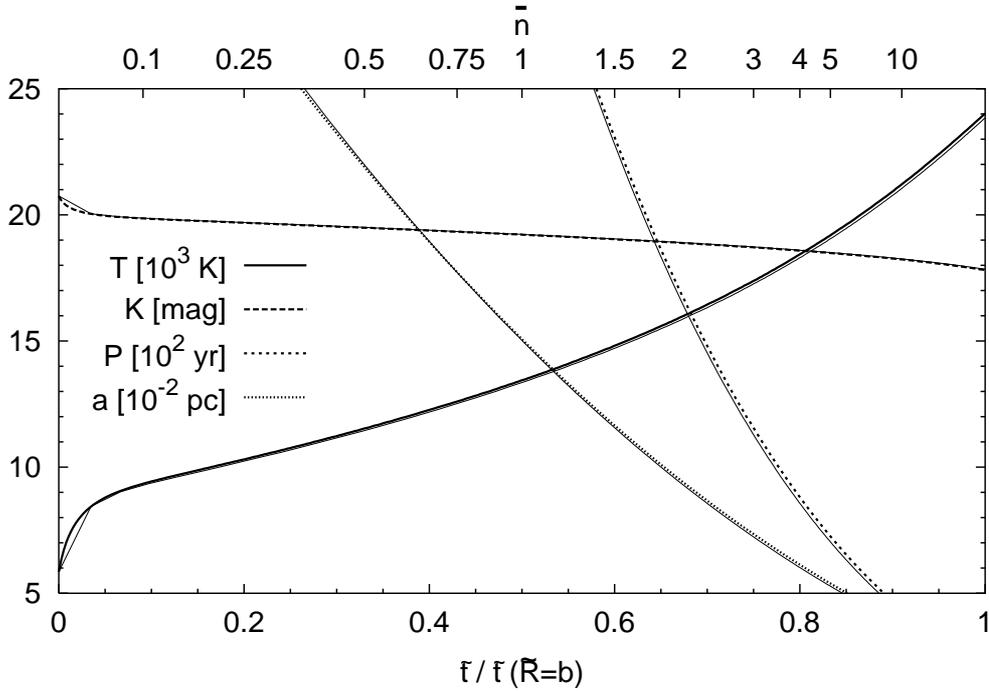}}

\caption{{\small \label{f:squeezar}The evolution of the orbit (period and
semi-major axis) and properties (temperature and $K$-band magnitude)
a $1\, M_{\odot}$ hot squeezar in the GC, initially deflected into
an orbit with $r_{p}\!=\!1.5r_{t}(t\!=\!0)$ and $P_{0}\!=\!1.4\!\times\!10^{4}$
yr, which corresponds to an inspiral time of $t_{0}\!=\!4.9\!\times\!10^{5}$
yr. Time is shown relative to $t_{\mathrm{disrupt}}\!=\!3.7\!\times\!10^{5}$,
the time when $r_{p}\!=\! r_{t}(t_{\mathrm{disrupt}})$ and the star
expands to the point where it is tidally disrupted. At disruption
the tidal luminosity is $642\,\Lo$, $P\!=\!210\,\mathrm{yr}$ and
$e\!=\!1-2.3\!\times\!10^{-4}$. For $\overline{n}$ squeezars in
the GC at any given time, the mean properties of the one in the most
advanced stage of its inspiral can be read off the top axis. (Alexander
\& Morris \citeyear{Ale03a}. Reprinted with permission from} \emph{\small The
Astrophysical Journal}{\small ).}}
\end{figure}

The end result of tidal inspiral (sometimes also called tidal capture)
is disruption and the accretion of a sizable fraction of the stellar
mass by the MBH. Tidal capture was hypothesized to be an important
channel of stellar mass supply to a MBH, equal to, or exceeding by
up to a factor of two the contribution of direct disruption (Frank
\& Rees \citeyear{Fra76}; Novikov, Pethick \& Polnarev \citeyear{Nov92};
Magorrian \& Tremaine \citeyear{Mag99}). Simulations indicate that
lower-mass MBHs ($m\!\lesssim\!10^{7}\,\Mo$) in low density galactic
cores obtain most of their mass from direct stellar disruption (Murphy,
Cohn \& Durisen \citeyear{Mur91}; Freitag \& Benz \citeyear{Fre02}).
If tidal capture were indeed so efficient, this would have implied
that such MBHs could be wholly constructed from the disruption of
stars in the radius of influence, in seeming contradiction with the
empirical $m/\sigma$ relation (Eq. \ref{e:msigma}), which holds
on the much larger scale of the bulge (\S\ref{ss:context}). The
contradiction is resolved by noting that tidal capture, like any slow
inspiral process, is strongly suppressed compared to direct infall.
When the effects of scattering are properly taken into account (\S\ref{sss:inspiral}),
it is found that tidal capture can contribute only an additional few
percent above the mass supplied by direct disruption (Alexander \&
Hopman \citeyear{Ale03b}).

\subsubsection{Tidal scattering }

\label{sss:tscatter}

Inspiral is very inefficient, and therefore most stars that are scattered
to {}``near-miss'' orbits pass by the MBH once and are then deflected
to wider orbits. When the encounter is very close, the star will undergo
an extreme non-disruptive tidal interaction and experience strong
distortion, spin-up, mixing, and possibly some mass loss (Fig. \hlink{f:BHtide}).
These effects may alter the star's subsequent appearance and evolution.
While it is difficult to predict the long-term observational signature
of such a {}``tidal scattering'' event, plausible arguments suggest
that the stellar luminosity will increase, photospheric abundances
will be enriched by hydrogen-burning products (enrichment of $^{4}\mathrm{He}$,
$^{14}\mathrm{N}$, $^{13}\mathrm{C}$ and $^{26}\mathrm{Al}$ and
depletion of $^{12}\mathrm{C}$, $^{16}\mathrm{O}$ and $^{15}\mathrm{N}$)
and the stellar colors will be bluer than expected for a normal star
of that mass (Alexander \& Livio \citeyear{Ale01b}). It is interesting
to note that precisely such abundance anomalies are observed in the
M supergiant IRS 7, $\sim0.2\,\mathrm{pc}$ from $\SgrA$ in projection
(Carr, Sellgren \& Balachandran \citeyear{Car00}). This suggests
that IRS 7 has undergone mixing in excess of values predicted by standard
models or observed in supergiants elsewhere, and has prompted Carr
et al. (\citeyear{Car00}) to hypothesize that {}``extra mixing induced
by rapid rotation may indeed be the fundamental difference between
the evolution of massive stars in the GC and those elsewhere in the
Galaxy.'' 

\begin{figure}[!t]
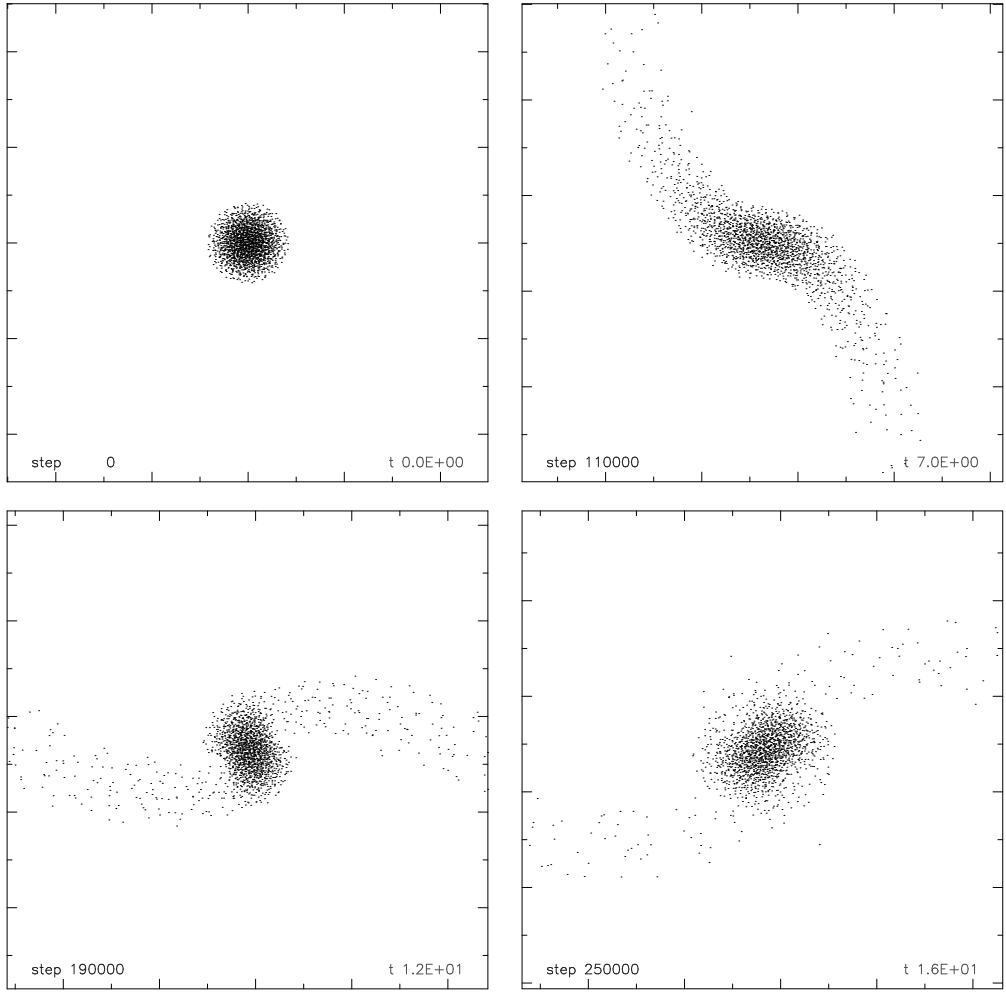

\centering{\htarget{f:BHtide}\begin{tabular}{cc}
\includegraphics[%
  width=0.40\textwidth,
  keepaspectratio,
  angle=270]{BHtide_rp100_m3e5_E0.0_bw_1.ps}&
\includegraphics[%
  width=0.40\textwidth,
  keepaspectratio,
  angle=270]{BHtide_rp100_m3e5_E0.0_bw_2.ps}\tabularnewline
\includegraphics[%
  width=0.40\textwidth,
  keepaspectratio,
  angle=270]{BHtide_rp100_m3e5_E0.0_bw_3.ps}&
\includegraphics[%
  width=0.40\textwidth,
  keepaspectratio,
  angle=270]{BHtide_rp100_m3e5_E0.0_bw_4.ps}\tabularnewline
\end{tabular}}

\caption{{\small \label{f:BHtide}A sequence of snapshots (left to right,
top to bottom) from an SPH simulation of a $10\,\Mo$, $\Rs\!=\!4.5\,\Ro$
star (represented by an $n=1.5$ ideal gas polytrope) undergoing an
extreme non-disruptive tidal interaction ({}``tidal scattering'')
as it passes near a $3\!\times\!10^{6}\,\Mo$ MBH. Time is measured
in units of the stellar dynamical time $\tau_{\star}\!=\!4840\,\mathrm{s}$.
The star passes near the black hole (located outside of the frame)
on a parabolic orbit with $r_{p}\!=\!100\Rs\!=\!1.5r_{t}\!=\!35.4r_{S}$.
After periapse passage (at $t\!=\!2.2$) the star appears to be on
the verge of breaking in two (bottom left panel). However, by the
end of the simulation, the two fragments coalesce, leaving a distorted,
mixed and rapidly rotating bound object. }}
\end{figure}

Whatever the exact nature of the after-effects of tidal scattering,
their magnitude must increase with decreasing periapse. The definition
of what constitutes a strong tidal interaction depends on the tidal
effect that is of interest, and this in turn corresponds to some maximal
periapse $r_{p}$, beyond which the effect becomes insignificant.
In spite of the smallness of the tidal radius, $r_{t}/r_{h}\!\sim\!10^{-6}$,
the number of tidally scattered stars that accumulate inside the MBH
radius of influence over the age of the Galaxy can be large (up to
$\lesssim\!0.1$ of the population). This can be shown qualitatively
by noting that most of the stars that are deflected into low angular
momentum orbits originate from $a\!\sim\! r_{h}$ orbits (Eq. \ref{e:Gtrh})
at the transition between the diffusive regime and full-loss cone
regimes. These orbits are approximately Keplerian and the DF is approximately
isotropic, so that stars are deflected with equal probability to all
directions. The cross-section for deflection into an eccentric orbit
is then $\Sigma(<\! r_{p})\!\propto\! r_{p}$, due to gravitational
focusing (Hills \citeyear{Hil75}; Eq. \ref{e:ThetaLC}). The rate
at which stars are initially deflected into orbits with periapse smaller
than $r_{p}$ but larger than $r_{t}$ is related to the direct disruption
rate by $\Gamma_{s}=(r_{p}/r_{t}-1)\Gamma_{t}$. This is also the
tidal scattering rate to good approximation, because inspiral is a
low probability process and most of these stars will be rescattered
to a wider orbit (the star can also be scattered into the loss-cone,
but the probability for that is smaller because loss-cone orbits occupy
a smaller volume in phase-space than wide orbits). Another effect
that will prevent once-scattered stars from interacting again with
the MBH is its Brownian motion (\S\ref{sss:distrDM}). The apoapse
of the scattered star's almost radial orbit will be $r_{a}\!\sim\!2r_{h}$,
where the potential is no longer dominated by the MBH. The MBH executes
Brownian motion relative to the stars inside $r_{h}$ with an amplitude
$\sim\!(r_{h}/\left\langle \Rs\right\rangle )(\left\langle \Ms\right\rangle /m)^{5/6}r_{t}\!\gg\! r_{t}$
and a period $\sim\! P(r_{h})$ (Bahcall \& Wolf \citeyear{Bah76}).
Therefore, while the star is near apoapse the MBH position will shift
relative to the stellar mas inside $r_{h}$ without changing much
the potential at $r_{a}$, so that on its return to the center the
star will miss the MBH. 

More accurate calculations (Alexander \& Hopman \citeyear{Ale03b})
show that the tidal scattering rate in a power-law cusp $n_{\star}\!\propto r^{-\alpha}$
in the diffusive limit is,\label{d:Gs}

\begin{equation}
\Gamma_{s}(<r_{p})\sim\left[\left(\frac{r_{p}}{r_{t}}\right)^{\delta}-1\right]P_{s}\Gamma_{t},\qquad\delta\equiv\frac{9-4\alpha}{8-2\alpha}\,,\label{e:Gs}\end{equation}
where $P_{s}\!\sim\!0.9$ is the survival probability against capture
by the MBH and where typical values for the power-law index are $\delta\!=\!0.60$
for $\alpha\!=\!3/2$, to $\delta\!=\!0.44$ for $\alpha\!=\!7/4$.
For the full loss cone regime $\delta\!=\!1$, and so the actual effective
power-law index is expected to lie in the range $0.5\!\lesssim\!\delta\!\lesssim\!1$. 

A substantial fraction $f_{t}$ of the total mass of a low-mass MBH
($10^{6}$--$10^{7}\,\Mo$) is supplied by tidally disrupted stars,
which contribute each a fraction $f_{m}\!\lesssim\!0.5$ of their
mass to the MBH (\S\ref{sss:tdafter}). Simulations indicate that
the total mass in tidally disrupted stars over the MBH lifetime lies
in the range $0.15m\!\lesssim\!(f_{t}/f_{m})m\!\lesssim\!0.65m$,
with the higher values typical of low-density nuclei where the population
is weighted toward low mass stars (Murphy, Cohn \& Durisen \citeyear{Mur91};
Freitag \& Benz \citeyear{Fre02}). Assuming for example $f_{t}/f_{m}\!=\!0.25$,
$\delta\!=\!0.75$ and a maximal periapse for strong tidal interaction
of $r_{p}\!=\!1.5r_{t}$, it then follows from Eq. (\ref{e:Gs}) that
the mass fraction of tidally scattered stars within the MBH volume
of influence is $(f_{t}/f_{m})(\Gamma_{s}/\Gamma_{t})\!\sim\!0.07$.
The possibility that some of the unusual stars observed in the GC
may be tidally scattered stars is intriguing. However, a better understanding
of the long-term observable consequences of tidal scattering is needed
to make progress on this issue.

\subsubsection{Star--disk interactions }

\label{sss:stardisk}

Drag against an accretion disk surrounding a MBH can dissipate orbital
energy, as well as affect the disk itself (e.g. Ostriker \citeyear{Ost83};
Syer, Clarke \& Rees \citeyear{Sye91}; Artymowicz, Lin \& Wampler
\citeyear{Art93}; Vilkoviskij \& Czerny \citeyear{Vil02}). There
is no evidence today for an accretion disk in the GC. It is conceivable,
however, that a dim, cold, low-viscosity {}``fossil'' disk still
exists around the MBH as a remnant of past periods of activity, perhaps
associated with the creation and fragmentation of the gaseous disks
that may have formed the observed star disks (\S\ref{sss:insituSF}),
or that a small disk is continuously fed by cooling stellar winds
from the young stars around the MBH (Cuadra et al \citeyear{Cua05};
\S\ref{ss:Windfeed}). Stellar interaction with the disk can reveal
its presence, or constrain its properties (extent, optical depth).
If it is optically thick, the disk can eclipse stars behind it. Stars
passing through the disk will likely emit X-ray and IR flares (Nayakshin
\& Sunyaev \citeyear{Nay03}; Nayakshin, Cuadra \& Sunyaev \citeyear{Nay04}).
The disk can reprocess the stellar UV light and re-emit it in the
IR (Cuadra et al. \citeyear{Cua03}). However, observations do not
support the presence of a disk. IR observations of the star S2 show
no eclipses or IR flares and are consistent with the absence of a
disk. Furthermore, the observed IR flares coincide to within a few
milli-arcseconds with $\SgrA$ and display a $\sim\!17$ min quasi-period
(Genzel et al. \citeyear{Gen03b}), which can not be explained by
passage through a disk. The temporal structure observed in the X-ray
flares (\citeyear{Asc04}) is likewise inconsistent with X-ray production
by passage through a disk.

\section{The riddle of the young stars}

\label{s:OBriddle}

The existence of young stars in the extreme environment so close to
a MBH (\S\ref{ss:1pc}) poses a challenge for theories of star formation
and stellar dynamics (Genzel et al \citeyear{Gen03a}; Ghez et al
\citeyear{Ghe03a}).  It has proved difficult to find a satisfactory
explanation of how they could have formed so close to the MBH, or
alternatively, of how they could have migrated inward from farther
away in the course of their short lifespans. This is the so-called
{}``paradox of youth''. The question applies to any of the young
stars in the inner parsec, but particularly so to the central cluster
of the {}``S-stars'', which exist a mere few hundredths of a parsec
from the MBH. The problem of the young stars has become one of the
major outstanding issues in GC research. 

The young population in the inner $\sim\!1$ pc is often loosely described
as the {}``OB-stars''. This general designation can be misleading,
as it fails to convey the significant systematic differences that
exist in the population. An important open question is the nature
of the connection, if any, between the S-stars inside $\sim\!0.04$
pc and the luminous emission line stars further out, on the $0.04$--$0.4$
pc scale. While it is plausible to assume that these are different
components of the same parent population, it should be noted the two
groups have distinct locations, kinematics and stellar properties.
The stars detected so far in the two young star disks at $p\!\sim\!1"$--$10"$
are luminous OB supergiants, giants and WR stars of various types
(Genzel et al. \citeyear{Gen03a}; Paumard et al. \citeyear{Pau01};
Paumard et al. \citeyear{Pau04a}). In contrast, orbital solutions
(Ghez et al. \citeyear{Ghe05}; Eisenhauer et al. \citeyear{Eis05})
show that the S-star orbits are randomly aligned and uncorrelated
with the planes of the disks. Spectral identifications (Eisenhauer
et al. \citeyear{Eis05}) reveal the S-stars to be B dwarfs (S2, the
brightest, is a transitional O8V-B0V star; Ghez et al. \citeyear{Ghe03a}).
This distinction is important, because B-stars are more numerous,
less massive (by about an order of magnitude) and longer lived (by
1--2 orders of magnitude) than the luminous O-stars and WRs (Fig.
\hlink{f:StellarKt}). This significantly relaxes the constraints
on models attempting to explain the central cluster. Thus there may
actually be two separate issues: how to explain the presence of the
OB stars on the $>\!0.04$ scale, and an even more acute problem,
how to explain the presence of the B-stars on the $<\!0.04$ pc scale.
Some of the scenarios discussed below (\S\ref{ss:OBsols}) attempt
to address both populations with a single solution, others focus only
on one or the other.

The stars in the central parsec can be described by stellar population
synthesis models. Such models assume an IMF and a star formation history
and use theoretically calculated stellar evolution tracks to follow
the stellar population in time. Population synthesis models do not
in themselves explain how the stars are born, or why the IMF and star
formation history have a particular form. However, by fitting the
models to the observed stellar population, it is possible to constrain
these input parameters and estimate the total mass in the stars. Krabbe
et al. (\citeyear{Kra95}) find that the luminous early- and late-type
stars can be modeled by a $7\!\pm\!1\,\mathrm{Myr}$ old short burst
of star formation with an IMF of $\mathrm{d}\Ns/\mathrm{d}\Ms\!\propto\!\Ms^{-2}$
between $1\,\Mo$ and $120\,\Mo$ (that is, skewed to high masses
as compared to the generic Salpeter IMF, $\mathrm{d}\Ns/\mathrm{d}\Ms\!\propto\!\Ms^{-2.35}$),
which formed $\sim\!3000$ stars containing in total$\sim\!1.5\!\times\!10^{4}\,\Mo$.
An additional, older star burst $\sim\!10^{8}\,\mathrm{yr}$ ago is
indicated by the intermediate-mass asymptotic branch giants. This
picture is consistent with the fact that the overall population in
the central few parsecs is well represented by a continuous star formation
model (Alexander \& Sternberg \citeyear{Ale99b}; \S\ref{ss:1pc}).
It appears that there is ongoing episodic star formation in the GC.
This suggests that we are not observing the GC at a very special epoch,
but rather that the presence of young stars there is a quasi-steady
state situation. 

Any explanation of the luminous OB-stars on the $>\!0.04$ pc scale
should account for their distinctive, disk-like distribution and kinematics
and for the presence of the cluster-like structure IRS13. Any explanation
of the S-stars should account for three principal properties. 

\begin{enumerate}
\item \emph{The stellar properties.} The stars appear to be \emph{entirely
normal} MS O8V/B0V to B9V stars, in terms of their (extinction corrected)
luminosities, their absorption lines equivalent widths and line ratios
(Eisenhauer et al. \citeyear{Eis05}). In particular, their rotational
velocities are similar to those of nearby Galactic disk B-stars. The
central cluster does not contain any star earlier than O8V, although
such stars do exist outside the central arcsecond.
\item \emph{The spatial concentration.} The relative fraction of the young
stars increases toward the center, to the near exclusion of any old
giant stars stars in the inner $\sim\!0.02$ pc (90\% of all stars
with $K\!<\!16$ mag in the central $0.02\,\mathrm{pc}$ are young
main-sequence stars, Eisenhauer et al. \citeyear{Eis05}). 
\item \emph{The orbital properties.} The orientations of the stellar orbits
appear overall random, in marked contrast to the ordered planar rotation
observed for the much more luminous emission line stars farther out.
There are some statistically marginal trends observed in the orbital
properties: higher than random eccentricity (Sch\"odel et al. \citeyear{Sch03})
and a lower bound on the orbital apoapse of $\sim\!0.01$ pc.
\end{enumerate}
The apparent normalcy of the B-stars is intriguing. This would seem
to argue against any process that involves strong perturbations of
the star, such as mergers (Genzel et al. \citeyear{Gen03a}), tidal
heating (Alexander \& Morris \citeyear{Ale03a}) or stripping (Hansen
\& Milosavljevi\'c \citeyear{Han03}). However, this is by no means
a conclusive argument, because it is not clear how efficiently a star
relaxes after a major perturbation, and in particular how angular
momentum is lost from the star or redistributed in it.

A relevant comparison can be made with the properties of blue stragglers
in clusters (MS stars that appear younger than the cluster's single
age population), whose formation is thought to involve mergers, binary
coalescence or 3-body interactions (Bailyn \citeyear{Bai95}). In
spite of their violent birth, observations show that in many cases,
late-B blue stragglers in intermediate age globular clusters, and
early-B and O blue stragglers in open clusters do not seem to rotate
faster than normal stars of the same mass in the field, in spite of
the fact that, unlike lower-mass stars (later than F5V), they do not
have efficient magnetic breaking (Leonard \& Livio \citeyear{Leo95}
and references therein; \S\ref{sss:spinup}). In this context it
is interesting to note the claimed detection of a circumstellar disk
around a blue straggler (de Marco et al. \citeyear{deM04}), which
may provide a breaking mechanism via magnetic anchoring. Neither is
it clear whether collisional merger products should be mixed, and
consequently, whether they should display unusual photospheric abundances.
Lombardi, Rasio \& Shapiro (\citeyear{Lom95},\citeyear{Lom96}) find
that the mixing by the collision itself is minimal (at least for lower
mass stars near a globular cluster's turnoff-mass). Sandquist, Bolte
and Hernquist (\citeyear{San97}) find similar results. However, subsequent
convection or meridional circulation can still induce mixing (Leonard
\& Livio \citeyear{Leo95}; Lombardi et al \citeyear{Lom95}), although
this was not found to be the case in detailed numerical work by Sills
et al. (\citeyear{Sil97}) and Ouellette \& Pritchet (\citeyear{Oue98}).
In summary, {}``absence of proof is not proof of absence''; if the
analogy to blue stragglers is justified, then it appears that the
lack of unusual spectral features in the S-stars does not place strong
constraints on their origin. 

\texttt{\textbf{}}

\subsection{The difficulties of forming or importing stars near a MBH }

\label{ss:SFMBH}

There is little evidence for star formation in the inner parsec at
this time%
\footnote{Some resolved infrared sources were initially explained as newly born
stars still enshrouded in their dust cocoons (Ott, Eckart \& Genzel
\citeyear{Ott99}). However, more detailed observations later revealed
these to be massive stars interacting with dust lanes extruding from
the circum-nuclear molecular disk inward (Tanner et al \citeyear{Tan02};
Geballe et al. \citeyear{Geb04}; Paumard et al. \citeyear{Pau04b};
Tanner et al. \citeyear{Tan05}).%
}. At present only low-density ionized gas is observed interior to
the circum-nuclear molecular disk, which encircles the MBH at a radius
of $\sim\!1.5$ pc. Estimates of the molecular hydrogen density and
the total mass in the disk vary from $n_{\mathrm{H}_{2}}\!\lesssim\!10^{6}\,\mathrm{cm^{-3}}$
(tidally unstable, see Eq. \ref{e:minnt2} below) and $M\sim\mathrm{few\!\times}\!10^{4}\,\Mo$
(based on molecular line ratios; Genzel et al. \citeyear{Gen85};
Jackson et al. \citeyear{Jac93}; Marr, Wright \& Backer \citeyear{Mar93};
Marshall, Lasenby \& Harris \citeyear{Mar95}) to $n_{\mathrm{H}_{2}}\!\sim\!10^{7}\,\mathrm{cm^{-3}}$
(tidally stable) and $M\sim\mathrm{3\!\times}\!10^{5}\,\Mo$ (assuming
that the molecular clumps are gravitationally bound; Shukla, Yun \&
Scoville \citeyear{Shu04}). An additional $\lesssim\!10^{3}\,\Mo$
\texttt{\textbf{}}may be contained in the gas and dust lanes (the
{}``spiral'') that extrude from circum-nuclear molecular disk inward
(Liszt \citeyear{Lis03}; Paumard et al. \citeyear{Pau04b}; Paumard
et al. \citeyear{Pau04a}). Even if $>\!10^{4}\,\Mo$ of cold molecular
gas existed in the past in the central parsec, as is implied by the
mass in the observed blue and red giants (Krabbe et al. \citeyear{Kra95}),
it is not clear how stars could condense from the gas so close to
the MBH. The minimal proto-stellar cloud density that can resist the
MBH tide is extremely high compared to values typically encountered
in molecular clouds elsewhere in the Galaxy ($n\!\lesssim\!10^{4}\,\mathrm{cm^{-3}}$),
where star formation normally takes place. For $r\lesssim\!1$ pc,

\begin{equation}
\min n\sim\frac{(m/m_{u})}{r^{3}}\sim10^{8}\left(\frac{m}{3\!\times\!10^{6}\,\Mo}\right)\left(\frac{r}{1\,\mathrm{pc}}\right)^{-3}\,\mathrm{cm^{-3}}\,,\label{e:minnt1}\end{equation}
where $m_{u}$ is the atomic mass unit and where the gravitational
potential of extended stellar mass is ignored. For $r\!\gtrsim1$
pc, \begin{equation}
\min n\sim\frac{3(\alpha-1)M_{0}}{4\pi r_{0}^{3}m_{u}}\left(\frac{r}{r_{0}}\right)^{-\alpha}\sim4\!\times\!10^{7}(\alpha-1)\left(\frac{M_{0}(1\,\mathrm{pc})}{4\!\times\!10^{6}\,\Mo}\right)\left(\frac{r}{1\,\mathrm{pc}}\right)^{-\alpha}\,\mathrm{cm^{-3}}\,,\label{e:minnt2}\end{equation}
where $M_{0}$ is the total mass enclosed within $r_{0}$ and it is
assumed that all the mass is distributed in a $r^{-\alpha}$ cusp.
These high minimal densities exceed, or are just at the inferred densities
in the circum-nuclear molecular disk. In addition, the molecular clouds
in the GC have large turbulent velocities ($\sim\!10\,\mathrm{km\, s^{-1}}$),
which may provide pressure support against fragmentation, and unusually
strong magnetic fields ($\sim\!1$ mG, e.g. Aitken, Moore \& Roche
\citeyear{Ait98}) are observed in the GC, which, if they permeate
the clouds, would provide additional support against fragmentation
and collapse (Morris \citeyear{Mor93}). The unavoidable conclusion
is that if the young stars well inside the central parsec were formed
locally, then they must have done so by a different mechanism than
the collapse of self gravitating cold molecular gas clouds that occurs
in normal star forming regions. 

The young stars are also too short-lived and too light to have formed
far from the MBH, where the tidal field is weak, and then to have
migrated in by dynamical friction. The {}``collection basin'' for
the migration of young MS stars of mass of $\Ms\!=\!15\,\Mo$ and
lifespan $t_{\star}\!=\!2\!\times\!10^{7}\,\mathrm{yr}$ (\S\ref{ss:1pc})
is only $\max r_{\mathrm{df}}\!\sim\!0.2\,\mathrm{pc}$ (Eq. \ref{e:rdf}),
and for stars of mass $\Ms\!=\!3\,\Mo$ and lifespan $t_{\star}\!=\!4\!\times\!10^{8}$
it is only $\max r_{\mathrm{df}}\!\sim\!0.5\,\mathrm{pc}$.

\subsection{Proposed solutions}

\label{ss:OBsols}

The solutions proposed so far for the riddle of the young stars (see
reviews by Genzel et al. \citeyear{Gen03a}; Ghez et al. \citeyear{Ghe05})
fall into three main categories: unusual modes of star formation near
the MBH; rejuvenation of old stars from the local population; and
dynamic migration or capture from farther out, where stars can form.
While each has some attractive features, none is quite satisfactory.
The paradox of youth remains unsolved at this time.

\subsubsection{Unusual \emph{in-situ} star formation }

\label{sss:insituSF}

One class of \emph{in-situ} star formation models invokes external
pressure to trigger cloud collapse. Because the clouds are stabilized
against collapse by their high turbulent velocities and magnetic fields,
star formation in such clouds will be skewed toward massive stars,
as are observed in the GC. Cloud--cloud collisions and the resulting
shocks and cooling could in principle initiate fragmentation (Morris
\citeyear{Mor93}). The near identical age of the two star disks could
perhaps be explained by two colliding clouds, each fragmenting to
form a disk (Genzel et al. \citeyear{Gen03a}). However, it is not
clear that the required very high compression ratio (Eq. \ref{e:minnt1})
can be achieved in such collisions. An additional objection to any
scenario that postulates the existence of dense clouds near the MBH
is that those clouds that are dense enough to resist the tidal forces
at $r\!\lesssim\!2$ pc will also be Jeans-unstable and will fragment
and presumably form stars on that large scale before reaching the
center (Vollmer \& Duschl \citeyear{Vol01}). 

Alternatively, Morris, Ghez \& Becklin (\citeyear{Mor99}) propose
a scenario of recurrent nuclear activity regulated by a limit cycle.
At the current phase of the cycle, the circum-nuclear molecular gas
disk, which has a central cavity inside $\sim\!1.5$ pc, is prevented
from filling the cavity by the radiation pressure from the hot massive
stars in the center. However, once these stars die (in $\lesssim\!10^{7}$
yr), the internal viscosity of the disk will set an inflow. The central
cavity will be filled and the gas will reach the center, setting off
a strong burst of luminous accretion. The radiation pressure will
shock and compress the inner parts of the gas disk, presumably triggering
an intense phase of star formation. Stellar winds and the combined
radiation from the young massive stars and the accretion on the MBH
will evacuate a central cavity, which will remain empty of gas as
long as there are enough massive luminous stars inside, thereby starting
the limit cycle again. This scenario has yet to be studied in detail.
Here again, the problem is that it is unclear whether radiation pressure
can lead to the very high compression ratio that is required for fragmentation. 

Another class of models for \emph{in-situ} star formation propose
that the young stars were formed by the fragmentation of a gaseous
disk around the MBH that gradually grew in mass (perhaps fed by tidally
disrupted molecular clouds, Sanders \citeyear{San98}) to the point
it became self gravitating (Levin \& Beloborodov \citeyear{Lev03};
\texttt{\textbf{}}Milosavljevi\'c \& Loeb \citeyear{Mil04}; Nayakshin
\& Cuadra \citeyear{Nay05b}). The outer parts of thin massive disks
are susceptible to fragmentation (Paczy\'nski \citeyear{Pac78}),
so much so that it is, in fact, difficult to explain theoretically
the observed existence of extended disks (Goodman \citeyear{Goo03}).
While such a disk does not exist in the GC today, Milosavljevi\'c
\& Loeb (\citeyear{Mil04}) point out that masering disks observed
around MBHs in other galaxies (disks emitting coherent radio emission
due to population inversion by shocks) have properties similar to
those needed to create the star disks in the GC.

The criterion for disk fragmentation due to self-gravity (Paczy\'nski
\citeyear{Pac78}) can be expressed, up to factors of order unity,
as a tidal limit, $\rho\!=\!\Sigma/h\!>\! m/r^{3}\!\equiv\!\rho_{c}$,
where $\Sigma$ is the disk's surface density. The disk's scale height
$h\!\sim\! c_{s}/\Omega$ is set by the maximal height $z$ an atom
can reach when launched from the disk mid-plane with the sound speed
$c_{s}$ against a vertical acceleration $a_{z}\simeq\Omega^{2}z$
($z\!\ll\! r$), where $\Omega\!=\!\sqrt{Gm/r^{3}}$ is the local
Keplerian frequency. At the critical density, $\Omega$ equals the
disk's gravitational free-fall rate, $\Omega\!=\!\sqrt{G\rho_{c}}\!\equiv\! t_{\mathrm{ff}}^{-1}$.
The Jeans (minimal) radius, $R_{J}$, and mass, $M_{J}$, of a collapsing
fragment are set by the condition that the free-fall time be shorter
than the sound crossing time (the time for the propagation of a gravity-resisting
pressure adjustment), $R_{J}\!>\! c_{s}t_{\mathrm{ff}}$, and $M_{J}\!>\!\rho_{c}c_{s}^{3}/\Omega^{3}\!=\! c_{s}^{4}/(G^{2}\Sigma_{c})$.
In addition, the gas must cool faster than the dynamical time to offset
the increased pressure support from the contracting gas. For the mass
and distance scales of the GC and for reasonable assumptions about
the disk properties, $M_{J}\!\sim\!\mathrm{few\!\times\!1\,\Mo}$
(Levin \& Beloborodov \citeyear{Lev03}). The collapsing proto-stellar
cloud then begins to accrete gas from the disk within its tidal radius,
$\sim\!(M_{J}/m)^{1/3}r$. The available mass there exceeds $M_{J}$
by a few orders of magnitude (Levin \citeyear{Lev03}; Goodman \&
Tan \citeyear{Goo04}). Due to the differential rotation in the disk,
the proto-stellar cloud acquires its own mini-accretion disk, which
may itself become unstable and fragment to form groups of stars. Milosavljevi\'c
\& Loeb (\citeyear{Mil04}) propose that IRS13 is such a gravitationally
bound group of $\sim\!2500\,\Mo$ (including low luminosity stars
not yet observed), and that the S-stars were internally scattered
from such groups by very hard encounters with binaries. 

Nayakshin \& Cuadra (\citeyear{Nay05b}) derive limits on the initial
total masses (stars and gas) of the two star disks observed today:
a lower limit of $\sim\!10^{4}\,\Mo$from the requirement of self-gravitation
instability, and an upper limit of $10^{5}\,\Mo$ based on the magnitude
of velocity dispersion in the outer disk that is caused by the potential
of the inner disk. They are thus able to rule out the possibility
that the young stars were formed by gas accretion on low mass stars
that got captured in the disk. This would have required disk masses
well in excess of the upper limit.

The disk fragmentation model is a promising scenario for the formation
of the star disks and the massive young stars on the $\gtrsim\!0.1$
pc scale, but it is less clear whether it can account for the S-cluster
as well. According to their stellar contents, the two star disks are
contemporaneous to better than 1 Myr, and only $\sim5$ Myr old (Krabbe
et al. \citeyear{Kra95} ; Genzel et al 2003; Paumard et al. 2005,
in prep.). Any model that associates the S-stars with the star disks
has to explain why it is only B-stars that are found in the S-cluster,
and not any of the more massive stars that exist in the disks. Furthermore,
if the B-stars are associated with the disks, then the relevant time
constraint for redistributing the orbits is the age of the disks,
and not the considerably longer MS lifespan of a B-star.

\subsubsection{Rejuvenation }

\label{sss:rejuvenation}

The inner $\sim\!0.02$--$0.03$ pc of the GC, the domain of the S-cluster,
roughly coincides with the region where the stellar density is so
high that stellar collisions occur more than once over the lifetime
of a MS dwarf (Alexander \citeyear{Ale99a}; \S\ref{ss:coll}). The
strong tidal field of the MBH may also affect the stellar structure
(\S\ref{ss:dissipation}). Rejuvenation models seek to explain the
S-cluster in terms of low mass, long-lived stars that had enough time
to migrate to the center from their formation sites far from the MBH
(however, the central number density of such stars will be suppressed
by mass segregation). 

All rejuvenation models proposed so far have serious problems in explaining
the S-stars. The physical processes were already discussed above in
some detail, and are listed here briefly. Envelope stripping collisions
(Hansen \& Milosavljevi\'c \citeyear{Han03}) will reveal a hot core,
but much fine-tuning is required for the bare core to masquerade as
an apparently normal B-star (\S\ref{sss:RGcoll}). Tidal heating
(\S\ref{sss:squeezar}) requires a very small periapse, $r_{p}\!\lesssim\!2r_{t}$
to be effective (Alexander \& Morris \citeyear{Ale03a}). Even the
smallest measured periapse to date, that of star S0-16 with $r_{p}\!\sim\!600r_{S}\!\sim\!30r_{t}$
(Ghez et al. \citeyear{Ghe05}) is much too large for tidal heating
to be of any relevance (tidal heating and scattering to a wider orbit
can be ruled out as an explanation for its present luminosity since
the scattering timescale is much longer than the stellar cooling time).
Successive mergers of low-mass stars (Genzel et al. \citeyear{Gen03a})
are inefficient because the mass retention in a high velocity collision
is typically very small (\S\ref{sss:mergers}). Exotic objects formed
by the capture of a compact remnant inside a normal star will probably
not look like B-stars (\S\ref{sss:mergers}).

\subsubsection{Dynamical migration}

\label{sss:dynfric}

Stellar dynamics are generally better understood than star formation,
or the effects of far from equilibrium conditions on stellar structure
and properties. Therefore models that invoke dynamical processes to
explain the S-stars tend to have more definite, falsifiable predictions.
The orbits and the stellar spatial distribution thus appear to be
a more promising way of discriminating between models. One class of
dynamical models seeks to accelerate the dynamical friction (\S\ref{ss:SFMBH})
by attaching the young stars to a massive {}``anchor''. This allows
the stars to form far from the MBH, where star formation is not strongly
inhibited, and yet sink into the center within their short lifespans.

A natural candidate for such an anchor is a dense, young, star-forming
cluster (Gerhard \citeyear{Ger01}; \S\ref{ss:100pc}) like the Arches
and Quintuplet clusters, which lie within tens of parsecs from the
center, and contain a young population closely resembling that near
the MBH, in particular He stars (Figer et al. \citeyear{Fig99}).
Such a cluster, being a compact massive object ($M\!\sim\!10^{4}\,\Mo$),
will undergo dynamical friction and can sink to the center in $10^{7}\,\mathrm{yr}$
from a distance of $\sim\!5$ pc (Eq. \ref{e:rdf}). Much more massive
clusters than are seen today in the GC, with $M\!\sim\!10^{6}\,\Mo$
will be able to sink in from a distance of $\sim\!50$ pc. The problem
with this scenario is that the cluster will be stripped and dissolved
by the tidal field of the GC before it reaches the central parsec
(Portegies Zwart, McMillan \& Gerhard \citeyear{Por03}), unless it
has an extremely high central density, $\rho\!\sim\!10^{8}\,\Mo\,\mathrm{pc^{-3}}$
(Kim \& Morris \citeyear{Kim03}). Alternatively, the center of the
cluster may be stabilized against tidal stripping by a central IBH
(Hansen \& Milosavljevi\'c \citeyear{Han03}). 

The scenario of the dissolving cluster with an IBH proposes that a
very dense stellar cluster containing an IBH, perhaps formed by runaway
mergers in the cluster core (e.g. Portegies Zwart et al. \citeyear{Por04}),
sinks rapidly to the center. The stars most tightly bound to the IBH
avoid tidal stripping until the IBH reaches $\sim\!0.1$ pc from the
MBH, where they are deposited in a disk-like configuration. Some of
them are subsequently scattered to tighter eccentric orbits by repeated
interactions with the orbiting IBH. An analogous mechanism operates
in the Solar System, where Jupiter scatters comets from the outer
Solar System to tightly bound inner orbits. Subsequent, more detailed
simulations (Levin, Wu \& Thommes \citeyear{Lev05}) showed that a
dissolving cluster with an IBH can deposit its last remaining stars
in a fairly thin ring (if moving on a circular orbit), but that even
repeated scattering of these stars by the IBH do not lead to the formation
of a tightly bound cluster such as the S-cluster. Hansen \& Milosavljevi\'c
(\citeyear{Han03}) hypothesize that the S-stars are in fact O-stars
from the dissolving cluster that were captured in very bound orbits
by a close, nearly disruptive encounter with another star, which stripped
their outer envelopes and whittled them down to B-star masses. The
general concept of this model is supported by recent claims for an
IBH embedded in the IRS13 {}``cluster'' (Maillard et al. \citeyear{Mai04};
but see dissenting view by Sch\"odel et al. \citeyear{Sch05}). The
weakness of this scenario for explaining the S-stars is that, lacking
quantitative calculations, it is unclear whether this method can actually
form such a tightly bound cluster, and it is also unclear whether
collisional stripping is efficient enough and consistent with the
apparent normalcy of the B-stars. 

Another objection to this scenario is the large number of massive
young stars that are implied by a dissolving $M\!>\!10^{5}\,\Mo$
cluster (Kim, Figer \& Morris \citeyear{Kim04}; G\"urkan \& Rasio
\citeyear{Gur04a}), and which are not observed in the GC. G\"urkan
\& Rasio (\citeyear{Gur04a}) show that a very massive ($M\!>\!10^{6}\,\Mo$)
and dense cluster starting at $10\,\mathrm{pc}$ can sink in less
than $10^{6}$ yr to the central $0.5$ pc, bringing with it $10^{4}\,\Mo$
of $>\!10\,\Mo$ stars while undergoing core collapse and forming
a $5000\,\Mo$ IBH. However, if a lower mass cluster is assumed (and
a correspondingly smaller initial radius, $r_{\mathrm{df}}\!\sim\!5$pc)
to mitigate the problem of the over-abundance of massive stars, then
in order to carry stars into the inner parsec, the IBH must have a
large mass relative to the cluster, $m\!\gtrsim\!10^{4}\,\Mo$ and
$m/M\!>\!0.1$. This is about 2 orders of magnitude larger than can
feasibly grow in runaway mergers, $m/M\sim10^{-3}$ (G\"urkan, Freitag
\& Rasio \citeyear{Gur04b}; G\"urkan \& Rasio \citeyear{Gur04a}).

\subsubsection{Exchange capture}

\label{sss:exchange}

Another class of dynamical scenarios are those that invoke 3-body
interactions to capture the B-star on a tightly bound orbit around
the MBH.

The massive binary exchange scenario (Gould \& Quillen \citeyear{Gou03})
postulates that the B-star originally had a very massive binary companion
($\sim\!100\, M_{\odot}$). The binary presumably originated in a
radially infalling, disintegrating massive cluster. Its radial orbit
brought it close to the MBH, where in the course of a 3-body exchange
interaction the B-star switched partners and became bound to the MBH.
Gould \& Quillen (\citeyear{Gou03}) find that the probability per
binary for capture in an orbit like that of the star S2 is of the
order of a few percent. The weakness of this scenario is the very
low joint probability for (i) having a cluster on a radial orbit,
(ii) containing a very massive star, (iii) paired in a binary with
a much lighter secondary (iv) with the right orbital parameters for
3-body exchange with the MBH. It also does not provide an explanation
why it is that only B-stars are captured, and not more massive stars,
and why there is a lower bound on the apoapse. 

The exchange capture with SBHs scenario (Alexander \& Livio \citeyear{Ale04})
proposes that the B-stars were originally formed far away from the
MBH where normal star formation is possible (not necessarily in a
massive cluster) and were then deflected into eccentric orbits that
intersect the dense central concentration of SBHs (with masses of
$\sim7$--$10\, M_{\odot}$), which sank to the center from the inner
$\sim\!5\,\mathrm{pc}$ due to mass segregation over the lifetime
of the GC (\S\ref{ss:Mseg}). The high concentration of the SBHs
is also responsible for the collisional destruction of any tightly
bound old red giants at the very center. Occasionally one of the B-stars
passes very close to a SBH, knocks it out and replaces it in a bound
orbit around the MBH (a 3-body exchange involving a MBH--SBH {}``binary''
and the single B-star). This mechanism can naturally explain several
of the properties of the S-cluster. The central concentration of the
S-cluster follows that of the highly segregated cluster of stellar
BHs. The S-cluster is composed of B-stars because they are most closely
matched in mass to the stellar BHs and so have the maximal probability
for exchange. The lower bound on the apoapse, $\sim\!0.01$ pc corresponds
to the point where an exchange requires that the B-star pass so close
to the SBH that it is tidally disrupted. The exchange mechanism also
predicts a trend toward high eccentricities in the orbits of the captured
stars. The weakness of this scenario is the high required central
concentration of SBHs and the large required number of B-stars on
eccentric orbits. Detailed calculations of the exchange capture cross-section
and rate indicate that the SBH concentration must be close to the
drain limit (Eq. \ref{e:drainlim}) in order to sustain a steady state
population of captured B-stars. These requirements may be relaxed
if the source of B-stars is a dissolving cluster that approaches the
SBH cluster (G\"urkan \& Rasio \citeyear{Gur04a}), however, detailed
calculations of this possibility have yet to be performed. 

The stripped AGB giant capture scenario (Davies \& King \citeyear{Dav05})
suggests that the S-cluster stars are the stripped core of very luminous
asymptotic giant branch stars, which are captured at typical S-star
periapses and periods by the dissipation of orbital energy required
for tidally stripping the extended giant envelopes. The criticism
leveled at this scenario is that it requires an unrealistically large
number of AGB giants to sustain a steady-state S-cluster population
(total mass consumed in AGB stars larger than the stellar mass in
$r_{h}$); that much fine tuning is required for the stripped cores
to masquerade as B stars; and that the work done to strip the envelope
will be taken out of the envelope's orbital energy rather than that
of the core (Goodman \& Paczy\'nski \citeyear{Goo05}).

\subsection{Feeding the MBH with stellar winds}

\label{ss:Windfeed}

Given the yet unexplained fact that young massive stars do exist very
close to the MBH, it is natural to consider the role that the mass
ejected by their strong stellar winds may play in the MBH accretion.
The massive He stars with their copious mass-loss winds can supply
mass for accretion at a rate of $\dot{M}_{w}\!\gtrsim\!10^{-3}\,\Mo\,\mathrm{yr^{-1}}$.
Estimates of the gas density and temperature from \emph{Chandra} X-ray
observations (Baganoff et al. \citeyear{Bag03}) imply a Bondi accretion
rate (Bondi \citeyear{Bon52}; Melia \citeyear{Mel92}) of $\dot{M}_{B}\!\sim\!10^{-5}\,\Mo\,\mathrm{yr^{-1}}\!\ll\!\dot{M}_{w}$,
\emph{if} the accretion is spherically symmetric. Accretion at this
rate by a geometrically thin, optically thick accretion disk, which
has a typical radiative efficiency of $\eta\!\sim\!0.1$ (Shakura
\& Sunyaev \citeyear{Sha73}; \S\ref{sss:tdafter}), can be ruled
out since it would produce a luminosity of $L_{B}\!\sim\!\eta\dot{M}_{B}c^{2}\!\sim\!10^{41}\,\mathrm{erg\, s^{-1}}$,
about $10^{5}$ higher than observed. The implications are that (1)
the He stars alone can easily supply all the mass required for accretion,
and (2) $\SgrA$ is extremely under-luminous compared to available
gas supply rate. It is believed that the dimness problem of $\SgrA$,
which is common to many MBHs in quiescent galactic nuclei, can be
explained by radiatively inefficient accretion. The detection of linearly
polarized sub-mm emission from $\SgrA$ (Aitken et al. \citeyear{Ait00};
Bower et al. \citeyear{Bow03}) as well as the observed spectral energy
distribution of $\SgrA$ suggest that only $\dot{M}\!\sim\!10^{-8}\,\Mo\,\mathrm{yr^{-1}}\!\ll\!\dot{M}_{B}$
is accreted on the MBH, with a low efficiency of $\eta\!\sim\!10^{-3}$.
The rest of the mass is presumably carried away by a global outflow
(see review by Quataert \citeyear{Qua03}). Note that $\dot{M}t_{H}\!\ll\! m$,
so the present accretion rate cannot explain the mass of the MBH.
Either the MBH was accreting at much higher rates (at the level of
the Bondi rate or higher ) for most of its past, or else it grew by
channels other than gas accretion, for example, a substantial fraction
can be supplied by tidal disruption of stars, which can account in
steady state for $\sim\!0.5\,\Mo\Gamma_{t}t_{H}\!\sim\!2.5\!\times\!10^{5}\,\Mo\!\sim\!0.1m$
(\S\ref{sss:tdrate}). 

A more realistic picture of the flow of the stellar winds into the
MBH must take into account the discrete distribution of stars (Coker
\& Melia \citeyear{Cok97}; Rockefeller et al. \citeyear{Roc04}).
Cuadra et al. (\citeyear{Cua05}) note that stellar mass lost by slow
winds ($v\!\lesssim\!300\,\mathrm{km\, s^{-1}}$) or in a faster wind,
but in the direction opposite to the star's orbital motion, will not
be shocked to very high temperatures and can cool on a dynamical timescale.
They simulate the gas flow from the winds of an ensemble of orbiting
stars in the inner $\sim\!0.35$pc (roughly the region where the 2
stellar rings lie, \S\ref{ss:1pc}) and find that a 2-phase gas distribution
develops, with hot, X-ray emitting gas in the inner $1"$ around the
MBH and a cold, fragile mini-disk further out (containing $\sim\!10\,\Mo$).
Cold blobs occasionally detach from the disk and flow inward, leading,
if they do not evaporate, to a variable accretion rate with a variability
timescale of $10^{2}$--$10^{3}$ yr.

While the low level, quasi-steady state accretion of $\SgrA$ may
be supplied by the shocked winds of massive stars in the inner parsec,
a time dependent component may be contributed directly by stars on
eccentric orbits that approach within $\sim\!10^{3}r_{S}$ of the
MBH. Loeb (\citeyear{Loe04}) suggest that a fraction of the stellar
wind in the direction oriented toward the MBH can fall ballistically
into the MBH, thereby possibly introducing variability that correlates
with the orbital phase.

\section{Outlook}

\label{s:outlook}

\subsection{Progress report}

\label{ss:progress}

\begin{table}

\caption{\label{t:score}Score card and progress report (mid 2005)}

\centerline{\htarget{t:score}\begin{tabular}{lp{4ex}p{4ex}p{4ex}p{3.25in}}
&
\multicolumn{3}{c}{}&
\tabularnewline
\hline 
{\small Category}&
\multicolumn{3}{c}{{\small Score $^{a}$}}&
{\small Remarks}\tabularnewline
&
\multicolumn{3}{c}{{\small Theory/Predict./Obs.}}&
\tabularnewline
\hline
&
&
&
&
\tabularnewline
\multicolumn{5}{l}{\textbf{\small Primary parameters of the dark mass }}\tabularnewline
{\small Nature of DM}&
{\small $\sqrt{}$}&
{\small $\sqrt{}$}&
{\small $\sqrt{}$}&
{\small MBH, from lower limit on density (\S\ref{ss:DM}). Dynamical
upper limits on extended component (\S\ref{sss:distrDM}).}\tabularnewline
{\small Position of DM}&
{\small $!$}&
{\small $!$}&
{\small $\sqrt{}$}&
{\small Apparent position from orbital solutions (\S\ref{sss:orbsol}).
Line-of-sight distance from radial velocities (\S\ref{sss:R0}).}\tabularnewline
{\small BH mass}&
??&
{\small !}&
{\small $\sqrt{}$}&
{\small No basic theory, but mass consistent with $m/\sigma$ prediction
(\S\ref{ss:mass}).}\tabularnewline
{\small BH spin $^{b}$}&
{\small $\sqrt{}$}&
{\small !}&
{\small !}&
{\small Quasi-period in IR accretion flares (Genzel et al. \citeyear{Gen03b};
Aschenbach et al. \citeyear{Asc04}).}\tabularnewline
{\small Binary BH}&
{\small ??}&
{\small !}&
{\small ?}&
{\small IRS13? Bounds on secondary mass, distance (\S\ref{ss:2MBH}).}\tabularnewline
&
&
&
&
\tabularnewline
\multicolumn{5}{l}{\textbf{\small Properties of the stellar system}}\tabularnewline
{\small Density distribution}&
{\small $\sqrt{}$}&
{\small $\sqrt{}$}&
{\small $\sqrt{}$}&
{\small High density stellar cusp observed (\S\ref{ss:cusp}).}\tabularnewline
{\small Dynamical properties}&
{\small !}&
{\small !}&
{\small $\sqrt{}$}&
{\small Vel. dispersion, anisotropy measured on all scales (\S\ref{sss:stat}).}\tabularnewline
{\small Population composition}&
??&
{\small ?}&
{\small $\sqrt{}$}&
{\small Mixed old and young stellar populations (\S\ref{s:GCstars}).}\tabularnewline
{\small Star formation mode}&
{\small ??}&
{\small ??}&
{\small ?}&
{\small Many ideas, but still no definite conclusion (\S\ref{ss:100pc}). }\tabularnewline
&
&
&
&
\tabularnewline
\multicolumn{5}{l}{\textbf{\small Stellar dynamical processes}}\tabularnewline
{\small Relaxation}&
{\small $\sqrt{}$}&
{\small !}&
{\small $!$}&
{\small Observed cusp slope and velocity dispersion of old population
consistent with relaxation (\S\ref{ss:cusp}). }\tabularnewline
{\small Mass segregation}&
{\small $\sqrt{}$}&
{\small !}&
{\small ?}&
{\small Strong segregation expected. Some evidence for over-abundance
of compact objects (\S\ref{ss:Mseg}).}\tabularnewline
{\small Collisions/mergers}&
{\small $\sqrt{}$}&
{\small ?}&
{\small !}&
{\small Depletion of bright giants toward center (\S\ref{ss:coll}).}\tabularnewline
{\small Tidal spin-up}&
{\small !}&
{\small ? }&
{\small ??}&
{\small Evidence of (rotational?) mixing (\S\ref{sss:spinup})}\tabularnewline
{\small Star disks}&
{\small ?}&
{\small ??}&
{\small $\sqrt{}$}&
{\small Star formation in fragmenting gas disks? (\S\ref{ss:1pc}).}\tabularnewline
&
&
&
&
\tabularnewline
\multicolumn{5}{l}{\textbf{\small Stellar interactions with the MBH}}\tabularnewline
{\small Tidal disruption}&
{\small $\sqrt{}$}&
{\small ?}&
{\small ?}&
{\small Evidence inconclusive (\S\ref{ss:tide}).}\tabularnewline
{\small Tidal scattering/heating}&
{\small !}&
{\small ?}&
??&
{\small Process inevitable, signature uncertain (\S\ref{sss:squeezar},
\S\ref{sss:tscatter}).}\tabularnewline
{\small Interactions with cold disk}&
{\small ??}&
{\small !}&
{\small $\sqrt{}$}&
{\small No evidence of cold disk (\S\ref{sss:stardisk}).}\tabularnewline
&
&
&
&
\tabularnewline
\multicolumn{5}{l}{\textbf{\small Post Newtonian physics}}\tabularnewline
{\small Gravitational redshift}&
{\small $\sqrt{}$}&
{\small $\sqrt{}$}&
{\small ---}&
{\small Awaiting more orbital data (\S\ref{sss:GRz}).}\tabularnewline
{\small Orbital precession}&
{\small $\sqrt{}$}&
{\small !}&
{\small ---}&
{\small Awaiting short period, better astrometry orbits (\S\ref{ss:GRorbit}).}\tabularnewline
{\small Frame dragging}&
{\small $\sqrt{}$}&
{\small !}&
{\small ??}&
{\small Proposed detection depends on untestable assumptions about
the star S2 (\S\ref{ss:GRorbit}).}\tabularnewline
{\small Gravitational lensing}&
{\small $\sqrt{}$}&
{\small !}&
{\small ?}&
{\small Expected lensing probability low (\S\ref{ss:GL}).}\tabularnewline
{\small Gravitational waves}&
{\small !}&
{\small !}&
{\small ---}&
{\small Awaiting future detectors (\S\ref{sss:GW}).}\tabularnewline
\hline
\multicolumn{5}{l}{{\scriptsize $^{a}\,$`$\sqrt{}$' known with confidence; `!' promising
but uncertain; `?' uncertain; `??' speculative; `---' unknown or not
applicable}}\tabularnewline
\multicolumn{5}{l}{{\scriptsize $^{b}\,$Listed here for completeness. This estimate
was derived from gas accretion properties, not stellar processes. }}\tabularnewline
\hline
\end{tabular}}
\end{table}

Table (\hlink{t:score}) summarizes the progress achieved to date
(mid 2005) by the study of stars and stellar processes near the MBH
in the GC, as well as the many remaining challenges. Several subjects
are listed, broken down to sub-topics. For each topic, progress in
three categories, theory, predictions and observations, is graded
on a qualitative scale, from {}``unknown'' to {}``known with confidence''.
The theory category refers to the level of basic understanding of
the particular object or process and the degree of confidence that
it should exist or play a role in the GC. The predictions category
refers to the ability to predict the observable consequences in the
GC, either from a basic theory or from established empirical relations,
\emph{assuming} that the object or process is indeed relevant. The
observations category refers to the extent at which observations were
able to provide information on the matter (either supporting or refuting
existence). 

For example, concerning the nature of the dark mass, the theoretical
expectation is that the dark masses in galactic centers are MBHs (based,
among others, on the close analogies with Galactic {}``micro-quasars''
(SBHs accreting from a binary companion) which are not composite objects
such as clusters and most probably are not made of exotic dark mass).
Furthermore, a basic theory of BHs exists. The theoretical predictions
of the effect of a MBH on stellar orbits are firm, and there are empirical
expectations for its mass based on the $m/\sigma$ relation. Even
for the less-well defined exotic dark mass alternatives it is possible
to predict that there should be deviations from Keplerian motion.
The observations of stellar orbits provide a compelling case for a
MBH. Consequently, theory, predictions and observations for this topic
are all graded as {}``known with confidence''. Another example,
where the situation is less clear, concerns stellar collisions and
mergers. While it is theoretically known with confidence that stellar
collisions should occur in a high density environment, the predicted
observational signature is uncertain, as it depends on the details
of stellar physics. Nevertheless, there are promising, albeit still
uncertain observational indications that giants are collisionally
destroyed very near the MBH.

It should be emphasized that it is often the case that there isn't
one definitive theoretical prediction, either because of the complexity
of the process, or because the predictions depends on various unknown
parameters. The scores listed here reflect an admittedly subjective
assessment of the overall robustness of the theoretical understanding
and predictions. Likewise, the question whether or not available observations
can come to bear on a particular topic is tied to the theoretical
progress. It is quite possible that relevant observations are already
available, but remain {}``hidden in plain sight'' due to the lack
of a theoretical framework.

\subsection{Future directions}

\label{ss:future}

The main achievements resulting to date from the study of stars near
the Galactic MBH (table \hlink{t:score}) include the determination
of the primary parameters of the central dark object (its nature,
mass, location and distance), the detection of an extremely high density
stellar cusp, and the observational characterization of the puzzling
stellar population near the MBH. There are still many challenges in
making the connection between these discoveries and the underlying
theories. This review concludes with a brief survey of some of the
next research frontiers and the theoretical and observational advances
that are required for making progress on these issues.

A yet unrealized goal is the detection of post-Newtonian effects in
the stellar orbits (\S\ref{ss:GRorbit}). The prospects of progress
will be much improved if stars on even tighter and more eccentric
orbits than those observed today are detected. Such stars, if they
exist, must be less massive and less luminous than the faintest stars
already detected near the MBH, late B-stars ($\Ms\gtrsim3\,\Mo$).
The question whether such stars are present near the MBH is closely
linked to the issue of mass segregation (\S\ref{sss:distrDM}). If,
as is anticipated, most of the objects near the MBH are massive remnants,
then the density of low mass stars there will be significantly suppressed.
Furthermore, collisional destruction will limit the lifespan of tightly
bound stars (\S\ref{sss:RGcoll}). 

To detect and track stellar orbits in a field that becomes increasingly
cluttered as more faint stars are revealed ({}``source confusion''),
it will be necessary to increase the photometric, astrometric and
spectroscopic sensitivities. This can be achieved by much larger telescopes
than the existing 8--10 m class telescopes, which will have a correspondingly
smaller diffraction limit (e.g. the proposed Thirty Meter Telescope
(TMT), for GC applications see Weinberg, Milosavljevi\'c \& Ghez
\citeyear{Wei04}). Alternatively, optical/IR interferometers such
as the Very Large Telescope Interferometer (VLTI), the Keck interferometer
and the Large Binocular Telescope (LBT) will achieve very high angular
resolution and may also reach high photometric sensitivity with adaptive
optics. Monitoring a short period star will require frequent sampling
to avoid losing track of the star in the dense field, and to obtain
good coverage of the orbit, especially near periapse, where the post-Newtonian
deviations are largest (\S\ref{ss:GRorbit}).

A closely related problem is the detection and characterization of
the extended dark mass around the MBH. Orbital monitoring can also
detect deviations from Keplerian orbits due to a smooth distribution
of dark mass or perturbations due to interactions with discrete objects
(compact remnants). It may also reveal gravitational lensing events
(\S\ref{ss:GL}) involving both the MBH and the cluster of compact
remnants or faint stars around it (Alexander \& Loeb \citeyear{Ale01c};
Chanam\'e, Gould \& Miralda-Escud\'e \citeyear{Cha01}). On a scale
larger than probed by the orbits, X-ray surveys have already discovered
an over-abundance of transient X-ray sources in the central parsec,
which are thought to be NSs or SBHs accreting from a binary companion
(Muno et al. \citeyear{Mun04}). Deep surveys at short radio wavelengths
may probe the distribution of millisecond pulsars, which, as long-lived
light test particles, are expected to be depleted in the center (Chanam\'e
\& Gould \citeyear{Cha02}). The distribution of other long-lived
light test particles, such as red clump giants, should also be sensitive
to the degree of mass segregation. Theoretical predictions of the
details of mass segregation around a MBH are only available for highly
idealized situations (Bahcall \& Wolf \citeyear{Bah77}), and the
application to real systems is unclear. Dynamical simulations of stellar
systems around a MBH (e.g. Freitag \& Benz \citeyear{Fre02}; Preto,
Merritt \& Spurzem \citeyear{Pre04}; Baumgardt, Makino \& Ebisuzaki
\citeyear{Bau04}) that take stellar evolution and continuous star
formation into account will be needed to confirm and refine these
predictions.

The detection of evidence of past strong interactions between stars
and the MBH and between stars themselves (\S\ref{s:inter}, \S\ref{ss:coll})
will require large area surveys, in particular spectroscopic surveys,
to look for tell-tale peculiar stellar properties. There are however
still no robust theoretical predictions of the appearance of stars
that were subjected to major perturbations. Any progress on this complex
problem will be very relevant for such searches, and conversely, the
stellar population in the inner GC can be used test these ideas. 

The attempts to solve the riddle of the young stars (\S\ref{s:OBriddle})
raise many interesting possibilities about the contents and history
of the inner GC. A deep spectroscopic survey of the inner $\sim\!15"$
will help disentangle the different stellar populations there and
provide clues about the history and origin of the young stars and
the properties of the star disks. Such a survey can establish whether
or not apparent concentrations of stars, such as IRS13, are indeed
self-bound objects, perhaps harboring an IBH.

The Galactic MBH provides a uniquely accessible laboratory for studying
in detail the connections and interactions between a massive black
hole and the stellar system in which it grows; for investigating the
effects of extreme density, velocity and tidal fields on stars; and
for using stars to probe the central dark mass and post-Newtonian
gravity in the weak- and strong-field limits. These issues are relevant
for understanding the MBH phenomenon in general. The wealth of observed
phenomena in the GC provides the impetus to study stellar processes
in the extreme environment near a MBH. As shown in this review, such
studies prove to be very fruitful, yielding valuable insights even
in cases where they do not fully succeed to explain the observations.
Many questions remain---it is quite likely that the most exciting
science still lies ahead.

\subsubsection*{Acknowledgements\addcontentsline{toc}{section}{Acknowledgements}}

Helpful discussions and comments by F. Eisenhauer, R. Genzel, S. Gillessen,
C. Hopman and S. Zucker are gratefully acknowledged. Useful comments
by the referee, Y. Levin, are much appreciated. This work was supported
by ISF grant 295/02-1, Minerva grant 8484 and a New Faculty grant
by Sir H. Djangoly, CBE, of London, UK. 

{\small 
\bibliographystyle{my-hypertext}

}
{\small \par}

\marginpar{%
}
\end{document}